\pdfoutput=1
\documentclass[aps,pre,twocolumn,amssymb,showpacs,amsmath,superscriptaddress,nofootinbib]{revtex4-2}
\usepackage{graphics}
\usepackage{color}
\usepackage{empheq}
\usepackage[scr]{rsfso}
\usepackage{times}
\usepackage[colorlinks=true,linkcolor=blue,urlcolor=blue,citecolor=blue]{hyperref}
\usepackage{mathrsfs}

\usepackage{amsfonts}
\usepackage[normalem]{ulem}

\newcommand{\sref}[1]{Sec.~\ref{#1}}
\newcommand{\fref}[1]{Fig.~\ref{#1}}
\newcommand{\aref}[1]{App.~\ref{#1}}

\newcommand{\mquad}{\hspace{-1em}}
\newcommand{\ca}{\mathcal}

\newenvironment{remark}[1][Remark:]{\begin{trivlist}
\item[\hskip \labelsep {\bfseries #1}]}{\end{trivlist}}

\newcommand{\qed}{\nobreak \ifvmode \relax \else
      \ifdim\lastskip<1.5em \hskip-\lastskip
      \hskip1.5em plus0em minus0.5em \fi \nobreak
      \vrule height0.75em width0.5em depth0.25em\fi}

\usepackage{color}
\definecolor{Blue}{rgb}{0.00, 0.00, 1.00}
\definecolor{Red}{rgb}{1.00, 0.00, 0.00}

\newcommand{\sgn}{{\mathrm{sgn}}}
\newcommand{\rme}{{\mathrm{e}}}

\newcommand{\nn}{\nonumber}\newcommand {\eq}[1]{(\ref{#1})}
\newcommand {\Eq}[1]{Eq.\hspace{0.55ex}(\ref{#1})}
\newcommand {\Eqs}[1]{Eqs.\hspace{0.55ex}(\ref{#1})}

\newcommand{\fig}[2]{\includegraphics[width=#1]{./#2}}
\newcommand{\Fig}[1]{\includegraphics[width=8.7cm]{./#1}}

\renewcommand{\epsilon}{\varepsilon}

\def\be{\begin{equation}}
\def\ee{\end{equation}}

\def\bal{\begin{align}}
\def\eal{\end{align}}

\def\bea{\begin{eqnarray}}
\def\eea{\end{eqnarray}}
\renewcommand{\log}{\ln }

\newcommand{\braket}[1]{\left \langle #1 \right \rangle}

\makeatletter
\newsavebox{\@brx}
\newcommand{\llangle}[1][]{\savebox{\@brx}{\(\m@th{#1\langle}\)}%
  \mathopen{\copy\@brx\kern-0.7\wd\@brx\usebox{\@brx}}}
\newcommand{\rrangle}[1][]{\savebox{\@brx}{\(\m@th{#1\rangle}\)}%
  \mathclose{\copy\@brx\kern-0.7\wd\@brx\usebox{\@brx}}}
\makeatother

\begin{document}

\title{Functionals of fractional Brownian motion and the three arcsine laws}

 \author{Tridib Sadhu}
\affiliation{Department of Theoretical Physics, Tata Institute of Fundamental Research, Dr. Homi Bhabha Road, Mumbai 400005, India.}

\author{Kay J\"org Wiese}
\affiliation{Laboratoire de Physique de l'Ecole normale sup\'erieure, ENS, Universit\'e PSL, CNRS,\\
 Sorbonne Universit\'e, Universit\'e Paris-Diderot, Sorbonne Paris Cit\'e, 24 rue Lhomond, 75005 Paris, France.}

\begin{abstract}
 Fractional Brownian motion is a non-Markovian Gaussian process indexed by the Hurst exponent $H\in [0,1]$,  generalizing 
standard Brownian motion to account for  anomalous diffusion.  Functionals of this process are important for practical applications as a standard reference point for non-equilibrium dynamics.  We describe a perturbation expansion allowing us to evaluate many non-trivial observables analytically: We  generalize the celebrated three {\em arcsine-laws} of   standard Brownian motion.  The functionals are: (i) the fraction of time the process remains positive, (ii) the  time when the process last visits the origin, and (iii) the time when it achieves its maximum (or minimum).  We derive expressions for the probability of these three functionals as an expansion in $\epsilon = H-\tfrac{1}{2}$,   up to second order.   We find that the three probabilities are different,  except for $H=\tfrac{1}{2}$ where they coincide. Our results are confirmed to high precision by numerical simulations. 
\end{abstract}


\date{\today}
\maketitle

\tableofcontents

\section{Introduction}
\subsection{Fractional Brownian motion}
\begin{figure}[b]
\Fig{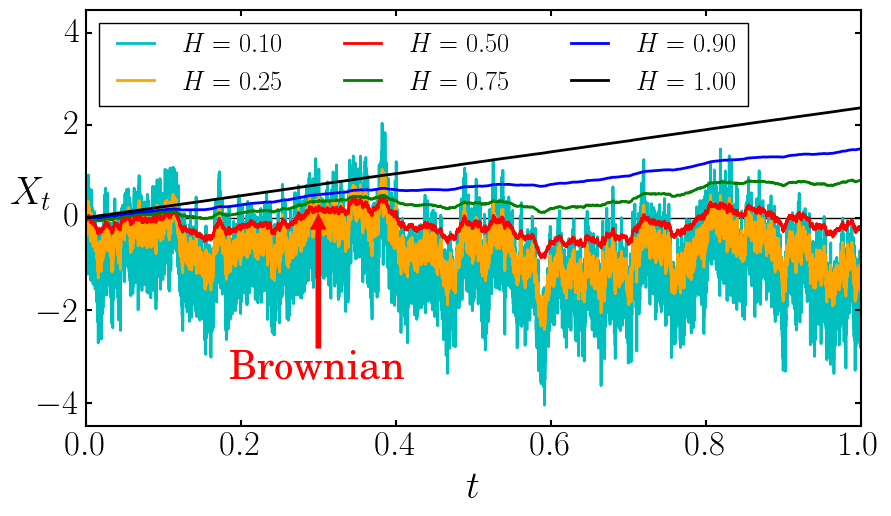}
\caption{(color online) Sample trajectories of an fBm corresponding to different Hurst exponent ($H$). Anti-correlation of increments for $H<\frac{1}{2}$ can be seen from larger fluctuations of the trajectories. In comparison, smoother trajectories for $H>\frac{1}{2}$ reflect positive correlations, which become a straight line for $H\to 1$.}
\label{fig:fbm sample}
\end{figure}

In the theory of stochastic processes fractional Brownian motion (fBm) plays as important a role as   standard Brownian motion \cite{DecreusefondUstunel1998,fbmReview2,NourdinBook,Shevchenko2015}. It was introduced \cite{Kolmogorov1940,MandelbrotVanNess1968} to incorporate  anomalous diffusive transport \cite{MetzlerKlafter2000}, which is abundant in  nature, but  not describable by  standard Brownian motion. FBm has  several key mathematical structures   to qualify it as the most fundamental stochastic process for anomalous diffusion: translation invariance in both time and space (stationarity), invariance under rescaling, and Gaussianity \cite{CohenIstas2013}. The current mathematical formulation of fBm was given by Mandelbrot and Van Ness \cite{MandelbrotVanNess1968} to describe correlated time-series   in natural processes. It is defined as a Gaussian stochastic process $X_t$ with $X_0=0$, mean $\left<X_{t}\right>=0$  and  covariance
\begin{equation}
\langle X_t X_s \rangle = t^{2H} + s^{2H} - |t-s|^{2H} \ . \label{eq:covariance}
\end{equation}
The parameter $H\in (0,1)$ is the Hurst exponent. An example is given in Fig.~\ref{fig:fbm sample}.
Standard Brownian motion corresponds to $H=\tfrac{1}{2}$ where the covariance reduces to $\langle X_t X_s \rangle = 2 \min(s,t) $. 

FBm is important as it successfully models  a variety of natural processes \cite{DecreusefondUstunel1998,fbmReview2}: A tagged particle in   single-file diffusion ($ H\,{=}\,0.25 $) \cite{KuklaKornatowskiDemuthGirnusal1996,WeiBechingerLeiderer2000,LizanaAmbjornssonTaloniBarkaiLomholt2010,KrapivskyMallickSadhu2015,SadhuDerrida2015}, the integrated current in diffusive transport ($ H\,{=}\,0.25 $) \cite{SadhuDerrida2016}, polymer translocation through a narrow pore ($ H\,{\simeq}\,0.4 $) \cite{ZoiaRossoMajumdar2009,DubbeldamRostiashvili2011,PalyulinAlaNissilaMetzler2014}, anomalous diffusion \cite{BouchaudGeorges1990}, values of the log return of a stock  ($H\,{\simeq}\,0.6\; {\rm to }\; 0.8 $) \cite{Peters1996,CutlandKoppWillinger1995,Biagini2008,Sottinen2001}, hydrology ($H\,{\simeq}\,0.72\;{\rm to}\;0.87 $) \cite{Hurst1951,MandelbrotWallis1968}, a tagged monomer in a polymer  chain ($ H\,{=}\,0.25$) \cite{GuptaRossoTexier2013}, solar flare activity ($H\,{\simeq}\,0.57\;{\rm to}\;0.86$) \cite{Moreno2014}, the price of electricity in a liberated market ($ H\,{\simeq}\,0.41 $) \cite{Simonsen2003}, telecommunication networks ($H\,{\simeq}\,0.78\;{\rm to}\;0.86$) \cite{Norros2006}, telomeres inside the nucleus of human cells ($H\,{\simeq}\,0.18\;{\rm to}\;0.35$) \cite{Burnecki2012}, sub-diffusion of lipid granules in yeast cells \cite{JeonTejedorBurovBarkaiSelhuber-UnkelBerg-SorensenOddershedeMetzler2011}, and diffusion inside crowded fluids ($ H\,{\simeq}\,0.4 $) \cite{Ernst2012}, are few such examples. Due to the simplicity of its definition,    fBm has a fundamental importance, as well as a multitude of potential applications. The pressing questions are how the celebrated properties of   standard Brownian motion generalize for  fBm, and how can one analyze them? In this paper we aim to address some of these questions. 

The anomalous diffusion in fBm comes from the long-range correlations in time, which makes the process non-Markovian, \textit{i.e.}\ its increments are not independent, unless $H=\tfrac{1}{2}$; this can be seen from the correlation of increments,
\be
\langle \partial_{t} X_t \,\partial_{s} X_s \rangle = 2H (2H-1) |s-t|^{2(H-1)} \label{eq:covariance dxdx}\ .
\ee 
The positivity of correlations for $H>\tfrac{1}{2}$ means that the process is correlated and the paths appear to be more regular  than for standard Brownian motion.  The converse   holds  for $H<\tfrac{1}{2}$, where increments are anti-correlated, making the process rough on short scales. This can be seen in \fref{fig:fbm sample} for the sample trajectory of a fBm generated in our computer simulation, using the same random numbers for the Fourier modes, which  renders the resulting curves comparable. 

The non-Markovian dynamics makes a theoretical analysis of   fBm difficult. Until now, few exact results are available in the literature \cite{Molchan1999,KrugKallabisMajumdarCornellBraySire1997,GuerinLevernierBenichouVoituriez2016}. In this paper, we describe  a systematic theoretical approach to  fBm, by constructing  a perturbation theory  in
\be
\epsilon = H - \frac12
\ee
around the Markovian dynamics.
 We describe this approach with a focus on observables that are functionals of the fBm trajectory $X_t$, and thereby depend on the entire history of the process. The fraction of time $X_t$ remains positive, the area under $X_t$, the position of the last maximum, or the time where $X_t$ reaches its maximum are examples of such functionals.

Functionals of stochastic processes are a topic of general   interest \cite{Majumdar2006,Hida1977}. Beside their relevance in addressing practical problems, they appear in path-ensemble generalizations of traditional statistical mechanics \cite{Ruelle1999,Ruelle2004}. Beyond   equilibrium statistical mechanics, the dynamics plays a crucial role in the statistical theory of non-equilibrium systems. Observables that are functionals of a stochastic trajectory, \textit{e.g.} entropy production, empirical work, integrated current, or activity, are   relevant dynamical observables for a   thermodynamic description of non-equilibrium systems \cite{LebowitzSpohn1999}.

The statistics of functionals is non-trivial already for Markovian processes, and is much harder for non-Markovian ones like   fBm. In our work, we overcome the inherent difficulty of the non-Markovian dynamics of   fBm by using a perturbation expansion around   standard Brownian motion ($H=\tfrac{1}{2}$), which is a Markovian process. This allows us to use many tricks available for Brownian motion, such as the method of images.  

\subsection{The three arcsine laws}

\begin{figure}[t]
\Fig{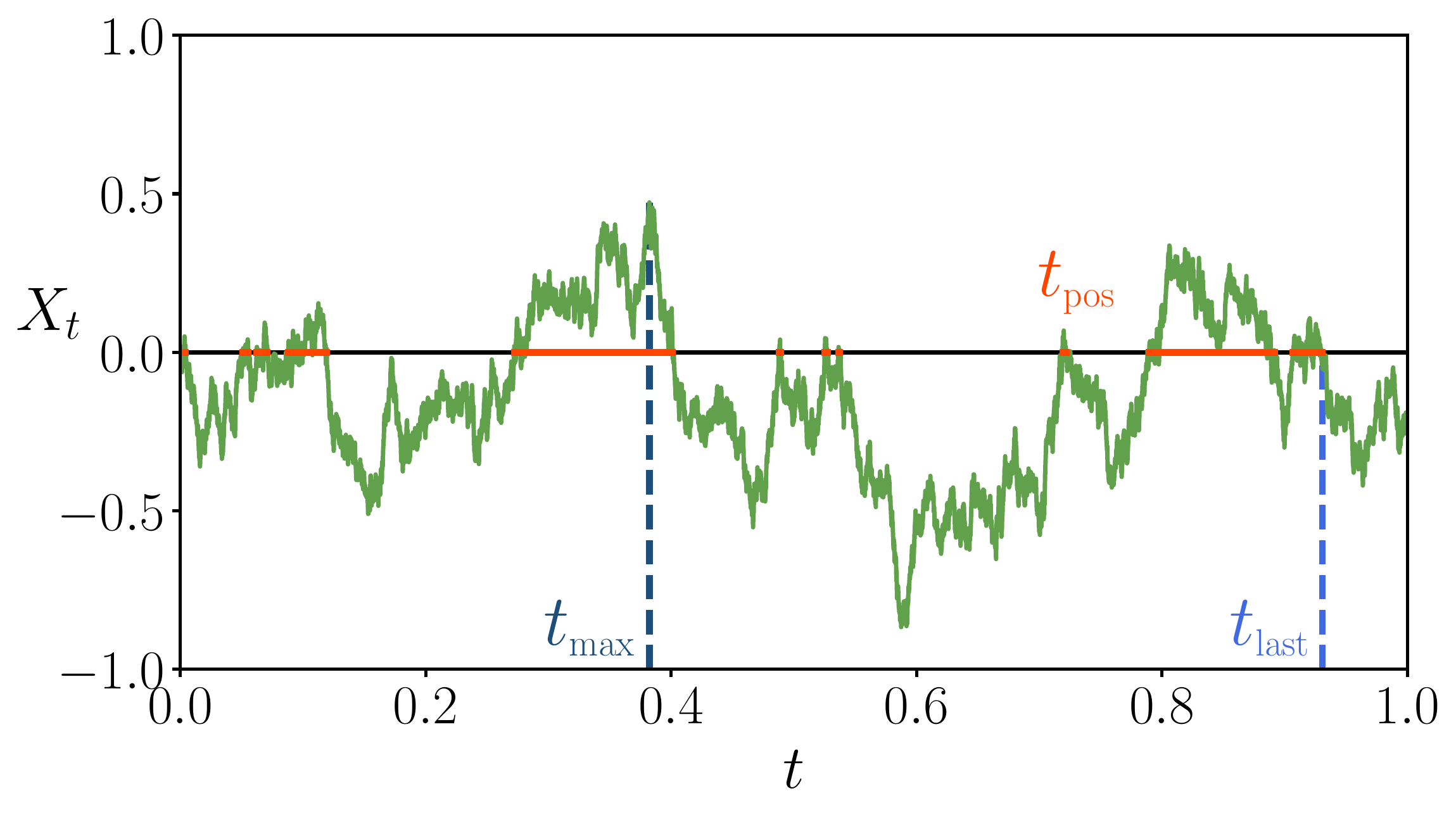}
\caption{(color online) The three observables $t_{\rm pos}$, $t_{\rm last}$, and $t_{\rm max}$ for a stochastic process starting at the origin. For the standard Brownian motion, all three have the same cumulative probability distribution expressed in terms of arcsine function \cite{Levy1940,FellerBook,Morters2010,Yen2013}.}
\label{f:traj}
\end{figure}

We illustrate this approach by considering a generalization of a famous result for standard Brownian motion: the three arcsine-laws \cite{Levy1940,FellerBook,Morters2010,Yen2013}. This result is about the following three functionals of a Brownian motion $B_t$ starting from the origin $B_0=0$, and evolving during time $T$  (see Fig.~\ref{f:traj}): 
\begin{enumerate}
\item[(i)] the total duration $t_{\rm pos}$ when the process is positive, 
\item[(ii)] the last time $t_{\rm last}$ the process visits the origin, and 
\item[(iii)] the time $t_{\rm max}$ it achieves its maximum (or minimum). 
\end{enumerate}
Remarkably, all three functionals have the same probability distribution as a function of $\vartheta := t/T$,  given by \cite{Levy1940,FellerBook,Morters2010,Yen2013}
\be\label{Levey-law}
p( \vartheta)= \frac{1}{{\pi \sqrt{\vartheta(1-\vartheta)}} }\ .
\ee 
As the cumulative distribution contains an arcsine function, these laws are commonly referred to as the {\it first, second}, and {\it third} arcsine-law. These laws apply quite generally to   Markov processes, {\em i.e.}\ processes where the increments are uncorrelated \cite{FellerBook}. Their counter-intuitive form with a divergence at $\vartheta=0$ and $\vartheta=1$ has sparked a lot of interest, and they are considered  among the most important results for stochastic processes. Recent studies   led to many extensions,  in constrained Brownian motion \cite{MajumdarRandon-FurlingKearneyYor2008,Bouchaud2008,Randon-FurlingMajumdar2007}, for general stochastic processes \cite{Majumdar2010,SchehrLeDoussal2010,HOCHBERG1994,Pitman1992,Carmona1994,Lamperti1958}, and even in higher dimensions \cite{BARLOW1989,Bingham1994,Ernst2017}. 
The laws are realized in a  plethora of real-world examples, from finance \cite{Dale1980,Baz2004} to competitive team sports \cite{Clauset2015}.

Using our perturbative approach, we show how the three arcsine-laws generalize for   fBm. Our results show that unlike for standard Brownian motion, all three functionals have different probability distributions, which coincide only when $\epsilon=0$, \textit{i.e.} for Brownian motion. As for two of the laws the difference is first seen at   second order in $\epsilon$, we have to develop the technology beyond what was done at leading order \cite{WieseMajumdarRosso2010,DelormeWiese2015,DelormeWiese2016b,DelormeWiese2016,DelormeRossoWiese2017,DelormeThesis,Wiese2018,ArutkinWalterWiese2020}.
 Using our perturbation results up to second order, and a scaling ansatz, we propose   expressions for all three probability densities. These expressions agree well with our numerical results, even for large values of $\epsilon$, i.e.\ including the full range of Hurst exponents  reported in the literature   cited above \cite{KuklaKornatowskiDemuthGirnusal1996,WeiBechingerLeiderer2000,LizanaAmbjornssonTaloniBarkaiLomholt2010,KrapivskyMallickSadhu2015,SadhuDerrida2015,SadhuDerrida2016,ZoiaRossoMajumdar2009,DubbeldamRostiashvili2011,PalyulinAlaNissilaMetzler2014,BouchaudGeorges1990,Peters1996,CutlandKoppWillinger1995,Biagini2008,Sottinen2001,Hurst1951,MandelbrotWallis1968,GuptaRossoTexier2013,Moreno2014,Simonsen2003,Norros2006,Burnecki2012,JeonTejedorBurovBarkaiSelhuber-UnkelBerg-SorensenOddershedeMetzler2011}. A short account of our main results was reported in \cite{SadhuDelormeWiese2017}.

This article is organized  in the following order: In \sref{sec:basics} we discuss   basics of an fBm and introduce the perturbation expansion of the action. 
As a consistency check 
we derive the free propagator for an fBm in \sref{sec:free propagator}, which is checked against the exact result. In the rest of the sections we discuss the three functionals for the arcsine-law. In \sref{sec:results}, we summarize our main analytical results for the generalization of the arcsine laws for an fBm, and compare them with our numerical simulations. How these results are derived is first sketched in \sref{sec:general approach}, and   thoroughly discussed in   later sections. Many algebraic details and a  description of our numerical algorithm are given in the appendices.

\section{Perturbation theory} 
\label{sec:basics}
\subsection{The action to second order in $\epsilon$}
Our analysis is based on a perturbation expansion of the action for an fBm trajectory around   standard Brownian motion ($H=\frac{1}{2}$). This expansion   was discussed and used earlier in \cite{WieseMajumdarRosso2010,DelormeWiese2015,DelormeWiese2016,DelormeWiese2016b,DelormeRossoWiese2017,DelormeThesis,Wiese2018,
SadhuDelormeWiese2017,ArutkinWalterWiese2020}  at  linear order. Here, we give additional details at   second order, which is essential to show the difference between the  generalizations of the three arcsine-laws.

An ensemble of trajectories for fBm in a time window $[0,T]$ is characterized by the Gaussian action
\begin{equation}
S[X_t]=\frac{1}{2}\int_0^{T}dt_1\int_{0}^{T}dt_2 X_{t_1}G(t_1,t_2)X_{t_2}
\label{eq:sH 16}
\end{equation}
with   covariance $G^{-1}(t_1,t_2)=\langle X_{t_1}X_{t_2}\rangle $ as given in \Eq{eq:covariance}. The probability of a trajectory, up to a normalization,  is given by
\begin{equation}
P[X_t]\sim e^{-S[X_t]}.\label{eq:P weight first}
\end{equation}
For $H=\frac{1}{2}$ one   recovers the Feynman-Kac formula \cite{Kac1949} for   standard Brownian motion.

Writing $H=\frac{1}{2}+\epsilon$ and expanding \Eq{eq:sH 16} in powers of $\epsilon$ we obtain (a derivation is in \aref{sec:expansion app})
\begin{equation}
S=\frac{1}{D}\left[S_0-\frac{\epsilon}{2}S_1+\epsilon^2 S_2+\cdots\right],
\label{eq:expansion of S}
\end{equation}
where
\begin{subequations}
\begin{align}
S_0=&\frac{1}{4}\int_{0}^{T}dt \;\dot{x}(t)^2, \label{eq:S0}\\
S_1=&\int_{0}^{T}dr_1\int_{r_1+\omega}^{T}dr_2 \; \frac{\dot{x}(r_1)\dot{x}(r_2)}{r_2-r_1}, \label{eq:S1}\\
S_2=&\frac{1}{2}\int_{0}^{T}dr_1\int_{r_1+\omega}^{T}dr_2 \; \dot{x}(r_1)\dot{x}(r_2) \times \label{eq:S2}  \\
&\bigg[\int_{0}^{r_1-\omega}\!\!\!\!\!\!\frac{ds}{(r_1-s)(r_2-s)}+\int_{r_2+\omega}^{T}\frac{ds}{(s-r_1)(s-r_2)}\bigg].~~~~~~\nn
\end{align}
The pre-factor, the diffusion constant, reads 
\label{eq:S0S1S2}\end{subequations}
\begin{align}
D\equiv D(\epsilon,\omega)=e^{\epsilon\, 2(1+\log\omega)-\epsilon^2\, 2( 1-\frac{\pi^2}{6})+ \ca O(\epsilon^3)} .
\label{eq:c0}
\end{align}
The  small-time (ultraviolet) cutoff $\omega> 0$  is introduced to regularize the integrals in the action. Our final results  are in the limit of $\omega\rightarrow 0$, and   independent of $\omega$.  The second-order term in the exponential in \Eq{eq:c0} is independent of $\omega$, since from dimensional arguments $D\sim\omega^{\epsilon}$, 

\begin{remark}
To keep our formulas simple, we explicitly write the ultraviolet cutoff in Eqs.~(\ref{eq:S1})-(\ref{eq:S2}) only for     integrals which would otherwise diverge. 

\end{remark}

\subsection{Integral representation of the action, and normal-ordered form of the weight}
\label{Integral representation of the action, and normal-ordered form of the weight}
For our  explicit calculations we   use an alternative representation of \Eqs{eq:S1} and \eq{eq:S2}:
\begin{subequations}
\begin{equation}
S_1=\int_0^{\Lambda}dy\int_{0}^{T}dr_1\int_{r_1}^{T}dr_2 \,\dot{x}(r_1)\dot{x}(r_2) e^{y(r_1-r_2)}, \label{eq:S1 1 integral}
\end{equation}
and
\begin{align}\nn
S_2=\frac{1}{2}&\int_0^{\Lambda}dy_1\int_0^{\Lambda}dy_2\int_{0}^{T}dr_1\int_{r_1}^{T}dr_2 \,\dot{x}(r_1)\dot{x}(r_2)\times  \\ 
&
\left[\int_{0}^{r_1}ds\;e^{-y_1(r_1-s)-y_2(r_2-s)}\;+\right.\cr &\left.\qquad\qquad \int_{r_2}^{T}ds\;e^{-y_1(s-r_1)-y_2(s-r_2)} \right], \qquad  
 \label{eq:S1 2 integral}
\end{align}
where the ultraviolet cutoff $\omega$ in time is replaced by an upper limit $\Lambda$ for the $y$ variables. A vanishing $\omega$ is equivalent to $\Lambda\to \infty$, which is always taken in the final results.
\label{eq:integral reperesentation S1S2}
\end{subequations} 

Their relation can be inferred as follows:
for small $\omega$
\be
\int_\omega^\infty \frac{dt}{t} e^{-s t}\simeq -\log (s\,\omega)-\gamma_{\rm E} + \ca O(\omega),
\ee  where $\gamma_{\rm E}=0.57721\dots$ is the Euler constant. On the other hand,  the integral representation for large $\Lambda$ reads
\be
\int_0^\Lambda dy \int_0^\infty dt\;e^{-t s-ty}\simeq \log \left(\frac{\Lambda}{s}\right) + \ca O(\Lambda^{-1}).
\ee
Demanding that they agree, we get
\begin{equation}
\Lambda=\frac{1}{\omega}e^{-\gamma_{\rm E}}. \label{eq:Lambda omega relation}
\end{equation} 
In \sref{sec:free propagator} we further  check  \Eq{eq:Lambda omega relation} by constructing the free diffusion propagator for fBm.
In terms of $\Lambda$, \Eq{eq:c0} reads
\begin{equation}
D=e^{\epsilon \; 2(1-\log\Lambda-\gamma_{\rm E})-\epsilon^2\, 2(1-\frac{\pi^2}{6})+\ca O(\epsilon^3)}.
\label{eq:D exp}
\end{equation}

\begin{remark}
Keeping in mind the ultraviolet cutoff $\omega$ present in \Eqs{eq:S1}--\eq{eq:S2},   integrals arising from  \Eq{eq:integral reperesentation S1S2} are interpreted such that
\begin{equation} \int_{r_1}^{T}dr_2 \; \delta(r_2-r_1):= \lim_{\omega \to 0 } \int_{r_1+\omega}^{T}dr_2 \; \delta(r_2-r_1) =0 .
\label{eq:int dr1dr2 delta}
\end{equation}
This convention for the expression of the action is used   throughout our analysis.

\end{remark}

\begin{remark}

A subtle point is that at second order  for the probability in \Eq{eq:P weight first}  one encounters terms of order $S_1^2$ in which one  contracts two of  the $\dot x$ (contracting all four gives a constant entering into the normalization of the probability, thus ignored), 
\bea
\frac {S_1^2}8 &=&\frac18  \int_{0}^{T}dr_1\int_{r_1+\omega}^{T}dr_2 \; \frac{\dot{x}(r_1)\dot{x}(r_2)}{r_2-r_1} \times \nn\\
&&  \qquad \int_{0}^{T}dr_3\int_{r_3+\omega}^{T}dr_4 \; \frac{\dot{x}(r_3)\dot{x}(r_4)}{r_4-r_3}\nn\\[10pt]
&&\longrightarrow \begin{cases} \frac14 \int_{r_1<r_2,r_4 }\; \frac{ \dot{x}(r_2)}{r_2-r_1}\frac{ \dot{x}(r_4)}{r_4-r_1} \nn\\[10pt]
\frac14\int_{r_3< r_1<r_2} \frac{ \dot{x}(r_2)}{r_2-r_1}  \frac{\dot{x}(r_3)}{r_1-r_3}\nn\\[10pt]
\frac14 \int_{r_1<r_2<r_4} \; \frac{\dot{x}(r_1)}{r_2-r_1} \frac{\dot{x}(r_4)}{r_4-r_2}\nn\\[10pt]
\frac14\int_{r_1,r_3< r_2}
 \frac{\dot{x}(r_1) }{r_2-r_1}   \frac{\dot{x}(r_3) }{r_2-r_3}. 
 \end{cases}\label{16}
\eea
The cutoffs in the integrations are implicit and the right arrow indicates contraction of a pair of $\dot{x}$.
The four terms come, in the given order, from the contraction of $\dot x(r_1)\dot x(r_3)$, $\dot x(r_1)\dot x(r_4)$, $\dot x(r_2)\dot x(r_3)$, and $\dot x(r_2)\dot x(r_4)$.  They have the same structure as those of $S_2$ in  \Eq{eq:S2}, and we can group them together: the first contracted term in \Eq{16} cancels the first term of \Eq{eq:S2}, the fourth contracted term in \Eq{16} cancels the last term in \Eq{eq:S2} (note that $S_2$ comes with a minus sign in the expansion, and the points $r_1$ and $r_2$ are ordered); the remaining two contracted terms are identical and   can be incorporated  into a redefinition of $S_2$ as discusses below. 

These cancellations make it advantageous to exclude self-contractions, i.e.\ the terms on the r.h.s.\ of \Eq{16}, from $e^{-S}$,  which in field theory is  noted as a {\em normal-ordered} \cite{Zinn} weight,   
\be
e^{-S} ~~\longrightarrow ~~ :\!e^{-\ca S^{\rm (n)}}\!\!: \label{eq:S normal ordered}
\ee
In this normal-ordered form, the second-order term $S_2$ is replaced by 
\begin{align}\nn
\ca S_2^{\rm (n)}=\frac{1}{2}&\int_0^{\Lambda}dy_1\int_0^{\Lambda}dy_2\int_{0}^{T}dr_1\int_{r_1}^{T}dr_2 \dot{x}(r_1)\dot{x}(r_2) \\ 
&
 \times \int_{r_1}^{r_2}ds\;e^{-y_1(r_1-s)-y_2(r_2-s)},
 \label{eq:S1 2 integral normal}
\end{align}
(cutoffs are implicit).
Using the normal-ordered weight makes our calculations simple and elegant.  However,  to keep our calculation accessible for a non-specialist, we present our analysis using the weight in 
\Eq{eq:integral reperesentation S1S2}. We shall mention at relevant stages of the calculation which can be simplified using normal ordered weight.
\end{remark}

\section{The free fBm propagator} 
\label{sec:free propagator}
In this section, we verify the perturbation expansion in Eqs.~\eqref{eq:expansion of S}-\eqref{eq:integral reperesentation S1S2} by deriving a known result about the propagator of an fBm. The probability for an fBm, starting at $X_0=0$, to be at $X_T=m$ at time $T$ is given by 
\begin{equation}
\mathcal{G}_H(m,T)=\frac{e^{-\frac{m^2}{4 T^{2H}}}}{\sqrt{4 \pi T^{2H}}},\label{eq:Free fBm propagator}
\end{equation}
which is straightforward to see for the Gaussian process with covariance \Eq{eq:covariance}.

In terms of the action in \Eq{eq:sH 16}, the same propagator can be expressed as
\begin{equation}
\mathcal{G}_H(m,T)=\frac{W_H(m,T)}{N_T},\label{eq:ZH}
\end{equation}
where
\begin{subequations}
\begin{equation}
W_H(m,T)=\int_{x(0)=0}^{x(T)=m}\mathcal{D}[x]e^{-S[x]} \label{eq: W and N a}  
\end{equation}
and normalization
\begin{equation}
N_T=\int_{-\infty}^{\infty}dm W_H(m,T).
\end{equation}\label{eq: W and N}%
\end{subequations}
\Eq{eq:Free fBm propagator} can be derived from \Eq{eq:ZH} using the perturbation expansion \eqref{eq:expansion of S}. For this, we Taylor expand \Eq{eq:Free fBm propagator} in $\epsilon$ as 
\begin{align}
\mathcal{G}_H(m,T)=&\mathcal{G}(m,T)+\epsilon\,2T(\log T)\partial^2_m\mathcal{G}(m,T)+\cr
& \epsilon^2\,\left[2(T\log T)^2\partial^4_m\mathcal{G}(m,T)\right.\cr & \qquad\quad  \left.+2T(\log T)^2\partial^2_m\mathcal{G}(m,T)\right]+\cdots,\qquad
\label{eq:ZH expansion}
\end{align}
where $\mathcal{G}(m,T)$ (without the subscript $H$) is the propagator for   standard Brownian motion ($H=\tfrac{1}{2}$) with unit diffusivity. In this section, we   restrict our analysis     to second order in $\epsilon$, which is enough to verify     formulas \eqref{eq:S0S1S2}-\eqref{eq:integral reperesentation S1S2}. An  all-orders analysis 
is deferred to   \aref{app:free propagator}.

Using Eqs.~\eqref{eq:expansion of S} and \eqref{eq:D exp} in \Eq{eq: W and N} we get
\begin{equation*}
W_H(m,T)=W_0(m,T)+\epsilon\, W_1(m,T)+\epsilon^2 W_2(m,T)+\cdots,
\end{equation*}
where 
\begin{subequations}
\begin{align}
W_0(m,T)&=\int\limits_{x(0)=0}^{x(T)=m}\mquad\mathcal{D}[x]e^{-\frac{S_0}{D}},\label{eq:W0 definition}\\
W_1(m,T)&=\frac{1}{2}\int\limits_{x(0)=0}^{x(T)=m} \mquad \mathcal{D}[x]e^{-\frac{S_0}{D}}S_1, \label{eq:W1 1}\\
W_2(m,T)&= \!\!\!\!\int\limits_{x(0)=0}^{x(T)=m}\mquad\!\mathcal{D}[x]e^{-\frac{S_0}{D}}\!\bigg[\frac{ S_1^2 }{8}-(1{-}\gamma_{\rm E}{-}\log\Lambda)S_1-S_2\bigg]. \label{eq:W2 definition}
\end{align}
\end{subequations}
The second term  comes from  the order-$\epsilon$ contribution to the diffusion constant \eq{eq:D exp} inserted into \Eq{eq:expansion of S}. 

Each term in the expansion of $W_H$ can now be evaluated as an average with a Brownian measure of diffusivity $D$. The path integral measure $\mathcal{D}[x]$ is defined such that the leading term
\begin{equation}
W_0(m,T)=Z_T(0,m):=\frac{e^{-\frac{m^2}{4 D T}}}{\sqrt{4 \pi D T}}\label{eq:W0 Z}
\end{equation}
is the normalized propagator $Z_T(0,m)$ for standard Brownian motion with diffusivity $D$, starting from $x=0$ at $t=0$, and ending in $x=m$ at $t=T$. (For $D=1$, $Z_T(0,m)\equiv\mathcal{G}(m,T)$ in \Eq{eq:ZH expansion}.)

For the linear-order term in \Eq{eq:W1 1} we use \Eq{eq:S1 1 integral} and the  identity  \eq{eq:xx dot simple} derived in  \aref{sec:correlation x dot} to obtain
\begin{align}
&\int_{x(0)=0}^{x(T)=m}\mquad \mathcal{D}[x]\dot{x}(r_1)\dot{x}(r_{2}) e^{-\frac{S_0}{D}}=\Delta(r_1-r_2)Z_T(0,m), \label{eq:two-point correlation}
\\
&\Delta(r_1-r_2):=2^{2}D^{2}\partial^{2}_{m}+2D\,\delta(r_1-r_2).
\end{align}
Using the convention in \Eq{eq:int dr1dr2 delta} we get
\begin{align}
W_1(m,T)&=f_1(T)D^{2}\;2\;\partial^{2}_{m}Z_T(0,m),\label{eq:W1 result} \\
f_1(T)&=\int_0^{\Lambda}dy\int_0^Tdr_1\int_{r_1}^{T}dr_2e^{-y(r_2-r_1)}\cr
&\simeq T\left[\log\left(T \Lambda e^{\gamma_{\rm E}}\right)-1\right] + \ca O(\Lambda^{-1}).
\label{eq:f1 result}
\end{align}
For the quadratic   term in \Eq{eq:W2 definition} we use   Wick's theorem to obtain
\begin{align}
&\int\limits _{x(0)=0}^{x(T)=m} \mquad \mathcal{D}[x]\dot{x}(r_1)\dot{x}(r_{2})\dot{x}(r_{3})\dot{x}(r_{4}) e^{-\frac{S_0}{D}}=\cr
&\left(\sum_{\sigma}\Delta(r_{\sigma(1)}-r_{\sigma(2)})\Delta(r_{\sigma(3)}-r_{\sigma(4)})\right)Z_T(0,m),
\qquad 
\label{eq:four-point correlation}
\end{align}
where $\sigma$ denotes the set of all pairs. Then, using \Eqs{eq:S1 2 integral}  and   \eq{eq:int dr1dr2 delta} leads to
\begin{align}
W_2(m,T)&=2f_1^2(T)D^4\partial^{4}_{m}Z_T(0,m)+\cr
&\bigg[f_5(T)D-4(1-\gamma_{\rm E}-\log\Lambda )f_1(T)-2f_3(T)\bigg]\cr
& \times D^2\partial^{2}_{m}Z_T(0,m)+\frac{1}{2}f_6(T)D^2Z_T(0,m). \label{eq:W2 result}
\end{align}
Here
\begin{align*}
&f_5(T)=\int\limits_0^{\Lambda}\!\! dy_1\!\! \int\limits_0^{\Lambda}\!\! dy_2\!\!\int\limits_0^T\!\! dr_1\!\!\int\limits_{r_1}^{T} \!\!  dr_2\!\!\int\limits_0^T  \!\! dr_3 \!\!\int\limits_{r_3}^{T} \!\!  dr_4e^{y_1(r_1-r_2)+y_2(r_3-r_4)}\cr
&~~~~~~~~~~~~~\times \bigg[\delta(r_1{-}r_3)+\delta(r_2{-}r_4)+\delta(r_1{-}r_4)+\delta(r_3{-}r_2)\bigg],
\end{align*}
which simplifies to
\begin{subequations}
\begin{align}
f_5(T)=&\,2\!\!\int\limits_0^{\Lambda}\!\! dy_1\!\! \int\limits_0^{\Lambda}\!\! dy_2\!\!\int\limits_0^T\!\! dr_1\!\!\int\limits_{r_1}^T\!\! dr_2\!\!\int\limits_0^{T}\!\! ds \,e^{-y_1\vert s-r_1\vert-y_2\vert s-r_2\vert}. \label{eq:f5}
\end{align}
The remaining terms are 
\begin{align}
\!\!\!\!f_3(T)=&\int\limits_0^{\Lambda}dy_1 \int\limits_0^{\Lambda} dy_2 \int\limits_{0}^{T}dr_1\int_{r_1}^{T}dr_2\cr 
& \left\{\int\limits_{0}^{r_1} ds+\int\limits_{r_2}^{T} ds\right\}e^{-y_1\vert s-r_1\vert-y_2\vert s-r_2\vert},\qquad
\\
\!\!\!\!f_6(T)=&\int\limits_0^{\Lambda} dy_1 \int\limits_0^{\Lambda} dy_2 \int\limits_0^Tdr_1\int\limits_{r_1}^{T}dr_2\;e^{-(y_1+y_2)(r_2-r_1)}. \qquad ~~
\end{align}
\end{subequations}
In a similar calculation, the normalization in \Eq{eq: W and N} is obtained from Eqs.~(\ref{eq:W0 Z}), (\ref{eq:W1 result}), and (\ref{eq:W2 result}) as
\begin{equation*}
N_T=1+\epsilon^2\frac{1}{2}f_6(T)D^2+\mathcal{O}(\epsilon^3).
\end{equation*}
Note that the linear-order term vanishes. 

Altogether, from \Eq{eq:ZH} we get
\begin{align}
\mathcal{G}_H(m,T)= Z_T&(0,m)+\epsilon \; 2f_1(T)D^2\partial^{2}_{m}Z_T(0,m)\cr 
+\epsilon^2 \bigg\{2f_1^2(T)&D^4\partial^{4}_{m}Z_T(0,m)+\bigg[f_5(T)D-2f_3(T)\cr
-4(1-\gamma_{\rm E} -\log&\Lambda )f_1(T)\bigg]D^2\partial^{2}_{m}Z_T(0,m)\bigg\}+\mathcal{O}(\epsilon^3).\qquad
\label{eq:ZH expand}
\end{align}
To see that \Eq{eq:ZH expand} agrees with \Eq{eq:ZH expansion} we use \Eq{eq:D exp} in \Eq{eq:W0 Z} and write
\begin{align*}
&Z_T(0,m)=\mathcal{G}(m,T)+\epsilon \; 2(1-\gamma_{\rm E}-\log\Lambda) T \partial^2_m \mathcal{G}(m,T)\cr
&\qquad +\epsilon^2 \bigg[2\left( (1-\gamma_{\rm E}-\log\Lambda)^2-1+\frac{\pi^2}{6}\right)T\partial^2_m \mathcal{G}(m,T)\cr
&\qquad \qquad \quad  +2(1-\gamma_{\rm E}-\log\Lambda)^2T^2\partial^4_m \mathcal{G}(m,T)\bigg]+\cdots .
\end{align*}
Substituting the above expression of $Z_T(0,m)$ in \Eq{eq:ZH expand} and then using \Eq{eq:D exp} yields
\begin{align}
\mathcal{G}_H&(m,T)=\mathcal{G}(m,T)+\epsilon \; 2 T K_1^2\partial^2_m \mathcal{G}(m,T)+\nn\\[1mm]
\epsilon^2 &\left[2 T^2K_1^2\partial^4_m \mathcal{G}(m,T)+2 T K_2\partial^2_m \mathcal{G}(m,T)\right]+\cdots,
\label{eq:ZH expand 2}
\end{align}
where
\begin{equation}
K_1=\frac{f_1(T)}{T}+1-\gamma_{\rm E}-\log\Lambda,
\end{equation}
and
\begin{align}
K_2=\frac{f_5(T)-2f_3(T)}{2 T}&+2(1-\gamma_{\rm E}-\log\Lambda)\frac{f_1(T)}{T}\cr
&+(1-\gamma_{\rm E}-\log\Lambda)^2-1+\frac{\pi^2}{6}.~~~~~~~
\end{align}
It is then easy to see from the expression of $f_1(T)$ in \Eq{eq:f1 result} and
\begin{align}
\frac{1}{2}f_5(T)-&f_3(T)= \int_0^{\Lambda} dy_1\int_0^{\Lambda} dy_2\int_{0}^{T} dr_1\int_{r_1}^{T} dr_2  \cr
& \qquad \qquad \times \int_{r_1}^{r_2} ds e^{-y_1(s-r_1)-y_2(r_2-s)} \cr
\simeq  T\bigg\{&\Big[\log T-(1-\gamma_{\rm E}-\log\Lambda)\Big]^2+1-\frac{\pi^2}{6}\bigg\} \label{eq:conflicting integral}
\end{align}
for large $\Lambda$, that \Eq{eq:ZH expand 2} agrees with \Eq{eq:ZH expansion}.

\begin{figure}[t]
\fig{\columnwidth}{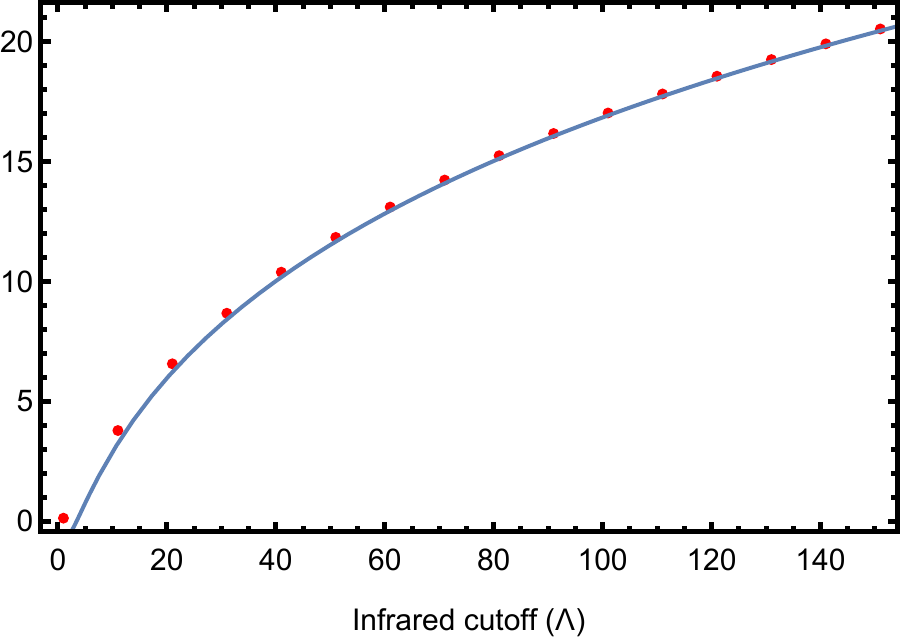}
\caption{(color online) A comparison of the integral in \Eq{eq:conflicting integral} (indicated by red points) with its asymptotic (indicated by solid line) for large $\Lambda$ and $T=1$.\label{fig:NumAnalyticComparisonEq28}}
\end{figure}

\begin{remark}
The
asymptotics of the integral in \Eq{eq:conflicting integral} is numerically verified in Mathematica, with results   shown in \fref{fig:NumAnalyticComparisonEq28}. 
\end{remark}

\begin{remark}
The analogue of the integral in \Eq{eq:conflicting integral} with ultraviolet cutoff $\omega$ in time is 
\begin{equation*}
\int_0^{T-2\omega}dr_1\int_{r_1+2\omega}^Tdr_2\int_{r_1+\omega}^{r_2-\omega}ds\; \frac{1}{(s-r_1)(r_2-s)}.
\end{equation*}
As a consistency check we verified that for small $\omega$, and using  the identification \Eq{eq:Lambda omega relation}, the integral yields the asymptotics in \Eq{eq:conflicting integral}.
\end{remark}

\begin{remark}
In our derivation of \Eq{eq:ZH expansion} using \Eq{eq:expansion of S} we   interchanged the small-$\epsilon$ and large-$\Lambda$ limits. Agreement of the final result in \Eq{eq:ZH expansion} shows that this step is justified. We   assume the same property in our perturbation analysis in the   observables of the three arcsine-laws.
\end{remark}

\begin{remark}
The analysis would be simpler with the normal-ordered Action in \Eq{eq:S normal ordered},  because then  terms $f_5$ and $f_6$  in \Eq{eq:W2 result} vanish.
\end{remark}

\section{A generalization of the three arcsine-laws} 
\label{sec:results}
Unlike for standard Brownian motion, the probabilities for the three observables $t_{\rm last}$, $t_{\rm max}$, and  $t_{\rm pos}$ all differ. Self-affinity of an fBm (invariance under rescaling of space with $T^{H}$) means that the three probabilities are a function of the rescaled variable $\vartheta=t/T$ ($t$ being $t_{\rm last}$, $t_{\rm max}$, $t_{\rm pos}$). They can be written as 
\bea 
\label{Plast}
p_{\rm last}(\vartheta)&=&\frac{{\cal N}_{\rm last}}{\pi\,\vartheta^H(1-\vartheta)^{1-H}} \,\rme^{\mathcal{F}^{\rm last}(\vartheta, H)}, ~~~\\
p_{\rm max}(\vartheta)&=&\frac{{\cal N}_{\rm max}}{\pi\,[\vartheta(1-\vartheta)]^H}\, \rme^{\mathcal{F}^{\rm max}(\vartheta, H)},
\label{Pmax}\\
\label{P+}
p_{\rm pos}(\vartheta) &=& \frac{{\cal N}_{\rm pos}}{\pi\,[\vartheta(1-\vartheta)]^H}\, \rme^{\mathcal{F}^{\rm pos}(\vartheta, H)}.
\eea
The divergences in the prefactor of the exponential terms are predicted using a scaling argument (discussed in \sref{sec:scaling}) for $\vartheta\to 0$ and $\vartheta\to 1$. They are linked to earlier results about the persistence exponent $\Theta=1-H$ \cite{Molchan1999,KrugKallabisMajumdarCornellBraySire1997,WieseMajumdarRosso2010}. The terms $\mathcal{F}$ in the exponential are non-trivial and {\em remain finite over the full range of $\vartheta$.}
We use the convention that the integral of each ${\cal F}$ function over $\vartheta$ vanishes, which together with the normalization $\int_0^1 d\vartheta\, p(\vartheta)=1$ fixes the constants ${\cal N}$.

For $H=\tfrac{1}{2}$, all three $\mathcal{F}$ functions vanish, $H=1-H$, and the   expressions \eq{Plast} to \eq{P+} reduce to the same well-known result of   standard Brownian motion (``arcsine law"). For   $H \neq \frac12$, they can be written as a perturbation expansion in $\epsilon=H-\frac{1}{2}$,
\begin{subequations}
\begin{align}
\mathcal{F}^{\rm last}(\vartheta, H)=&\epsilon \, {\cal F}^{\rm last}_1(\vartheta) +\epsilon^2 {\cal F}^{\rm last}_2(\vartheta)  + \cdots, \label{eq:F expansion last}\\
\mathcal{F}^{\rm max}(\vartheta, H)=&\epsilon \, {\cal F}^{\rm max}_1(\vartheta) +\epsilon^2 
{\cal F}^{\rm max}_2(\vartheta) + \cdots, \\
\mathcal{F}^{\rm pos}(\vartheta, H)=&\epsilon \,{\cal F}^{\rm pos}_1(\vartheta) +\epsilon^2 {\cal F}^{\rm pos}_2(\vartheta) + \cdots .\label{eq:F expansion +}
\end{align}\label{eq:F expansion}
\end{subequations}
For the leading-order terms we find 
\begin{subequations}
\begin{equation}
\mathcal{F}_1^{\rm last} (\vartheta)=  0, \label{eq:F1 last final result}
\end{equation}
and
\begin{align}
\mathcal{F}_1^{\rm max}& (\vartheta)= \mathcal{F}_1^{\rm pos} (\vartheta)=2-\frac{\pi ^2}{2}+\psi\left( \sqrt{\frac{\vartheta}{1-\vartheta}}\right), 
\label{eq:F1 max final result}
\end{align}
with  
\begin{equation}
\psi(x)=\frac{2}{x}\arctan (x)+2\,x\arctan \!\left( \frac{1}{x} \right).
\label{eq:psi}
\end{equation}%
\label{8}%
\end{subequations}%
\begin{figure}[t]
\fig{8.7cm}{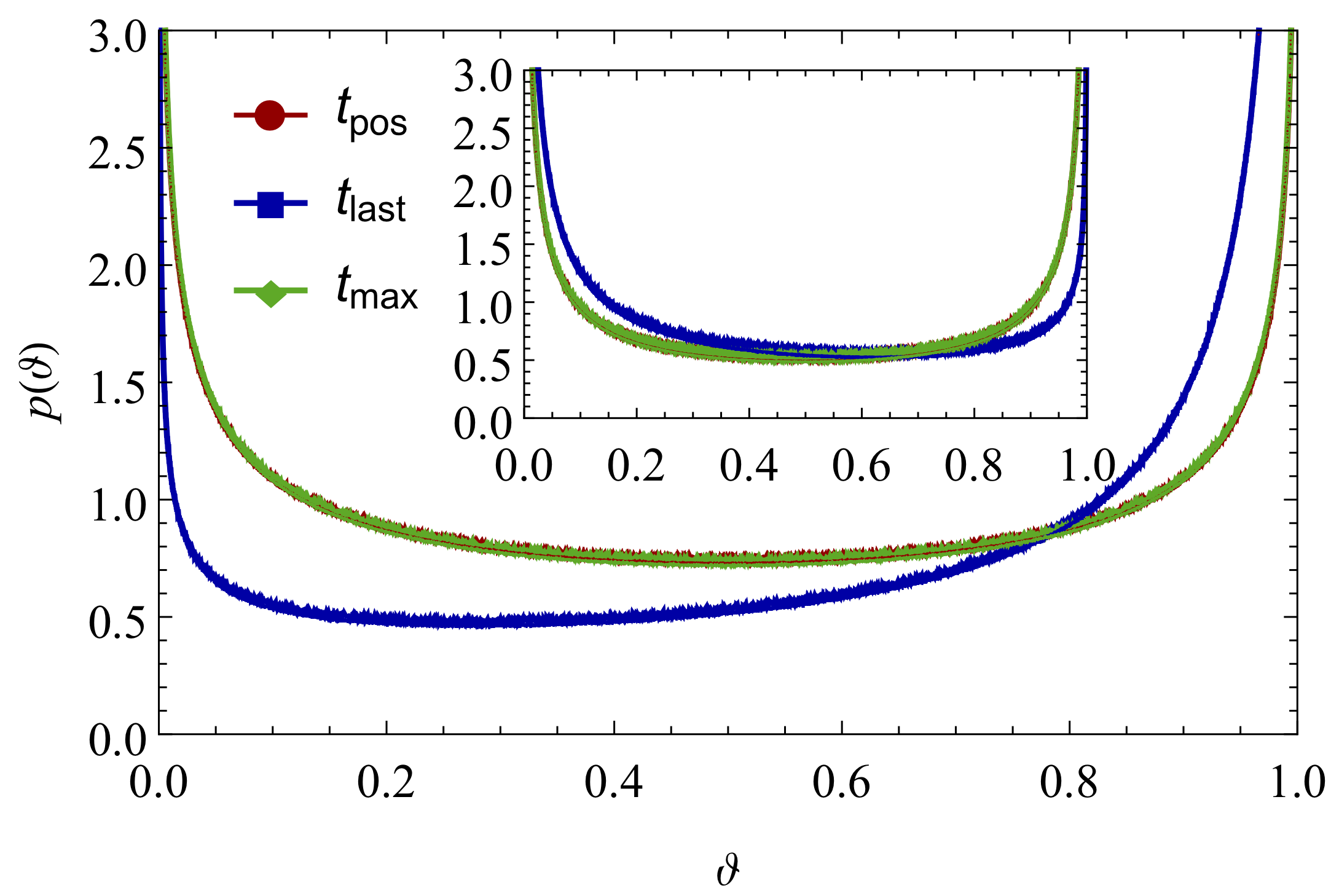}
\caption{(color online) Numerical simulation results for the probability of the three observables $t_{\rm last}$, $t_{\rm max}$, and $t_{\rm pos}$ for an fBm with $H=0.33$. The inset shows the probabilities for $H=0.66$. Note that the distributions of $t_{\rm pos}$ and $t_{\rm max}$ are almost indistinguishable.\label{f:raw}}
\end{figure}%
This is the simplest   form we found. Alternative expressions were given in \cite{DelormeWiese2015,DelormeWiese2016,DelormeWiese2016b,DelormeThesis}, using that 
$
\arctan (x)+\arctan \!\left( \frac{1}{x} \right)=\frac \pi 2. 
$
Yet another equivalent form is given in Eq.~(6) of \cite{SadhuDelormeWiese2017}.
We note that the expression \eqref{eq:F1 max final result} is symmetric under $\vartheta \to 1-\vartheta$. This can be understood from the symmetry of the problem. We do not have an intuitive understanding of the equality of  $\mathcal{F}_1^{\rm pos}$ and $\mathcal{F}_1^{\rm max}$, while the    vanishing of $\mathcal{F}_1^{\rm last}$ in \Eq{8} can easily be understood from perturbation theory \cite{DelormeThesis}.

Expressions for the sub-leading terms ${\cal F}_2$ can be written as   integrals, which are hard to evaluate analytically. For $t_\text{last}$, it is given up to an additive constant by
\begin{subequations}\label{eq:F2 last exact}
\begin{align}
{\cal F}^{\rm last}_2(\vartheta)= & \int_{0}^{\infty}\frac{dy_1dy_2}{y_1^2y_2^2} \; \Psi^{\rm last}\left(y_1,y_2,\frac{1-\vartheta}{\vartheta}\right),
\end{align}
where $\Psi^{\rm last}(y_1,y_2,z)$ is symmetric in $(y_1,y_2)$ and given by
\begin{align}
\Psi^{\rm last}\left(y_1,y_2,z\right)&=2\sqrt{(1+y_1+y_2)z}\nn\\[1mm]
\times\Big(&1-\sqrt{1+y_1}-\sqrt{1+y_2}+\sqrt{1+y_1+y_2}\Big)\nn\\[1mm]
\times\Big(\sqrt{z}\,-\,&\Theta(z-y_1)\sqrt{z-y_1}-\Theta(z-y_2)\sqrt{z-y_2}\nn\\[1mm]
&+\Theta(z-y_1-y_2)\sqrt{z-y_1-y_2} \Big),
\end{align}
with $\Theta(x)$ being the Heaviside step function. 
Expressions for ${\cal F}_2^{\rm max}$ and ${\cal F}_2^{\rm pos}$  are cumbersome and  given     later. 
\end{subequations}

\begin{figure}[t]
\fig{8.7cm}{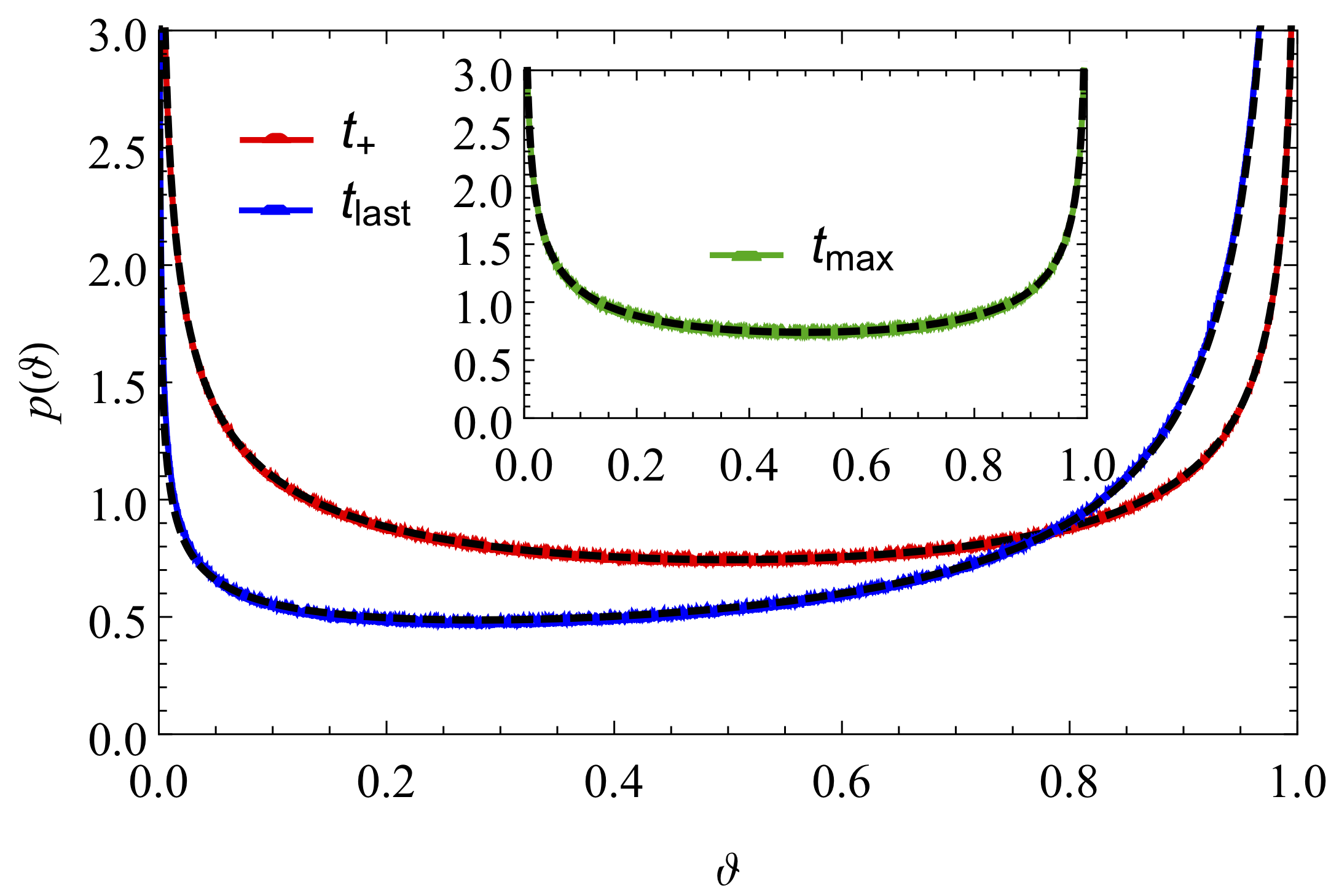}
\caption{A comparison of the data shown in \fref{f:raw} with their theoretical formula \eqref{Plast}-\eqref{P+}. The dashed lines are for theoretical results.  The distribution $p_{\rm max}(\vartheta)$ is shown in the inset as it is almost indistinguishable from the distribution $p_{\rm pos}(\vartheta)$.}
\label{f:all}
\end{figure}

In order that the reader can use our results, we give simple but rather precise approximations for the results obtained after numerical integration.
\begin{align}
{\cal F}_2^{\rm last}(\vartheta)\simeq & -17.92401+13.30207 \sqrt{\vartheta}\cr &-2.16604 \sqrt{1-\vartheta} + 8.30059 \vartheta +11.59529 \vartheta^{\frac{3}{2}}\cr
&+13.23121 (1-\vartheta)^{\frac{3}{2}} - 10.74274 \vartheta^2,\label{eq:F last numerical approximation}
\\
{\cal F}_2^{\rm max}(\vartheta)\simeq &-0.431001+1.69259 \left[ \vartheta(1-\vartheta)\right]^{\frac{1}{2}}\cr
&-1.93367 \left[ \vartheta(1-\vartheta)\right] +1.3572 \left[ \vartheta(1-\vartheta)\right]^{\frac{3}{2}} \nn\\ 
& -0.33995  \left[ \vartheta(1-\vartheta)\right]^{2},\label{eq:F max numerical approximation}
\\
{\cal F}_2^{\rm pos}(\vartheta) \simeq & -0.842235+1.76479 \left[ \vartheta(1-\vartheta)\right]^{\frac{1}{2}}\cr
&+3.70810 \left[ \vartheta(1-\vartheta)\right] -9.71973 \left[ \vartheta(1-\vartheta)\right]^{\frac{3}{2}}\cr
&+7.40511 \left[ \vartheta(1-\vartheta)\right]^{2} .\label{eq:F + numerical approximation}
\end{align}
These approximate functions are estimated respecting symmetries in the problem, i.e.\ that ${\cal F}_2^{\rm pos}(\vartheta)$ and ${\cal F}_2^{\rm max}(\vartheta)$ are symmetric under the exchange of $\vartheta \rightarrow 1-\vartheta$ while ${\cal F}_2^{\rm last}(\vartheta)$ is not.

\begin{remark}
We stated above that $p_{\rm pos}(\vartheta)$ and $p_{\rm max}(\vartheta)$ are symmetric around $\vartheta=\frac{1}{2}$, while $p_{\rm last}(\vartheta)$ is not (except for $H=\frac{1}{2}$). Symmetry of the first two probabilities is due to the observation that  $x_t$ and $X_T-x_{T-t}$ have the same law. For $p_{\rm last}$ the asymmetry is easy to see from the almost-straight-line trajectories for $H\simeq 1$ in \fref{fig:fbm sample}, which makes $\vartheta= 0$ the most probable value. This is reflected in the small-$\vartheta$ divergence of the distribution \eq{Plast} in the limit of $H\to 1$. 
\end{remark}

\subsection{Comparison with numerical results}
An efficient implementation of   fBm on a computer   is non-trivial due to its long-range correlations in time. For   this paper, we use the Davis-Harte  algorithm \cite{DaviesHarte1987,DiekerPhD}, which generates sample trajectories drawn from a Gaussian probability with covariance \eqref{eq:covariance} in a time of order $N \ln (N)$, given $N$ equally spaced discretization points. Details of this algorithm are given in   \aref{app:computer simulation}.
Interestingly, for the first-passage time, recently an algorithm was introduced which grows   as $\ln( N)^3$, albeit accepting a small error probability \cite{ArutkinWalterWiese2020,WalterWiese2019a,WalterWiese2019b}, allowing for even more precise estimates. 

Results for the three probabilities from our computer simulations are shown in figure \ref{f:raw} for $H=0.33$. They are obtained by averaging over $5\times 10^9$ sample trajectories, each generated  with $2^{13}$ discrete-time steps. The two distributions  $p_{\rm max}(\vartheta)$ and $p_{\rm pos}(\vartheta)$ are almost indistinguishable, as predicted in their theoretical expressions  in Eqs.~\eqref{Pmax} and \eqref{P+}. 
 
\begin{figure*}
\parbox{0mm}{\raisebox{1.2cm}[0mm][0mm]{\hspace{1.3cm}$H=0.5$}}\fig{9.0cm}{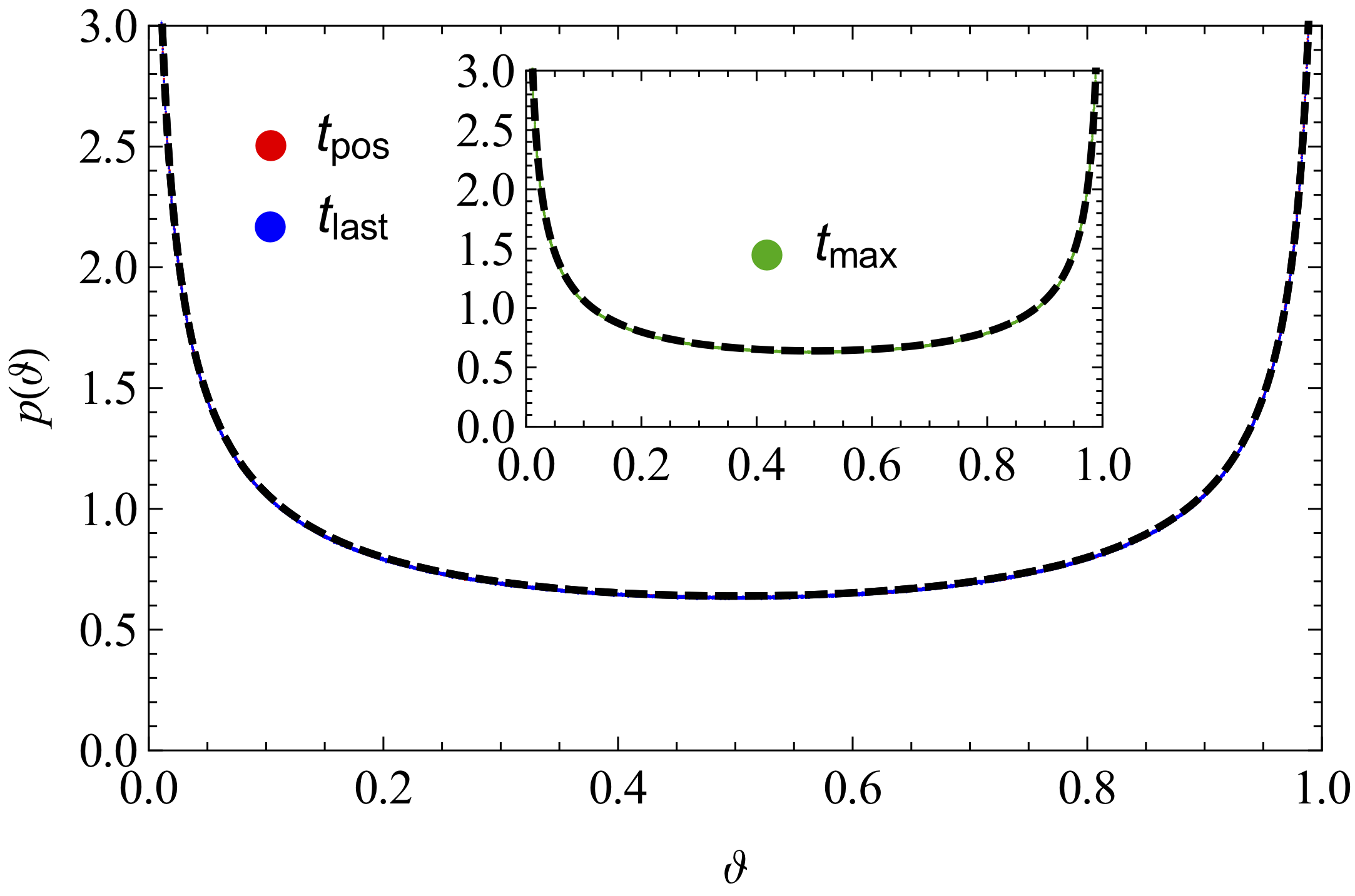}~%
\parbox{0mm}{\raisebox{1.2cm}[0mm][0mm]{\hspace{1.3cm}$H=0.6$}}\fig{9.0cm}{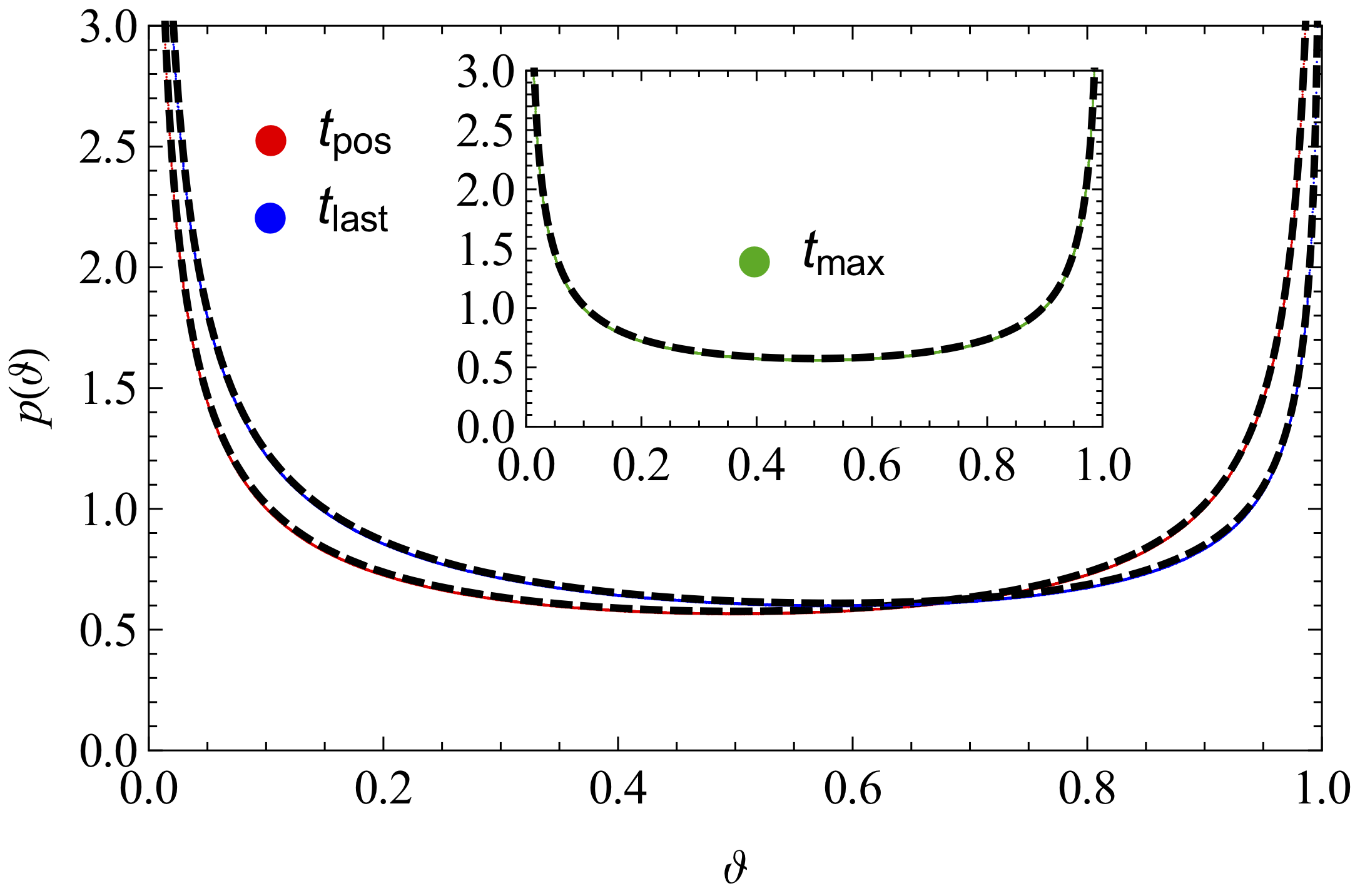}
\parbox{0mm}{\raisebox{1.2cm}[0mm][0mm]{\hspace{1.3cm}$H=0.75$}}\fig{9.0cm}{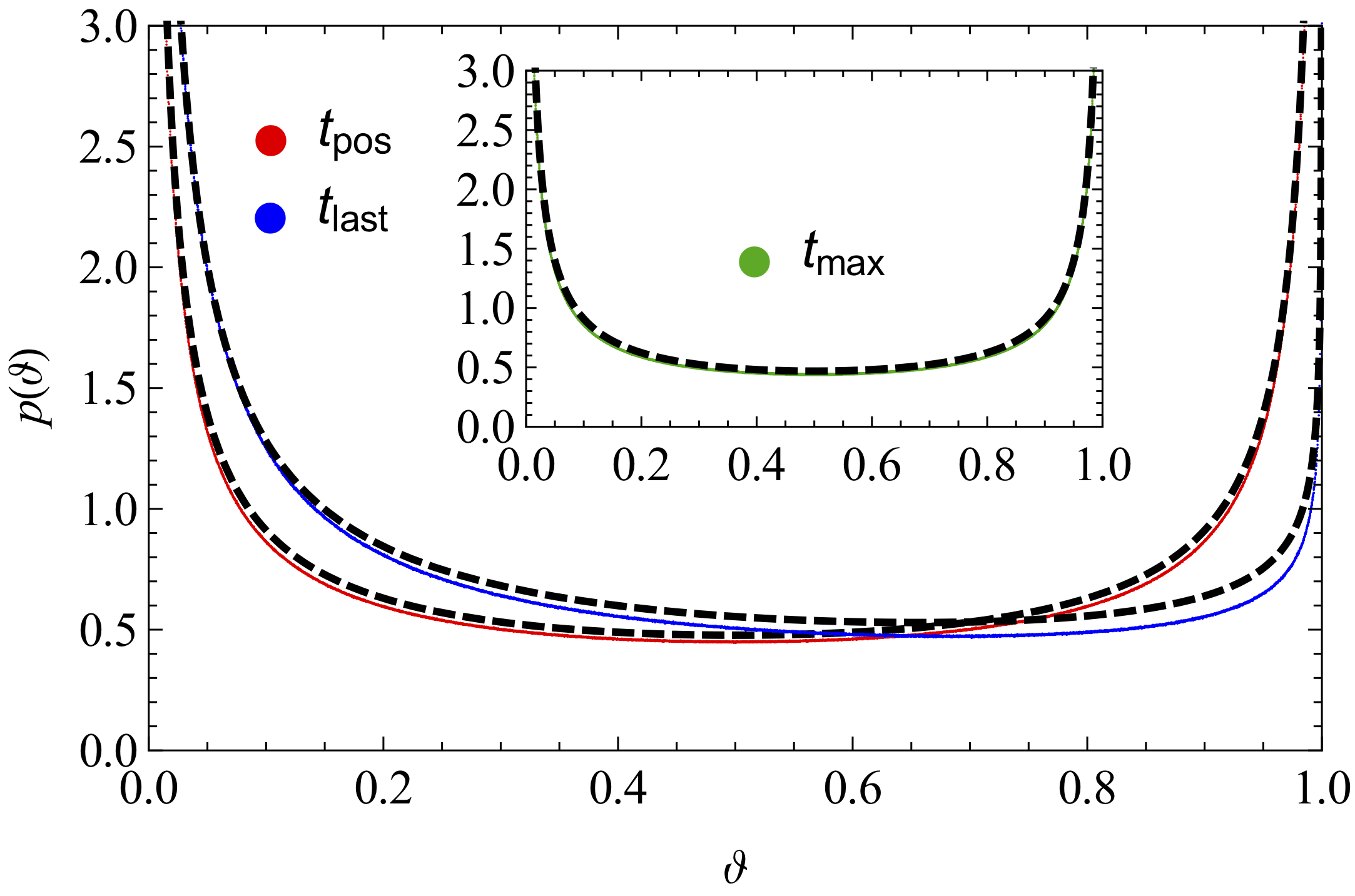}~%
\parbox{0mm}{\raisebox{1.2cm}[0mm][0mm]{\hspace{1.3cm}$H=0.9$}}\fig{9.0cm}{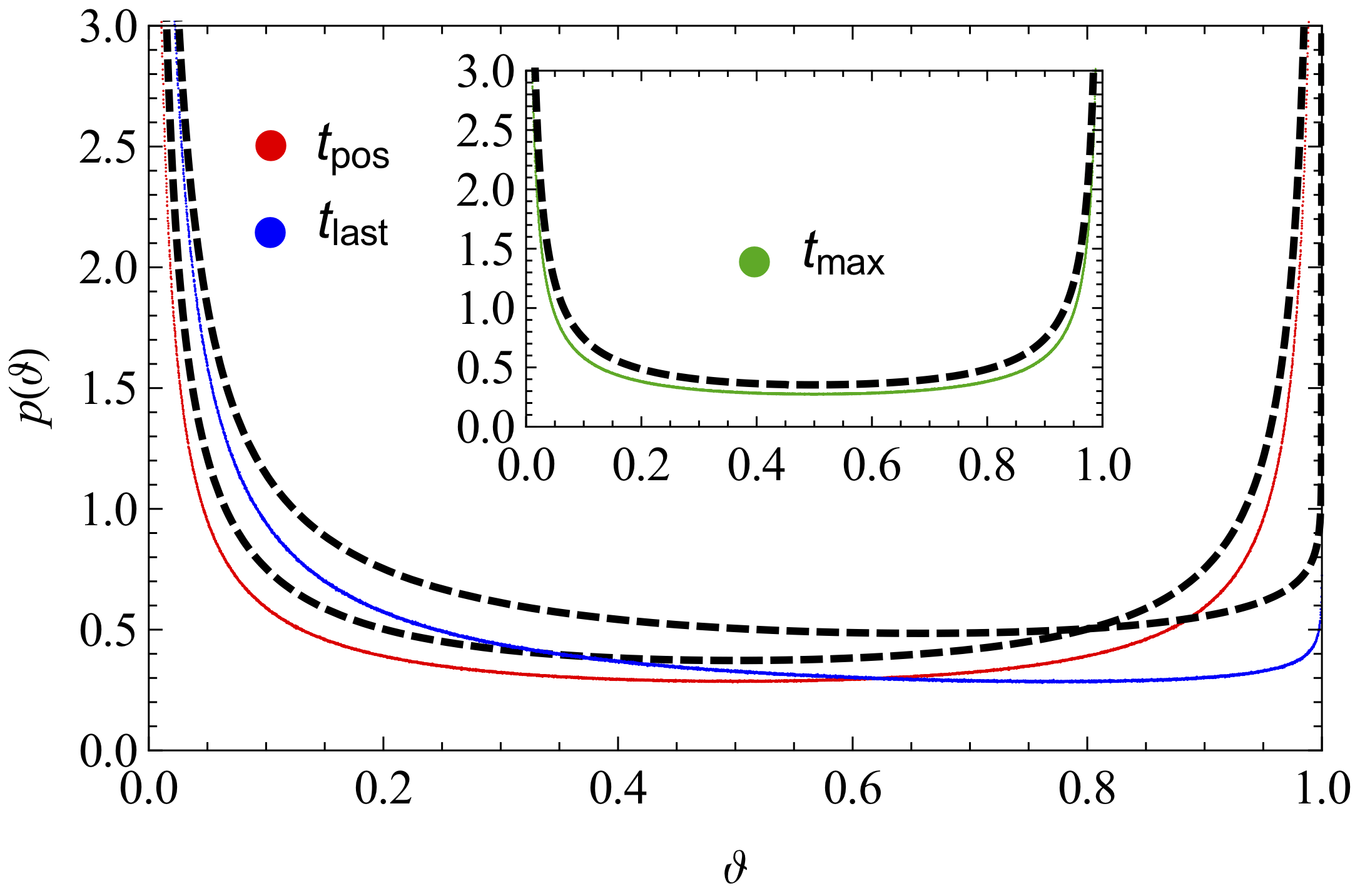}
\caption{A comparison of the theoretical formulas in Eqs.~\eqref{Plast}-\eqref{P+} with their corresponding numerical simulation result of an fBm at diffrent values of $H\ge \frac{1}{2}$:  $H=0.5$, $0.6$, $0.75$ and $0.9$. The dashed lines are the theoretical results,   the continuous lines   the numerical results.}
\label{f:all diff H}
\end{figure*}

Figure \ref{f:raw} also shows that $p_{\rm last}(\vartheta)$ behaves markedly differently from the other two distributions; especially, it is asymmetric under the exchange $\vartheta\to 1-\vartheta$.  This  asymmetry in exponents is reversed around $H=\tfrac{1}{2}$,  as shown in the inset of figure \ref{f:raw}.  This can be seen in the scaling form in \Eq{Plast}. 

A comparison of  numerical data for $H=0.33$ with their corresponding theoretical result in Eqs.~\eqref{Plast}-\eqref{P+} are shown in \fref{f:all}. They are in   excellent agreement. Deviations  are visible for higher values of $H$ as shown in \fref{f:all diff H} for a set of increasing values of $H\ge \frac{1}{2}$.
We see a perfect agreement between theoretical and numerical results for $H=\frac{1}{2}$, (i.e. $\epsilon=0$). The agreement is very good for small $\epsilon=H-\tfrac{1}{2}$, but deviations can be seen as $\epsilon$ is increased beyond $|\epsilon| \approx 0.25$, i.e.\ $H\le  0.25$ or $H\ge 0.75$. 

\begin{figure*}
\newcommand{\figsize}{5.9cm}\fig{\figsize}{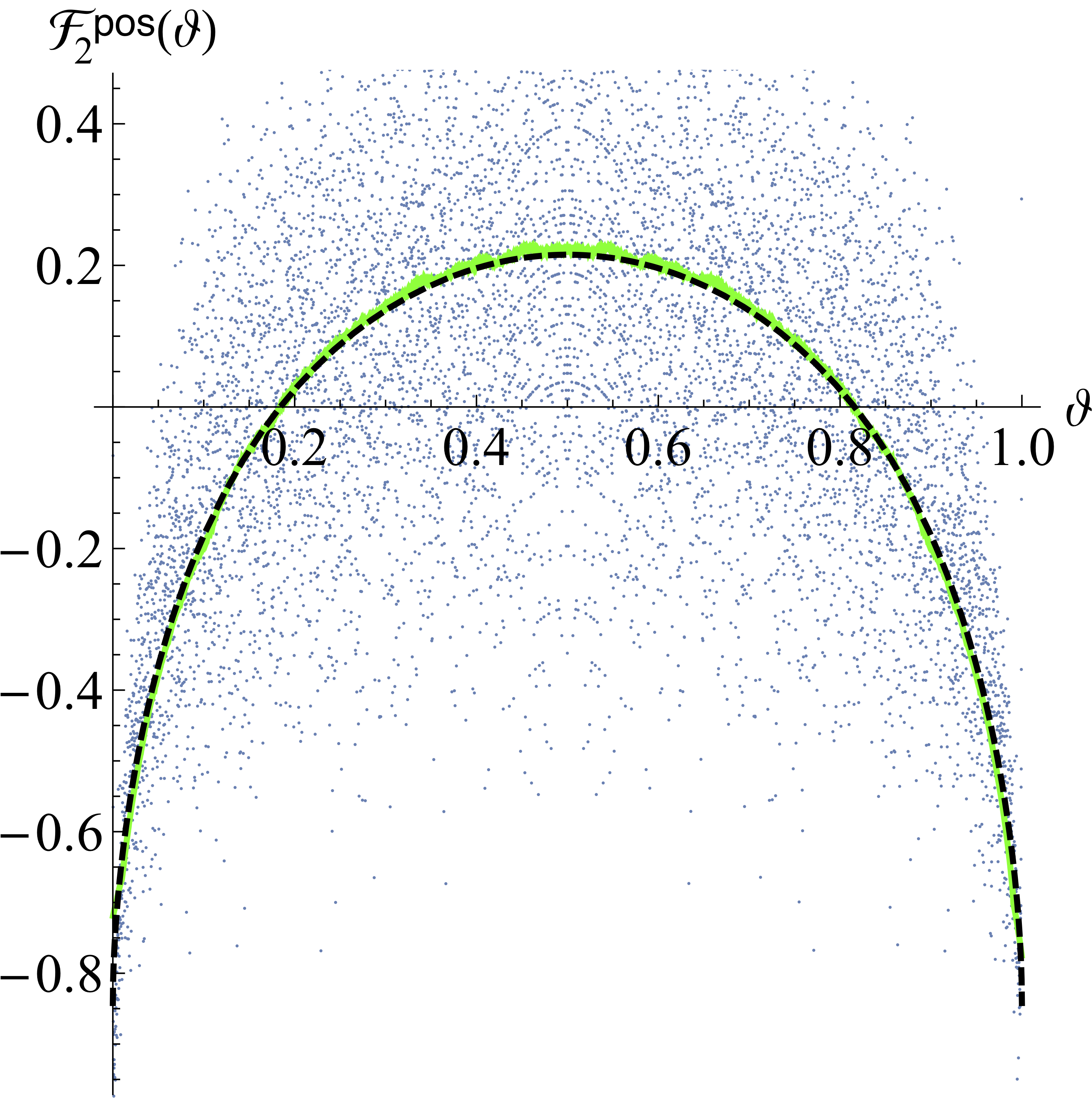} \fig{\figsize}{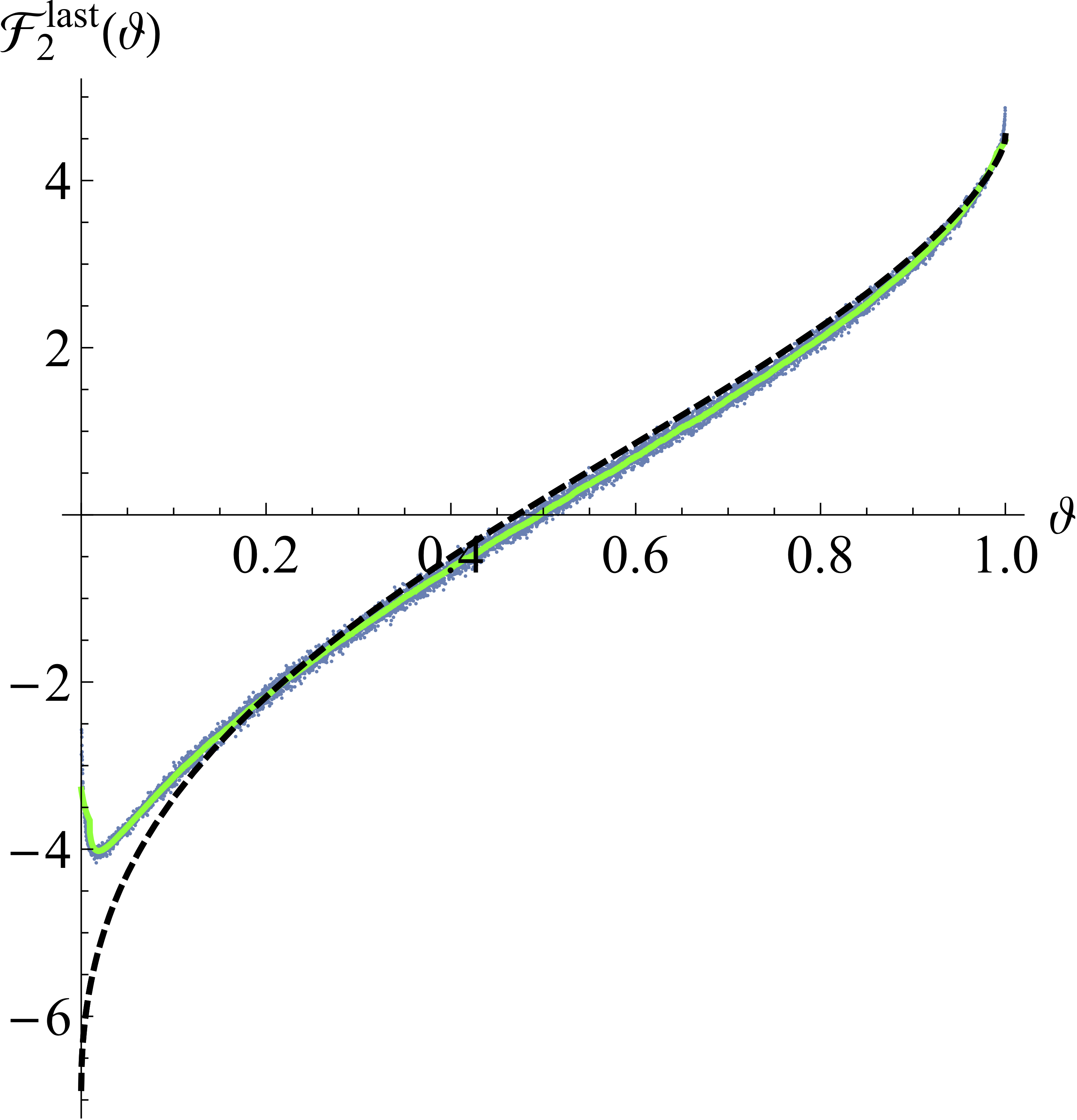}
\fig{\figsize}{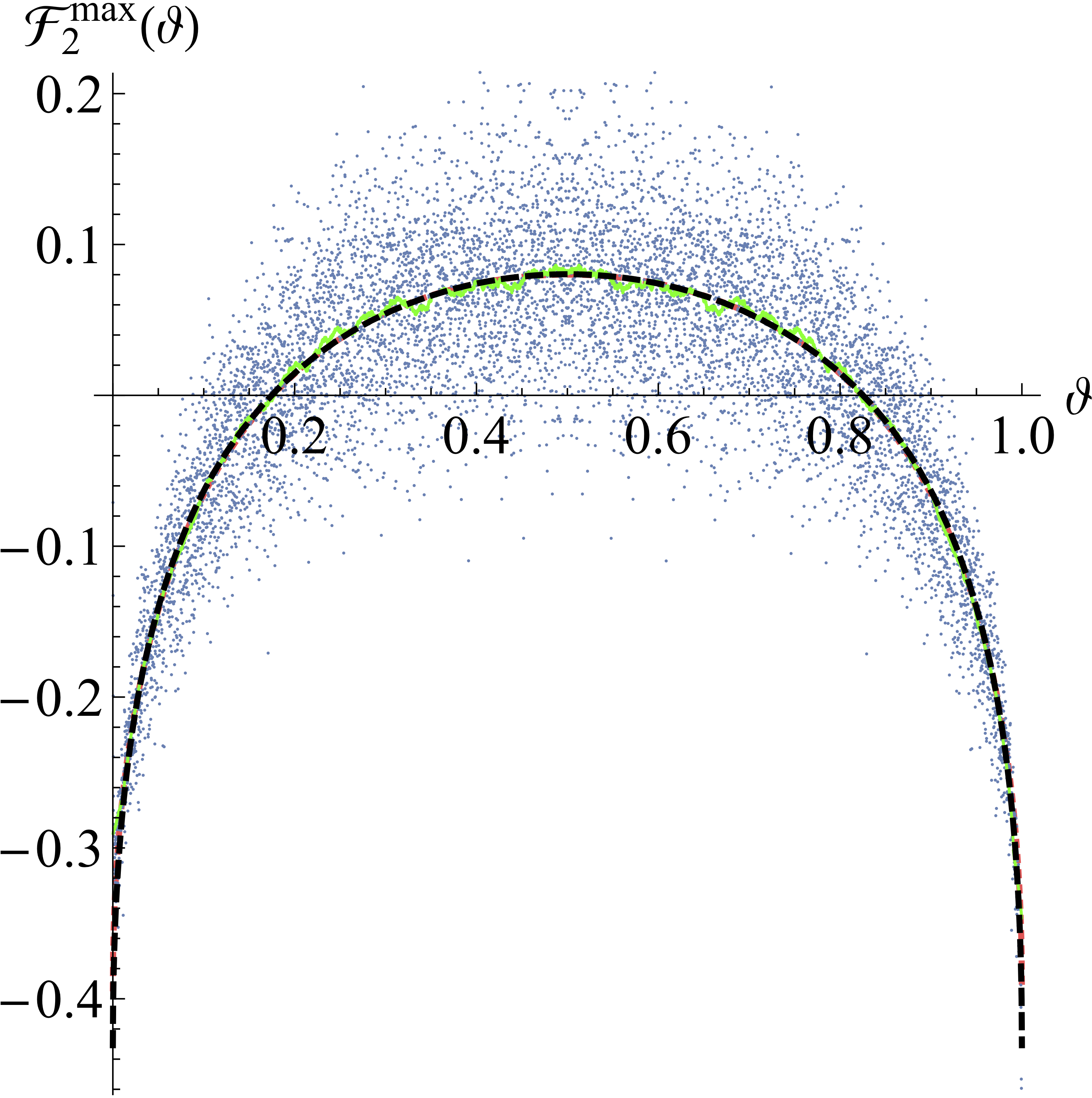}
\put(-425,30){$(a)$}
\put(-250,30){$(b)$}
\put(-85,30){$(c)$}
\caption{A comparison  of the three  ${\cal F}_2(\vartheta)$ obtained analytically (black dashed lines) and their measurement using formula (\ref{9})  with $\epsilon=\pm\tfrac{1}{6}$. From left to right: $(a)$ positive time, $(b)$ time for the last visit to the origin, and $(c)$ time for the maximum. The scattered dots are the raw data  from trajectories of $N=2^{13}$ time steps,  averaged over $5 \times 10^9$ samples, which are coarse grained by a factor of 100 to give the green curve. \label{f:F2s}}
\end{figure*}

The difference between $p_{\rm last}$ and $p_{\rm max}$ first appears in the second-order term $\mathcal{F}_2$ in \Eq{eq:F expansion}. In \fref{f:F2s} we plot our theoretical results of ${\cal F}_2(\vartheta)$ alongside the   results   from computer simulations. This give a finer verification of our  theory. To illustrate this procedure, we use \Eq{P+} to define 
\be
{\cal F}_{2,\epsilon}^{\rm pos}(\vartheta):= \frac1\epsilon\left[\frac1\epsilon \ln \!\bigg( p_{\rm pos}(\vartheta)  \frac{[\vartheta (1-\vartheta)]^H}{\cal N}  \bigg) -{\cal F}_1^{\rm pos}(\vartheta)\right].
\ee
Then, ${\cal F}_{2,\epsilon}^{\rm pos}(\vartheta)  = {\cal F}_{2}^{\rm pos}(\vartheta) + {\cal O}(\epsilon)$ and it contains all terms in the exponential in \Eq{P+} except ${\cal F}_{1}^{{\rm pos}}(\vartheta)$. We can further improve this estimate by observing that the sub-leading term in ${\cal F}_{2,\epsilon}^{\rm pos}(\vartheta)$ is odd in $\epsilon$. Define
\be \label{9}
\overline{\cal F}_{2,\epsilon}^{\rm pos}(\vartheta):=\frac12\Big[ 
{\cal F}_{2,\epsilon}^{\rm pos}(\vartheta) +{\cal F}_{2,-\epsilon}^{\rm pos}(\vartheta)  \Big], 
\ee
then $\overline{\cal F}_{2,\epsilon}^{\rm pos}(\vartheta)$ differs from the theoretical ${\cal F}_{2}^{\rm pos}(\vartheta)$ by order $\epsilon^2$ or higher, for small $\epsilon$, equivalent to an order $\epsilon^4$ correction to $p_{\rm last}(\vartheta)$.

A comparison of $\overline{\cal F}_{2,\epsilon}^{\rm pos}(\vartheta)$ extracted from   numerical simulations   of $p_{\rm pos}(\vartheta)$ to the theoretical result of ${\cal F}_{2}^{\rm pos}(\vartheta)$ is shown in \fref{f:F2s} for $\epsilon =\pm \frac16$ (\textit{i.e.} for $H=\frac{2}{3}$ and $\frac{1}{3}$). The figure also contains a similar comparison for ${\cal F}_{2}^{\rm last}(\vartheta)$ and ${\cal F}_{2}^{\rm max}(\vartheta)$, with their corresponding numerical results.
One   sees the excellent agreement between results from  our theory and numerical simulations. We remind that these are {\em sub-sub-leading} corrections, almost indiscernible in the probability density $p(\vartheta)$ shown on \fref{f:all}.

\begin{figure}
\newcommand{\figsize}{8.7cm} \fig{\figsize}{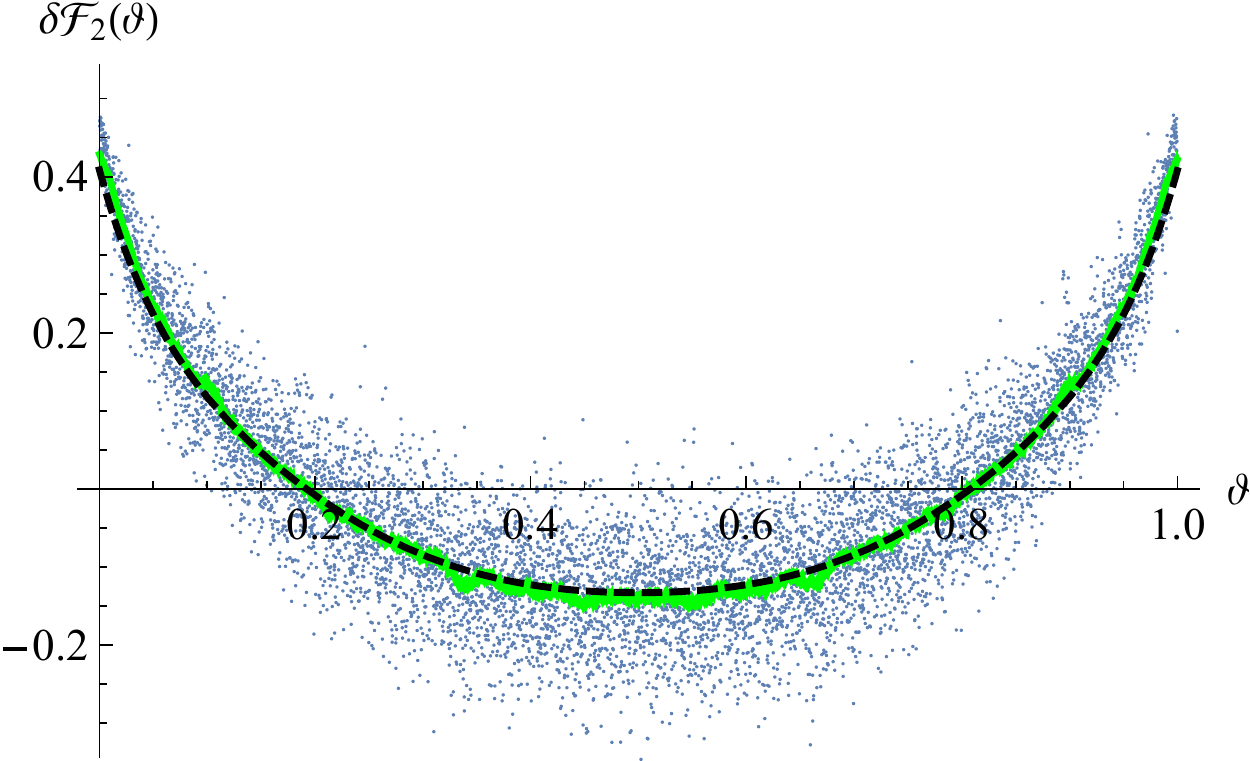} 
\caption{The difference $\delta {\cal F}_2(\vartheta) = {\cal F}_2^{\rm max}(\vartheta)-{\cal F}_2^{\rm pos}(\vartheta)$ using the same conventions as in Fig.~\ref{f:F2s}. This plot quantifies the difference between the distribution of $t_{\rm max}$ and $t_{\rm pos}$.}
\label{f:F2diffs}
\end{figure}
 
An important observation from \fref{f:F2s} is that for all three observables $\mathcal{F}_2(\vartheta)$ is finite in the entire range of $\vartheta$. We note that the amplitude of ${\cal F}_2^{\rm last}(\vartheta)$ is about ten times larger than ${\cal F}_2^{\rm pos}(\vartheta)$ and ${\cal F}_2^{\rm max}(\vartheta)$.  The former also shows the largest deviations from our theoretical result, especially for $\vartheta \to 0$. These indicate the presence of sub-leading terms of order $\epsilon^4$, or higher in $p$.

The difference between $p_{\rm pos}(\vartheta)$ and $p_{\rm max}(\vartheta)$ first appears  at second order in   perturbation theory. 
To underline that $\mathcal{F}_2^{\rm pos}(\vartheta)$ and $\mathcal{F}_2^{\rm max}(\vartheta)$ in Eqs.~\eqref{eq:F max numerical approximation} and \eqref{eq:F + numerical approximation} are distinct functions, we show in \fref{f:F2diffs} their difference 
\bea
\delta \mathcal{F}_2(\vartheta) &=& {\cal F}_{2}^{\rm max}(\vartheta)-{\cal F}_{2}^{\rm pos}(\vartheta)  \cr
&=& \lim_{\epsilon\to 0}\frac1{\epsilon^2} \ln \!\bigg(\frac{p_{\rm max}(\vartheta)}{p_{\rm pos}(\vartheta)}\Big) .
\eea 
 The theoretical result of the difference shows excellent agreement with the numerical data for $\overline{\cal F}_{2,\epsilon}^{\rm max}(\vartheta)-\overline{\cal F}_{2,\epsilon}^{\rm pos}(\vartheta)$ defined following the same conventions as in \Eq{9}. This proves that the laws for $t_{\rm max}$ and $t_{\rm pos}$ are indeed different.

\subsection{Scaling analysis \label{sec:scaling}}
The prefactor of the exponential in formula Eqs.~\eqref{Plast}-\eqref{P+} can be predicted using   scaling arguments. The simplest one is $p_{\rm last}(\vartheta)$, which is the probability that the fBm  is at the origin at time $\vartheta$ and   does not   return for the remaining time $1-\vartheta$. (We put the total time $T=1$, s.t.\ $\vartheta=t$.) The probability for the first part of the event scales as $\vartheta^{-H}$, see   \Eq{eq:Free fBm propagator}. The second part  scales as $\vartheta^{-\theta}$, where $\theta=1-H$ is the persistent exponent \cite{Molchan1999,KrugKallabisMajumdarCornellBraySire1997,WieseMajumdarRosso2010}. Combining the two gives the prefactor in \Eq{Plast}.

The scaling argument for $p_{\rm max}(\vartheta)$ is more involved, and   was first discussed in Refs.~\cite{WieseMajumdarRosso2010,DelormeWiese2016,DelormeThesis}.
One starts with the relation
\begin{equation}
{\cal P}_T(m)=\frac{d\mathscr{S}_T(m)}{dm},
\end{equation}
where ${\cal P}_T(m)$ is the probability for the position of the maximum $m$ for an fBm in a time interval $T$ started at the origin;   $\mathscr{S}_{T}(m)$ is the survival probability up to time $T$ for an fBm started at $m>0$, in presence of an absorbing wall at the origin.
Self-affinity of an fBm suggests the scaling form
\begin{equation}
 {\cal P}_T(m)=\frac{1}{T^H}g_1\!\left( \frac{m}{T^H} \right); \qquad \mathscr{S}_{T}(m)=g_2\!\left( \frac{m}{T^H} \right),
\end{equation}
which leads to
\begin{equation}
g_1(x)=g_2^{\prime}(x).
\end{equation}
To be consistent with the result for the persistence exponent \cite{Molchan1999,KrugKallabisMajumdarCornellBraySire1997}, one must have $g_2(x)\sim x^{\frac{\theta}{H}}$ for small $x$. This leads to $g_1(x)\sim x^{\frac{\theta}{H}-1}$, equivalent to
\begin{equation}
{\cal P}_T(m)\sim \frac{m^{\frac{\theta}{H}-1}}{T^\theta}\qquad \textrm{for small $m$.} 
\end{equation}
To relate to the distribution $P_T(t_{\rm max})$ of $t_{\rm max}$ we use that at small  $t_{\rm max}$ the   maximum $m$ is also small and $m\sim t_{\rm max}^H$. This leads to
\begin{equation}
P_T(t_{\rm max})={\cal P}_T(m)\frac{dm}{dt_{\rm max}}\sim \frac{1}{t_{\rm max}}\left(\frac{t_{\rm max}}{T}\right)^{\!\!\theta}.
\end{equation}
Substituting $\theta=1-H$ one gets
\begin{equation}
P_T(t_{\rm max})\sim \frac{1}{T}\left(\frac{t_{\rm max}}{T}\right)^{\!\!-H},
\end{equation}
and equivalently
\begin{equation}
p_{\rm max}(\vartheta)=T\,P_T(\vartheta\, T)\sim {\vartheta^{-H}}\qquad \textrm{for small $\vartheta$.} \label{eq:pmax}
\end{equation}
Using the symmetry of the probability $p_{\rm max}(\vartheta)$ under $\vartheta \to 1-\vartheta$ one gets   $(1-\vartheta)^{-H}$ for $\vartheta \rightarrow 1$. This gives the prefactor in \Eq{Pmax}.

A similar argument relating to the persistent exponent \cite{Dhar1999} can be constructed for the distribution of $t_{\rm pos}$. For $t_{\rm pos}\ll T$, probability $P_T(t_{\rm pos})$ for an fBm to remain positive of net $t_{\rm pos}$ time, relates to persistence probability for the fBm to stay negative for most of its total duration $T$. This means, for $1\ll t_\text{pos}\ll T$, 
\begin{equation}
P_T(t_{\rm pos}) \sim T^{-\theta},
\end{equation}
with the persistent exponent $\theta$. For this $T$-dependence to be consistent with the re-scaled probability $P_T(\vartheta\,T)=\frac{1}{T}p_{\rm pos}(\vartheta)$, one must have
\begin{equation}
p_{\rm pos}(\vartheta)\sim \vartheta^{\theta-1}\qquad \text{for $\vartheta \to 0$,} 
\end{equation}
giving the small $\vartheta$ divergence in \Eq{P+}. The symmetry under $\vartheta \to 1-\vartheta$ gives the divergence near $\vartheta\to 1$.

\subsection{Comparison to an exact result}
In Ref.~\cite{KrapivskyMallickSadhu2015} the first few moments of $t_{\rm pos}$ were calculated analytically for an fBm of $H=\frac{1}{4}$. It is straightforward to generalize this analysis for arbitrary $H$. For the fraction of positive time $\vartheta= {t_{\rm pos}}/{T}$, we obtain the first three moments: $\langle\vartheta\rangle=\frac{1}{2}$ (obvious from the symmetry of the distribution),
\begin{subequations}
\begin{align}
\langle\vartheta^2\rangle & =\frac{1}{4}+\frac{1}{2\pi}\int_0^1dr\, \arcsin R(r),\\
\langle\vartheta^3\rangle & =\frac{1}{8}+\frac{3}{4\pi}\int_0^1dr\, \arcsin R(r),
\end{align}
where
\begin{equation}
R(r)=\frac{1}{2 r^H}\left[1+r^{2H}-(1-r)^{2H} \right].
\end{equation}
It is  hard to determine higher moments. The problem maps to the orthant probability problem for a multivariate Gaussian,   which is still unsolved \cite{Owen2014}.
\label{eq:two moments exact}
\end{subequations}

A perturbation expansion of \Eq{eq:two moments exact} in $\epsilon=H-\frac{1}{2}$ gives
\begin{subequations}
\begin{align}
\langle\vartheta^2\rangle & =\frac{3}{8}+\frac{\epsilon}{4} \left( \log 4-1\right)+\frac{\epsilon^2}{24}\left( 6\log ^2 4-\pi^2\right)+\cdots,\\
\langle\vartheta^3\rangle & =\frac{5}{16}+\frac{3\epsilon}{8} \left( \log 4-1\right)+\frac{\epsilon^2}{16}\left( 6\log ^2 4-\pi^2\right)+\cdots .
\end{align}\label{eq:two moments exact series}%
\end{subequations}
Terms up to   linear order are   reproduced using our perturbation result \Eq{P+}. The $\epsilon^2$  order terms ($0.0693$ for $\langle\vartheta^2\rangle$ and $0.1040$ for $\langle\vartheta^3\rangle$) obtained using the numerical approximation \Eq{eq:F + numerical approximation} agree  with the exact result in \Eq{eq:two moments exact series} up to the third decimal place. (This is a $0.2\%$ disagreement, as apposed to a $40\%$ disagreement if $\mathcal{F}_2^{\rm pos}$ is ignored in \Eq{eq:F expansion +}.)

\begin{figure}
\newcommand{\figsize}{8.7cm} \fig{\figsize}{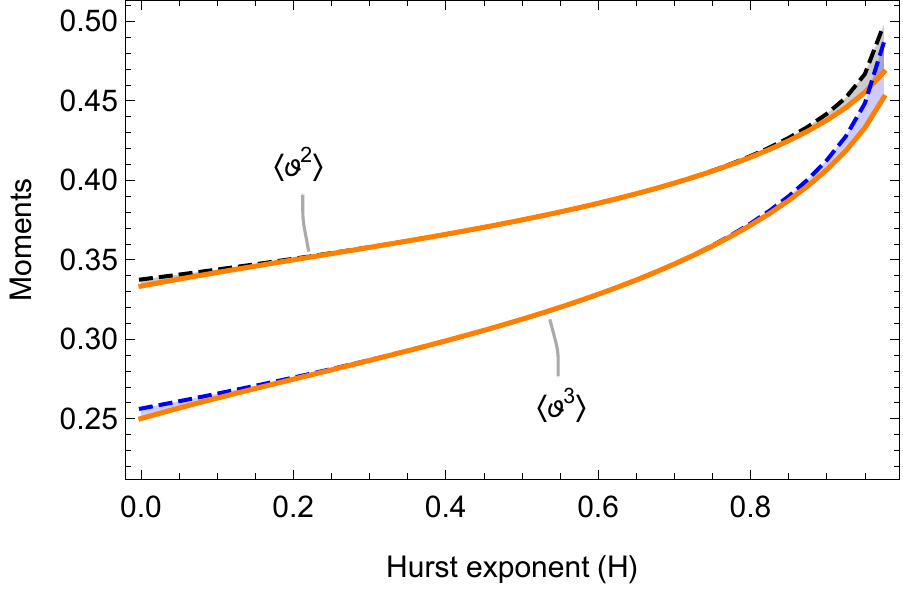} 
\caption{Second and third moment for the fraction of positive time $\vartheta=t_{\rm pos}/T$ as a function of the Hurst exponent $H$. The solid lines are   the exact result in \Eq{eq:two moments exact},  whereas the dashed lines denote their result obtained using \Eq{P+} with $\mathcal{F}^{\rm pos}$ in \Eq{eq:F expansion} up to   second order. The difference is noticeable for $H$ far from $\frac{1}{2}$, indicating corrections from   higher-order terms in \Eq{eq:F expansion}.}
\label{fig:Two Moments}
\end{figure}

A comparison of the exact result for the moments with their results obtained using \Eq{P+} is shown in \fref{fig:Two Moments}.

\section{Overview of theoretical analysis} 
\label{sec:general approach}
Before we present details of the derivation for Eqs.~\eqref{Plast}-\eqref{P+}, we give an overview of our approach. Our calculation is done using a double Laplace transformation $\ca D$ for the probability $P_T(\tau)$,  defined by
\begin{eqnarray}
\widetilde{P}(\lambda,s)&=&\mathscr{D}_{\substack{\tau \to \lambda \\ T \to s}}\odot P_T(\tau), \qquad \text{with} \\[1mm]
\mathscr{D}_{\substack{\tau \to \lambda \\ T \to s}}\odot P_T(\tau)&:=&\int_{0}^{\infty}dT\int_{0}^{T}d\tau \; e^{-sT-\lambda \tau}\;P_T(\tau). \qquad \label{eq:D transformation}
\end{eqnarray}
For the re-scaled probability $p(\vartheta):=T\, P_T(\vartheta\, T)$ and it's Laplace transform  
\begin{equation}
\widetilde{p}(\kappa)=\int_0^\infty d\vartheta \, e^{-\kappa\, \vartheta}p(\vartheta),
\end{equation}
the $\mathscr{D}$-transformation gives
\begin{equation}
\widetilde{P}(\lambda,s)=\frac{1}{s}\widetilde{p}\left(\frac{\lambda}{s}\right), ~~ \text{with} ~~ \widetilde{p}\left(\kappa\right)=\int_0^{1}d\vartheta \, \frac{p(\vartheta)}{1+\kappa \,\vartheta}. \label{eq:ptilde p}
\end{equation}
Complex analysis using the residue theorem gives the corresponding inverse transformation (see \aref{app:inverse} for a derivation),
\begin{equation}
p(\vartheta)=\frac{1}{2\pi i}\lim_{\delta\rightarrow 0^+}\frac{\widetilde{p}(-\frac{1}{\vartheta}-i\delta)-\widetilde{p}(-\frac{1}{\vartheta}+i\delta)}{\vartheta}.
\label{eq:inverse KT1}
\end{equation}
Equivalently, one can write
\begin{equation}
p(\vartheta)=\frac{1}{2\pi i}\lim_{\phi \to \pi^-}\Big[\kappa\,\widetilde{p}(\kappa)-\kappa^\star\widetilde{p}(\kappa^\star)\Big]_{\kappa=\frac{e^{i\phi}}{\vartheta}},
\label{eq:inverse KT2}
\end{equation}
where the limit is taken from $\phi$ below $\pi$, and the star ($\star$) denotes complex conjugation.

The analysis can be simplified by considering the  form of results in Eqs.~\eqref{Plast}-\eqref{P+} expected from scaling arguments. We write
\begin{equation}
p(\vartheta)=\frac{e^{\mathcal{F}(\vartheta,H)-(H-\frac{1}{2})\mathcal{R}(\vartheta)}}{\pi \sqrt{\vartheta(1-\vartheta)}}
\label{eq:gF def}
\end{equation}
with $\mathcal{R}(\vartheta)=\log\tfrac{\vartheta}{(1-\vartheta)}$ for $t_{\rm last}$ and $\mathcal{R}(\vartheta)=\log\vartheta(1-\vartheta)$ for $t_{\rm max}$ and $t_{\rm pos}$. (In writing \Eq{eq:gF def} the normalization constant $\mathcal{N}$ from Eqs.~\eqref{Plast}-\eqref{P+} is absorbed in $\mathcal{F}$.) Then, from Eqs.~\eqref{eq:inverse KT2} and \eqref{eq:gF def} we write
\begin{equation}
\widetilde{p}(\kappa)=\frac{e^{\widetilde{\mathcal{F}}(\kappa,H)}}{\sqrt{1+\kappa}}\label{eq:last mathcal F}, 
\end{equation}
such that
\begin{equation}
e^{\mathcal{F}(\vartheta,H)-(H-\frac{1}{2})\mathcal{R}(\vartheta)}=\mathcal{K}_{\kappa \rightarrow \vartheta}^{-1}\odot e^{\widetilde{\mathcal{F}}(\kappa,H)}.
\label{eq:main inverse KT}
\end{equation}
Here we define the transformation
\begin{equation}
\mathcal{K}_{\kappa \rightarrow \vartheta}^{-1}\odot f(\kappa)\equiv\lim_{\phi\rightarrow \pi- }\mathscr{R}\left[f\left(\frac{e^{i\phi}}{\vartheta}\right)\right], \label{eq:Kay inverse transformation}
\end{equation}
with $\mathscr{R}$ denoting the real part.

In our derivation of the probabilities in Eqs.~\eqref{Plast}-\eqref{P+}, we   first calculate   $\widetilde{\mathcal{F}}(\kappa,H)$, and then use \Eq{eq:main inverse KT} to obtain  $\mathcal{F}(\vartheta,H)$. To  do this order by order in a perturbation expansion in $\epsilon=H-\frac{1}{2}$,   write
\begin{subequations}
\begin{equation}
\widetilde{\mathcal{F}}(\kappa,H)=\epsilon\;\widetilde{\mathcal{F}}_1(\kappa)+\epsilon^2\widetilde{\mathcal{F}}_2(\kappa)+\mathcal{O}(\epsilon^3).\label{eq:F kappa expansion}
\end{equation}
Using this expansion in \Eq{eq:main inverse KT} we get \Eq{eq:F expansion} with
\begin{align}
\mathcal{F}_{1}(\vartheta)=&\, \mathcal{R}(\vartheta)+\mathcal{K}_{\kappa \rightarrow \vartheta}^{-1}\odot \widetilde{\mathcal{F}}_{1}(\kappa), \label{eq:main inverse KT 1}\\
\mathcal{F}_{2}(\vartheta)=&-\frac{1}{2}\left[\mathcal{F}_{1}(\vartheta)-\mathcal{R}(\vartheta)\right]^2\cr&+\mathcal{K}_{\kappa \rightarrow \vartheta}^{-1}\odot \left[\widetilde{\mathcal{F}}_{2}(\kappa)+\frac{1}{2}\widetilde{\mathcal{F}}_{1}(\kappa)^2\right]. \label{eq:main inverse KT 2}
\end{align}
\label{eq:F last inversion formula}
\end{subequations}

\begin{remark}
For completeness and for verification purposes, let us write the inverse transformation of \Eq{eq:Kay inverse transformation},
\begin{equation}
\mathcal{K}_{\vartheta \to k}\odot f(\vartheta):=\frac{1}{\pi}\int_0^1 d\vartheta \; \frac{\sqrt{1+\kappa}}{1+\kappa \, \vartheta}\,\frac{f(\vartheta)}{\sqrt{\vartheta(1-\vartheta)}}.\label{eq:Kay transformation}
\end{equation}
A list of the used inverse $\ca K$-transforms is  given in \aref{sec:list of K transformations}.
\end{remark}
\begin{remark}
From the normalization condition $\int_0^1d\vartheta \,p(\vartheta)=1$ one can see in \Eq{eq:ptilde p} that $\widetilde{p}(0)=1$ and therefore in \Eq{eq:last mathcal F},
\begin{equation}
\widetilde{\mathcal{F}}(\kappa,H)=0\qquad \text{for $\kappa=0$.} \label{eq:F tilde zero}
\end{equation}
\end{remark}

\begin{remark}
There are two reasons for performing our analysis using Laplace transform. The first is  that convolutions in time are factorized, the second that integrations over space can be done over the Laplace-transformed propagator, but not the propagator in time. This will become clear in the analysis in the following sections.
\end{remark}

\section{Distribution of time $t_{\rm last}$ for the last visit to the origin}
\label{sec:last}
\begin{figure}
\includegraphics[scale=0.75]{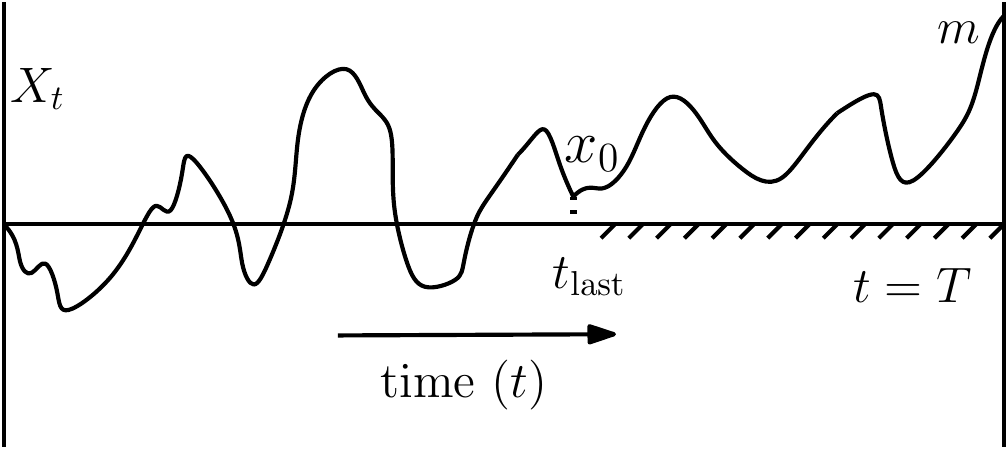}
\caption{A schematic of an fBm trajectory contributing to the time $t_{\rm last}$ of last visit to the origin. The striped line indicates an absorbing boundary.}
\label{fig:lastVisit}
\end{figure}
The 
analysis for the distribution of $t_{\rm last}$ is the simplest among the three observables, and we   present it first. The probability of $t_{\rm last}=\tau$ for an fBm in a time window $[0,T]$ can be determined by
\begin{equation}
P_T(t_{\rm last}=\tau)=\frac{W(\tau,T)}{N(T)} \quad \text{for $x_0\to 0$,} \label{eq:PT last limit formal}
\end{equation}
where $W(\tau,T)$ is {\em twice }  the     weight of fBm trajectories that start at $X_0=0$, pass through $X_\tau=x_0>0$, and remain positive for the rest of the time (see \fref{fig:lastVisit} for an illustration). 
Note that the factor of $2$ accounts for the possibility that the final position is either $m>0$, or $m<0$.
Here
$N(T)$ is the normalization
\begin{equation}
N(T)=\int_{0}^{T}d\tau W(\tau,T).
\label{eq:norm 2}
\end{equation}
(To keep   notations simple, we   avoid explicit reference to $x_0$, unless necessary.)

Formally, we write 
\begin{align}
W(\tau,T)=&\,2\int\limits_{0}^{\infty}dm\int_{x(0)=0}^{x(T)=m}\mathcal{D}[x]\;\delta\big(x(\tau)-x_0\big)\cr &\qquad \times\prod_{t= \tau}^{T}\Theta\big(x(t)\big)\;e^{-S} .
\label{eq:NW last}
\end{align}
The perturbative expansion in \Eq{eq:expansion of S} of the action leads to a similar expansion for $W$, given by
\begin{equation}
W(\tau,T)= W_0(\tau,T)+\epsilon W_1(\tau,T)  + \epsilon^2 W_2(\tau,T)+\ldots \label{eq:W last expansion}
\end{equation}
with
\begin{align}
W_0(\tau,T)& = 2\int_{0}^{\infty}dm\,\llangle 1 \rrangle_{m}, \label{eq:w0 last}\\
W_1(\tau,T)& = \int_{0}^{\infty}dm \,\llangle[\Big]\frac{S_1}{D}\rrangle[\Big]_{m} ,\label{eq:w1 last}\\
W_2(\tau,T)& = \int_{0}^{\infty}dm\,\llangle[\bigg]\frac{S_1^2}{4D^2}-2\frac{S_2}{D}\rrangle[\bigg]_{m} .\label{eq:w2 last}
\end{align}
The double-angular brackets denote (for $m> 0$) the average over trajectories as sketched in \fref{fig:lastVisit} with a standard Brownian measure,
\begin{equation}
\llangle O[x]\rrangle_{m}:=\!\!\!\!\!\!\!\int\limits_{x(0)=0}^{x(T)=m}\mquad
\mathcal{D}[x]\delta\big(x(\tau)-x_0\big)\prod_{t=\tau}^{T}\Theta\big(x(t)\big)\,e^{-\frac{S_0}{D}}O(x(t)). 
\label{eq:angular bracket last}
\end{equation}
This definition of double-angular brackets is specific to the trajectories used here, its definition in other sections will include the corresponding boundary conditions needed there.

\subsection{Zeroth order term}
In terms of the free Brownian propagator \Eq{eq:W0 Z} and the propagator in presence of an absorbing wall,
\begin{equation}
Z^{+}_t(x_1,x_2)=\int_{x(0)=x_1}^{x(t)=x_2}\mathcal{D}[x]\prod_{r=0}^{t}\Theta(x(r))e^{-\frac{S_0}{D}}\label{eq:propagator free absorbing}
\end{equation}
we write \Eq{eq:w0 last} as
\begin{equation}
W_0(\tau,T)=2 \int_{0}^{\infty}dm\, Z_{\tau}(0,x_0)\,Z^{+}_{T-\tau}(x_0,m).
\end{equation}
Its double Laplace transformation \Eq{eq:D transformation} denoted by
\begin{equation}
\widetilde{W}_0(\lambda,s)=\mathscr{D}_{\substack{\tau \to \lambda \\ T \to s}}\odot W_0(\tau,T)
\label{eq:double L of W0}
\end{equation}
is
\begin{equation}
\widetilde{W}_0(\lambda,s)=2 \int_{0}^{\infty}dm\,\widetilde{Z}_{s+\lambda}(0,x_0) \;\widetilde{Z}^{+}_s(x_0,m). \label{eq:W0 last factorization}
\end{equation}
Here $\widetilde{Z}_s$ and $\widetilde{Z}_s^+$ are the Laplace transforms of $Z_t$ and $Z_t^{+}$, given by
\begin{subequations}
\begin{align}
\widetilde{Z}_s(x_1,x_2)&=\int_0^\infty dt \, e^{-s t}\; Z_t(x_1,x_2)\nn \\[1mm] &=\frac{e^{-\sqrt{\frac{s}{D}}\vert x_1-x_2\vert}}{2\sqrt{s D}} \label{eq:Z tilde final},
\end{align}
and
\begin{align}
\widetilde{Z}^{+}_s(x_1,x_2)&= \int_0^\infty dt \, e^{-s t} \; Z^{+}_t(x_1,x_2)\nn \\[1mm]
&=\frac{e^{-\sqrt{\frac{s}{D}}\vert x_1-x_2\vert}-e^{-\sqrt{\frac{s}{D}}\vert x_1+x_2\vert}}{2\sqrt{s D}} . \label{eq:Z + tilde final}
\end{align}%
\label{eq:L of Z Z+}%
\end{subequations}
Using these results in \Eq{eq:W0 last factorization} and evaluating the integral for small $x_0$ we get, (see \Eq{eq:id Z plus int}) \begin{equation}
\widetilde{W}_0(s\,\kappa,s)\simeq \frac{x_0}{Ds}\times \frac{1}{\sqrt{1+\kappa}} .\label{eq:w0 last result}
\end{equation}

\begin{remark}
The factorization in \Eq{eq:W0 last factorization}  results from the identity
\begin{equation}
\mathscr{D}_{\substack{\tau \to \lambda \\ T \to s}}\odot  \Big[g(\tau)f(T-\tau)\Big]=\widetilde{g}(s+\lambda)\;\widetilde{f}( s) ,
\label{eq:id L-transform}
\end{equation}
where $\widetilde{g}(s)$ and $\widetilde{f}(s)$ are the Laplace transforms of $g(t)$ and $f(t)$, respectively. 
\end{remark}
\begin{remark}
From \Eq{eq:w0 last result} it is straightforward to verify the arcsine-law \eqref{Levey-law} for   Brownian motion. One can use $D=1$ for $\epsilon=0$ in \Eq{eq:w0 last result}, and verify that $W_0(\vartheta T,T)\simeq x_0[\pi T \sqrt{\vartheta(1-\vartheta)}]^{-1}$. Then, Eqs.~\eqref{eq:PT last limit formal} and \eqref{eq:norm 2} lead to the distribution \eqref{Levey-law}.
\end{remark}

\subsection{Linear order: 1-loop diagrams}

\begin{figure}
\includegraphics[scale=0.7]{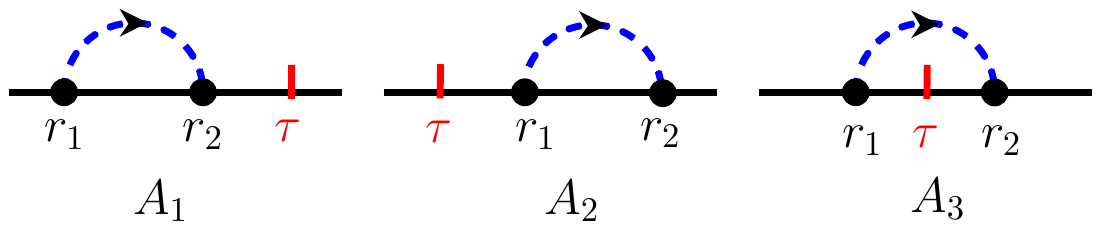}
\caption{One-loop diagrams: a graphical representation of the terms in \Eq{eq:W1 last A123} for the linear order in our perturbation expansion. For all diagrams $r_1<r_2$, staying on the same side of $\tau$ as indicated.  The dashed lines indicate coupling between points $r_1$ and $r_2$ with $r_1<r_2$ (indicated by an arrowhead) and a coupling strength $e^{y(r_1-r_2)}$. The solid disks indicate the `charge' $\dot{x}(r_1)$ and $\dot{x}(r_2)$ for the associated points. A similar convention will be used for diagrams in later parts of our analysis.}
 \label{fig:diagramELastLinear}
\end{figure}

Using $S_1$ from \Eq{eq:S1 1 integral} we explicitly write \Eq{eq:w1 last} as
\begin{align}
W_1(\tau,T) = \frac{1}{D}\int_{0}^{\infty}&dm\int_0^{\Lambda}dy \int_0^T dr_1\int_{r_1}^{T}dr_2\nn \\[1mm]
& \times e^{y(r_1-r_2)} \; \llangle[\big]\dot{x}(r_1)\dot{x}(r_2)\rrangle[\big]_{m} .\label{eq:W1 last 02} 
\end{align}
For   convenience we use a graphical representation of the expression in \Eq{eq:W1 last 02}. We write the amplitude in three parts, according to the relative order of times $r_1$, $r_2$, and $\tau$, as illustrated in the 1-loop diagrams in \fref{fig:diagramELastLinear}.

\begin{remark}
Diagrams in \fref{fig:diagramELastLinear} consists of couplings between a single pair of points, resulting in the $y$-integral in \Eq{eq:W1 last 02}. In analogy with field theory, we   refer to them as  1-loop diagrams, with $y$ representing the loop-variable to be integrated over. In \sref{sec:two loop last}, i.e.\ at   second order, amplitudes involve couplings between two pairs, resulting into  two $y$-integrations, and therefore referred to as  2-loop  diagrams.
\end{remark}

Following our convention for the diagrams in \fref{fig:diagramELastLinear} we write \Eq{eq:W1 last 02} as
\begin{subequations}
\begin{equation}
W_1=A_1+A_2+A_3 \label{eq:W1 last A123}
\end{equation}
with
\begin{align}
&A_1(\tau,T)=\frac{1}{D} \int\limits_{0}^{\infty}  dm \int\limits_{0}^{\Lambda}  dy\, J_\tau(0,x_0; -y,y)\,Z^{+}_{T-\tau}(x_0,m) ,\\
&A_2(\tau,T)=\frac{1}{D} \int\limits_{0}^{\infty}  dm \int\limits_{0}^{\Lambda}  dy\, Z_\tau(0,x_0) \; J_{T-\tau}^{+}(x_0,m;-y,y) ,\\
&A_3(\tau,T) \nn\\
&\! =\frac{1}{D} \int\limits_{0}^{\infty}  dm  \int\limits_{0}^{\Lambda}  dy\, J_\tau(0,x_0; -y)\; e^{-y\tau}J^+_{T{-}\tau}(x_0,m; y) .\label{eq:A3 last}
\end{align}
We defined
\end{subequations}
\begin{align}
&J_{t}(u,v;y_1,\ldots,y_n)\nn\\
&:=\int\limits_{0< r_1< \cdots< r_n}^{t} 
\prod_{i=1}^{n}  \; e^{-y_i r_i}  \int\limits_{x(0)=u}^{x(t)=v}\mathcal{D}[x] \dot{x}(r_1)\cdots \dot{x}(r_n) e^{-\frac{S_0}{D}}
\label{eq:J2}
\end{align}
and its analogue $J^{(+)}$ in presence of an absorbing wall at $x=0$. The integral over time in \Eq{eq:J2} is interpreted as in \Eq{eq:int dr1dr2 delta}, \textit{i.e.} with an ultraviolet cutoff $\Lambda$ on $y$. 

Using \Eq{eq:id L-transform} we write the double Laplace transform  \Eq{eq:D transformation} of 
the diagrams $A_i$ in terms of Laplace transforms of $Z$ and  $Z^+$ in \Eq{eq:L of Z Z+}, as well as  Laplace transforms for $J$ and $J^+$. Expressions are   obtained in \aref{app:B}, and summarized here,
\begin{align*}
\widetilde{A}_{1}(\lambda,s)=&\frac{1}{D}\int\limits_{0}^{\infty}  dm \int\limits_{0}^{\Lambda}  dy\,\widetilde{J}_{s+\lambda}(0,x_0;-y,y)\;\widetilde{Z}^{+}_s(x_0,m), \\
\widetilde{A}_{2}(\lambda,s)=&\frac{1}{D}\int\limits_{0}^{\infty}  dm \int\limits_{0}^{\Lambda}   dy\,\widetilde{Z}_{s+\lambda}(0,x_0)\; \widetilde{J}_s^{+}(x_0,m;-y,y) .
\end{align*}
Using Eqs.~\eqref{eq:L of Z Z+}, \eqref{eq:J2 0}, and \eqref{eq:IJPm1} gives, for small $x_0$,
\begin{align*}
\widetilde{A}_{1}(s\,\kappa,s) & \simeq -\frac{x_0}{Ds}\times\frac{\widetilde{\mathcal{A}}(1+\kappa)}{\sqrt{1+\kappa}},\cr
\widetilde{A}_{2}(s\,\kappa,s) & \simeq \frac{x_0}{Ds}\times\frac{\widetilde{\mathcal{A}}(1)}{\sqrt{1+\kappa}},
\end{align*}
with
\begin{align}
\widetilde{\mathcal{A}}(z)=&\int_0^{\Lambda/s}\frac{dy }{y ^2}\left(\sqrt{z+y }-\sqrt{z}\right)^2.
\label{eq:chi2 last}
\end{align}
A similar analysis for $A_3$ in \Eq{eq:A3 last}, using Eqs.~(\ref{eq:J1 tilde final}) and (\ref{eq:J + tilde 1 asymptotic}),   shows that the corresponding double Laplace transform $\widetilde{A}_{3}\sim x_0^2$, for small $x_0$. As a result, the double Laplace transform  of $W_1(\tau,T)$  defined in analogy to \Eq{eq:double L of W0}  reads, for small $x_0$,  
\begin{align}
\widetilde{W}_{1}(s\,\kappa,s)\simeq \frac{x_0}{Ds}\times\frac{\widetilde{\mathcal{A}}(1)-\widetilde{\mathcal{A}}(1+\kappa)}{ \sqrt{1+\kappa}}.
\label{eq:W1 tilde result}
\end{align}

\begin{remark}
The reason for $\widetilde{A}_{3}$ to vanish as $x_0^2$ or faster, for small $x_0$, can be understood from a simple observation. In the limit of $x_0\to 0$, $J_{t}(0,x_0;y_1,\ldots,y_n)$ in \Eq{eq:J2} vanishes for odd $n$. One way to see this is by noting that, in the limit of $x_0\to 0$, for each trajectory with a certain $\dot{x}(r)$, there is a mirror trajectory  $-\dot{x}(r)$, with equal probability. In comparison, $J^{+}(x_0,m;y_1,\ldots,y_n)$ vanishes for $x_0\to 0$ because of the absorbing boundary. This means that  in \Eq{eq:A3 last}, both $J$ and $J^+$ are at least of order $x_0$, and therefore $\widetilde{A}_{3}\sim x_0^2$, to the least. We shall see later that for a similar reason the amplitudes of the 2-loop diagrams $B$ and $C$ in \fref{fig:diagram} are of order $x_0^2$, for small $x_0$.
\end{remark}

\begin{remark}
We shall see that these diagrams $A_1$, $A_2$, and $A_3$ contribute to the propagator $W_1$ in \Eq{eq:W1 last A123}, thus to the scaling prefactor in \Eq{Plast}, but they do not feed into the exponential term $\ca F^ {\rm last}$.
\end{remark}

\subsection{Quadratic order: 2-loop diagrams \label{sec:two loop last}}

\begin{figure}
\includegraphics[width=0.48\textwidth]{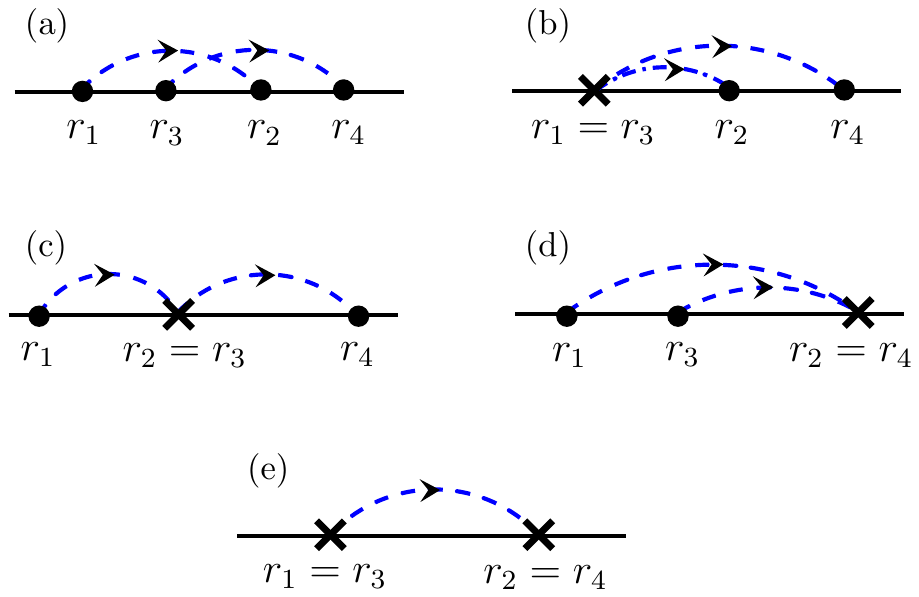}
\caption{A diagramatic representation for the amplitude in \Eq{eq:S1 sqr explicit} where all orders of time are allowed keeping $r_2>r_1$ and $r_4>r_3$ (as indicated by an arrowhead). We write the amplitude \Eq{eq:S1 sqr explicit} in five parts according to the contraction of times (indicated by cross). In (a) none of the times are equal (contracted). In (b,c,d) two times are contracted and in (e) all four times are contracted. \label{fig:S1sqr Last}}
\end{figure}
\begin{figure}
\includegraphics[width=0.48\textwidth]{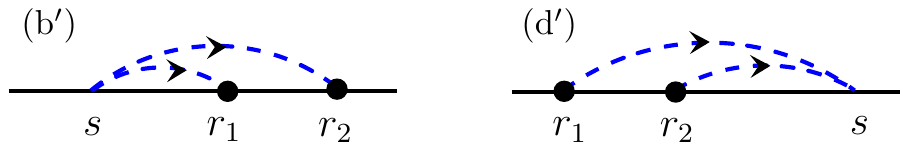}
\caption{A diagramatic representation of the formula in \Eq{eq:S2 explicit} where for (b$^\prime$) $s<r_1<r_2$ and for (d$^\prime$) $r_1<r_2<s$.}\label{fig:S2 Last First Formula}
\end{figure}

Using \Eq{eq:integral reperesentation S1S2} we explicitly write the terms  in \Eq{eq:w2 last} as
\begin{align}\label{eq:S1 sqr explicit}
&\int_{0}^{\infty}dm \, \llangle[\bigg]\frac{S_1^2}{4D^2}\rrangle[\bigg]_{m} \\
&=\frac{1}{4D^2}\int\limits_{0}^{\infty}dm\int\limits_0^{\Lambda}dy_1dy_2 \int\limits_{0}^{T}dr_1\int\limits_{r_1}^{T}dr_2 \int\limits_{0}^{T}dr_3\int\limits_{r_3}^{T}dr_4 \; \cr
& \qquad \times  e^{y_1(r_1-r_2)}  \;  e^{y_2(r_3-r_4)}\llangle[\Big]\dot{x}(r_1)\dot{x}(r_2)\dot{x}(r_3)\dot{x}(r_4)\rrangle[\Big]_{m} \nn
\end{align}
and
\begin{align}
\int_{0}^{\infty}dm \, \llangle[\bigg] &\frac{2S_2}{D}\rrangle[\bigg]_{m}=\frac{1}{D}\int\limits_{0}^{\infty} dm \int\limits_0^{\Lambda}dy_1dy_2 \cr 
\times\int\limits_{0}^{T}dr_1&\int\limits_{r_1}^{T}dr_2 
 \bigg[\int\limits_{0}^{r_1}ds\;e^{-y_1(r_1-s)-y_2(r_2-s)}\cr 
+\int\limits_{r_2}^{T}ds & \;e^{-y_1(s-r_1)-y_2(s-r_2)} \bigg] \llangle[\Big]\dot{x}(r_1)\dot{x}(r_2)\rrangle[\Big]_{m}. \qquad\label{eq:S2 explicit}
 \end{align}
A graphical illustration of the amplitudes in \Eq{eq:S1 sqr explicit} and \Eq{eq:S2 explicit} is shown in Figs.~\ref{fig:S1sqr Last} and \ref{fig:S2 Last First Formula}. Similar to the conventions in \fref{fig:diagramELastLinear}, a dashed line indicates an interaction between points $r_i$ and $r_j$ with an amplitude $e^{y(r_i-r_j)}$. The solid disks indicate the  field derivative $\dot{x}(r_i)$ at point $r_i$. For a contracted point, indicated by a cross, the associated amplitude is $2D$. A reason for this will be clear shortly. Empty points in \fref{fig:S2 Last First Formula} have an amplitude 1.

We shall see that among these diagrams,  only diagrams (a) and (c) contribute at the second order.  This can be directly seen using the normal-ordered weight in \Eq{eq:S normal ordered}.  Here,  we explicitly show why this happens. 

We find that the amplitudes of diagrams (b) and (b$^\prime$) are equal, as are those of (d) and (d$^\prime$). To see this we use that under Wick contraction between  $\dot x(r_1)$ and $\dot x(r_3)$
\begin{equation}
\llangle[\Big]\dot{x}(r_1)\dot{x}(r_2)\dot{x}(r_3)\dot{x}(r_4)\rrangle[\Big]_{m}\to 2 D\; \llangle[\Big]\dot{x}(r_2)\dot{x}(r_4)\rrangle[\Big]_{m}. \label{eq:contraction 4 to 2 last}
\end{equation}
(A similar result holds for contraction of any pair of times.) One can see this as a consequence of $\delta(r_i-r_j)$ term in \Eq{eq:id xxxx dot}, and its analogue in presence of an absorbing boundary.
Using the result \eqref{eq:contraction 4 to 2 last} in \Eq{eq:S1 sqr explicit} for diagram (b) we write its amplitude as
\begin{align*}
&\frac{1}{2D}\int_{0}^{\infty}dm\int_0^{\Lambda}dy_1dy_2 \int_{0}^{T}dr_1\int_{r_1}^{T}dr_2 \int_{r_1}^{T}dr_4   \nn \\[2mm]
& \qquad \qquad\times e^{y_1(r_1-r_2)} e^{y_2(r_1-r_4)}\llangle[\Big]\dot{x}(r_2)\dot{x}(r_4)\rrangle[\Big]_{m} \nn \\[2mm]
&=\frac{1}{D}\int_{0}^{\infty}dm\int_0^{\Lambda}dy_1dy_2 \int_{0}^{T}dr_2 \int_{r_2}^{T}dr_4 \int_{0}^{r_2}dr_1   \nn \\[2mm]
&\qquad \qquad \times e^{y_1(r_1-r_2)} e^{y_2(r_1-r_4)}\llangle[\Big]\dot{x}(r_2)\dot{x}(r_4)\rrangle[\Big]_{m}.
\end{align*}
Following a relabeling of the dummy variables $r$ we see that the integral is equal to the amplitude of diagram (b$^\prime$) from \Eq{eq:S2 explicit} and \fref{fig:S2 Last First Formula}. A similar analysis shows equal amplitude for diagrams (d) and (d$^\prime$). 

The amplitude of diagram (e), where all four times are contracted, is proportional to $W_0$ in \Eq{eq:w0 last}, which can be seen by using
\begin{equation}
\llangle[\Big]\dot{x}(r_1)\dot{x}(r_2)\dot{x}(r_3)\dot{x}(r_4)\rrangle[\Big]_{m}\to 4 D^2\llangle1\rrangle_{m},
\end{equation}
when all four points are contracted.
This means that the contribution of (e) can be included in the normalization \Eq{Plast}, and therefore ignored.

\begin{figure}
\includegraphics[width=0.48\textwidth]{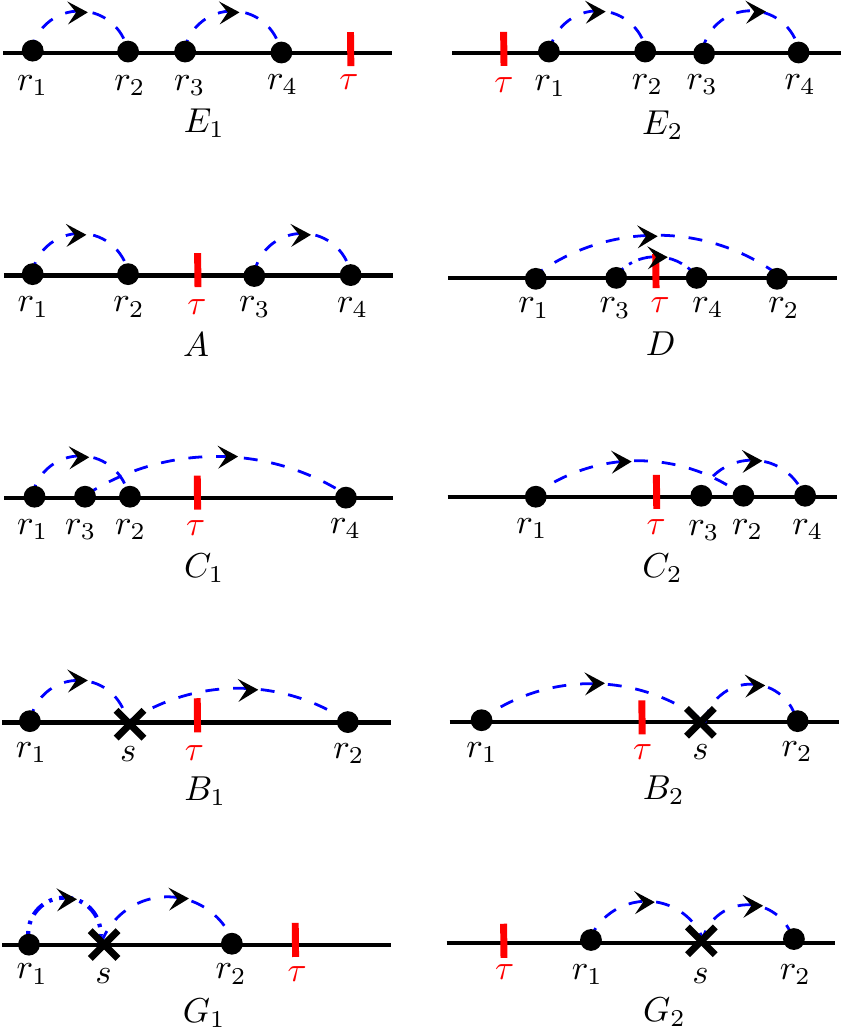}
\caption{Two-loop diagrams for the quadratic order term $W_2$ in \Eq{eq:w2 last}. The diagrams are categorized according to relative position of the loops with respect to $\tau$. For diagram $E_1$, the times $r_1<r_2<\tau$ and $r_3<r_4<\tau$, excluding cases where any two times are equal (contracted). Similar convention is adopted for the diagrams $E_2$, $A$, $D$, and $C$, where $r_1<r_2$, $r_3<r_4$, and their relative position with $\tau$ indicated in the diagrams. For diagrams $B$ and $G$ we consider, $r_2>s>r_1$ being on the same side of $\tau$ as indicated. A solid disk denotes a `charge' $\dot{x}(r)$ for the associated point $r$, and a cross denotes a `charge' $2D$. A dashed line indicates coupling between points $r_i$ and $r_j$ with a coupling strength $e^{y(r_i-r_j)}$. \label{fig:diagram}}
\end{figure}

Considering the contribution of the diagrams in Figs.~\ref{fig:S1sqr Last} and \ref{fig:S2 Last First Formula}, resulting into \Eq{eq:w2 last}, we see that the relevant contribution for $W_2$ comes   from the 2-loop diagrams (a) and (c) in \fref{fig:S1sqr Last}. Considering the relative position of the loops with respect to $\tau$, we write the amplitude $W_2$ as a sum of the following ten diagrams,  
\begin{align}
W_2&=a+c\cr
&=\left(E_1+E_2\right)+A+D+\left(C_1+C_2\right) \cr
& \quad +\left(B_1+B_2\right)+\left(G_1+G_2\right).\label{eq:last quadratic splitting W2}
\end{align}
This is  shown in \fref{fig:diagram}. Explicit formulas of their amplitudes are given in \aref{app:two loop diagram for last}.
We shall see that among these diagrams, only   diagram $D$ contributes to the non-trivial term $\mathcal{F}^{\rm last}$ in \Eq{Plast}, whereas the remaining diagrams contribute to the   power-law prefactor only.

Here, we present the double Laplace transformation \Eq{eq:D transformation} of the amplitude of these diagrams, for small $x_0$ limit. Their derivation is similar to those of the amplitude of zeroth and linear order terms in Eqs.~(\ref{eq:w0 last result},\,\ref{eq:W1 tilde result}). We defer their explicit calculation to the \aref{app:two loop diagram for last}.

For small $x_0$, we get
\begin{equation}
\widetilde{D}(s\,\kappa,s)\simeq \frac{x_0}{Ds}\times \frac{\widetilde{\mathcal{D}}(1+\kappa)}{\sqrt{1+\kappa}}\label{eq:D tilde last final}
\end{equation}
with 
\begin{align}\label{eq:var D last}
\widetilde{\mathcal{D}}(z)&= -2\int_{0}^{\Lambda/s}\frac{dy_1dy_2}{y_1^2 y_2^2}\sqrt{z}\sqrt{1+y_1+y_2}  \\[1mm]
& \quad \times \left(\sqrt{1+y_1+y_2}-\sqrt{1+y_1}-\sqrt{1+y_2}+1\right) \cr
& \quad \times \bigg(\sqrt{z+y_1+y_2}-\sqrt{z+y_1}-\sqrt{z+y_2}+\sqrt{z}\bigg).\nn
\end{align}
The amplitude of the diagrams $B$ and $C$ is of order $x_0^2$ for small $x_0$,
\begin{equation}
\widetilde{B}(s\,\kappa,s)\simeq \widetilde{C}(s\,\kappa,s)\sim x_0^2.
\end{equation}
This can be seen from the argument given in the remark below \Eq{eq:W1 tilde result}. Their explicit derivation is in Appendix \ref{app:B C last}.

The amplitude for the remaining diagrams is of order $x_0$, and given as follows.
For small $x_0$,
\begin{align}
\widetilde{E}_{1}(s\,\kappa,s)+\widetilde{E}_{2}(s\,\kappa,s)& \simeq \frac{x_0}{s}\times\frac{\widetilde{\mathcal{E}}(1+\kappa)+\widetilde{\mathcal{E}}(1)}{D\sqrt{1+\kappa}},
\label{eq:diagram E final}
\end{align}
where
\begin{align}
&\widetilde{\mathcal{E}}(z)=  -\frac{1}{2}\int_0^{\Lambda/s}\frac{ dy_1dy_2}{ y_1^2 y_2^2} \bigg\{(z+y_1) (z+y_2)\qquad  \nn \\[1mm]
&+\sqrt{z}\,\big(\sqrt{z}-\sqrt{z+y_1}-\sqrt{z+y_2}\big)\bigg[ \sqrt{z+y_1}  \nn \\[1mm]
&\times \big(\sqrt{z}-\sqrt{z+y_1}\big)+\sqrt{z+y_2} \left(\sqrt{z}-\sqrt{z+y_2}\right)  \nn \\[1mm]
&+\big(\sqrt{z}-\sqrt{z+y_1}-\sqrt{z+y_2}\big)^2- \cr
2 \bigg(&\sqrt{z+y_1+y_2}-
\sqrt{z+y_1}-\sqrt{z+y_2}+\sqrt{z}\bigg)^2\bigg]\bigg\}.\qquad\label{eq:var E}
\end{align}
Similarly, for small $x_0$,
\begin{equation}
\widetilde{A}(s\,\kappa,s)\simeq -\frac{x_0}{s}\times\frac{\widetilde{\mathcal{A}}(1+\kappa)\widetilde{\mathcal{A}}(1)}{D\sqrt{1+\kappa}}
\label{eq:2-loop A tilde final}
\end{equation}
with \Eq{eq:chi2 last}, and 
\begin{equation}
\widetilde{G}_1(s\, \kappa,s)+\widetilde{G}_2(s\, \kappa,s)\simeq\frac{x_0}{s}\times\frac{\widetilde{\mathcal{G}}(1+\kappa)+\widetilde{\mathcal{G}}(1)}{D\sqrt{1+\kappa}},
\label{eq:2-loop F tilde final last}
\end{equation}
where
\begin{align} \label{eq:var F}
\widetilde{\mathcal{G}}(z)
=&\int_0^{\Lambda/s}\frac{ dy_1dy_2}{y_1^2 y_2^2}\\[1mm]
&\times \left[\frac{(\sqrt{z+y_2}-\sqrt{z})^2y_1^2-(\sqrt{z+y_1}-\sqrt{z})^2y_2^2}{(y_1-y_2)}\right].\nn
\end{align}
Considering the amplitude of these 2-loop diagrams in \Eq{eq:last quadratic splitting W2} we get the double Laplace transform \Eq{eq:D transformation} of $W_2(\tau,T)$ in \Eq{eq:w2 last}. For small $x_0$ it reads
\begin{align}
\widetilde{W}_2 &(s\,\kappa,s) \simeq  \frac{ x_0
 }{s\,D\sqrt{1+\kappa}}\bigg[\widetilde{\mathcal{E}}(1+\kappa)+\widetilde{\mathcal{E}}(1)-\cr
 &\widetilde{\mathcal{A}}(1+\kappa)\widetilde{\mathcal{A}}(1)+\widetilde{\mathcal{D}}(1+\kappa)
 +\widetilde{\mathcal{G}}(1+\kappa)+\widetilde{\mathcal{G}}(1)\bigg].\quad
 \label{eq:w2 last tilde result}
\end{align}

\subsection{Result for $\mathcal{F}^{\rm last}(\kappa,H)$ \label{sec:F last}}
From the results in Eqs.~\eqref{eq:w0 last result}, \eqref{eq:W1 tilde result}, and \eqref{eq:w2 last tilde result} we obtain the double Laplace transform \Eq{eq:D transformation} of $W(\tau,T)$ in \Eq{eq:W last expansion} in an exponential form,
\begin{equation}
\widetilde{W}(s\, \kappa,s)=\mathscr{D}_{\substack{\tau \to s \,\kappa \\ T \to s}}\odot W(\tau,T)\simeq \frac{ x_0
 }{s}\times \frac{e^{\widetilde{\mathcal{W}}(\kappa)}}{\sqrt{1+\kappa}},\label{eq:w last scaled form}
\end{equation}
Here $x_0$ is small, and we used $D$ in \Eq{eq:D exp} to explicitly write the exponential term $\widetilde{\mathcal{W}}=\epsilon \widetilde{\mathcal{W}}_1+\epsilon^2 \widetilde{\mathcal{W}}_2+\cdots$, with
\begin{subequations}
\begin{align}
\widetilde{\mathcal{W}}_1(\kappa)=&\;\widetilde{\mathcal{A}}(1)-\widetilde{\mathcal{A}}(1+\kappa)+ 2(\log\Lambda+\gamma_{\rm E}-1),\\[1mm]
\widetilde{\mathcal{W}}_2(\kappa)=&\;\widetilde{\mathcal{E}}(1+\kappa)+\widetilde{\mathcal{E}}(1)+\widetilde{\mathcal{D}}(1+\kappa)\cr
 &-\widetilde{\mathcal{A}}(1+\kappa)\widetilde{\mathcal{A}}(1)
 +\widetilde{\mathcal{G}}(1+\kappa)+\widetilde{\mathcal{G}}(1)\cr &-\frac{1}{2}\left[ \widetilde{\mathcal{A}}(1)-\widetilde{\mathcal{A}}(1+\kappa) \right]^2+ 2\left[1-\frac{\pi^2}{6}\right].\qquad\qquad
\end{align}%
\label{eq:Ws exponentiated last}%
\end{subequations}
To relate to the exponential form in \Eq{eq:last mathcal F} we note that the Laplace transform of $N_T$ in \Eq{eq:norm 2}  is 
\begin{equation}
\widetilde{N}(s)=\widetilde{W}(0,s).
\end{equation}
The simple $s$-dependence in \Eq{eq:w last scaled form} (for $\Lambda\to \infty$) makes it easy to invert the Laplace transform, giving
\begin{equation}
N(T)\simeq x_0\; e^{\widetilde{\mathcal{W}}(0)} \qquad \text{for small $x_0$.}
\end{equation}

This means, for small $x_0$, $N(T)\equiv N$ is independent of $T$,  and the double Laplace transform of $P_T(\tau)$ in \Eq{eq:PT last limit formal} is
\begin{equation}
\widetilde{P}(\lambda,s)\simeq \frac{\widetilde{W}(\lambda,s)}{N} \qquad \text{for small $x_0$.}
\end{equation}
Then, using \Eq{eq:w last scaled form} and comparing with Eqs.~\eqref{eq:ptilde p} and \eqref{eq:last mathcal F} gives
\begin{equation}
\widetilde{\mathcal{F}}^{\rm last}(\kappa,H)=\widetilde{\mathcal{W}}(\kappa)-\widetilde{\mathcal{W}}(0),
\label{eq:F last W last}
\end{equation}
which we shall need to determine $\mathcal{F}(\vartheta,H)$ in \Eq{eq:main inverse KT}.
The leading terms in its perturbation expansion \Eq{eq:F kappa expansion} is given by
\begin{subequations}
\begin{align}
\widetilde{\mathcal{F}}_1^{\rm last}(\kappa)=&\;\widetilde{\mathcal{A}}(1)-\widetilde{\mathcal{A}}(1{+}\kappa),\label{eq:F12 tilde last a}\\
\widetilde{\mathcal{F}}_2^{\rm last}(\kappa)=&\;\left[\widetilde{\mathcal{D}}(1{+}\kappa)-\widetilde{\mathcal{D}}(1)\right]+\bigg[\widetilde{\mathcal{E}}(1{+}\kappa)+ \widetilde{\mathcal{G}}(1{+}\kappa)\cr
&-\frac{1}{2}  \widetilde{\mathcal{A}}^2(1{+}\kappa)-\left(\widetilde{\mathcal{E}}(1)+\widetilde{\mathcal{G}}(1)-\frac{1}{2}\widetilde{\mathcal{A}}^2(1) \right)\bigg].\label{eq:F12 tilde last b}
\end{align}
\end{subequations}
We have numerically verified that, for $\Lambda\to \infty$,  
\begin{align} 
&\widetilde{\mathcal{E}}(z)+\widetilde{\mathcal{G}}(z)-\frac{1}{2} \widetilde{\mathcal{A}}^2(z)=\left(1+\log(2)\right)^2-\frac{5 \pi^2}{12}. \label{eq:efa cancellation}
\end{align}
Therefore, the only non-vanishing contribution for $\Lambda\to \infty$ comes from the diagram $D$, leading to
\begin{eqnarray}
\widetilde{\mathcal{F}}_2^{\rm last}(\kappa)&=&\widetilde{\mathcal{D}}(1{+}\kappa)-\widetilde{\mathcal{D}}(1). \label{eq:F12 tilde last}
\end{eqnarray}

\begin{remark}
We see that \Eq{eq:F last W last} is consistent with the condition \Eq{eq:F tilde zero}. Moreover, we shall see that the integrals in \Eq{eq:F12 tilde last a} and \Eq{eq:F12 tilde last} converge in the $\Lambda\to \infty$ limit, as one would expect for   our theory to be correct.
\end{remark}
\begin{remark}
Note that in \Eq{eq:Ws exponentiated last} the contribution from diffusion constant $D$ in \Eq{eq:D exp} is constant, which cancels in \Eq{eq:F last W last}. This is expected as the distribution of $t_{\rm last}$ is independent of the diffusion constant, whereas as a distribution involving space would depend on $D$. The same applies for the distribution of $t_{\rm max}$ and $t_{\rm pos}$.
\end{remark}

For the leading-order term \Eq{eq:F12 tilde last a}, explicitly carrying out the integral in \Eq{eq:chi2 last} in the limit of $\Lambda\to \infty$, we get
\begin{equation}
\widetilde{\mathcal{F}}_1^{\rm last}(\kappa)=\log(1+\kappa),
\end{equation} 
whose $\mathcal{K}^{-1}$-transformation is (see \Eq{eq:inverse KT id four})
\begin{equation}
\mathcal{K}^{-1}_{\kappa\to \vartheta}\odot\widetilde{\mathcal{F}}_1^{\rm last}(\kappa)=-\log\tfrac{\vartheta}{(1-\vartheta)}=-\mathcal{R}^{\rm last}(\vartheta). \label{eq:F1 transform result}
\end{equation}
Using the result   \Eq{eq:main inverse KT 1} for $t_{\rm last}$ gives the leading-order result in \Eq{eq:F1 last final result}.

For the second-order term in \Eq{eq:main inverse KT 2} we use \Eq{eq:F12 tilde last}, \Eq{eq:F1 last final result}, and 
\begin{equation}
\mathcal{K}^{-1}_{\kappa\to \vartheta}\odot\widetilde{\mathcal{F}}_1^{\rm last}(\kappa)^2=\mathcal{R}^{\rm last}(\vartheta)^2-\pi^2
\end{equation}
(using the identity \Eq{eq:inverse KT id five}) to write
\begin{align}
\mathcal{F}_{2}^{\rm last}&(\vartheta) =\mathcal{K}_{\kappa \rightarrow \vartheta}^{-1}\odot \left[\widetilde{\mathcal{D}}(1+\kappa)-\widetilde{\mathcal{D}}(1)\right] -\frac{\pi^2}{2},
 \label{eq:F2 last final}
\end{align}
where we use linearity of the operator $\mathcal{K}^{-1}$.
  
\begin{figure}
\includegraphics[width=\columnwidth]{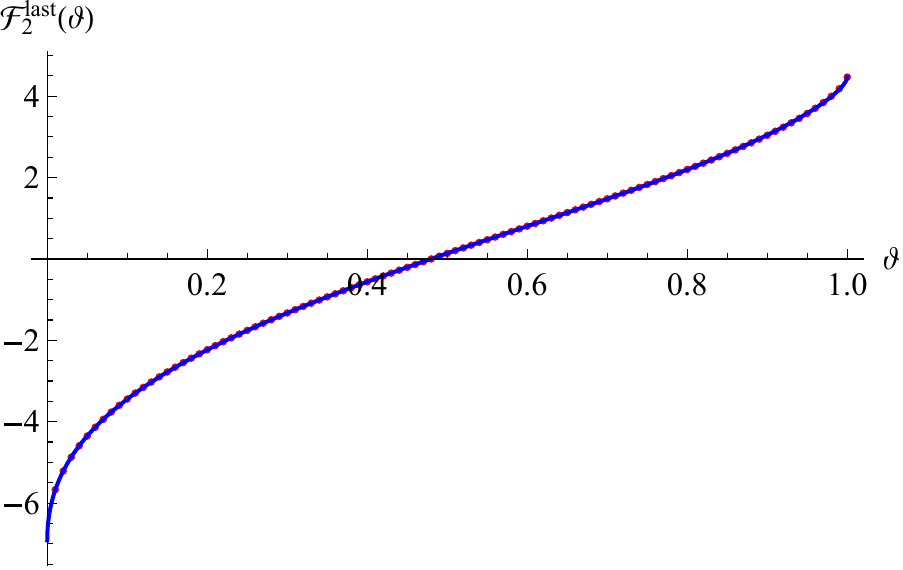}
\caption{The dotted points (colored red) show results of numerical integration for $\mathcal{F}_{2}^{\rm last}(\vartheta)$ in \Eq{eq:F2 last exact}. The solid line is the polynomial fit in  \Eq{eq:F last numerical approximation}, which gives a good estimation for $\mathcal{F}_{2}^{\rm last}(\vartheta)$. \label{fig:lastF2 numerical analytical comparison}}
\end{figure}

{\color{black}
The integral for $\widetilde{\mathcal{D}}(z)$ in \Eq{eq:var D last} is convergent in the limit of $\Lambda\to \infty$, but it is hard to evaluate analytically. The expression for $\mathcal{F}_{2}^{\rm last}(\vartheta)$ in \Eq{eq:F2 last exact} is obtained \cite{Mathematica} by exchanging the order of $\mathcal{K}_{\kappa \rightarrow \vartheta}^{-1}$ transformation and the $y$-integrals in  Eqs.~\eq{eq:F2 last final} and \eq{eq:var D last}. (For several other examples like in Eqs.~\eq{eq:F12 tilde last a} and \eq{eq:F1 transform result} where integration can be explicitly carried out, we have verified that this exchange of order gives the correct result.) The resulting function $\mathcal{F}_{2}^{\rm last}(\vartheta)$ in \Eq{eq:F2 last exact} is plotted in \fref{fig:lastF2 numerical analytical comparison} along with a polynomial estimation given in \Eq{eq:F last numerical approximation}. The expression \Eq{eq:F2 last exact} is in good agreement with our computer simulation result in \fref{f:F2s}.}

\section{Distribution of the time $t_{\rm max}$ when the fBm attains maximum} 
\label{sec:max}
The probability for an fBm, starting at $X_0=0$ and evolving till time $T$, to attain its maximum at time $t_{\rm max}=\tau$ can be expressed as 
\begin{equation}
P_T(t_{\rm max}=\tau)=\frac{W(\tau,T)}{N(T)} \label{eq:P max first formal}.
\end{equation}
Here $W(\tau,T)$ is the   weight 
of all contributing trajectories, and $N(T)$ is the corresponding normalization. 
We   use the same notations as   in \sref{sec:last}. Note, however, that the definition of these quantities ($W$, $N$, etc.) is specific  to the problem in this section. 

\begin{figure}
\includegraphics[width=\columnwidth]{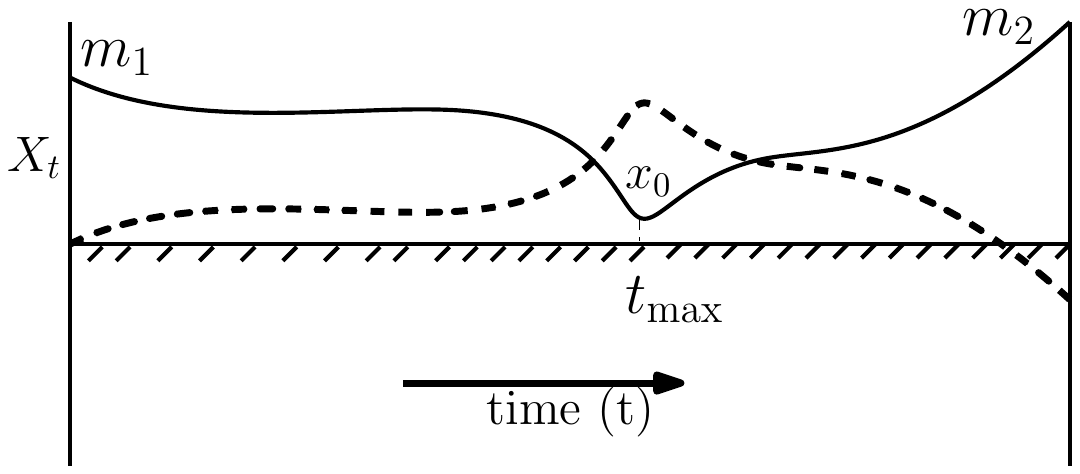}
\caption{The dark solid curve is a schematic of paths $X_t$ for \Eq{eq:W max formal}, where the stripped line indicates an absorbing boundary at the origin. For $x_0\to 0$, there is an one-to-one correspondence with an fBm path (indicated by gray dashed curve) that contributes for the process to attain its maximum $m_1$ at time $t_{\rm max}$.  \label{fig:lastMax}}
\end{figure}

Noting the symmetry of the problem (illustrated in \fref{fig:lastMax}), we write 
\begin{align}\label{eq:W max formal}
W(\tau,T)=\, &\int_{0}^{\infty}dm_1\int_{0}^{\infty}dm_2\int_{x(0)=m_1}^{x(T)=m_2}\mathcal{D}[x]\cr
&\qquad \times\delta(x(\tau)-x_0)\prod_{t=0}^{T}\Theta[x(t)]e^{-S[x]}.\qquad 
\end{align}
The probability density $P_T(\tau)$ in \Eq{eq:P max first formal} is obtained  by taking the limit of $x_0\to 0$.
(Like in the previous section, we do not write any explicit reference to $x_0$, unless necessary.)

The perturbation expansion \Eq{eq:expansion of S} of the fBm action $S$ leads to an expansion of $W$ similar to \Eq{eq:W last expansion} with
\begin{align}
W_0(\tau,T)& = \int_{0}^{\infty}\mquad dm_1\int_{0}^{\infty}\mquad dm_2\; \llangle 1 \rrangle_{(m_1,m_2)} ,\label{eq:w0 max}\\[1mm]
W_1(\tau,T)& = \int_{0}^{\infty}\mquad dm_1\int_{0}^{\infty}\mquad dm_2\;\llangle[\Big]\frac{S_1}{2D}\rrangle[\Big]_{(m_1,m_2)} ,\label{eq:w1 max}\\[1mm]
W_2(\tau,T)& = \int_{0}^{\infty}\mquad dm_1\int_{0}^{\infty}\mquad dm_2\;\llangle[\bigg]\frac{S_1^2}{8D^2}-\frac{S_2}{D}\rrangle[\bigg]_{(m_1,m_2)} .\label{eq:w2 max}
\end{align}
By the double-angular brackets we denote 
\begin{align}
\label{eq:angular bracket max}
&\llangle O[x]\rrangle_{(m_1,m_2)} \\
&:=\int\limits_{x(0)=m_1}^{x(T)=m_2}  \mathcal{D}[x]\delta\big(x(\tau)-x_0\big)\prod_{t=0}^{T}\Theta[x(t)]\,e^{-\frac{S_0}{D}}O[x(t)] .\nn
\end{align}
Here, both $m_1\ge 0$ and $m_2\ge 0$, and the average is over trajectories sketched in \fref{fig:lastMax} with the standard Brownian measure.
Note that this definition is different from the one in  \Eq{eq:angular bracket last}, due to the different boundary conditions employed there. 
We will now in turn study averages at different orders, expressed in terms of the Brownian propagator \Eq{eq:propagator free absorbing} in presence of an absorbing wall. This is similar to   the analysis of $t_{\rm last}$ in the previous \sref{sec:last}.

\subsection{Zeroth order}
Similar to \Eq{eq:W0 last factorization}, we write the double Laplace transformation of \Eq{eq:w0 max}   as
\begin{equation*}
\widetilde{W}_0(\lambda,s)= \int_{0}^{\infty}dm_1 \int_{0}^{\infty}dm_2 \, \widetilde{Z}^{+}_{s+\lambda}(m_1,x_0)\,\widetilde{Z}^{+}_s(x_0,m_2).
\end{equation*}
Using \Eq{eq:L of Z Z+} and integrating, it is easy to see that for small $x_0$,
\begin{equation}
\widetilde{W}_0(s\,\kappa,s)\simeq  \frac{x_0^2}{D  s}\times \frac{1}{\sqrt{1+\kappa}}.\label{eq:w0 tilde max result}
\end{equation}
The leading non-vanishing term is of order $x_0^2$, and its amplitude is same as in \Eq{eq:w0 last result}. This gives the well-known arcsine-law \eqref{Levey-law} for $t_{\rm max}$. 

\subsection{Linear order: 1-loop diagrams}
Similar to \Eq{eq:W1 last A123} we write $W_1$ in \Eq{eq:w1 max} in three parts according to the order of $(r_1,r_2,\tau)$. Their diagrammatic representation is similar to the 1-loop diagrams in \fref{fig:diagramELastLinear}, but their amplitude is different. They are given by
\begin{subequations}
\begin{align}
A_1(\tau,T)=\frac{1}{2D}&\int_{0}^{\infty} dm_1 dm_2 \int_{0}^{\Lambda} dy \nn\\[1mm]
\times &J_{\tau}^{+}(m_1,x_0;-y,y)Z_{T-\tau}^{+}(x_0,m_2), \\
A_2(\tau,T)=\frac{1}{2D}&\int_{0}^{\infty} dm_1 dm_2 \int_{0}^{\Lambda} dy \nn\\[1mm]
 \times &Z_{\tau}^{+}(m_1,x_0)J_{T-\tau}^{+}(x_0,m_2;-y,y), \\
A_3(\tau,T)=\frac{1}{2D}&\int_{0}^{\infty} dm_1 dm_2 \int_{0}^{\Lambda} dy \nn \\[1mm]
\times J_{\tau}^{+}(m_1&,x_0;-y)J_{T{-}\tau}^{+}(x_0,m_2;y)\; e^{-y\tau}.~~~
\end{align}
The function $J_t^+$ is the counterpart of \Eq{eq:J2} in presence of an absorbing wall at the origin.
\end{subequations}
Their double Laplace transform \Eq{eq:D transformation} gives
\begin{subequations}
\begin{align}
\widetilde{A}_{1}(\lambda,s)=\frac{1}{2D}&\int_{0}^{\infty} dm_1 dm_2 \int_{0}^{\Lambda} dy \nn\\[1mm]
\times & \widetilde{J}_{s+\lambda}^{+}(m_1,x_0;-y,y)\widetilde{Z}_{s}^{+}(x_0,m_2), \label{eq:w11 Max}\\
\widetilde{A}_{2}(\lambda,s)=\frac{1}{2D}&\int_{0}^{\infty} dm_1 dm_2 \int_{0}^{\Lambda} dy \nn\\[1mm]
\times & \widetilde{Z}_{s+\lambda}^{+}(m_1,x_0)\widetilde{J}_{s}^{+}(x_0,m_2;-y,y), \label{eq:w12 Max}\\
\widetilde{A}_{3}(\lambda,s)=\frac{1}{2D}&\int_{0}^{\infty} dm_1 dm_2 \int_{0}^{\Lambda} dy \nn\\[1mm]
\times &\widetilde{J}_{s+\lambda+y}^{+}(m_1,x_0;-y)\widetilde{J}_{s}^{+}(x_0,m_2;y).\label{eq:w13 Max}
\end{align}
\end{subequations}
These integrals can be evaluated explicitly using the results in Apps.~\ref{app:Identities} and \ref{app:B}, specifically Eqs.~\eqref{eq:J2 + tilde integral final full}, \eqref{eq:id Z plus int},  and their symmetry properties for evaluating  Eqs.~(\ref{eq:w11 Max}), (\ref{eq:w12 Max}), as well as  Eqs.~(\ref{eq:J1IntL}) and (\ref{eq:J1IntR}) for evaluating \Eq{eq:w13 Max}. For small $x_0$, we get
\begin{align}
\widetilde{A}_{1}(s\,\kappa,s) & \simeq \frac{x_0^2}{Ds}\times\frac{\widetilde{\mathcal{A}}(1+\kappa)}{\sqrt{1+\kappa}},\\
\widetilde{A}_{2}(s\,\kappa,s) & \simeq \frac{x_0^2}{Ds}\times\frac{\widetilde{\mathcal{A}}(1)}{\sqrt{1+\kappa}},\\
\widetilde{A}_{3}(s\,\kappa,s) & \simeq \frac{x_0^2}{Ds}\times\frac{\widetilde{{\mathcal{A}_3}}(1+\kappa)}{\sqrt{1+\kappa}},
\end{align}
with $\widetilde{\mathcal{A}}(z)$ defined in \Eq{eq:chi2 last} and
\begin{equation}
\widetilde{{\mathcal{A}_3}}(z)=-2\int_0^{\Lambda/s}\frac{d y}{y^2}\Big(\sqrt{z+y}-\sqrt{z}\Big)\Big(\sqrt{1+y}-1\Big).
\end{equation}
Summing all three contributions we get the double Laplace transform  \Eq{eq:D transformation} of the linear-order term $W_1(\tau,T)$ in \Eq{eq:w1 max}. It reads, for small $x_0$, 
\begin{equation}
\widetilde{W}_{1}(s\,\kappa,s)\simeq \frac{x_0^2}{s D}\times\frac{\widetilde{\mathcal{A}}(1+\kappa)+\widetilde{\mathcal{A}}(1)+\widetilde{{\mathcal{A}_3}}(1+\kappa)}{\sqrt{1+\kappa}}.
\label{eq:W1 tilde max result}
\end{equation}
 We note the simplification
\begin{align}
 \widetilde{\mathcal{A}}&(z)+\widetilde{\mathcal{A}}(1)+\widetilde{{\mathcal{A}_3}}(z)\cr & =\int_0^{\Lambda/s}\!\!\frac{d y }{ y ^2}\left[\left(\sqrt{z+ y }-\sqrt{z}\right)-\left(\sqrt{1+ y }-1\right)\right]^2.\qquad 
\label{eq:chi2 Max}
\end{align}

\subsection{Quadratic order}
Similar to \Eq{eq:last quadratic splitting W2}, we find that the second-order term $W_2$ in \Eq{eq:w2 max} is composed of the 2-loop diagrams in \fref{fig:diagram}.
The amplitudes of these diagrams are   different for this problem. Here we summarize their result for small $x_0$. Their derivation is given in \aref{app:two loop diagram for max}.

The list below contains  the double Laplace transform  of all 2-loop diagrams. All amplitudes are of order $x_0^2$ for small $x_0$. Note  that    many  diagrams  are the same as in the problem of $t_{\rm last}$ in \sref{sec:last}; this may  not be surprising as   the same power-law corrections for $\vartheta\to 0$ and $\vartheta\to 1$ are also  present in the distribution of  $t_{\rm last}$. 

The list of already calculated diagrams  reads ($x_0\ll1$): 
\begin{align}
\widetilde{E}_{1}(s\,\kappa,s)+\widetilde{E}_{2}(s\,\kappa,s)& \simeq \frac{x_0^2}{sD}\times\frac{\widetilde{\mathcal{E}}(1+\kappa)+\widetilde{\mathcal{E}}(1)}{\sqrt{1+\kappa}} \label{eq:E tilde final max sum},
\end{align}
with $\widetilde{\mathcal{E}}$ given in \Eq{eq:var E}.
\begin{equation}
\widetilde{A}(s\,\kappa,s)\simeq \frac{x_0^2}{sD}\times\frac{\widetilde{\mathcal{A}}(1+\kappa)\widetilde{\mathcal{A}}(1)}{\sqrt{1+\kappa}}\label{eq:A tilde final max},
\end{equation}
with $\widetilde{\mathcal{A}}(z)$ given in \Eq{eq:chi2 last}.
\begin{equation}
\widetilde{G}_1(s\, \kappa,s)+\widetilde{G}_2(s\, \kappa,s)\simeq \frac{x_0^2}{sD}\times\frac{\widetilde{\mathcal{G}}(1+\kappa)+\widetilde{\mathcal{G}}(1)}{\sqrt{1+\kappa}}
\label{eq:F tilde final max},
\end{equation}
with $\widetilde{\mathcal{G}}(z)$ given in \Eq{eq:var F}.

The amplitudes of the remaining diagrams are different. We get, for small $x_0$,
\begin{equation}
\widetilde{D}(s\,\kappa,s)=\frac{x_0^2}{sD}\times\frac{\widetilde{\mathcal{D}}(1+\kappa)}{\sqrt{1+\kappa}}\label{eq:diagram D max},
\end{equation}
with
\begin{align}
\widetilde{\mathcal{D}}(z)&= 2\int_{0}^{\Lambda/s}\frac{d y _1d y _2}{ y _1^2  y _2^2}\sqrt{z+y _1+y _2}\,\sqrt{1+y _1+y _2}\cr 
& \quad \times\left(\sqrt{1+y _1+y _2}-\sqrt{1+y _1}-\sqrt{1+y _2}+1\right)\cr
& \quad\times \bigg(\sqrt{z+y _1+y _2}-\sqrt{z+y _1}-\sqrt{z+y _2}+\sqrt{z}\bigg).
\label{eq:var D max}
\end{align}
The difference to \Eq{eq:var D last} is in the first term inside the integrals and the overall sign.

The leading non-vanishing amplitudes of diagrams $B$ and $C$ are of order $x_0^2$, and unlike in \sref{sec:last}, these diagrams are relevant here. Their Laplace transform, for small $x_0$ are
\begin{equation}
\widetilde{B}_1(s\,\kappa,s)+\widetilde{B}_2(s\,\kappa,s)\simeq \frac{x_0^2}{s}\times\frac{\widetilde{\mathcal{B}}(1+\kappa)}{D\sqrt{1+\kappa}},\label{eq:B tilde sub final expression rescaled}
\end{equation}
where
\bea
&&\widetilde{\mathcal{B}}(z)=2\int\limits_0^{\Lambda/s}\!\! \frac{dy _1dy _2 }{y _1^2 y _2^2(y _1-y _2)}\;\bigg[y _2^2(\sqrt{z{+}y _1}{-}\sqrt{z})(\sqrt{1{+}y _1}{-}1)\cr 
&&\qquad\qquad\qquad -y _1^2(\sqrt{z+y _2}-\sqrt{z})(\sqrt{1+y _2}-1)\bigg]
\eea
and
\begin{equation}
\widetilde{C}_1(s\,\kappa,s)+\widetilde{C}_2(s\,\kappa,s)\simeq \frac{x_0^2}{s}\times \frac{ \left(\widetilde{\mathcal{C}}(1,\,1+\kappa)+\widetilde{\mathcal{C}}(1+\kappa,\,1) \right)}{D\sqrt{1+\kappa}},\label{eq:C tilde final expression scaled}
\end{equation}
where
\begin{align}
 \widetilde{\mathcal{C}}&(z_1,z_2)=2\int_{0}^{\Lambda/s}\frac{dy _1dy _2}{\;y _1^2\;y _2^2}\left(\sqrt{z_1}-\sqrt{z_1+y _1}\right)\sqrt{z_2+y _1}\nn\\[1mm]
 &~~~~~\times \left(\sqrt{z_2+y _1+y _2}-\sqrt{z_2+y _1}-\sqrt{z_2+y _2}+\sqrt{z_2}\right)^2.
\end{align}
From the amplitude of all 2-loop diagrams in \Eq{eq:last quadratic splitting W2} we get the double Laplace transform of $W_2(\tau,T)$ in \Eq{eq:w2 max},  for small $x_0$, 
\begin{align}
&\widetilde{W}_2 (s\,\kappa,s) \simeq  \frac{ x_0^2
 }{s\,D\sqrt{1+\kappa}}\bigg[\widetilde{\mathcal{E}}(1+\kappa)+\widetilde{\mathcal{E}}(1)+\nn\\[1mm]
 &\widetilde{\mathcal{A}}(1+\kappa)\widetilde{\mathcal{A}}(1)+\widetilde{\mathcal{D}}(1+\kappa)
 +\widetilde{\mathcal{C}}(1,\,1+\kappa)+\widetilde{\mathcal{C}}(1+\kappa,\,1)\nn\\[1mm]
 &\qquad \qquad +\widetilde{\mathcal{B}}(1+\kappa)+\widetilde{\mathcal{G}}(1+\kappa)+\widetilde{\mathcal{G}}(1)\bigg].
 \label{eq:w2 tilde max result}
\end{align}

\subsection{Result for $\mathcal{F}^{\rm max}(\kappa,H)$}
\begin{figure}
\includegraphics[width=\columnwidth]{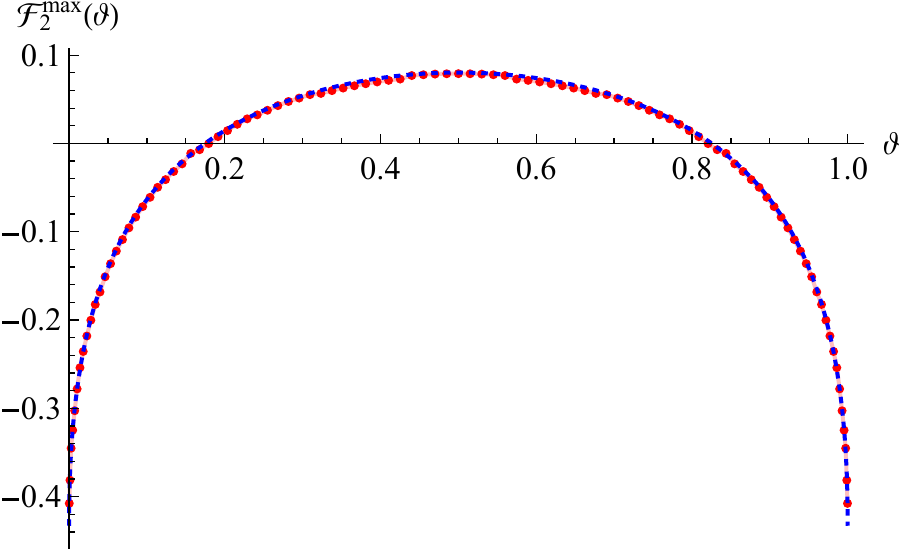}
\caption{The dotted points (colored red) show $\mathcal{F}_{2}^{\rm max}(\vartheta)$ evaluated by numerical integration from \Eq{eq:F2 max final before transformation}. The solid line is the polynomial in  \Eq{eq:F max numerical approximation}, which gives a good estimation for $\mathcal{F}_{2}^{\rm max}(\vartheta)$. \label{fig:MaxF2 numerical analytical comparison}}
\end{figure}
Taking the results in Eqs.~\eqref{eq:w0 tilde max result}, \eqref{eq:W1 tilde max result}, \eqref{eq:w2 tilde max result}, and the expansion \eqref{eq:D exp} we write in an exponential form analogous to \Eq{eq:w last scaled form}, where, for this problem,
\begin{align*}
\widetilde{\mathcal{W}}_1(\kappa)=&\widetilde{\mathcal{A}}(1+\kappa)+\widetilde{\mathcal{A}}(1)+\widetilde{{\mathcal{A}_3}}(1+\kappa)+ 2(\log\Lambda+\gamma_{\rm E}-1),\\[1mm]
\widetilde{\mathcal{W}}_2(\kappa)=&\Big[\widetilde{\mathcal{E}}(1+\kappa)+\widetilde{\mathcal{E}}(1)+\widetilde{\mathcal{A}}(1+\kappa)\widetilde{\mathcal{A}}(1)+\widetilde{\mathcal{D}}(1+\kappa)\cr
 &
 +\widetilde{\mathcal{C}}(1,\,1+\kappa)+\widetilde{\mathcal{C}}(1+\kappa,\,1) +\widetilde{\mathcal{B}}(1+\kappa)\cr
 &+\widetilde{\mathcal{G}}(1+\kappa)+\widetilde{\mathcal{G}}(1)\Big] -\frac{1}{2}\Big[ \widetilde{\mathcal{A}}(1+\kappa)+\widetilde{\mathcal{A}}(1)\cr
 &+\widetilde{{\mathcal{A}_3}}(1+\kappa) \Big]^2+ 2\left(1-\frac{\pi^2}{6}\right).
\end{align*}
The rest of the analysis is very similar to that in \sref{sec:F last}. To leading order we get
\begin{align}\label{eq:F1 max kappa before integral}
&\widetilde{\mathcal{F}}_1^{\rm max}(\kappa)=\widetilde{\mathcal{W}}_1(\kappa)-\widetilde{\mathcal{W}}_1(0)\\
&=\int_0^{\Lambda/s}\frac{dy }{y ^2}\left(\sqrt{1+\kappa+y }-\sqrt{1+\kappa}-\sqrt{1+y }+1\right)^{\!2}.\qquad
\nn
\end{align}
Explicitly carrying out the integral in the $\Lambda\to \infty$ limit yields
\begin{align}
\widetilde{\mathcal{F}}_1^{\rm max}&(\kappa)=-8\log 2-\left(1+\sqrt{1+\kappa} \right)\log(1+\kappa)\cr & +\frac{2}{\sqrt{1+\kappa}}\left(1+\sqrt{1+\kappa}\right)^2\log\left(1+\sqrt{1+\kappa}\right).~~~~~~~~
\label{eq:F1 max kappa after integral}
\end{align}
Its inverse transform \Eq{eq:Kay inverse transformation} is (see \Eq{eq:inverse KT id eight})
\begin{eqnarray}
\mathcal{K}_{\kappa \rightarrow \vartheta}^{-1}\odot\widetilde{\mathcal{F}}_1^{\rm max}(\kappa)
&=&-8\log 2+\psi \left( \sqrt{\frac{\vartheta}{1-\vartheta}}\right)\nn\\
&&-\log(\vartheta(1-\vartheta)), 
\end{eqnarray}
with $\psi(x)$ defined in \Eq{eq:psi}. Then \Eq{eq:main inverse KT 1} with $\mathcal{R}^{\rm max}(\vartheta)=\log(\vartheta(1-\vartheta))$ gives the leading-order term
\begin{equation}
\mathcal{F}^{\rm max}_1(\vartheta)=-8\log2 +\psi\left( \sqrt{\frac{\vartheta}{1-\vartheta}}\right).
\label{eq:F1 max theta final result}
\end{equation} 
The expression in \Eq{eq:F1 max theta final result} differs from \Eq{eq:F1 max final result} by a constant, which comes from our convention that for the latter the integral over $\vartheta$ vanishes.

At second order, we get
\begin{align}
\widetilde{\mathcal{F}}_2^{\rm max}(\kappa)&=\widetilde{\mathcal{W}}_2(\kappa)-\widetilde{\mathcal{W}}_2(0)\nn\\[1mm]
=\bigg[\widetilde{\mathcal{D}}(1&+\kappa)-\frac{1}{2}\widetilde{{\mathcal{A}_3}}(1+\kappa)^2-\left(\widetilde{\mathcal{D}}(1)-\frac{1}{2}\widetilde{{\mathcal{A}_3}}(1)^2\right)\bigg]\nn \\[1mm]
+\bigg[\widetilde{\mathcal{B}}&(1+\kappa)+\widetilde{\mathcal{C}}(1,\,1+\kappa)+\widetilde{\mathcal{C}}(1+\kappa,\,1)\cr
&\qquad \quad -\widetilde{{\mathcal{A}_3}}(1+\kappa)\left(\widetilde{\mathcal{A}}(1+\kappa)+\widetilde{\mathcal{A}}(1) \right)\cr 
&-\left(\widetilde{\mathcal{B}}(1) +2\widetilde{\mathcal{C}}(1,\,1)-2\widetilde{{\mathcal{A}_3}}(1)\widetilde{\mathcal{A}}(1) \right)\bigg]\cr
+\bigg[\widetilde{\mathcal{E}}&(1+\kappa)+\widetilde{\mathcal{G}}(1+\kappa)-\frac{1}{2} \widetilde{\mathcal{A}}^2(1+\kappa)\cr
&\qquad \qquad -\left(\widetilde{\mathcal{E}}(1)+\widetilde{\mathcal{G}}(1)-\frac{1}{2}\widetilde{\mathcal{A}}^2(1) \right)\bigg]. \label{eq:F2 tilde max final}
\end{align}
The terms are written such that each square bracket remains finite for $\Lambda\to \infty$ limit. In fact, we see that the expression in the last square bracket is same as in \Eq{eq:F12 tilde last b} and it vanishes for $\Lambda \to \infty$. Rest two square brackets give $\widetilde{\mathcal{F}}_2^{\rm max}(\kappa)$ for the $\Lambda\to \infty$.

\begin{remark}
We see that for $\kappa=0$, both $\widetilde{\mathcal{F}}_1^{\rm max}(\kappa)$ and $\widetilde{\mathcal{F}}_2^{\rm max}(\kappa)$ vanish, which is consistent with the condition \eqref{eq:F tilde zero}.
\end{remark}

From \Eq{eq:main inverse KT 2} and using linearity of the transformation $\mathcal{K}_{\kappa \rightarrow \vartheta}^{-1}$ we write
\begin{align}
\mathcal{F}_{2}^{\rm max}&(\vartheta)=\mathcal{K}_{\kappa \rightarrow \vartheta}^{-1}\odot \widetilde{\mathcal{F}}_{2}^{\rm max}(\kappa)+ \\[1mm]
\frac{1}{2}\Big\{&\mathcal{K}_{\kappa \rightarrow \vartheta}^{-1}\odot\widetilde{\mathcal{F}}_{1}^{\rm max}(\kappa)^2-\left[\mathcal{F}_{1}^{\rm max}(\vartheta)-\mathcal{R}^{\rm max}(\vartheta)\right]^2\Big\}.\nn
\end{align}
Using an identity \Eq{eq:inverse KT id nine} we see that
\begin{align}
\mathcal{K}_{\kappa \rightarrow \vartheta}^{-1}&\odot\widetilde{\mathcal{F}}_{1}^{\rm max}(\kappa)^2=\cr
&\Big[\mathcal{F}_{1}^{\rm max}(\vartheta)-\mathcal{R}^{\rm max}(\vartheta)\Big]^2
-\psi_2\left( \sqrt{\frac{\vartheta}{1-\vartheta}}\right)^{\!2},\qquad
\end{align}
where we define
\begin{align}
\psi_2(x)= & 2 \arctan x+ x\log\left(1+\frac{1}{x^2}\right)\cr
&-\left[2 \arctan \frac{1}{x}+ \frac{1}{x}\log\left(1+x^2\right)\right].
\label{eq:psi2}
\end{align}
This leads to our result
\begin{align}
\mathcal{F}_{2}^{\rm max}&(\vartheta)=\mathcal{K}_{\kappa \rightarrow \vartheta}^{-1}\odot \widetilde{\mathcal{F}}_{2}^{\rm max}(\kappa)-\frac{1}{2}\psi_2\left( \sqrt{\frac{\vartheta}{1-\vartheta}}\right)^2.\label{eq:F2 max final before transformation}
\end{align}
(We note that the last term is symmetric in $\vartheta \to 1-\vartheta$.)

It is hard to analytically evaluate the integrals in \Eq{eq:F2 tilde max final}. Similar to \Eq{eq:F2 last final}, we determine $\mathcal{F}_{2}^{\rm max}(\vartheta)$ by exchanging the order of $\mathcal{K}$-transformation and integration. This gives, up to an additive constant, 
\begin{align}
{\cal F}^{\rm max}_2(\vartheta) & = -\frac{1}{2}\psi_2\left( \sqrt{\frac{\vartheta}{1-\vartheta}}\right)^2\nn \\[1mm]
 + & 2\int_{0}^{\infty}\frac{dy_1dy_2}{y_1^2y_2^2} \; \Psi^{\rm max}\left(y_1,y_2,\frac{1-\vartheta}{\vartheta}\right),\label{eq:F2 max exact integral}
\end{align}
where $\Psi^{\rm max}(y_1,y_2,z)$ has a lengthy expression given in the Appendix \ref{app:R max}. The expression is also given in the supplemental Mathematica notebook \cite{Mathematica} for numerical evaluation.

Our result for ${\cal F}^{\rm max}_2(\vartheta)$ in \Eq{eq:F2 max final before transformation} is plotted in \fref{fig:MaxF2 numerical analytical comparison}, which agrees well with our computer simulation result in \fref{f:F2s}. For this, we evaluated both the ${\ca K}^{-1}$-transformation and the $y$-integration numerically.

\section{Distribution of  time $t_{\rm pos}$ where the process is positive}
\label{sec:positive}
\begin{figure}
\begin{center}
\includegraphics[width=\columnwidth]{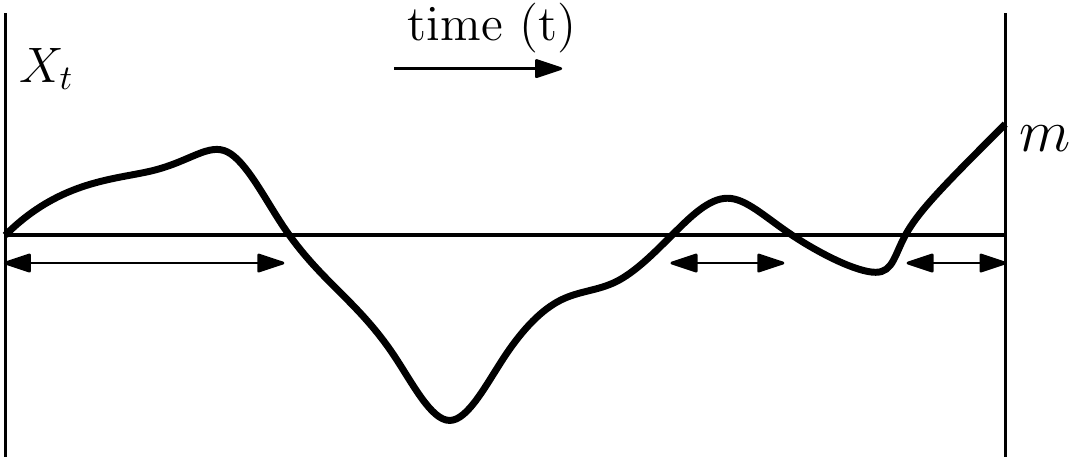}
\end{center}
\caption{Schematic of an fBm trajectory leading to positive time $t_{\rm pos}$. Times spent on the positive side is indicated by double-sided arrow. \label{fig:positive}}
\end{figure}

This analysis is more involved compared to the analysis for $t_{\rm last}$ and $t_{\rm max}$. The main reason is that the expressions at second order are very cumbersome, and a lot of ingeniosity is needed to reduce them to a manageable size. 

Analogous to \Eq{eq:P max first formal}, the probability that an fBm, starting at $X_0=0$ and evolving until time $T$, spends time $t_{\rm pos}=\tau$ being positive ($X_t>0$), can be expressed as 
\begin{equation}
P_T(t_{\rm pos}=\tau)=\frac{W(\tau,T)}{N(T)}, \label{eq:p pos first}
\end{equation}
where $W(\tau,T)$ is the weight of all fBm trajectories contributing to the event and $N(T)$ its normalization. Formally,
\begin{equation}
W(\tau,T)=\int\limits_{-\infty}^{\infty}  dm\int\limits_{x(0)=0}^{x(T)=m}  \!\!  \mathcal{D}[x]\, \delta \bigg(\!\tau{-}\!\int\limits_0^T   dt\,\Theta\big(x(t)\big)\!\bigg)e^{-S[x]}, \label{eq:W pos very first}
\end{equation}
where $\Theta(x)$ is the Heaviside step function.
A sketch of such a trajectory is given in \fref{fig:positive}. 
We follow the same notations as in sections \ref{sec:last} and \ref{sec:max}. The definition of the  quantities $W$, $N$, etc.,  is modified to measure the positive time. 

Using the perturbation expansion of the fBm action in \Eq{eq:expansion of S} we write \eqref{eq:W last expansion}, with
\begin{subequations}
\begin{align}
W_0(\tau,T)& = \int_{-\infty}^{\infty}dm\,\llangle 1 \rrangle_{(0,m)}, \label{eq:w0 +}\\[1mm]
W_1(\tau,T)& = \int_{-\infty}^{\infty}dm \,\llangle[\Big]\frac{S_1}{2D}\rrangle[\Big]_{(0,m)}, \label{eq:w1 +}\\[1mm]
W_2(\tau,T)& = \int_{-\infty}^{\infty}dm \, \llangle[\bigg]\frac{S_1^2}{8D^2}-\frac{S_2}{D}\rrangle[\bigg]_{(0,m)}, \label{eq:w2 +}
\end{align}
where   the double-angular brackets   denote 
\label{eq:w 012 +}
\end{subequations}
\bea\label{eq:average +}
&&\llangle O[x]\rrangle_{(m_1,m_2)}  \\
&&\quad:= \int\limits_{x(0)=m_1}^{x(T)=m_2}  \!\!  \mathcal{D}[x]\, \delta \bigg(\!\tau{-}\!\int\limits_0^T   dt\,\Theta\big(x(t)\big)\!\bigg)e^{-\frac{S_0}D} O[x(t)]. \nn
\eea
This is an average over trajectories with   Brownian measure.

\subsection{Conditional propagator \label{sec:conditional propagator}}
In \sref{sec:last} and \sref{sec:max}, the amplitudes in the expansion \eqref{eq:W last expansion} are expressed in terms of the free Brownian propagator $Z$ in \Eq{eq:W0 Z} and its analogue $Z^+$ in presence of an absorbing wall. For amplitudes \eqref{eq:w 012 +}, it is natural to express in terms of a conditional Brownian propagator, defined by
\begin{equation}
\mathbb{Z}_T( m_1,m_2 \vert \tau)=
\int\limits_{x(0)=m_1}^{x(T)=m_2}  \!\!  \mathcal{D}[x]\, \delta \bigg(\!\tau{-}\!\int\limits_0^T   dt\,\Theta\big(x(t)\big)\!\bigg)e^{-\frac{S_0}D} .
\label{eq:conditional propagator definition}
\end{equation}
This gives the weight  of all Brownian paths starting at $m_1$ and ending at $m_2$ at time $T$ conditioned to spending time $\tau$ on the positive half.

\begin{figure}
\includegraphics[width=\columnwidth]{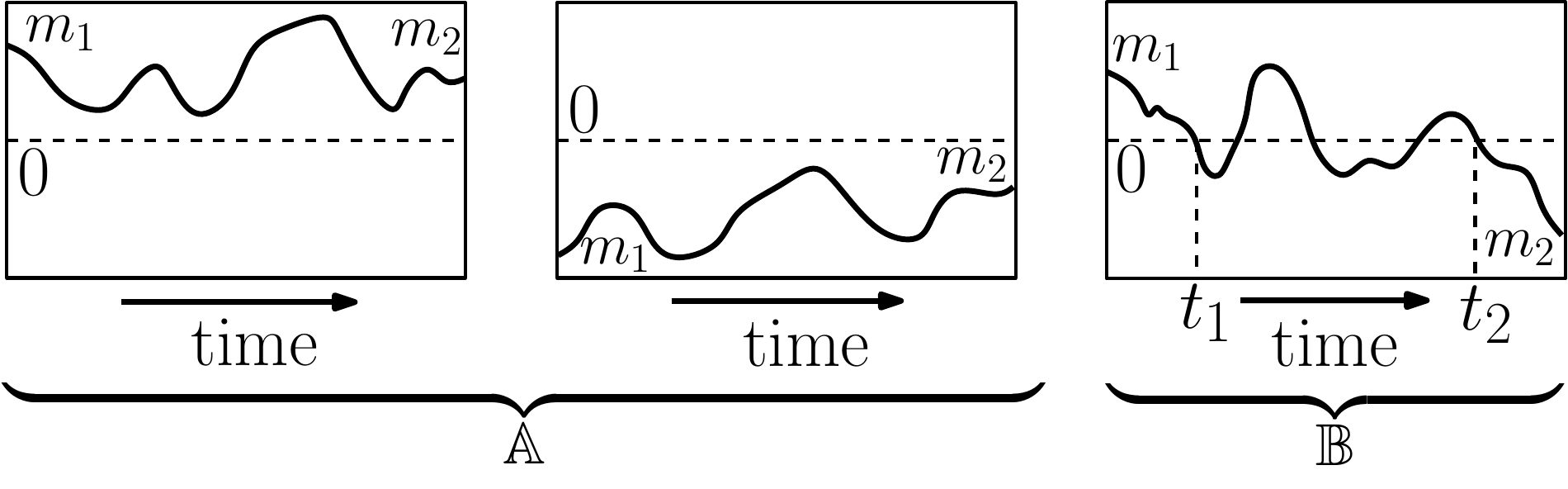}
\caption{Different Brownian paths for the conditional propagator in \Eq{eq:Za+Zb}. ($\mathbb{A}$) Includes paths which  have never crossed the origin, ($\mathbb{B}$) includes paths which have crossed the origin at least once.}
\label{fig:bridge}
\end{figure}

To find an explicit expression for the conditional propagator, we write the associated paths into two groups,
\begin{equation}
\mathbb{Z}_T(  m_1,m_2\vert \tau)=\mathbb{A}_T(  m_1,m_2\vert \tau)+\mathbb{B}_T(  m_1,m_2\vert \tau),
\label{eq:Za+Zb}
\end{equation}
 shown in the Fig.~\ref{fig:bridge}. The term $\mathbb{A}$ is non-zero only for $\tau=0$ or $T$. Using \Eq{eq:propagator free absorbing}, we write
\begin{align*}
\mathbb{A}_T(  m_1,m_2\vert \tau)=& \Theta(m_1)\Theta(m_2)\delta(\tau-T)Z_T^{+}(m_1,m_2)\\[1mm]
+ \Theta(&-m_1)\Theta(-m_2)\delta(\tau)Z_T^{+}(-m_1,-m_2).
\end{align*}
Its double Laplace transform can be written with  the help of  identity \eqref{eq:id L-transform} as
\bea
\widetilde{\mathbb{A}}_s( m_1,m_2\vert \lambda) &= & \Theta(m_1)\Theta(m_2)\widetilde{Z}_{s+\lambda}^{+}(m_1,m_2)  \\[1mm]
 & + & \Theta(-m_1)\Theta(-m_2)\widetilde{Z}_s^{+}(-m_1,-m_2),\nn
\eea
where   expression \eqref{eq:L of Z Z+} leads to
\begin{align}
\widetilde{\mathbb{A}}_s &( m_1,m_2\vert \lambda)=\frac{\Theta(m_1m_2)}{2\sqrt{D(s+\lambda \Theta(m_1))}}\label{eq:Za final}\\[1mm]
&\times\left[e^{-\vert m_1-m_2\vert\sqrt{\frac{s+\lambda \Theta(m_1)}{D}}}-e^{-\vert m_1+m_2\vert\sqrt{\frac{s+\lambda \Theta(m_1)}{D}}}\right].\qquad \nn
\end{align}
The second part of \Eq{eq:Za+Zb} is defined by (see \fref{fig:bridge})
\begin{align}
\mathbb{B}_T(m_1,m_2\vert \tau )=& \int\limits_{0}^{T}  dt_1\int\limits_{t_1}^{T}  dt_2 \llangle \delta\left(x(t_1)\right)\delta\left(x(t_2)\right) \rrangle_{(m_1,m_2)} 
\end{align}
with $\tau$ 
specified in the average \eqref{eq:average +}. One can estimate, for example, for $m_1>0$ and $m_2>0$, 
\begin{align}
\llangle \delta\left(x(t_1)\right)\delta\left(x(t_2)\right) \rrangle_{(m_1,m_2)}=&\ca N\; D^2 \; \left[\lim_{x_0\to 0}\frac{Z^{+}_{t_1}(m_1,x_0)}{x_0}\right]\nn \\[1mm]
 \times \mathscr{G}(\tau-t_1-t_2,T-t_1-t_2)&\times \left[\lim_{x_0\to 0}\frac{Z^{+}_{t_2}(x_0,m_2)}{x_0}\right],~~~
\end{align}
(here $D^2$ is from dimensional argument) up to a normalization $\ca N$, where $\mathscr{G}(\tau,T)$ is the   weight of Brownian paths  starting at the origin and returning there at time $T$, spending   time $\tau$ in the positive half.  

In general, using   identity \eqref{eq:id L-transform}, we write the double Laplace transform of $\mathbb{B}$ as
\begin{align*}
& \widetilde{\mathbb{B}}_s( m_1,m_2\vert \lambda) = \mathcal{N}\;D^2\;\widetilde{\mathscr{G}}(\lambda,s) \nn\\[1mm] & \times\left[\lim_{x_0\rightarrow 0}\frac{\Theta(m_1)\widetilde{Z}_{s+\lambda}^{+}(m_1,x_0)+\Theta(-m_1)\widetilde{Z}_{s}^{+}(-m_1,x_0)}{x_0}\right]\cr
&\times\left[\lim_{x_0\rightarrow 0}\frac{\Theta(m_2)\widetilde{Z}_{s+\lambda}^{+}(x_0,m_2)+\Theta(-m_2)\widetilde{Z}_{s}^{+}(x_0,-m_2)}{x_0}\right],
\end{align*}
The normalization $\mathcal{N}$ to be determined self-consistently, and $\widetilde{\mathscr{G}}(\lambda,s)$ is the double Laplace transform of $\mathscr{G}(\tau,T)$. 

We see that
\begin{equation*}
\mathscr{G}(\tau,T)= Z_T(0,0)P_{\rm bridge}(\tau,T),
\end{equation*} 
where $P_{\rm bridge}(\tau,T)$ is the probability of positive time $t_{\rm pos}=\tau$ for a Brownian bridge of duration $T$. One can show (a derivation is given in \aref{app:Brownian Bridge}) that for a Brownian bridge, all values of $\tau$ are equally probable, and therefore $P_{\rm bridge}(\tau,T)=1/T$. This, along with \Eq{eq:W0 Z}, gives
\begin{align*}
\widetilde{\mathscr{G}}(\lambda,s)=\frac{\sqrt{s+\lambda}-\sqrt{s}}{\lambda\sqrt{D}}.
\end{align*}
Using these results and \Eq{eq:L of Z Z+}, we find
\begin{align}
\widetilde{\mathbb{B}}_s( m_1,m_2\vert \lambda)&= e^{-\vert m_1 \vert \sqrt{\frac{s+\Theta(m_1)\lambda}{D}}}\label{eq:Zb final}\\[1mm] \times &\left[\frac{\sqrt{s+\lambda}-\sqrt{s}}{\lambda\sqrt{D}}\right]
 e^{-\vert m_2 \vert \sqrt{\frac{s+\Theta(m_2)\lambda}{D}}},\nn
\end{align}
where we used $\mathscr{N}=1$, determined using the self-consistency condition that
\begin{equation}
\int_{0}^{T}d\tau\; \mathbb{Z}_T(m_1,m_2\vert \tau )=Z_T(m_1,m_2), \label{eq:norm cond}
\end{equation} 
for \Eq{eq:Za+Zb}, and equivalently,
\begin{equation*}
\widetilde{\mathbb{Z}}_s( m_1,m_2\vert 0)=\widetilde{Z}_{s}(m_1,m_2)=\frac{e^{-\sqrt{\frac{s}{D}}\vert m_1-m_2\vert}}{2\sqrt{s D}},
\end{equation*}
where $\widetilde{\mathbb{Z}}_s( m_1,m_2\vert \lambda)$ is the Double Laplace transformation \Eq{eq:D transformation} of $\mathbb{Z}_T(m_1,m_2\vert \tau )$.
Results \eqref{eq:Za final} and \eqref{eq:Zb final} together give
\begin{equation}
\widetilde{\mathbb{Z}}_s( m_1,m_2\vert \lambda)=\widetilde{\mathbb{A}}_s( m_1,m_2\vert \lambda)+\widetilde{\mathbb{B}}_s( m_1,m_2\vert \lambda).
\label{eq:Z tilde conditional final}
\end{equation} 
This will be used extensively  in the following sections.

\subsection{Zeroth order term}

The leading term \eqref{eq:w0 +} is
\begin{equation*}
W_0(\tau,T)=\int_{-\infty}^{\infty}dm\, \mathbb{Z}_T( 0,m \vert \tau).
\end{equation*}
Its double Laplace transform is
\begin{align*}
\widetilde{W}_0(\lambda,s)& =\int_{-\infty}^{\infty}dm \; \widetilde{\mathbb{Z}}_s(0,m\vert \lambda),
\end{align*}
with $\widetilde{\mathbb{Z}}$ in \Eq{eq:Z tilde conditional final}.

The integration can be evaluated using \Eq{eq:Z tilde conditional final} with  $\widetilde{\mathbb{A}}$ and $\widetilde{\mathbb{B}}$ given in Eqs.~(\ref{eq:Za final}) and (\ref{eq:Zb final}). The result is given in \Eq{eq:Zbb tilde integral Right} using which we write\begin{equation}
\widetilde{W}_0(s\,\kappa,s)=\frac{1}{s}\times \frac{1}{\sqrt{1+\kappa}}.
\label{eq:W0 final positive}
\end{equation}
This is same as for distribution of $t_{\rm last}$ and $t_{\rm max}$, and confirms the arcsine-law \eqref{Levey-law}.

\subsection{Linear order: 1-loop diagram \label{sec:oneloop +}}
\begin{figure}
\includegraphics[scale=1.1]{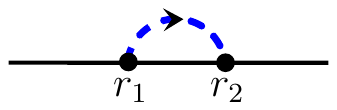}
\caption{A 1-loop diagram representation of the linear order term \eqref{eq:w1 positive first} for distribution of positive time. We follow a similar convention as earlier. A dashed line indicates coupling between points $(r_1,r_2)$ (their order indicated by an arrowhead) with a coupling strength $e^{y(r_1-r_2)}$ and a solid disk indicates a `charge' of amplitude $\dot{x}(r)$ for the associated point $r$. \label{fig:oneloop +}}
\end{figure}

Using \Eq{eq:S1 1 integral} we write the linear order term \eqref{eq:w1 +} as
\begin{align}
W_1(\tau,T)=&\, \frac{1}{2D} \int_{-\infty}^{\infty}  dm \int_0^{\Lambda} dy \int_0^T dr_1 \int_{r_1}^T  dr_2\,\nn\\[2mm]
& \qquad \times  e^{y(r_1-r_2)} \; \llangle \dot{x}(r_1)\dot{x}(r_{2}) \rrangle_{(0,m)},
\label{eq:w1 positive first}
\end{align}
where the integral over time $r$ is interpreted as in \Eq{eq:int dr1dr2 delta}.
A graphical representation of the amplitude as a 1-loop diagram is sketched in \fref{fig:oneloop +}.

To evaluate the conditional average in \Eq{eq:w1 positive first} we use a result for the correlation similar to \Eq{eq:id xx dot}. Generalizing the analysis in \aref{sec:correlation x dot} for the conditioned case, we see that for $r_2>r_1$, 
\begin{align}
&\llangle \dot{x}(r_1)\dot{x}(r_{2}) \rrangle_{(m_1,m_2)} =2^2 D^2\int\limits_0^{r_1}d\tau_1\int\limits_0^{r_2-r_1}  d\tau_2\nn \\
&\times\int\limits_0^{T-r_2}  d\tau_3\delta(\tau-\tau_1-\tau_2-\tau_3)\int\limits_{-\infty}^{\infty}  dx_1\, dx_2\, \mathbb{Z}_{r_1}(m_1,x_1\vert \tau_1)\nn \\[1mm]
& \qquad   \partial_{x_1}\mathbb{Z}_{r_2-r_1}(x_1,x_2\vert \tau_2) \partial_{x_2}\mathbb{Z}_{T-r_2}(x_2,m_2\vert \tau_3).
\label{eq:two-point correlation conditional}
\end{align}
This helps us to write $W_1(\tau,T)$ in terms of the conditional propagator $\mathbb{Z}$. By a change of variables and an integration by parts we obtain
\begin{align}
&W_1(\tau,T)=-2 D\int\limits_0^{\infty}dt_1dt_2dt_3\int\limits_0^{t_1}d\tau_1
\int\limits_0^{t_2}d\tau_2\int\limits_0^{t_3}d\tau_3\cr 
&\times \delta(T{-}t_1{-}t_2{-}t_3)\;\delta(\tau{-}\tau_1{-}\tau_2{-}\tau_3)\int\limits_0^{\Lambda}  dy\, e^{-y \, t_2}\int\limits_{-\infty}^{\infty}  dm\cr
&\times\int\limits_{-\infty}^{\infty}   dx_1 dx_2\;\partial_{x_1}\mathbb{Z}_{t_1}(0, x_1 \vert \tau_1 )\mathbb{Z}_{t_2}(x_1,x_2 \vert \tau_2) \cr
& \qquad \times \partial_{x_2}\mathbb{Z}_{t_3}(x_2,m \vert \tau_3).
\end{align}
A double Laplace transform \Eq{eq:D transformation} of the amplitude gives
\begin{align}
&\widetilde{W}_1(\lambda,s)=-2D\int\limits_0^{\Lambda}dy\int\limits_{-\infty}^{\infty}dx_1dx_2\partial_{x_1}\widetilde{\mathbb{Z}}_s(0,x_1 \vert \lambda)\cr 
& \qquad \times \widetilde{\mathbb{Z}}_{s+y}(x_1,x_2 \vert \lambda)\int\limits_{-\infty}^{\infty}dm\partial_{x_2} \widetilde{\mathbb{Z}}_s(x_2,m \vert \lambda),\label{eq:W1 tilde pos some intermediate expression}
\end{align}
with $\widetilde{\mathbb{Z}}$ defined in \Eq{eq:Z tilde conditional final}.

Using the result \eqref{eq:Z tilde conditional final} and integrating using  integrals (\ref{eq:B49}), (\ref{eq:B48}), we get
\begin{equation}
\widetilde{W}_1(\kappa \, s,s)=\frac{\widetilde{\mathcal{A}}(1+\kappa)}{s\sqrt{1+\kappa}},
\end{equation}
with
\begin{equation}
\widetilde{\mathcal{A}}(z)=\int\limits_0^{\Lambda/s}\frac{dy }{y ^2}\left[\sqrt{z+y }-\sqrt{z}-\sqrt{1+y }+1\right]^2,
\label{eq:A positive def}
\end{equation}
which by mere coincidence happens to be the same  integral as in \Eq{eq:chi2 Max},  although their corresponding diagrams are different.

\subsection{Quadratic order: 2-loop diagrams}

\begin{figure}
\includegraphics[scale=1]{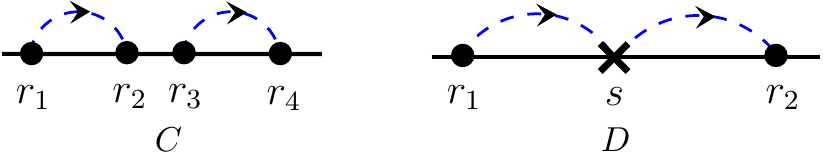}
\caption{Two-loop diagrams for the quadratic order term $W_2$ for the distribution of positive time $t_{\rm pos}$. In this illustration we choose $r_2>r_1$ and $r_4>r_3$ for diagram $C$, whereas $r_1<s<r_2$ for diagram $D$. A solid disc denotes a `charge' $\dot{x}(r)$ for the associated point $r$, whereas a cross denotes a contracted point with a `charge' $2D$. \label{fig:diagramPositive}}
\end{figure}

Following an analysis similar to that in \sref{sec:two loop last}, it is straightforward to see that for  $W_2$ in \Eq{eq:w2 +} contributions come only from the two diagrams shown in \fref{fig:diagramPositive}, 
\begin{equation}
W_2(\tau,T)=C(\tau,T)+D(\tau,T),\label{eq:W2 pos first equation}
\end{equation}
where the amplitudes are given by
\begin{align}
C(\tau,T)=&\frac{1}{8D^2} \int_{-\infty}^{\infty} dm \int_0^{\Lambda} dy_1 dy_2 \int_0^T dr_1 \nn \\[1mm]
&\times \int_{r_1}^T  dr_2  \int_0^T dr_3 \int_{r_3}^T  dr_4 e^{y_1(r_1-r_2)} e^{y_2(r_3-r_4)}\nn \\[1mm]
& \times \llangle[\big] \dot{x}(r_1)\dot{x}(r_{2})\dot{x}(r_3)\dot{x}(r_{4}) \rrangle[\big]_{(0,m)}
\label{eq:C positive formal}
\end{align}
and
\begin{align}
D(\tau,T)=\frac{1}{2D} &\int_{-\infty}^{\infty} dm \int_0^{\Lambda} dy_1 dy_2 \int_0^T dr_1 \int_{r_1}^T  ds  \cr
\times \int_{s}^T dr_2 & e^{y_1(r_1-s)} e^{y_2(s-r_2)} \llangle[\big] \dot{x}(r_1)\dot{x}(r_{2}) \rrangle[\big]_{(0,m)}.
\label{eq:D positive formal}~~
\end{align}
These amplitudes can be expressed in terms of the conditional propagator $\mathbb{Z}_T$ in \Eq{eq:Za+Zb}, and then an explicit result can be derived following an analysis similar to that of the linear-order term in \sref{sec:oneloop +}. Here we give their final expression, and defer their derivation to the \aref{app:two loop positive}.

The double Laplace transform of the amplitude of the diagram $D$ in \fref{fig:diagramPositive} can be written as
\begin{equation}
\widetilde{D}(\kappa \, s,s)=\frac{\widetilde{\mathcal{D}}(1+\kappa)}{s\sqrt{1+\kappa}},\label{eq:D tilde positive final rescaled}
\end{equation}
where 
\begin{align}
\widetilde{\mathcal{D}}(z)=\frac{2}{(1+\sqrt{z})}&\int_0^{\Lambda/s} \frac{dy _1dy _2}{y _1 y _2} \\[1mm]
\times \bigg\{&\frac{y _2\, \mathfrak{h}(1,z,y _1)}{(y _2-y _1)}+\frac{y _1\, \mathfrak{h}(1,z,y _2)}{(y _1-y _2)}\bigg\}\qquad \nn
\end{align}
with
\begin{align}
\mathfrak{h}&(s_1,s_2,y)= \label{eq:h positive}\\[1mm]
&\frac{\left(\sqrt{s_2+y}-\sqrt{s_1+y}\right)\left[\sqrt{s_2(s_1+y)}-\sqrt{s_1(s_2+y)}\right]}{(\sqrt{s_1+y}+\sqrt{s_1})(\sqrt{s_2+y}+\sqrt{s_2})}.\nn
\end{align}

The double Laplace transform for the diagram $C$ in \fref{fig:diagramPositive} is
\begin{equation}
\widetilde{C}(\kappa\, s,s)=\frac{\widetilde{\mathcal{C}}(1+\kappa)}{s\sqrt{1+\kappa}} \label{eq:C tilde positive final rescaled}
\end{equation}
with
\begin{align}
\widetilde{\mathcal{C}}&(z)=\frac{4}{(1+\sqrt{z})} \int_{0}^{\Lambda/s}\frac{dy _1dy _2}{y _1 y _2}\bigg\{\mathfrak{f}(1,z,y _1,y _2)+\nn\\[2mm] &\mathfrak{f}(z,1,y _1,y _2)+\frac{\mathfrak{g}(1,z,y _1,y _2)+\mathfrak{g}(z,1,y _1,y _2)}{y _1}\bigg\},
\end{align}
where we define
\begin{widetext}
\begin{align}
\mathfrak{g}(s_1,s_2,y_1,y_2)=&\sqrt{{s_1}+y_1+y_2}\; \Big(\sqrt{{s_2}+y_1+y_2}-\sqrt{{s_1}+y_1+y_2}\Big)\cr
\times & \Big(\sqrt{{s_1}+y_1}+\sqrt{{s_1}+y_2}-\sqrt{{s_1}}+\sqrt{{s_2}+y_1+y_2}-\sqrt{{s_2}+y_1}-\sqrt{{s_2}+y_2}+\sqrt{{s_2}}\Big)\cr
\times & \left(\frac{\left(\sqrt{{s_1}}+\sqrt{{s_2}}\right) \left(-\sqrt{{s_1}}-\sqrt{{s_2}+y_1}+\sqrt{{s_2}}\right)}{\left(\sqrt{{s_1}+y_1}+\sqrt{{s_2}+y_1}\right) \left(\sqrt{{s_1}+y_1+y_2}+\sqrt{{s_1}+y_1}\right)}+\frac{\sqrt{{s_1}}}{\sqrt{{s_1}+y_1+y_2}+\sqrt{{s_1}}}\right)\cr
+\left({s_1}+y_2\right) &\Big(\sqrt{{s_1}+y_2}-\sqrt{{s_2}+y_2}\Big) \left(\frac{\left(\sqrt{{s_1}}+\sqrt{{s_2}}\right) \left(-\sqrt{{s_1}}-\sqrt{{s_2}+y_1}+\sqrt{{s_2}}\right)}{\left(\sqrt{{s_1}+y_1}+\sqrt{{s_2}+y_1}\right) \left(\sqrt{{s_1}+y_1}+\sqrt{{s_1}+y_2}\right)}+\frac{\sqrt{{s_1}}}{\sqrt{{s_1}+y_2}+\sqrt{{s_1}}}\right)\cr & +\frac{1}{4} (s_2-s_1) \Big(\sqrt{{s_1}+y_1}+\sqrt{{s_1}}-\sqrt{{s_2}+y_1}-\sqrt{{s_2}}\Big)\cr
&+\frac{\sqrt{{s_1}} \Big(\sqrt{{s_1}+y_1}-\sqrt{{s_2}+y_1}\Big) \Big(\sqrt{{s_1}} \left(\sqrt{{s_2}+y_1}-\sqrt{{s_2}}\right)+2 {s_1}+y_1\Big)}{\sqrt{{s_1}+y_1}+\sqrt{{s_1}}}, \label{eq:g positive}
\end{align}
and
\begin{align}
\mathfrak{f}(s_1,s_2,y_1,y_2)=&\frac{({s_1}+y_2)}{y_2} \Big(\sqrt{{s_2}+y_2}-\sqrt{{s_1}+y_2}\Big) \left(\frac{\left(\sqrt{{s_1}}+\sqrt{{s_2}}\right) \left(\sqrt{{s_2}}-\sqrt{{s_1}}-\sqrt{{s_2}+y_1}\right)}{\left(\sqrt{{s_1}+y_1}+\sqrt{{s_2}+y_1}\right) \left(\sqrt{{s_1}+y_1}+\sqrt{{s_1}+y_2}\right)}+\frac{\sqrt{{s_1}}}{\sqrt{{s_1}+y_2}+\sqrt{{s_1}}}\right)\cr
+\frac{\left(\sqrt{{s_2}}-\sqrt{{s_1}}\right)}{4 y_2} & \left(1+\frac{2 \left(\sqrt{{s_1}}+\sqrt{{s_2}}\right) \left(\sqrt{{s_2}}-\sqrt{{s_1}}-\sqrt{{s_2}+y_1}\right)}{\left(\sqrt{{s_1}+y_1}+\sqrt{{s_1}}\right) \left(\sqrt{{s_1}+y_1}+\sqrt{{s_2}+y_1}\right)}\right) \Big(2 \sqrt{{s_1}} \sqrt{{s_2}+y_2}-2 \sqrt{{s_1}} \sqrt{{s_2}}-2 \sqrt{{s_1}} \sqrt{{s_1}+y_2}-y_2\Big)\cr
&\qquad \qquad +\frac{\sqrt{{s_1}} \Big(\sqrt{{s_2}+y_1}-\sqrt{{s_1}+y_1}\Big) \Big(\sqrt{{s_2}}-\sqrt{{s_1}}-\sqrt{{s_2}+y_1}\Big)}{2 \Big(\sqrt{{s_1}+y_1}+\sqrt{{s_1}}\Big)^2}. \label{eq:f positive}
\end{align}
\end{widetext}

Adding contribution of these two diagrams we get the double Laplace transform of the second order term
\begin{equation*}
\widetilde{W}_2(\lambda,s)=\widetilde{C}(\lambda,s)+\widetilde{D}(\lambda,s).
\end{equation*}
The expressions in Eqs.~\eqref{eq:g positive} and \eqref{eq:f positive} are given in the supplemental Mathematica notebook \cite{Mathematica} for their numerical evaluation.

\subsection{Result for $\mathcal{F}^{\rm pos}(\kappa,H)$}
Rest of the analysis is very similar to that for $t_{\rm last}$ and $t_{\rm max}$. We write the amplitude $\widetilde{W}(\lambda,s)$ in \Eq{eq:W pos very first} in an exponential form such that
\begin{equation}
\widetilde{W}(s\, \kappa,s)= \frac{e^{\widetilde{\mathcal{W}}(\kappa)}}{s\sqrt{1+\kappa}},\label{eq:w pos scaled form}
\end{equation}
where $\widetilde{\mathcal{W}}=\epsilon\, \widetilde{\mathcal{W}}_1+\epsilon^2 \, \widetilde{\mathcal{W}}_2+\cdots$, with
\begin{subequations}
\begin{align}
\widetilde{\mathcal{W}}_1(\kappa)=&\;\widetilde{\mathcal{A}}(1+\kappa), \\
\widetilde{\mathcal{W}}_2(\kappa)=&\;  \widetilde{\mathcal{C}}(1+\kappa)+\widetilde{\mathcal{D}}(1+\kappa)-\frac{1}{2}\widetilde{\mathcal{A}}(1+\kappa)^2.
\end{align}
\label{eq:Ws exponentiated pos}
\end{subequations}
Considering the normalization in \Eq{eq:p pos first} we get the Laplace transform of the distribution of $t_{\rm pos}$ in \Eq{eq:last mathcal F} with
\begin{equation}
\widetilde{\mathcal{F}}^{\rm pos}(\kappa,H)=\widetilde{\mathcal{W}}(\kappa)-\widetilde{\mathcal{W}}(0). \label{eq:F pos W pos}
\end{equation}
One can verify that $\widetilde{\mathcal{W}}(0)=0$ up to the second order in the perturbation expansion, and this means in the expansion \Eq{eq:F kappa expansion},
\begin{align}
\widetilde{\mathcal{F}}_1^{\rm pos}(\kappa)=\widetilde{\mathcal{W}}_1(\kappa) \quad \text{and}\quad \widetilde{\mathcal{F}}_2^{\rm pos}(\kappa)=\widetilde{\mathcal{W}}_2(\kappa).
\end{align}

Comparing with \Eq{eq:F1 max kappa before integral} we see that $\widetilde{\mathcal{F}}_1^{\rm pos}(\kappa)$ is exactly same as $\widetilde{\mathcal{F}}_1^{\rm max}(\kappa)$, and therefore we get 
\begin{equation}
\mathcal{F}_1^{\rm pos}(\vartheta)=\mathcal{F}_1^{\rm max}(\vartheta)
\end{equation} 
given in \Eq{eq:F1 max theta final result}.

The difference with the distribution for $t_{\rm max}$ comes in the second order term. This is given by
\begin{align}
\mathcal{F}_{2}^{\rm pos}(\vartheta)=&-\frac{1}{2}\left[\mathcal{F}_{1}^{\rm pos}(\vartheta)-\mathcal{R}^{\rm pos}(\vartheta)\right]^2\cr&+\mathcal{K}_{\kappa \rightarrow \vartheta}^{-1}\odot \left[\widetilde{\mathcal{F}}_{2}^{\rm pos}(\kappa)+\frac{1}{2}\widetilde{\mathcal{F}}_{1}^{\rm pos}(\kappa)^2\right].
\end{align}
Following a similar analysis as used for \Eq{eq:F2 max final before transformation} we get our result
\begin{align}
\mathcal{F}_{2}^{\rm pos}(\vartheta)&=-\frac{1}{2}\psi_2\left( \sqrt{\frac{\vartheta}{1-\vartheta}}\right)^2 + \cr \mathcal{K}_{\kappa \rightarrow \vartheta}^{-1}&\odot  \left[\widetilde{\mathcal{C}}(1+\kappa)+\widetilde{\mathcal{D}}(1+\kappa)-\frac{1}{2}\widetilde{\mathcal{A}}(1+\kappa)^2\right] \quad \label{eq:F2 pos final before transformation}
\end{align}
with \Eq{eq:psi2}. 

It is difficult to analytically do the integration for the amplitudes in the second term in \Eq{eq:F2 pos final before transformation}. We have numerically verified that the term remains finite for $\Lambda\to \infty$. For an explicit formula in terms of $\vartheta$ we exchange the order of $\mathcal{K}_{\kappa \rightarrow \vartheta}^{-1}$-transformation and the integration. This allows us to write 
\begin{align}
{\cal F}^{\rm pos}_2(\vartheta)  = & -\frac{1}{2}\psi_2\left( \sqrt{\frac{\vartheta}{1-\vartheta}}\right)^2 \label{eq:F2 pos final exact}\\[1mm]
 &+  2\int_{0}^{\infty}\frac{dy_1dy_2}{y_1^2y_2^2} \; \Psi^{\rm pos}\left(y_1,y_2,\frac{1-\vartheta}{\vartheta}\right).\nn
\end{align}
Expression for $\Psi^{\rm pos}$ is lengthy and it is given in the Appendix \ref{app:Psi pos}.  Our result for ${\cal F}^{\rm pos}_2(\vartheta)$ is plotted in \fref{f:F2s}, which agrees well with our computer simulation result.  For this we evaluated both the ${\cal K}^{-1}$-transformation and the $y$-integration numerically. 

\section{Summary}
We found a generalization of the three arc-sine laws of Brownian motion for an fBm. Unlike in the Brownian motion, the probabilities are different and given in Eqs.~\eqref{P+}-\eqref{Pmax}. These results are obtained using a perturbation expansion around the Brownian motion, and by a scaling argument for divergences near  $\vartheta \to 0$ and $1$. Our numerical simulations confirm these highly non-trivial predictions accurately. We find a very good convergence to the numerical results for the entire range of $\vartheta$ even for large $\epsilon$. 
Most realizations of fBm found in practical applications fall within the range $H\simeq \tfrac{1}{2} \pm 0.25$ where our formulas yield  high-precision predictions.

Our perturbation approach offers a systematic framework to obtain analytical results for other  observables of an fBm, of which very few are available so far. For example, distribution of Area under a Brownian excursion is known to have an Airy distribution \cite{MajumdarComtet2005}. Corresponding generalization for an fBm is yet unavailable. On simpler examples, a closed form expression for an fBm propagator with absorbing and reflecting boundary is desirable.

\acknowledgements
TS acknowledges support of the Department of Atomic Energy, Government
of India, under Project Identification No. RTI-4002. KJW thanks PSL for support by grant ANR-10-IDEX-0001-02-PSL.
It is a pleasure to thank M.~Delorme for his contributions to the early stages of this work,  and J.U.~Klamser for her help with figures.

\appendix
\section{Perturbation expansion of the fBm action}
\label{sec:expansion app}
Writing $H=\frac{1}{2}+\epsilon$ in the expression for $G^{-1}(t_1,t_2)=\langle X_{t_1}X_{t_2}\rangle $ given in \Eq{eq:covariance} and expanding in powers of small $\epsilon$ we get
\begin{equation*}
G^{-1}(t_1,t_2)=K_0(t_1,t_2)+\epsilon K_1(t_1,t_2)+\epsilon^2 K_2(t_1,t_2)+\cdots,
\end{equation*}
where
\begin{align*}
&K_0(t_1,t_2)\equiv G_0^{-1}(t_1,t_2)=2\min(t_1,t_2),
\end{align*}
and, for $n\ge 1$,
\begin{align}
K_n(t_1,t_2)=\frac{2^n}{n!}\bigg[ t_1 \log^n t_1&+t_2 \log^n t_2\cr
-&\vert t_1-t_2\vert \log^n \vert t_1-t_2\vert \bigg].\qquad
\label{eq:Kn}
\end{align}

For $G$ related by $G^{-1}\, G(t_1,t_2)=G\, G^{-1}(t_1,t_2)=\delta(t_1-t_2)$, this is equivalent\footnote{To see this one can verify that $K_0\cdot G_0(r,s)=G_0\, K_0(r,s)=\delta(r-s)$ and then use $\sum_{q=0}^{n}K_q \, G_{n-q}=0$ for all $n\ge 1$, which can be seen from  \Eq{eq:Gn expansion}.} 
to a perturbation expansion
\begin{equation*}
G(t_1,t_2)=G_0(t_1,t_2)+\epsilon\, G_1(t_1,t_2)+\epsilon^2\, G_2(t_1,t_2)+\cdots
\end{equation*}
with
\begin{subequations}\label{eq:G_n together}
\begin{equation}
G_0(t_1,t_2)=-\frac{1}{2}\delta^{\prime\prime}(t_1-t_2),
\end{equation}
and for $n\ge 1$,
\begin{align}
G_n(t_1,t_2) =-\sum_{q=1}^{n}G_0  K_q  G_{n-q}(t_1,t_2). \label{eq:Gn expansion}
\end{align}
\end{subequations}
(Here we denote
\begin{equation}
A B (t_1,t_2)=\int_0^Tds\;A(t_1,s)B(s,t_2),
\end{equation}
for any two bivariate functions $A$ and $B$.)

It will be convenient for our analysis to write $G_n$ in \Eq{eq:G_n together} as 
\begin{equation}
G_n=G_0 \Sigma_n G_0
\end{equation}
for all positive integers $n$, such that
\begin{align}
\Sigma_0&= K_0, \quad \Sigma_1= - K_1,\quad \Sigma_2= -K_2 + K_1 G_0 K_1, 
 \nn \\[1mm] \Sigma_3=& - K_3 + K_2 G_0 K_1 + K_1 G_0 K_2- K_1 G_0 K_1 G_0 K_1,\label{eq:Sigma K}
\end{align}
and so on. In terms of this perturbation expansion, action \eqref{eq:sH 16} is written as
\begin{equation}
S=S_0+\epsilon \,\mathcal{L}_1+\epsilon^2 \mathcal{L}_2+\cdots, \label{eq:expansion S intro}
\end{equation}
where $S_0$ is in \Eq{eq:S0} and for $n\ge 1$,
\begin{align}
\mathcal{L}_n=\int_0^Tdt_1\int_{0}^Tdt_2\; \dot{x}(t_1)\left\{\frac{1}{8}\partial_{t_1}\partial_{t_2}\Sigma_n(t_1,t_2)\right\}\, \dot{x}(t_2) \label{eq:Lagrangians}
\end{align}
obtained by integration by parts.

For their explicit expression we use the following results obtained from \Eq{eq:Kn}: for $t_2\ge t_1$
\begin{subequations}
\begin{align}
\frac{1}{4}\partial_{t_1}\partial_{t_2}&K_1=(1+\ln \omega)\delta(t_1-t_2)+\frac{1}{2}\frac{1}{ (t_2-t_1)},\\
\frac{1}{4}\partial_{t_1}\partial_{t_2}&K_2=\left(\frac{\pi^2}{6}+2\ln \omega+\ln^2 \omega\right)\delta(t_1-t_2)\cr &+\frac{1+\ln \omega}{(t_2-t_1)} +\frac{1}{2}\int_{t_1+\omega}^{t_2-\omega}\frac{ds}{(t_2-s)(s-t_1)}, \label{eq:pd K2}
\end{align}
where singularities are regularized by introducing an infinitesimally small ultraviolet cutoff $\omega> 0$ in time, such that terms like $ \delta(t_1-t_2)\ln ( t_2-t_1)\simeq \delta(t_1-t_2)\ln \omega$ and 
\begin{align}
\frac{\ln( t_2-t_1)}{( t_2-t_1)}\simeq &\frac{\ln \omega}{(t_2-t_1)}+\frac{\pi^2}{6}\delta(t_1-t_2)\cr
&+\frac{1}{2}\int_{t_1+\omega}^{t_2-\omega}\frac{ds}{(t_2-s)(s-t_1)}, \label{eq:int strange}
\end{align}
which are used for writing \Eq{eq:pd K2}. Similarly, for $t_2\ge t_1$, 
\begin{align}
\frac{1}{4}\partial_{t_1}\partial_{t_2}&K_1G_0 K_1=2\left(1+\ln \omega\right)^2\delta(t_1-t_2)\cr
& +\frac{2(1+\ln \omega)}{(t_2-t_1)}+\frac{1}{2}\int_0^T\frac{ds}{\vert t_1-s\vert\vert t_2-s\vert }.
\end{align}
\label{eq:K identities}
\end{subequations}

Using \Eq{eq:K identities} in \Eq{eq:Sigma K} and \Eq{eq:Lagrangians} it is easy to see that
\begin{subequations}
\begin{align}
\mathcal{L}_1=& -2(1+\ln \omega) S_0-\frac{1}{2}S_1, \\
\mathcal{L}_2=& \left[2\left(1-\frac{\pi^2}{6}\right)+2\left(1+\ln \omega\right)^2\right] S_0\nn \\[1mm]
& \qquad \qquad +\left(1+\ln \omega\right)S_1+S_2,
\end{align}
where $S_0$, $S_1$, and $S_2$ are defined in \Eq{eq:S0S1S2}. The expansion \eqref{eq:expansion S intro} with \Eq{eq:L1 L2 results} gives \Eq{eq:expansion of S}.
\label{eq:L1 L2 results}
\end{subequations}

\section{Alternate derivation of the action}\label{app:S contactless}
Here we give an elegant and short derivation of the action in Eqs.~\eqref{eq:expansion of S}-\eqref{eq:S0S1S2} in a normal-ordered form.  Using integration by parts, \Eq{eq:sH 16} gives
\begin{equation}
S[X_t]=\frac{1}{2}\int_0^Tdt_1\int_0^Tdt_2\;\dot{X}_{t_1}C^{-1}(t_1,t_2)\dot{X}_{t_2}\label{eq:S X dot}
\end{equation}
with the correlation
\begin{align}
C(t_1,t_2)=&\langle \dot{X}_{t_1}\dot{X}_{t_2}\rangle=4H\vert t_1-t_2\vert^{2H-1}\delta(t_1-t_2)\nn \\[1mm]
&\qquad+2H(2H-1)\vert t_1-t_2\vert^{2(H-1)}.
\end{align}
An expansion in $\epsilon=H-\frac{1}{2}$ gives
\begin{align}
C(t_1,t_2)=2\widehat{D} & \left[ \delta(t_1-t_2)+\epsilon\frac{1}{\vert t_1-t_2\vert}\right.\nn \\[1mm]
&\qquad \quad \left.+\epsilon^2\frac{2\ln\frac{\vert t_1-t_2}{\omega}}{\vert t_1-t_2\vert}+\cdots\right],
\end{align}
with $\widehat{D}=2 H \omega^{2H-1}=(1+2\epsilon)\omega^{2\epsilon}$,  and $\omega$ being an ultraviolet cutoff in time.   This implies
\begin{align*}
C^{-1}&(t_1,t_2)=\frac{1}{2\widehat{D}}\left[ \delta(t_1-t_2)-\frac{\epsilon}{\vert t_1-t_2\vert} \right.\nn \\[1mm]
&\left.-\epsilon^2 \frac{2\ln\frac{\vert t_1-t_2\vert}{\omega}}{\vert t_1-t_2\vert}+\epsilon^2 \int ds\frac{1}{\vert s-t_1\vert \vert s- t_2\vert}+\cdots\right].
\end{align*}
Substituting in \Eq{eq:S X dot} and defining a normal-ordered form (non-contact terms only) in  \Eq{eq:S normal ordered}
we get
\begin{align}
\ca S^{(n)}[X_t]=\frac{1}{2\widehat{D}}&\int_{t1<t_2}dt_1 dt_2 \dot{X}_{t_1}\dot{X}_{t_2}\bigg[\delta(t_1-t_2)\nn \\[1mm]
&- \frac{\epsilon}{\vert t_1-t_2\vert}-2\epsilon^2\frac{\ln\frac{\vert t_1-t_2\vert}{\omega}}{\vert t_1-t_2\vert}+\cdots\bigg].
\end{align}
Using the integral representation \Eq{eq:int strange} this gives
\begin{align}
\ca S^{(n)}&[X_t]=\frac{1}{2D}\int_{t1<t_2}dt_1 dt_2 \dot{X}_{t_1}\dot{X}_{t_2}\bigg[\delta(t_1-t_2)\nn \\[1mm]
&- \frac{\epsilon}{\vert t_1-t_2\vert}-\epsilon^2\int_{t_1}^{t_2} ds\frac{1}{\vert s-t_1\vert \vert s- t_2\vert}+\cdots\bigg],\label{eq:S contactless}
\end{align}
with $D$ given in \Eq{eq:c0}.  Comparing with Eqs.~\eqref{eq:expansion of S}-\eqref{eq:S0S1S2} one can see that the both leading and sub-leading terms are same whereas the $\epsilon^2$ order term includes only contact-less terms.  An integral representation of the normal-ordered second-order term is in \Eq{eq:S1 2 integral normal}.

\section{The fBm propagator} 
\label{app:free propagator}
Here, we verify \Eq{eq:Free fBm propagator} using the perturbation expansion of the action \eqref{eq:expansion S intro} to all orders. In terms of this expansion, \Eq{eq: W and N a} can be written as
\begin{equation}
W_H(m,T)= \braket{ e^{-\sum_{n\ge 1}\epsilon^n \mathcal{L}_n}},
\end{equation}
where by the angular brackets we denote (definition restricted only for this Appendix)
\begin{equation}
\braket{O[x]}\equiv \int_{x(0)=0}^{x(T)=m}\mathcal{D}[x]\;e^{-S_0} O[x].
\end{equation}
Then, using a result for the multi-time correlation given later in \Eq{eq:main identity} for $D=1$ and the propagator \Eq{eq: W and N a} leads to
\begin{equation}
\mathcal{G}_H(m,T)=e^{F(T)\partial_m^2}\mathcal{G}(m,T) \label{eq:GH der m}
\end{equation}
with
\begin{equation}
F(T)=\frac{1}{2}\int_0^Tdt_1 \int_0^Tdt_1 \sum_{n\ge 1}\epsilon^n \partial_{t_1}\partial_{t_2}K_n(t_1,t_2), \label{eq:F}
\end{equation}
where we used \Eq{eq:Sigma K} and \Eq{eq:Lagrangians}.

\begin{remark}
In \Eq{eq:GH der m}, the contribution from terms like $K_1 G_0 K_1$ \textit{etc} in \Eq{eq:Sigma K} are canceled from the terms in normalization $N_T$ in \Eq{eq:Free fBm propagator}. One may explicitly verify this at lower orders in perturbation expansion. 
\end{remark}

Using \Eq{eq:Kn} in \Eq{eq:F}, it is easy to see that
\begin{equation}
F(T)=T\sum_{n\ge 1}\frac{(2\epsilon \ln T)^n}{n!}=T(T^{2\epsilon}-1),
\end{equation}
which in \Eq{eq:GH der m} leads to
\begin{equation}
\mathcal{G}_H(m,T)=e^{(T^{2H}-T)\partial_m^2}\mathcal{G}(m,T),
\end{equation}
where we used $1+2\epsilon=2H$. Using the expression of $\mathcal{G}(m,T)$ in \Eq{eq:W0 Z}, it is now easy to obtain \Eq{eq:Free fBm propagator}.

\section{Numerical simulation of an fBm} 
\label{app:computer simulation}
Efficient computer simulation of an fBm trajectory is a delicate task. A vast literature has been published on this subject. For a comparative study of many of the sampling methods for an fBm see the review \cite{Coeurjolly2000} and references therein.
In general these algorithms generate the full trajectory. If one is only interested in a specific observable, as the first-passage time, not all points need to be generated, allowing for tremendous gains both in memory usage and execution speed \cite{WalterWiese2019a,WalterWiese2019b,ArutkinWalterWiese2020}.

In our work, we use a discrete-time sampling method following the Davis and Harte procedure \cite{DaviesHarte1987} (also known as the Wood and Chan procedure \cite{WoodChan}) as described in Ref.~\cite{DiekerPhD}. The basic idea is to construct fBm paths from a discrete-time sampling of stationary, Gaussian-distributed, increments $\Delta X_n$ for integers $n=0, 1, \cdots, N-1$, with mean $\langle \Delta X_n\rangle=0$ and covariance 
\begin{align}
&\langle \Delta X_m \Delta X_n \rangle = \gamma(m-n) \label{eq:covariance discrete Delta X}\\[1mm]
&\quad=( m-n+1)^{2H}  + ( m-n-1 )^{2H} -2( m-n )^{2H}, \nn
\end{align}
for positive integers $n\le m< N$. For large $N$ with $t=n/N$, one can see that $N^{2-2H}\gamma(N t-N s)$ converge to the covariance \eqref{eq:covariance dxdx}. This means, the cumulated sum $N^{-H}\sum_{i=0}^{n}\Delta X_{i}$ for large $N$ gives an fBm path $X_t$ with $X_0=0$ in a time window $[0,1]$.

The Davis and Harte procedure is an efficient algorithm for generating samples of $\Delta X_n$ with a computational efficiency $\mathcal{O}(N \log N)$ (compared to $\mathcal{O}( N^3)$ for Choleski decomposition method \cite{Coeurjolly2000}). The algorithm involves the following simple steps. We construct two linear arrays $\{W_n\}$ and $\{ \lambda_n \}$ of length $2N$ with index $n=0,1,\cdots, 2N-1$. Elements of the first array are generated from a set of $2N$ independent Gaussian random numbers ${V_0, V_1,\cdots,V_{2N-1}}$, with $\langle V_n\rangle =0$ and $\langle V_m V_n \rangle=\delta_{m,n}$. We define
\begin{equation}
W_0=V_0, \qquad W_n=\frac{1}{\sqrt{2}}\left(V_n+i\; V_{2N-n}\right),
\end{equation} 
for $n=1,\cdots,N-1$, whereas 
\begin{equation}
W_N=V_N, \qquad W_n=\frac{(-i)}{\sqrt{2}}\left(V_n+i\; V_{2N-n}\right),
\end{equation}
for $n=N+1,\cdots,2N-1$. This construction ensures that $\langle W_n \rangle=0$ and 
\begin{equation}
\langle W_k W_{k'}\rangle=\delta_{k,0}\delta_{k',0}+\delta_{k+k',2N},\label{eq:WW covariance}
\end{equation}
for indices $0\le k\le 2N-1$.

Elements of the second array are defined by
\begin{equation}
\lambda_n=\sum_{k=0}^{2N-1}\Gamma_k \; e^{i\, \pi \frac{n k}{N}}
\end{equation}
for integers $0\le n\le 2N-1$, where $\Gamma_k=\gamma(k)$ for $0\le k \le N$ and $\Gamma_k=\gamma(2N-k)$ for $N+1\le k \le 2N-1$ with covariance in \Eq{eq:covariance discrete Delta X}. This means,
\begin{equation}
\lambda_{2N-n}=\lambda_n \label{eq:lambda symmetry}
\end{equation}
and the inversion formula
\begin{equation}
\Gamma_k=\frac{1}{2N}\sum_{n=0}^{2N-1}\lambda_n \; e^{-i\, \pi \frac{n k}{N}}.\label{eq:Gamma inverse formula}
\end{equation}

The set of increments for a discrete fBm are obtained from
\begin{equation}
\Delta X_n=\frac{1}{\sqrt{2N}}\sum_{k=0}^{2N-1}W_k\;\sqrt{\lambda_k}\;  e^{i \pi \frac{n k}{N}}
\label{eq:Delta Xn formula}
\end{equation}
for $0\le n \le N-1$. In comparison, we shall see that the set of increments for $N\le n \le 2N-1$ do not have the covariance \eqref{eq:covariance discrete Delta X} and they are discarded.

It is simple to verify that this construction \eqref{eq:Delta Xn formula} indeed generates Gaussian random numbers with covariance \eqref{eq:covariance discrete Delta X}. The simplest is to see that $\langle \Delta X_n \rangle=0$ from $\langle W_n \rangle=0$. Moreover, $X_n$ is a linear combination of Gaussian random variables $W_n$, and therefore it's distribution remains Gaussian. For the covariance, using \Eq{eq:Delta Xn formula} we write
\begin{equation*}
\langle \Delta X_m \Delta X_n \rangle=\frac{1}{2N}\sum_{k,k'=0}^{2N-1}\langle W_k W_{k'}\rangle\;\sqrt{\lambda_k\lambda_{k'}}\;  e^{\frac{i\pi}{N} \left(n k + m k'\right)},
\end{equation*}
which using \Eq{eq:WW covariance} gives
\begin{equation}
\langle \Delta X_m \Delta X_n \rangle=\frac{1}{2N}\left\{\lambda_0+\sum_{k=1}^{2N-1}\sqrt{\lambda_k\lambda_{2N-k}}\;  e^{-i \frac{\pi}{N} \left(m-n\right)k}\right\},
\end{equation}
for $n\le m$. Using the symmetry in \Eq{eq:lambda symmetry} the above expression simplifies to
\begin{align}
\langle \Delta X_m \Delta X_n \rangle=&\frac{1}{2N}\sum_{k=1}^{2N-1}\lambda_k\;  e^{-i \frac{\pi}{N} \left(m-n\right)k},\cr
=&\Gamma_{m-n}\label{eq:DeltaXX Gamma}
\end{align}
for $m\ge n$, where in the last step we used the inverse Fourier transformation \eqref{eq:Gamma inverse formula}. It is clear from \Eq{eq:DeltaXX Gamma} that,
\begin{equation}
\langle \Delta X_m \Delta X_n \rangle= \gamma(m-n) \quad \text{for }m-n\le N,
\end{equation}
which includes all $0\le n\le m\le N-1$. For indices $\ge N$, such that $m-n>N$, the covariance is $\gamma(2N-m+n)$, and therefore $\Delta X_n$ for $n\ge N$ are discarded.

The mathematics behind this algorithm is clearly explained in \cite{Coeurjolly2000,DiekerPhD}. It involves calculating square root of a positive matrix by embedding it in a circulant matrix. We shall not repeat the discussion this here. Reader may find details in \cite{Coeurjolly2000,DiekerPhD}.

\section{A derivation of the inverse transform} 
\label{app:inverse}
\begin{figure}[t]
\includegraphics[scale=1]{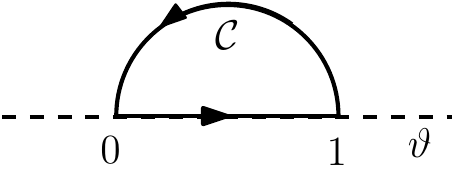}
\caption{Contour $C$ for the complex integral \eqref{eq:contour integral}.}
\label{f:SPContour}
\end{figure}

The inverse transformation in \Eq{eq:inverse KT1} can be derived using complex analysis by writing \Eq{eq:ptilde p} as
\begin{equation*}
\widetilde p(z)=\oint_{\mathcal{C}}d\ell \frac{p(\ell)}{1+z\ell}, \label{eq:contour integral}
\end{equation*}
where $C$ is a simple closed contour drawn in \fref{f:SPContour}. In an alternative representation
\begin{equation}
\frac{1}{2\pi i}\oint_{\mathcal{C}}d\ell \frac{p(\ell)}{\ell-z}=f(z):=\frac{1}{2\pi i}\left(-\frac{1}{z}\right)\widetilde p \left(-\frac{1}{z}\right).
\label{eq:SPContour}
\end{equation}
The \textit{Sokhotski-Plemlj} formula of complex analysis gives the inverse transformation 
\begin{equation}
p(\ell)=f_{+}(\ell)-f_{-}(\ell) \label{eq:SP formula}
\end{equation}
for any point $\ell$ on the contour $C$, where $f_{\pm}(\ell)=\lim_{z\rightarrow \ell}f(z)$ with the limit taken from the domain \textit{inside} (+) and \textit{outside} (-) the contour $C$, respectively. For $\ell=\vartheta$ on the real axis,
\begin{equation}
f_{\pm}(\vartheta)=-\frac{1}{2\pi i \, \vartheta}\lim_{\delta \rightarrow 0} \widetilde{p}\left(-\frac{1}{\vartheta}\pm i \delta \right)
\end{equation}
and this gives \Eq{eq:inverse KT1}.

\section{A list of useful $\mathcal{K}^{-1}$ transforms}
\label{sec:list of K transformations}
Here, we give functions, which are related by the transformation \Eq{eq:Kay inverse transformation} and its inverse transformation \Eq{eq:Kay transformation}. These relations, indicated below by $\leftrightarrow$, are useful for our analysis. They can be numerically verified in Mathematica. A trivial, but useful result is $ 1\leftrightarrow 1$.

Among others,
\begin{equation}
-2\log\left[1+\sqrt{1+\kappa}\right]\leftrightarrow\log \vartheta, \label{eq:inverse KT id one}
\end{equation}
\begin{equation}
-2\log\left[1+\frac{1}{\sqrt{1+\kappa}}\right]\leftrightarrow\log (1-\vartheta ),\label{eq:inverse KT id two}
\end{equation}
which using linearity of the transformation leads to
\begin{equation}
\log(1+\kappa)-4\log\left[1+\sqrt{1+\kappa}\right]\leftrightarrow\log \vartheta (1-\vartheta ),\label{eq:inverse KT id three}
\end{equation}
and
\begin{equation}
-\log(1+\kappa)\leftrightarrow\log \frac{\vartheta}{1-\vartheta }. \label{eq:inverse KT id four}
\end{equation}

Additionally,
\begin{equation}
\left(\log(1+\kappa)\right)^2\leftrightarrow \left(\log \frac{\vartheta}{1-\vartheta }\right)^2-\pi^2. \label{eq:inverse KT id five}
\end{equation}

We get,
\begin{equation}
\frac{\log(1+\sqrt{1+\kappa})}{\sqrt{1+\kappa}}\leftrightarrow x\arctan \frac{1}{x}, \label{eq:inverse KT id six}
\end{equation}
and
\begin{equation}
\sqrt{1+\kappa}\log(1+\frac{1}{\sqrt{1+\kappa}})\leftrightarrow \frac{1}{x}\arctan x, \label{eq:inverse KT id seven}
\end{equation}
where $x=\sqrt{\frac{\vartheta}{1-\vartheta}}$. A linear combination of Eqs.~(\ref{eq:inverse KT id two},\,\ref{eq:inverse KT id six},\,\ref{eq:inverse KT id seven}) gives
\begin{align}
-&\left(1+\sqrt{1+\kappa} \right)\log(1+\kappa) +\frac{2\left(1+\sqrt{1+\kappa}\right)^2}{\sqrt{1+\kappa}}\nn \\[1mm]
&\qquad \times \log\left(1+\sqrt{1+\kappa}\right)\leftrightarrow \psi(x)-\log \vartheta (1-\vartheta) \label{eq:inverse KT id eight}
\end{align}
with \Eq{eq:psi} and the same definition for $x$.
A related result about square of the above function is
\begin{align}
&\Big[-\left(1+\sqrt{1+\kappa} \right)\log(1+\kappa) +\frac{2\left(1+\sqrt{1+\kappa}\right)^2}{\sqrt{1+\kappa}}\times \cr
& \log\left(1+\sqrt{1+\kappa}\right) \Big]^2\leftrightarrow \left[\psi(x)-\log \vartheta (1-\vartheta)\right]^2-\psi_2(x)^2.\qquad  \label{eq:inverse KT id nine}
\end{align}
with Eqs.~\eqref{eq:psi} and \eqref{eq:psi2}.

Other results, useful for verifying Eqs.~(\ref{eq:F last numerical approximation},\,\ref{eq:F max numerical approximation},\,\ref{eq:F + numerical approximation}), are
\begin{equation}
\frac{\Gamma\left(\frac{n}{2}\right)}{\sqrt{\pi}}\;{}_2F_1\left(\frac{1}{2},\frac{n-1}{2},\frac{n+1}{2},-\kappa\right)\leftrightarrow (\vartheta)^{\frac{n-1}{2}}, \label{eq:inverse KT id eleven}
\end{equation}
\begin{equation}
\frac{\sqrt{1+\kappa}\;\Gamma\left(\frac{n}{2}\right)}{\sqrt{\pi}} \; {}_2F_1\left(\frac{1}{2},1,\frac{n+1}{2},-\kappa\right)\leftrightarrow (1-\vartheta)^{\frac{n-1}{2}}, \label{eq:inverse KT id twelve}
\end{equation}
and their product
\begin{equation}
\frac{\sqrt{1+\kappa}\; \Gamma\left(\frac{n}{2}\right)^2}{\sqrt{\pi}} \;{}_2F_1\left(1,\frac{n}{2},n,-\kappa\right)\leftrightarrow \left[\vartheta(1-\vartheta)\right]^{\frac{n-1}{2}}, \label{eq:inverse KT id thirteen}
\end{equation}
for $n\ge 1$, where ${}_2 F_1(a,b,c,z)/\Gamma(c)$ is regularized hypergeometric function and it can be evaluated to arbitrary numerical precision in Mathematica.

\section{Amplitude of the Two-loop diagrams for $t_{\rm last}$}
\label{app:two loop diagram for last}
Here, we give a detailed derivation of the amplitudes of 2-loop diagrams shown in \fref{fig:diagram}.

\subsection{Non-trivial diagram D contributing to $\mathcal{F}^{\rm last}$. \label{app:D last}}
\begin{figure}
\includegraphics[width=\columnwidth]{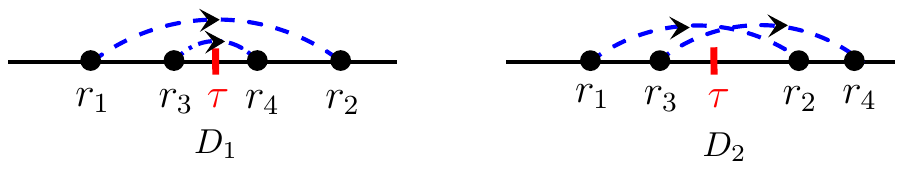}
\caption{Two distinct cases of the 2-loop diagram $D$ in \fref{fig:diagram} for distribution of  $t_{\rm last}$, categorized according to whether loops intersect (for $D_1$) or not (for $D_2$). The time variables $r$'s remain on the same side of $\tau$ as indicated. \label{fig:diagramLastDThree}}
\end{figure}

Amplitude of the diagram $D$ in \fref{fig:diagram} is given by
\begin{align*}
&D(\tau,T)=\frac{1}{4D^2}\int_{0}^{\infty}dm\int_0^{\Lambda}dy_1dy_2 \int_{0}^{\tau}dr_1\int_{\tau}^{T}dr_2 \int_{0}^{\tau}dr_3 \cr
&\times \int_{\tau}^{T}dr_4 
\, e^{y_1(r_1-r_2)}  e^{y_2(r_3-r_4)}\llangle[\Big]\dot{x}(r_1)\dot{x}(r_2)\dot{x}(r_3)\dot{x}(r_4)\rrangle[\Big]_{m}
\end{align*}
with the angular brackets defined in \Eq{eq:angular bracket last}.
Considering order of the time variables, the possible cases are illustrated in \fref{fig:diagramLastDThree}. Their amplitude can be expressed in terms of $J$ and $J^+$ defined in \Eq{eq:J2}. Adding them, we write 
\begin{align*}
D(\tau,T)
=&\frac{2}{4D^2}\!\int_{0}^{\infty}\mquad dm \!\! \int_{0}^{\Lambda}\mquad dy_1 dy_{2}e^{-\tau(y_1+y_2)}J_\tau(0,x_0;-y_{1},-y_{2})\nn \\[1mm] &\times \Big[J_{T-\tau}^{+}(x_0,m;y_{2},y_{1})+ J_{T-\tau}^{+}(x_0,m;y_{1},y_{2})\Big],
\end{align*}
where the pre-factor $2$ is due to interchange of pairs $(r_1,r_2)$ with $(r_3,r_4)$.

Its double Laplace transformation in \Eq{eq:D transformation} gives
\begin{align*}
\widetilde{D}(\lambda,s)
=&\frac{1}{2D^2}\!\int_{0}^{\infty}\mquad dm \!\! \int_{0}^{\Lambda}\mquad dy_1 dy_{2}\widetilde{J}_{s+\lambda+y_1+y_2}(0,x_0;-y_{1},-y_{2})\nn \\[1mm] &\times \left[\widetilde{J}_{s}^{+}(x_0,m;y_{2},y_{1})+ \widetilde{J}_{s}^{+}(x_0,m;y_{1},y_{2})\right].
\end{align*}
It is convenient to write the expression in a form such that the integrand is symmetric in $y_1$ and $y_2$. We write
\begin{align}
&\widetilde{D}(\lambda,s)= \frac{1}{4D^2}\int_{0}^{\Lambda}dy_{1}  dy_{2}\bigg[ \widetilde{J}_{s+\lambda+y_1 +y_2}(0,x_0;\cr
&-y_{1},-y_{2})  +\widetilde{J}_{s+\lambda+y_1+y_2}(0,x_0;-y_{2},-y_{1}) \bigg]\times\cr &  \int_{0}^{\infty}dm\bigg[\widetilde{J}_s^{+}(x_0,m;y_{2},y_{1})+\widetilde{J}_s^{+}(x_0,m;y_{1},y_{2})\bigg].
\label{eq:D tilde 1 app}
\end{align}

We show that (a derivation given in \aref{app:B})
\begin{align}
\widetilde{J}_{s}(m_1,m_2 &;y_1,y_2)=\frac{2\sqrt{D}}{y_1 y_2 (y_1+y_2)}\Bigg[ y_1\sqrt{s}\; e^{-z \sqrt{s}}\nn \\[1mm]
& \qquad + y_2\sqrt{s+y_1+y_2}\; e^{-z \sqrt{s+y_1+y_2}}\cr
& \qquad \quad -(y_1+y_2)\sqrt{s+y_2}\;e^{-z\sqrt{s+y_2}}\Bigg],
\label{eq:J2 tilde final full}
\end{align}
where $z=\tfrac{\vert m_1-m_2\vert}{\sqrt{D}}$ and
\begin{align}
\int_{0}^{\infty}&dm_2\widetilde{J}_{s}^{+}(m_1,m_2;y_1,y_2)\cr &= \frac{4D}{y_1y_2}\sqrt{\frac{s+y_2}{s}}\left[e^{-m_1\sqrt{\frac{s+y_2}{D}}}-e^{-m_1\sqrt{\frac{s+y_1+y_2}{D}}}\right]\cr
&\quad +\frac{4D}{(y_1+y_2)y_2}\left[e^{-m_1\sqrt{\frac{s+y_1+y_2}{D}}}-e^{-m_1\sqrt{\frac{s}{D}}}\right].\qquad 
\label{eq:J2 + tilde integral final full}
\end{align}

Using the asymptotic of \eqref{eq:J2 tilde final full} for small $x_0$, we obtain
\begin{align*}
 \widetilde{J}_{s}&(0,x_0;-y_{1},-y_{2})+ \widetilde{J}_{s}(0,x_0;-y_{2},-y_{1})\simeq \nn \\[1mm]
&\frac{2\sqrt{D}}{y_1 y_2}\bigg(\sqrt{s}-\sqrt{s-y_2} -\sqrt{s-y_1}+\sqrt{s-y_1-y_2}\bigg)
\end{align*}
and similarly from \Eq{eq:J2 + tilde integral final full} we get for small $x_0$,
\begin{align}
\int_{0}^{\infty}& dm \bigg\{\widetilde{J}_s^{+}(x_0,m;y_{2},y_{1}) + \widetilde{J}_s^{+}(x_0,m;y_{1},y_{2})\bigg\}\nn \\[1mm]
& \simeq -\frac{4x_0\sqrt{D}}{y_1 y_2}\times \frac{\sqrt{s+y_1+y_2}}{\sqrt{s}}\times \cr
&  \bigg(\sqrt{s+y_1+y_2}-\sqrt{s+y_1}- \sqrt{s+y_2}+\sqrt{s}\bigg).\quad\label{eq:J + int right sum asymptotic}
\end{align}

Using these asymptotics in \Eq{eq:D tilde 1 app} we get, for small $x_0$,
\begin{align*}
\widetilde{D}(\lambda,s) \simeq  -\frac{2x_0}{D}&\int_{0}^{\Lambda} \frac{dy_{1} dy_{2}}{y_1^2 y_2^2}\times \frac{\sqrt{s+y_1+y_2}}{\sqrt{s}}\times \cr \bigg( \sqrt{s+y_1+y_2} & -\sqrt{s+y_1}  -\sqrt{s+y_2}+\sqrt{s}\bigg)\cr
\times \bigg(\sqrt{s+\lambda+y_1+y_2} & -\sqrt{s+\lambda+y_1}\cr
& -\sqrt{s+\lambda+y_2}+\sqrt{s+\lambda}\bigg).
\end{align*}
This leads to the result in terms of re-scaled arguments in \Eq{eq:D tilde last final} and \Eq{eq:var D last}.

\subsection{Two-loop diagrams contributing to simple scaling}
\begin{figure}
\includegraphics[width=0.48\textwidth]{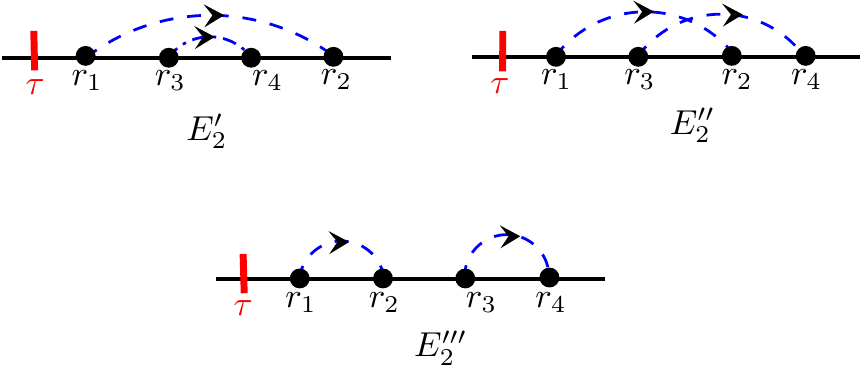}
\caption{Diagram $E_2$ of \fref{fig:diagram} is made of three cases according to relative order of time variables with $r_1<r_2$ and $r_3<r_4$, remaining on the side of $\tau$ as indicated. \label{fig:diagram E Last}}
\end{figure}

\subsubsection{Diagrams $E_1$ and $E_2$}
We begin with the diagram $E_2$ in \fref{fig:diagram}, whose amplitude is given by
\begin{align*}
&E_2(\tau,T)=\frac{1}{4D^2}\int_{0}^{\infty}\mquad dm\int_0^{\Lambda}dy_1dy_2 \int_{\tau}^{T}dr_1\int_{r_1}^{T}dr_2 \int_{\tau}^{T}dr_3 \cr
&\times \int_{r_3}^{T}dr_4 \; 
 e^{y_1(r_1-r_2)}  e^{y_2(r_3-r_4)}\llangle[\Big]\dot{x}(r_1)\dot{x}(r_2)\dot{x}(r_3)\dot{x}(r_4)\rrangle[\Big]_{m}
\end{align*}
with the angular brackets defined in \Eq{eq:angular bracket last}.

The expression can be written in three parts according to relative order of times $r$.
\begin{equation*}
E_2(\tau,T)= E_2^{\prime}(\tau,T)+E_2^{\prime\prime}(\tau,T)+E_2^{\prime\prime\prime}(\tau,T)
\end{equation*}
as shown in \fref{fig:diagram E Last}. Their amplitude can be written in terms of propagator $Z$ in \Eq{eq:W0 Z} and $J^+$ in \Eq{eq:J2}. Adding their amplitudes, we write
\begin{align*}
E_2(\tau,T)
= &\,\frac{2}{4D^2}\;Z_{\tau}(0,x_0)\int_{0}^{\Lambda}dy dy^{\prime}\int_{0}^{\infty}dm \times \cr
\bigg[J_{T-\tau}^{+}(x_0,m;-y,&-y^{\prime},y^{\prime},y) + 
J_{T-\tau}^{+}(x_0,m;-y,-y^{\prime},y,y^{\prime})\cr &  +J_{T-\tau}^{+}(x_0,m;-y,y,-y^{\prime},y^{\prime})\bigg].
\end{align*}
(The pre-factor $2$ comes from interchange of pairs $(r_1,r_2)$ and $(r_3,r_4)$.) 

Corresponding double Laplace transformation gives
\begin{align*}
\widetilde{E}_2(\lambda,s)=&\frac{1}{2D^2}\;\widetilde{Z}_{s+\lambda}(0,x_0)\int_{0}^{\Lambda}dy dy^{\prime}\int_{0}^{\infty}dm \times \cr 
\bigg[\widetilde{J}^{+}_{s}(x_0,m;-y,&y,-y^{\prime},y^{\prime})+\widetilde{J}^{+}_{s}(x_0,m;-y,-y^{\prime},y^{\prime},y)\cr
&\qquad \qquad +\widetilde{J}^{+}_{s}(x_0,m;-y,-y^{\prime},y,y^{\prime})\bigg].
\end{align*}

To evaluate the expressions we use $\widetilde{Z}_{s}(0,x_0)$ from \eqref{eq:Z tilde final}, and
\begin{widetext}
\begin{align}
\int_{0}^{\infty}dm \, \widetilde{J}^{+}_{s} &(x_0,m;y_1,y_2,y_3,y_4)=\frac{16 D^2}{s} \bigg[\frac{s^3 \left(e^{-s_4 z}-e^{-s z}\right)}{\left(s^2-s_1^2\right) \left(s^2-{s_2}^2\right) \left(s^2-{s_3}^2\right) \left(s^2-{s_4}^2\right)}\cr
&+\frac{{s_1}^3 \left(e^{-{s_4} z}-e^{-{s_1} z}\right)}{\left({s_1}^2-s^2\right) \left({s_1}^2-{s_2}^2\right) \left({s_1}^2-{s_3}^2\right) \left({s_1}^2-{s_4}^2\right)}+\frac{{s_2}^2 \left(s {s_1}+{s_2}^2\right) \left(e^{-{s_4} z}-e^{-{s_2} z}\right)}{(s+{s_1}) \left({s_2}^2-s^2\right) \left({s_2}^2-{s_1}^2\right) \left({s_2}^2-{s_3}^2\right) \left({s_2}^2-{s_4}^2\right)}\cr
&\qquad \qquad \qquad +\frac{{s_3} \left(\left({s_1} {s_2}+{s_3}^2\right) \left(s^2 {s_1}+{s_2} {s_3}^2\right)+s {s_3}^2 ({s_1}+{s_2})^2\right)\left(e^{-{s_4} z}-e^{-{s_3} z}\right) }{(s+{s_1}) (s+{s_2})  ({s_1}+{s_2}) \left({s_3}^2-s^2\right)\left({s_3}^2-{s_1}^2\right) \left({s_3}^2-{s_2}^2\right) \left({s_3}^2-{s_4}^2\right)}\bigg]\label{eq:J4 + integrate}
\end{align}
\end{widetext}
derived later in \Eq{eq:B39}, where we denote $z=\frac{x_0}{\sqrt{D}}$, $s_1=\sqrt{s+y_4}$, $s_2=\sqrt{s+y_3+y_4}$, $s_3=\sqrt{s+y_2+y_3+y_4}$, $s_4=\sqrt{s+y_1+y_2+y_3+y_4}$.

Using these two results for small $x_0$, we get the asymptotics  
\begin{equation*}
\widetilde{E}_2(\lambda,s)\simeq \frac{1}{2D}\times \frac{x_0}{\sqrt{s(s+\lambda)}}\int_0^{\Lambda} \frac{dy_1dy_2}{y_1^2 y_2^2}\;e(s,y_1,y_2)
\end{equation*}
where we define

\begin{widetext}
\begin{align}
e(s,y_1,y_2)=& -(s+y_1) (s+y_2)+ \sqrt{s} \left(\sqrt{s}-\sqrt{s+y_1}-\sqrt{s+y_2}\right)\bigg[2 \left(\sqrt{s+y_1+y_2}-\sqrt{s+y_1}-\sqrt{s+y_2}+\sqrt{s}\right)^2\cr
&-\left(\sqrt{s}-\sqrt{s+y_1}-\sqrt{s+y_2}\right)^2-\sqrt{s+y_1} \left(\sqrt{s}-\sqrt{s+y_1}\right)
-\sqrt{s+y_2} \left(\sqrt{s}-\sqrt{s+y_2}\right)\bigg].
\label{eq:small e}
\end{align}
\end{widetext}

Comparing the two diagrams $E_1$ and $E_2$  in \fref{fig:diagram}, one can see that, for small $x_0$,
\begin{equation*}
\widetilde{E}_1(\lambda,s)\simeq \frac{1}{2 D}\times \frac{x_0}{\sqrt{s(s+\lambda)}}\int_0^{\Lambda}\frac{dy_1dy_2}{y_1^2 y_2^2}\;e(s+\lambda,y_1,y_2)
\end{equation*}
with \Eq{eq:small e}. (This we have also explicitly verified.) Adding the two amplitudes we get \Eq{eq:diagram E final}.

\begin{remark}
Interestingly the integral in $\widetilde{E}_{1(2)}(\lambda,s)$ can be evaluated.
\begin{align*}
\int_0^{\Lambda} &\frac{dy_1dy_2}{y_1^2 y_2^2}\;e(s,y_1,y_2)= -\log^2\left(\frac{\Lambda}{s}\right)\cr
&+2\left[1+2\log 2\right]\log\left(\frac{\Lambda}{s}\right)+(1+2\log 2)^2-2-\frac{3}{2}\pi^2.
\end{align*}
(we have verified this numerically.) This result along with results \Eq{eq:A integration} and \Eq{eq:g definition} given later, helps recognize the linear combination of diagrams in \Eq{eq:efa cancellation} where divergences for $\Lambda\to \infty$ cancels.
\end{remark}

\subsubsection{Diagram $A$}
Amplitude of the diagram $A$ in \fref{fig:diagram} is given by
\begin{align*}
&A(\tau,T)=\frac{1}{4D^2}\int_{0}^{\infty}dm\int_0^{\Lambda}dy_1dy_2 \int_{0}^{\tau}dr_1\int_{r_1}^{\tau}dr_2 \int_{\tau}^{T}dr_3 \cr
&\times \int_{r_3}^{T}dr_4 \;
 e^{y_1(r_1-r_2)}  e^{y_2(r_3-r_4)}\llangle[\Big]\dot{x}(r_1)\dot{x}(r_2)\dot{x}(r_3)\dot{x}(r_4)\rrangle[\Big]_{m}
\end{align*}
with the angular brackets defined in \Eq{eq:angular bracket last}.
In terms of $J$ in \Eq{eq:J2} and its analogue $J^+$ in presence of absorbing boundary, we write
\begin{align*}
A(\tau,T)
=\frac{2}{4D^2}\int_{0}^{\Lambda} & dy_1 dy_2 J_{\tau}(0,x_0;-y_1,y_1) \cr & \times \int_{0}^{\infty}dm J_{T-\tau}^{+}(x_0,m;-y_2,y_2),
\end{align*}
where the prefactor $2$ is the degeneracy from the interchange of pair of indices $(1,2)$ and $(3,4)$. The double Laplace transformation \Eq{eq:D transformation} gives
\begin{align*}
\widetilde{A}(\lambda,s)
=\frac{1}{2D^2}\int_{0}^{\Lambda}&dy_1 dy_2\widetilde{J}_{s+\lambda}(0,x_0;-y_1,y_1) \cr & \times \int_{0}^{\infty}dm \widetilde{J}_{s}^{+}(x_0,m;-y_2,y_2).
\end{align*}
Using Eqs.~\eqref{eq:J2 tilde final full} and \eqref{eq:J2 + tilde integral final full} for small $x_0$, we get
\begin{align*}
\widetilde{A}(\lambda,s)
\simeq &-\frac{x_0}{D\sqrt{s(s+\lambda)}}\int_{0}^{\Lambda}\frac{dy_1 dy_2}{y_1^2y_2^2}\left(\sqrt{s+y_2}-\sqrt{s}\right)^2\cr 
& \qquad \qquad \qquad\times \left( \sqrt{s+\lambda}-\sqrt{s+\lambda+y_1} \right)^2.
\end{align*}
In terms of re-scaled variables this gives \Eq{eq:2-loop A tilde final}.
\begin{remark}
The $y$ integration in $\widetilde{A}$ can be evaluated explicitly using
\begin{align}
\int_{0}^{\Lambda}\frac{dy}{y^2}\left(\sqrt{s+y}-\sqrt{s}\right)^2=\ln\left(\frac{\Lambda}{s}\right)-1-2\ln 2.\label{eq:A integration}
\end{align} 
\end{remark}

\subsubsection{Diagrams $G_1$ and $G_2$}
Diagrams $G_1$ and $G_2$  in \fref{fig:diagram} has a contracted point $s$. Their amplitude is given by
\begin{align*}
G_1(\tau,T)=&\frac{1}{D}\int_{0}^{\infty}dm\int_0^{\Lambda}dy_1 dy_2\int_0^{\tau}dr_1\int_{r_1}^{\tau}ds\int_{s}^{\tau}dr_2 \cr
&\times e^{y_1(r_1-s)}  e^{y_2(s-r_2)}\llangle[\Big]\dot{x}(r_1)\dot{x}(r_2)\rrangle[\Big]_{m}
\end{align*}
and
\begin{align*}
G_2(\tau,T)=&\frac{1}{D}\int_{0}^{\infty}dm\int_0^{\Lambda}dy_1 dy_2\int_{\tau}^{T}dr_1\int_{r_1}^{T}ds\int_{s}^{T}dr_2\cr
&\times e^{y_1(r_1-s)}  e^{y_2(s-r_2)}\llangle[\Big]\dot{x}(r_1)\dot{x}(r_2)\rrangle[\Big]_{m}
\end{align*}
with the angular brackets defined in \Eq{eq:angular bracket last}.  (Their difference is in the range of integration for time variables.)

We write these amplitudes in terms of the fBm propagators defined in Eqs.~(\ref{eq:W0 Z},\,\ref{eq:propagator free absorbing}).
\begin{align*}
G_1(\tau,T)=\frac{1}{D}& \int_0^{\Lambda}dy_1dy_2 \; \mathscr{L}_{\tau}(0,x_0;-y_1,y_1-y_2,y_2) \cr & \times \int_0^{\infty}dm Z_{T-\tau}^{+}(x_0,m)
\end{align*}
and
\begin{align*}
G_2(\tau,T)=&\frac{1}{D}\int_0^{\Lambda}dy_1dy_2 Z(0,x_0,\tau)\cr & \times \int_0^{\infty}dm \; \mathscr{L}_{T-\tau}^{+}(x_0,m;-y_1,y_1-y_2,y_2),
\end{align*}
where we define
\begin{align}
\mathscr{L}_{t}(m_1,m_2;&y_1,z,y_2)=\int_0^{t}dr_1\int_{r_1}^{t}ds\int_{s}^{t}dr_2
\nn \\[1mm]
\times & e^{-y_1r_1-z\,s-y_2r_2} \langle \dot{x}(r_1) \dot{x}(r_2)\rangle_{(m_1,m_2)} \label{eq:L prop}
\end{align}
and its analogue $\mathscr{L}_{t}^+$ in presence of an absorbing line.  The angular brackets denote average with standard Brownian measure $e^{-\frac{S_0}{D}}$ starting at position $m_1$ and ending at position $m_2$.

Their double Laplace transformation \Eq{eq:D transformation} is given by
\begin{align}
\widetilde{G}_1(\lambda,s)=&\frac{1}{D}\int_0^{\Lambda}dy_1dy_2 \int_0^{\infty}dm  \cr 
 \times \widetilde{\mathscr{L}}_{s+\lambda}&(0,x_0;-y_1,y_1-y_2,y_2)\,\widetilde{Z}_{s}^{+}(x_0,m) \label{eq:app F1 tilde 1}
\end{align}
and
\begin{align}
\widetilde{G}_2(\lambda,s&)=\frac{1}{D}\int_0^{\Lambda}dy_1dy_2  \int_0^{\infty}dm \cr 
\times & \widetilde{Z}_{s+\lambda}(0,x_0) \, \widetilde{\mathscr{L}}_{s}^{+}(x_0,m;-y_1,y_1-y_2,y_2) .\quad \label{eq:app F2 tilde 1}
\end{align}

Expressions of $\widetilde{Z}$ and $\widetilde{Z}^{+}$ are in \Eq{eq:L of Z Z+} and the integral of the latter is in \Eq{eq:id Z plus int}. 

For Laplace transform $\widetilde{\mathscr{L}}$ of \Eq{eq:L prop} we note that
\begin{align}
\widetilde{\mathscr{L}}_{s}(m_1,m_2;y_1,z,y_2)=&\frac{1}{z}\bigg\{ \widetilde{J}_{s}(m_1,m_2;y_1+z,y_2)\cr & -\widetilde{J}_{s}(m_1,m_2;y_1,y_2+z)\bigg\} \label{eq:mathscr L in J}
\end{align}
with \Eq{eq:J2}, and a similar relation for $\widetilde{\mathscr{L}}^{+}$ in terms of $\widetilde{J}_{s}^{+}$.  This is easy to see from Eqs.~(\ref{eq:L prop},\;\ref{eq:J2}) and taking their Laplace transformation. 

Then, using Eqs.~\eqref{eq:J2 tilde final full} and \eqref{eq:J2 + tilde integral final full} we get
\begin{equation}
\widetilde{\mathscr{L}}_{s}(0,x_0;-y_1,y_1-y_2,y_2)=\sqrt{\frac{D}{s}}\times\frac{h_{s}\left(\frac{x_0}{\sqrt{D}},y_1,y_2 \right)}{y_1^2y_2^2}
\end{equation}
and
\begin{align}
\int_0^{\infty} dm \, \widetilde{\mathscr{L}}_{s}^{+}&(x_0,m;-y_1,y_1-y_2,y_2) \cr &= \frac{2 D}{\sqrt{s}}\times \frac{ h^{+}_{s}\left(\frac{x_0}{\sqrt{D}},y_1,y_2 \right)}{y_1^2y_2^2}, \label{eq:L scr 3 tilde integral right}
\end{align}
where we define
\begin{align}
h_{s}&\left(z,y_1,y_2 \right)= e^{-z\sqrt{s}}\,y_1y_2\cr
& -\frac{2\sqrt{s}}{(y_1-y_2)}\Bigg\{y_1^2\left(\sqrt{s+y_2}\;e^{-z\sqrt{s+y_2}}-\sqrt{s}\;e^{-z\sqrt{s}} \right)\cr
& \qquad -y_2^2\bigg(\sqrt{s+y_1}\;e^{-z\sqrt{s+y_1}}-\sqrt{s}\;e^{-z\sqrt{s}} \bigg) \Bigg\}, \label{eq:h full expr}
\end{align}
and
\begin{align}
h^{+}_{s}\left(z,y_1,y_2\right)=
&-z \; y_1 y_2 e^{-z\sqrt{s}}\nn \\[1mm]
+\frac{2}{y_1-y_2}&\bigg\{y_2^2 \sqrt{s+y_1}\left(e^{-z\sqrt{s+y_1}}-e^{-z\sqrt{s}}\right) \nn \\[1mm]  -  y_1^2 & \sqrt{s+y_2} \left(e^{-z\sqrt{s+y_2}}-e^{-z\sqrt{s}}\right)\bigg\}. \label{eq:h + full expr}
\end{align}

In terms of these functions in \Eq{eq:app F1 tilde 1} and \Eq{eq:app F2 tilde 1}, we write
\begin{equation*}
\widetilde{G}_{1}(\lambda,s)=\frac{ 1-e^{-x_0\sqrt{\frac{s}{D}}} }{s\sqrt{D(s+\lambda)}}  \int_0^{\Lambda}\frac{dy_1dy_2}{y_1^2 y_2^2}\;h_{s+\lambda}\left(\frac{x_0}{\sqrt{D}},y_1,y_2 \right)
\end{equation*}
and
\begin{equation*}
\widetilde{G}_{2}(\lambda,s)=\frac{ e^{-x_0\sqrt{\frac{s+\lambda}{D}}} }{\sqrt{D s(s+\lambda)}}\int_0^{\Lambda}\frac{dy_1dy_2}{y_1^2 y_2^2}h^{+}_{s}\left(\frac{x_0}{\sqrt{D}},y_1,y_2 \right).
\end{equation*}

For small $z$, the expressions in \Eq{eq:h full expr} and \Eq{eq:h + full expr} have the asymptotics
\begin{equation*}
h_{s}\left(z,y_1,y_2 \right)\simeq \frac{(\sqrt{s+y_2}-\sqrt{s})^2y_1^2-(\sqrt{s+y_1}-\sqrt{s})^2y_2^2}{(y_1-y_2)} 
\end{equation*}
and
\begin{equation}
h^{+}_{s}\left(z,y_1,y_2\right)\simeq  z \; h_{s}\left(z,y_1,y_2 \right). \label{eq:h+ h asymptotics}
\end{equation}

Substituting this in the expression for $\widetilde{G}_{1}$ and $\widetilde{G}_{2}$ in the small $x_0$ limit, we get
\begin{equation}
\widetilde{G}_{1}(\lambda,s)\simeq \frac{x_0}{D \sqrt{s(s+\lambda)}}\;g(s+\lambda)
\label{eq:G1 final explicit expressions}
\end{equation}
and
\begin{equation}
\widetilde{G}_{2}(\lambda,s)\simeq \frac{x_0}{D \sqrt{s(s+\lambda)}}\; g(s),\label{eq:G2 final explicit expressions}
\end{equation}
where 
\begin{align}
g(s)&=\int_0^{\Lambda}\frac{ dy_1dy_2}{y_1^2 y_2^2}\times\label{eq:small g definition} \\[1mm]
&\left[\frac{(\sqrt{s+y_2}-\sqrt{s})^2y_1^2-(\sqrt{s+y_1}-\sqrt{s})^2y_2^2}{(y_1-y_2)}\right].\nn
\end{align}
In terms of rescaled variables, we get \Eq{eq:2-loop F tilde final last}.
\begin{remark}
The integral in \Eq{eq:small g definition} can be evaluated analytically, 
\begin{equation}
g(s)=\left[\ln\left(\frac{\Lambda}{s}\right)-1-2\log 2\right]^2+1+\frac{\pi^2}{3}. \label{eq:g definition}
\end{equation}
\end{remark}

\subsubsection{Diagrams $B$ and $C$}
\label{app:B C last}
Amplitude of $B_1$ and $B_2$ in \fref{fig:diagram} is given by
\begin{align*}
B_1(\tau,T)=&\frac{1}{D}\int_{0}^{\infty}dm\int_0^{\Lambda}dy_1 dy_2\int_0^{\tau}dr_1\int_{r_1}^{\tau}ds\int_{\tau}^{T}dr_2 \nn \\[2mm]
&\times e^{y_1(r_1-s)}  e^{y_2(s-r_2)}\llangle[\Big]\dot{x}(r_1)\dot{x}(r_2)\rrangle[\Big]_{m}
\end{align*}
and
\begin{align*}
B_2(\tau,T)=&\frac{1}{D}\int_{0}^{\infty}dm\int_0^{\Lambda}dy_1 dy_2\int_{0}^{\tau}dr_1\int_{\tau}^{T}ds\int_{s}^{T}dr_2  \nn \\[2mm]
&\times e^{y_1(r_1-s)}  e^{y_2(s-r_2)}\llangle[\Big]\dot{x}(r_1)\dot{x}(r_2)\rrangle[\Big]_{m}.
\end{align*}
Their difference is in the limit of the time integrals. 

Amplitude of these diagrams are of order $x_0^2$ or higher, for small $x_0$, and therefore they do not contribute in the leading order amplitude in \Eq{eq:w2 last tilde result}. To see this let us consider $B_2$, which we write as
\begin{align*}
B_2(\tau,T)=\frac{1}{D}\int_0^{\Lambda}& dy_1dy_2 J_{\tau}(0,x_0;-y_1)e^{-y_1\tau}
\cr \times & \int_0^{\infty}dm \mathscr{L}_{T-\tau}^{+}(x_0,m;y_1-y_2,y_2),
\end{align*}
where, similar to \Eq{eq:L prop}, we define 
\begin{align*}
\mathscr{L}_{t}^{+}(m_1,m_2;y_1,y_2)=\int_0^{t}dr_1\int_{r_1}^{t}dr_2e^{-y_1r_1-y_2r_2}\langle \dot{x}(r_2)\rangle^{+}.
\end{align*}
The double Laplace transformation of $B_2$ is then given by
\begin{align}
\widetilde{B}_2(\lambda,s)=\frac{1}{D}\int_0^{\Lambda}& dy_1dy_2 \widetilde{J}_{s+\lambda+y_1}(0,x_0;-y_1)\cr  \times \int_0^{\infty} & dm \widetilde{\mathscr{L}}_s^{+}(x_0,m;y_1-y_2,y_2) .
\label{eq:B2 tilde last formal}
\end{align}

From the definition in \Eq{eq:J2} it is easy to see that
\begin{align*}
\mathscr{L}_{t}^{+}&(m_1,m_2;y_1,y_2)\cr  & =\frac{1}{y_1}\left[ J_t^{+}(m_1,m_2;y_2)-J_t^{+}(m_1,m_2;y_1+y_2)\right] 
\end{align*}
and similar for their Laplace transformation. Then using \Eq{eq:J1IntR} we see that, for small $x_0$,
\begin{align*}
\int_0^{\infty}dm \; \widetilde{\mathscr{L}}_s^{+}&(x_0,m;y_1-y_2,y_2) \sim x_0 
\end{align*}
and similarly, $\widetilde{J}_{s}(0,x_0;y)\sim x_0$  from \Eq{eq:J1 tilde final}. This means $\widetilde{B}_2\sim x_0^2$ for small $x_0$. 

Following a very similar calculation one can verify that $\widetilde{B}_1$ is also of order $x_0^2$ for small $x_0$. These are easy to see using the argument given in the remark below \Eq{eq:W1 tilde result}. 

The argument can be used to show that the diagram $C$ is also of order $x_0^2$. We have as well verified this explicitly using their amplitude 
\begin{align*}
&C_1(\tau,T)=\frac{1}{4D^2}\int_{0}^{\infty}\!\!\!\!dm\int_0^{\Lambda}\!\!dy_1dy_2 \int_{0}^{\tau}dr_1\int_{r_1}^{\tau}dr_2 \int_{0}^{\tau}dr_3 \cr
&\times \int_{\tau}^{T}dr_4 \; 
 e^{y_1(r_1-r_2)}  e^{y_2(r_3-r_4)}\llangle[\Big]\dot{x}(r_1)\dot{x}(r_2)\dot{x}(r_3)\dot{x}(r_4)\rrangle[\Big]_{m}
\end{align*}
and
\begin{align*}
&C_2(\tau,T)=\frac{1}{4D^2}\int_{0}^{\infty}\!\!\!\!dm\int_0^{\Lambda}\!\!dy_1dy_2 \int_{0}^{\tau}dr_1\int_{\tau}^{T}dr_2 \int_{\tau}^{T}dr_3 \cr
&\times \int_{r_3}^{T}dr_4 \; 
 e^{y_1(r_1-r_2)}  e^{y_2(r_3-r_4)}\llangle[\Big]\dot{x}(r_1)\dot{x}(r_2)\dot{x}(r_3)\dot{x}(r_4)\rrangle[\Big]_{m}
\end{align*}
as indicated in the diagram \fref{fig:diagram}.

\section{Amplitude of 2-loop diagrams for $t_{\rm max}$}
\label{app:two loop diagram for max}
All diagrams in \fref{fig:diagram} for distribution of $t_{\rm max}$ are of order $x_0^2$ for small $x_0$. Among these, the diagrams $E$ and $A$ contribute to the scaling term in \Eq{Pmax}, and the rest $D$, $B$, $C$, and $G$ contribute to the non-trivial function $\mathcal{F}^{\rm max}$.

\subsection{Diagrams for scaling term}

\subsubsection{Diagrams $E_1$ and $E_2$}
We begin with the diagram $E_2$ in \fref{fig:diagram}, whose amplitude for the problem of $t_{\rm max}$ is given by
\begin{align}
E_2(&\tau,T)=\frac{1}{8D^2}\int_{0}^{\infty}dm_1 dm_2\int_0^{\Lambda}dy_1dy_2 \cr
&\times \int_{\tau}^{T}dr_1\int_{r_1}^{T}dr_2 \int_{\tau}^{T}dr_3 \int_{r_3}^{T}dr_4 e^{y_1(r_1-r_2)}  \nn \\[1mm]
& \times  e^{y_2(r_3-r_4)}\llangle[\Big]\dot{x}(r_1)\dot{x}(r_2)\dot{x}(r_3)\dot{x}(r_4)\rrangle[\Big]_{(m_1,m_2)}\label{eq:E2 max formal}
\end{align}
with the angular brackets defined in \Eq{eq:angular bracket max}. Considering relative order of times $r$ we write the amplitude in three parts as indicated in \fref{fig:diagram E Last}. Their net amplitude can be written together as
\begin{align*}
&E_2(\tau,T)= \frac{2}{8D^2}\int_{0}^{\Lambda}dy_1 dy_2\int_{0}^{\infty}dm_1  Z^{+}_{\tau}(m_1,x_0) \times \cr & \int_{0}^{\infty}dm_2 \bigg[ J_{T-\tau}^{+}(x_0,m_2;-y_1,-y_2,y_2,y_1)+
 J_{T-\tau}^{+}(x_0,\cr & m_2;-y_1,-y_2,y_1,y_2)+J_{T-\tau}^{+}(x_0,m_2;-y_1,y_1,-y_2,y_2)\bigg],
\end{align*}
where the propagator $Z^+$ is in \Eq{eq:propagator free absorbing} and $J^+$ is an analogue of \eqref{eq:J2} with absorbing boundary. The prefactor $2$ is the degeneracy from interchange of pair of indices $(1,2)$ and $(3,4)$ in \fref{fig:diagram E Last}. 

A double Laplace transformation \Eq{eq:D transformation} of the amplitude is
\begin{align*}
&\widetilde{E}_2(\lambda,s)= \frac{1}{4D^2}\int_{0}^{\Lambda}dy_1 dy_2\int_{0}^{\infty}dm_1  \widetilde{Z}_{s+\lambda}^{+}(m_1,x_0) \times \cr & \int_{0}^{\infty}dm_2 \bigg[\widetilde{J}_{s}^{+}(x_0,m_2;-y_1,-y_2,y_2,y_1)+
\widetilde{J}_{s}^{+}(x_0,\cr & m_2;-y_1,-y_2,y_1,y_2)+\widetilde{J}_{s}^{+}(x_0,m_2;-y_1,y_1,-y_2,y_2)\bigg].
\end{align*}

Expression of $\tilde{Z}^+$ is in \Eq{eq:Z + tilde final} and integral of $\widetilde{J}^+$ is in \Eq{eq:J4 + integrate}. Using these results we get,  for small $x_0$,
\begin{equation*}
\widetilde{E}_2(\lambda,s)\simeq \frac{1}{2D}\frac{x_0^2}{\sqrt{s(s+\lambda)}}\int_0^{\Lambda} \frac{dy_1dy_2}{y_1^2 y_2^2}\;e(s,y_1,y_2)
\end{equation*}
with $e(s,y_1,y_2)$ in \Eq{eq:small e}.

Amplitude of the diagram $E_1$ for $t_{\rm max}$ is 
\begin{align}
E_1(\tau,T&)=\frac{1}{8D^2}\int_{0}^{\infty}dm_1 dm_2\int_0^{\Lambda}dy_1dy_2  \cr
\times &\int_{0}^{\tau}dr_1\int_{r_1}^{\tau}dr_2 \int_{0}^{\tau}dr_3 \int_{r_3}^{\tau}dr_4 \; 
  e^{y_1(r_1-r_2)}  \nn \\[1mm] \times & e^{y_2(r_3-r_4)}\llangle[\Big]\dot{x}(r_1)\dot{x}(r_2)\dot{x}(r_3)\dot{x}(r_4)\rrangle[\Big]_{(m_1,m_2)}.
\end{align}
Comparing with \Eq{eq:E2 max formal}, we see that for small $x_0$, the double Laplace transformation of the amplitude $E_1$ is
\begin{equation*}
\widetilde{E}_1(\lambda,s)\simeq \frac{1}{2D}\times \frac{x_0^2}{\sqrt{s(s+\lambda)}}\int_0^{\Lambda}\frac{ dy_1dy_2}{y_1^2 y_2^2}\;e(s+\lambda,y_1,y_2).
\end{equation*}

We note that amplitude of $\widetilde{E}_1$ and $\widetilde{E}_2$ for small $x_0$ are almost identical for both problems ($t_{\rm last}$ and $t_{\rm max}$). In terms of rescaled variables we get \Eq{eq:E tilde final max sum}.

\subsubsection{Diagram A}
Amplitude of the diagram $A$ in \fref{fig:diagram} for $t_{\rm max}$ is given by
\begin{align}
A(\tau,T)&=\frac{1}{8D^2}\int_{0}^{\infty}dm_1 dm_2\int_0^{\Lambda}dy_1dy_2 \cr
&\times \int_{0}^{\tau}dr_1\int_{r_1}^{\tau}dr_2 \int_{\tau}^{T}dr_3 \int_{r_3}^{T}dr_4 
 \; e^{y_1(r_1-r_2)}  \cr 
 \times & e^{y_2(r_3-r_4)}\llangle[\Big]\dot{x}(r_1)\dot{x}(r_2)\dot{x}(r_3)\dot{x}(r_4)\rrangle[\Big]_{(m_1,m_2)}
\end{align}
with the angular brackets defined in \Eq{eq:angular bracket max}. In terms of $J^+$ in \Eq{eq:J2}, we write
\begin{align*}
A(\tau,T)
=\frac{2}{8D^2}\int_{0}^{\Lambda} & dy_1 dy_2 \int_0^{\infty}dm_1 J_{\tau}^{+}(m_1,x_0;-y_1,y_1) \cr & \times \int_{0}^{\infty}dm_2 J_{T-\tau}^{+}(x_0,m_2;-y_2,y_2),
\end{align*}
where the prefactor $2$ is the degeneracy from the interchange of pair of indices $(1,2)$ and $(3,4)$. 

 The double Laplace transformation \Eq{eq:D transformation} of the amplitude can be written as
\begin{align*}
\widetilde{A}(\lambda,s)=\frac{1}{4 D^{2}}&\int_0^{\Lambda}dy_1dy_2\int_0^{\infty}dm_1 \widetilde{J}_{s+\lambda}^{+}(m_1,x_0;-y_1,y_1)\cr & \times \int_0^{\infty}dm_2 \, \widetilde{J}_{s}^{+}(x_0,m_2;-y_2,y_2).
\end{align*}

We use the results of integrals in \Eq{eq:J2 + tilde integral final full} and 
\begin{align}
\int_{0}^{\infty}& dm_1\widetilde{J}_{s}^{+}(m_1,m_2;y_1,y_2)=\frac{4D}{y_1y_2}\times \cr
&\sqrt{\frac{s+y_2}{s+y_1+y_2}}\left[e^{-m_2\sqrt{\frac{s+y_2}{D}}}-e^{-m_2\sqrt{\frac{s}{D}}}\right]\cr
&+\frac{4D}{(y_1+y_2)y_1}\left[e^{-m_2\sqrt{\frac{s}{D}}}-e^{-m_2\sqrt{\frac{s+y_1+y_2}{D}}}\right].\qquad 
\label{eq:B27}
\end{align}
Their derivation is in \aref{app:B}. Substituting the results, we get, for small $x_0$,
\begin{align*}
\widetilde{A}(\lambda,s)&\simeq \frac{x_0^2}{ D}\times \frac{1}{\sqrt{s(s+\lambda)}}\int_0^{\Lambda}\frac{dy_1dy_2}{y_1^2y_2^2}\times \nn \\[1mm] & \left(\sqrt{s+\lambda+y_1}-\sqrt{s+\lambda}\right)^2
\left(\sqrt{s+y_2}-\sqrt{s}\right)^2.
\end{align*}
In terms of re-scaled variables this give \Eq{eq:A tilde final max}.

\subsection{Non-trivial diagrams contributing to $\mathcal{F}^{\rm max}$}

\subsubsection{Diagram $D$}
Amplitude of the diagram  $D$ in \fref{fig:diagram} for $t_{\rm max}$ is given by
\begin{align}
D(\tau,&T)=\frac{1}{8D^2}\int_{0}^{\infty}dm_1 dm_2\int_0^{\Lambda}dy_1dy_2 \cr
& \times \int_{0}^{\tau}dr_1\int_{\tau}^{T}dr_2 \int_{0}^{\tau}dr_3 \int_{\tau}^{T}dr_4 
 \; e^{y_1(r_1-r_2)}  \nn \\[1mm]
 \times & e^{y_2(r_3-r_4)}  \llangle[\Big]\dot{x}(r_1)\dot{x}(r_2)\dot{x}(r_3)\dot{x}(r_4)\rrangle[\Big]_{(m_1,m_2)}
\end{align}
with the angular brackets defined in \Eq{eq:angular bracket max}.

Analysis for this amplitude is similar to the analysis in \aref{app:D last}. It is straightforward to get
\begin{align*}
D(\tau,T)
=& \frac{1}{4D^2}\int_{0}^{\Lambda}dy_1 dy_2 \; e^{-y_1\tau-y_2\tau} \cr & \times \int_{0}^{\infty}dm_1  J_{\tau}^{+}(m_1,x_0;-y_1,-y_2)\times \cr  \int_{0}^{\infty}dm_2 \bigg[J_{T-\tau}^{+}&(x_0,m_2;y_2,y_1)+J_{T-\tau}^{+}(x_0,m_2;y_1,y_2)\bigg]
\end{align*}
with $J^{+}$ in \Eq{eq:J2}. Taking the double Laplace transformation \Eq{eq:D transformation} we get
\begin{align*}
\widetilde{D}&(\lambda,s)
=\frac{1}{4D^2}\int_{0}^{\Lambda}dy_1 dy_2 \cr & \times \int_{0}^{\infty}dm_1  \widetilde{J}_{s+\lambda+y_1+y_2}^{+}(m_1,x_0;-y_1,-y_2) \cr & \times \left[\int_{0}^{\infty}dm_2 \widetilde{J}_{s}^{+}(x_0,m_2;y_2,y_1)+\widetilde{J}_{s}^{+}(x_0,m_2;y_1,y_2)\right].
\end{align*}
It is more convenient to write the expression in a symmetric form
\begin{align}
&\widetilde{D}(\lambda,s)=\frac{1}{8D^2}\int_{0}^{\Lambda}dy_1 dy_2\int_{0}^{\infty}dm_1 \bigg[\widetilde{J}_{s+\lambda+y_1+y_2}^{+}\cr & (m_1,x_0;-y_1,-y_2) +\widetilde{J}_{s+\lambda+y_1+y_2}^{+}(m_1,x_0;-y_2,-y_1) \bigg]\times \cr
&\int_{0}^{\infty}\!\!\!dm_2 \left[\widetilde{J}_{s}^{+}(x_0,m_2;y_1,y_2)+\widetilde{J}_{s}^{+}(x_0,m_2;y_2,y_1)\right].\quad \label{eq:D expr app formal}
\end{align}

For evaluating the expression we use the results for integrals in Eqs.~\eqref{eq:J2 + tilde integral final full}
and \eqref{eq:B27}. This leads to, for small $x_0$,
\begin{align*}
\int_{0}^{\infty}  dm_1 & \left[\widetilde{J}_{s}^{+}(m_1,x_0;y_1,y_2)+\widetilde{J}_{s}^{+}(m_1,x_0;y_2,y_1) \right]\cr 
\simeq & -\frac{x_0\;4\sqrt{D}}{y_1 y_2}\times \frac{\sqrt{s}}{\sqrt{s+y_1+y_2}}\times \nn \\[2mm] & \Big(\sqrt{s+y_1+y_2}-\sqrt{s+y_1}-\sqrt{s+y_2}+\sqrt{s}\Big)
\end{align*}
and an analogous formula \Eq{eq:J + int right sum asymptotic}.

More explicitly, for the integrals in \Eq{eq:D expr app formal} we get for small $x_0$,
\begin{align*}
\int_{0}^{\infty}& dm_1 \bigg\{\widetilde{J}_{s+\lambda+y_1+y_2}^{+}(m_1,x_0;-y_1,-y_2) + \cr &\qquad \qquad \widetilde{J}_{s+\lambda+y_1+y_2}^{+}(m_1, x_0;-y_2,-y_1) \bigg\}\nn \\[2mm]
&\simeq -\frac{2^{2}\sqrt{D}\;x_0}{y_1 y_2}\times \frac{\sqrt{s+\lambda+y_1+y_2}}{\sqrt{s+\lambda}}\times \bigg(\sqrt{s+\lambda}-\cr &\sqrt{s+\lambda+y_2} -\sqrt{s+\lambda+y_1} +\sqrt{s+\lambda+y_1+y_2}\bigg). 
\end{align*}
Using this with \Eq{eq:J + int right sum asymptotic} we get an explicit expression for $\widetilde{D}$ in \Eq{eq:D expr app formal}. For small $x_0$ limit, 
\begin{equation*}
\widetilde{D}(\lambda,s)\simeq \frac{1}{D}\times \frac{x_0^2}{\sqrt{s(s+\lambda)}}\int_{0}^{\Lambda}\frac{dy_1dy_2}{y_1^2 y_2^2}\;d(s,s+\lambda,y_1,y_2),
\end{equation*}
where we define
\begin{align}
 d(&s_1,s_2,y_1,y_2)=2\;\sqrt{s_1+y_1+y_2}\; \sqrt{s_2+y_1+y_2}\nn \\[1mm] \times & \Big(\sqrt{s_1+y_1+y_2}-\sqrt{s_1+y_1}-\sqrt{s_1+y_2}+\sqrt{s_1}\Big)
\cr \times&\Big(\sqrt{s_2+y_1+y_2}-\sqrt{s_2+y_1}-\sqrt{s_2+y_2}+\sqrt{s_2}\Big).\quad 
\end{align}

In terms of re-scaled variables, this gives the amplitude in \Eq{eq:diagram D max}.

\subsubsection{Diagram C}
One can see that for $t_{\rm max}$, amplitude of the diagrams $C_1$ in \fref{fig:diagram} is
\begin{align}
C_1(\tau,&T)=\frac{2}{8D^2}\int_{0}^{\Lambda}dy_1dy_2\int_{0}^{\infty}dm_1dm_2 \cr 
&\times \int_{0}^{\tau}dr_1\int_{r_1}^{\tau}dr_2\int_0^{\tau}dr_3 \int_{\tau}^{T}dr_4 \, e^{y_1(r_1-r_2)}\cr 
& \times e^{y_2(r_3-r_4)} \llangle[\Big]\dot{x}(r_1)\dot{x}(r_2)\dot{x}(r_3)\dot{x}(r_4)\rrangle[\Big]_{(m_1,m_2)}\qquad
\end{align}
with the angular brackets defined in \Eq{eq:angular bracket max}. (The prefactor $2$ is the degeneracy from interchange of pair of indices (1,2) and (3,4).)
The amplitude can be expressed in terms of $J^+$ in \Eq{eq:J2}, giving,
\begin{align}
C_1(\tau,T&) =\frac{1}{4D^2}\int_{0}^{\Lambda}dy_1dy_2\int_{0}^{\infty}dm_1 dm_2 \, e^{-y_2\tau} \nn \\[1mm]  \mathscr{I}_{\tau}^{+}&(m_1,x_0;-y_1,y_1,-y_2) \, J_{T-\tau}^{+}(x_0,m_2;y_2) ,\label{eq: C1 formal intermediate}\qquad
\end{align}
where we define
\begin{align}
\mathscr{I}_{\tau}^{+}(&m_1,m_2;y_1,y_2,y_3)=\int_{0}^{\tau}dr_1\int_{r_1}^{\tau}dr_2\int_0^{\tau}dr_3 \nn \\[1mm] &  e^{-y_1r_1-y_2r_2-y_3r_3}\langle \dot{x}(r_1)\dot{x}(r_2) \dot{x}(r_3)\rangle_{(m_1,m_2)}^{+},
\label{eq:I3+}
\end{align}
for $m_1>0$ and $m_2>0$. For an explicit evaluation one can use that $\mathscr{I}_{\tau}^{+}$ is related to $J^+$ (an absorbing-boundary-analogue of \Eq{eq:J2}) by
\begin{align}
\mathscr{I}_{\tau}^{+}(m_1,m_2;&y_1,y_2,y_3)=J_{\tau}^{+}(m_1,m_2;y_1,y_2,y_3) \nn \\[1mm] +J_{\tau}^{+}(m_1,&m_2;y_1,y_3,y_2)+J_{\tau}^{+}(m_1,m_2;y_3,y_1,y_2).\qquad\qquad 
\label{eq:150}
\end{align}

A double Laplace transform \Eq{eq:D transformation} of the amplitude in \Eq{eq: C1 formal intermediate} gives
\begin{align}
 \widetilde{C}_1&(\lambda,s)=\frac{1}{4D^2}\int_{0}^{\Lambda}dy_1dy_2\int_{0}^{\infty}dm_1dm_2\cr  \times & \widetilde{\mathscr{I}}_{s+\lambda+y_2}^{+}(m_1,x_0;-y_1,y_1,-y_2)\widetilde{J}_{s}^{+}(x_0,m_2;y_2).\quad \label{eq:C max formal expression intermediate}
\end{align}

To evaluate the integrals, we use a result from \Eq{eq:J1IntR} which, for small $x_0$, gives
\begin{equation}
\int_{0}^{\infty}dm_2\, \widetilde{J}_{s}^{+}(x_0,m_2;y_2)\simeq \frac{2x_0}{\sqrt{s}}\left(\frac{\sqrt{s+y_2}-\sqrt{s}}{y_2} \right).\quad \label{eq:G13}
\end{equation}
Similarly, using \Eq{eq:150} and the integration result \Eq{eq:left integral J 3 small x0}, for small $x_0$, we get
\begin{align}
\int_{0}^{\infty}dm_1 \widetilde{\mathscr{I}}_{s+\lambda+y_2}^{+}&(m_1,x_0;-y_1,y_1,-y_2)\cr  \simeq  - & \frac{4 D x_0}{\sqrt{s+\lambda}\; y_1^2 y_2} \times c(s+\lambda,y_2,y_1),\qquad
\label{eq:154}
\end{align}
where we define 
\begin{align}
c(s,y_1,y_2)=\sqrt{s+y_1}\; &\bigg(\sqrt{s+y_1+y_2}-\sqrt{s+y_1}\cr
&\qquad\quad -\sqrt{s+y_2}+\sqrt{s}\bigg)^2.
\label{eq:fsy1y2}
\end{align}

Using Eqs.~\eqref{eq:G13} and \eqref{eq:154} for the integrals in the expression \Eq{eq:C max formal expression intermediate} we get the amplitude
\begin{align*}
\widetilde{C}_1(\lambda,s)=\frac{2 x_0^2}{D}&\times \frac{1}{\sqrt{s(s+\lambda)}}\int_{0}^{\Lambda}\frac{dy_1dy_2}{\;y_1^2\;y_2^2}\cr & \bigg(\sqrt{s}-\sqrt{s+y_1}\bigg) \, c(s+\lambda,y_1,y_2),
\end{align*}
for small $x_0$, where we exchanged the dummy variables $y_1$ and $y_2$.

Analysis for the diagram $C_2$ in \fref{fig:diagram} is similar. It's amplitude
\begin{align}
C_2(\tau,&T)=\frac{2}{8D^2}\int_{0}^{\Lambda}dy_1dy_2\int_{0}^{\infty}dm_1dm_2 \cr 
 \times &\int_{0}^{\tau}dr_1\int_{\tau}^{T}dr_2\int_{\tau}^{T}dr_3 \int_{r_3}^{T}dr_4 \; e^{y_1(r_1-r_2)}\cr 
\times & e^{y_2(r_3-r_4)}  \llangle[\Big]\dot{x}(r_1)\dot{x}(r_2)\dot{x}(r_3)\dot{x}(r_4)\rrangle[\Big]_{(m_1,m_2)}\quad 
\end{align}
and the asymptotics for the corresponding double Laplace transformation for small $x_0$ is 
\begin{align}
 \widetilde{C}_2(\lambda,s)\simeq &\frac{2 x_0^2}{D}\times \frac{ 1}{\sqrt{s(s+\lambda)}}\int_{0}^{\Lambda}\frac{dy_1dy_2}{\;y_1^2\;y_2^2}\cr \times &  \bigg(\sqrt{s+\lambda}-\sqrt{s+\lambda+y_1}\bigg)c(s,y_1,y_2).\quad \label{eq:G17}
\end{align}

Adding the results for $\widetilde{C}_1$ and $\widetilde{C}_2$ gives \Eq{eq:C tilde final expression scaled} in terms of re-scaled variables.

\subsubsection{Diagram B}
For $t_{\rm max}$, amplitude of $B_1$ and $B_2$ in \fref{fig:diagram} is
\begin{align}
B_1(&\tau,T)=\frac{1}{2D}\int_{0}^{\infty}dm_1 dm_2\int_0^{\Lambda}dy_1 dy_2\int_0^{\tau}dr_1\int_{r_1}^{\tau}ds \cr
 \times &\int_{\tau}^{T}dr_2  \;e^{y_1(r_1-s)}  e^{y_2(s-r_2)}\llangle[\Big]\dot{x}(r_1)\dot{x}(r_2)\rrangle[\Big]_{(m_1,m_2)}\quad 
\end{align}
and
\begin{align}
B_2(&\tau,T)=\frac{1}{2D}\int_{0}^{\infty}dm_1 dm_2\int_0^{\Lambda}dy_1 dy_2\int_{0}^{\tau}dr_1\int_{\tau}^{T}ds \cr
\times & \int_{s}^{T}dr_2 e^{y_1(r_1-s)}  \; e^{y_2(s-r_2)}\llangle[\Big]\dot{x}(r_1)\dot{x}(r_2)\rrangle[\Big]_{(m_1,m_2)}\label{eq:G9}
\end{align}
with the angular brackets defined in \Eq{eq:angular bracket max}. 
Their difference is in the limit of the time integrals.

These expressions can be written in terms of $J^+$ in \Eq{eq:J2}. We write
\begin{align}
B_1(\tau,T)&=\frac{1}{2D}\int_0^{\Lambda}dy_1dy_2 \int_{0}^{\infty}dm_1 dm_2\times \label{eq:B1 max formal expression intermediate} \\[1mm]   \mathbb{L}_{\tau}^{+}&(m_1,x_0;-y_1,y_1-y_2)\; e^{-y_2\tau} J_{T-\tau}^{+}(x_0,m_2;y_2),\nn
\end{align}
where we define
\begin{align*}
\mathbb{L}_{\tau}^{+}(m_1,m_2;y_1,y_2)=&\int_0^{\tau}dr_1\int_{r_1}^{\tau}dr_2e^{-y_1r_1-y_2r_2}\langle \dot{x}(r_1)\rangle^{+}.
\end{align*}
This function can be evaluated in terms of $J^+$  in \Eq{eq:J2},
\begin{align}
\mathbb{L}_{t}^{+}(m_1,m_2;y_1,y_2) =\frac{1}{y_2}&\bigg\{ J_{t}^{+}(m_1,m_2;y_1+y_2)\cr
-&e^{-y_2\tau}J_{t}^{+}(m_1,m_2;y_1)\bigg\}.\quad
\end{align}

In a similar way, we write \Eq{eq:G9} by
\begin{align}
B_2&(\tau,T)=\frac{1}{2D}\int_0^{\Lambda} dy_1 dy_2  \int_0^{\infty}dm_1 dm_2\times \label{eq:B2 max formal expression intermediate} \\[1mm]
&   J_{\tau}^{+}(m_1,x_0;-y_1)\; e^{-y_1\tau} \;  \mathscr{L}^{+}_{T-\tau}(x_0,m_2;y_1-y_2,y_2) \nn
\end{align}
with $J^{+}$ defined in \Eq{eq:J2} and $\mathscr{L}^{+}$ defined in \Eq{eq:L prop}. The last quantity can also be expressed in terms of $J^{+}$ by their analogue of \Eq{eq:mathscr L in J} with absorbing boundary.

A double Laplace transformation \eqref{eq:D transformation} of the amplitudes Eqs.~\eqref{eq:B1 max formal expression intermediate} and \eqref{eq:B2 max formal expression intermediate} are
\begin{align*}
\widetilde{B}_1(\lambda,&s)=\frac{1}{2D}\int_0^{\Lambda}dy_1dy_2 \int_{0}^{\infty}dm_1 dm_2 \nn \\[1mm]
&\times  \widetilde{\mathbb{L}}^{+}_{s+\lambda+y_2}(m_1,x_0;-y_1,y_1-y_2) \; \widetilde{J}_{s}^{+}(x_0,m_2;y_2)
\end{align*}
and
\begin{align*}
\widetilde{B}_2(\lambda,s)&=\frac{1}{2D}\int_0^{\Lambda}dy_1dy_2 \int_0^{\infty}dm_1 dm_2 \nn \\[1mm] \times & \widetilde{J}_{s+\lambda+y_1}^{+}(m_1,x_0;-y_1) \; \widetilde{\mathscr{L}}^{+}_{s}(x_0,m_2;y_1-y_2,y_2),
\end{align*}
where
\begin{align*}
\widetilde{\mathbb{L}}^{+}_{s}(m_1,m_2;&y_1,y_2)\cr  = \frac{1}{y_2}\bigg\{ \widetilde{J}_{s}^{+}&(m_1,m_2;y_1+y_2)-\widetilde{J}_{s+y_2}^{+}(m_1,m_2;y_1)\bigg\}
\end{align*}
and
\begin{align*}
\widetilde{\mathscr{L}}^{+}_{s}&(m_1,m_2;y_1,y_2)\cr & =\frac{1}{y_1}\bigg\{ \widetilde{J}_1^{+}(m_1,m_2,y_2,s)-\widetilde{J}_1^{+}(m_1,m_2,y_1+y_2,s)\bigg\}.
\end{align*}

For an explicit evaluation of the amplitudes we use the formula \eqref{eq:J1IntL} that for small $x_0$, leads to
\begin{align*}
 \int_0^{\infty}&dm_1\widetilde{\mathbb{L}}^{+}_{s+y_2}(m_1,x_0;-y_1,y_1-y_2)\cr &\simeq \frac{2 x_0}{\sqrt{s}}\times \frac{(\sqrt{s+y_1}-\sqrt{s})y_2-(\sqrt{s+y_2}-\sqrt{s})y_1}{y_1y_2(y_1-y_2)}.
\end{align*}

Similarly, using \Eq{eq:J2 + tilde integral final full} we get, for small $x_0$,
\begin{align*}
\int_0^{\infty}&dm_2  \; \widetilde{\mathscr{L}}^{+}_{s}(x_0,m_2;y_1-y_2,y_2)\cr &\simeq \frac{2 x_0}{\sqrt{s}}\times \frac{(\sqrt{s+y_2}-\sqrt{s})y_1-(\sqrt{s+y_1}-\sqrt{s})y_2}{y_1y_2(y_1-y_2)}.
\end{align*}

Using these asymptotics, along with Eqs.~\eqref{eq:J1IntL} and \eqref{eq:J1IntR} we get the amplitudes, for small $x_0$,
\begin{align*}
 &\widetilde{B}_1(\lambda,s)\simeq \frac{2 x_0^2}{D\sqrt{s(s+\lambda)}}\int_0^{\Lambda}dy_1dy_2\; \frac{(\sqrt{s+y_2}-\sqrt{s})}{y_1 y_2^2 (y_1-y_2)}\times \nn \\[1mm] &\left[(\sqrt{s+\lambda+y_1}-\sqrt{s+\lambda})y_2-(\sqrt{s+\lambda+y_2}-\sqrt{s+\lambda})y_1\right]\quad
\end{align*}
and
\begin{align*}
\widetilde{B}_2(\lambda,s)\simeq &\frac{2 x_0^2}{D\sqrt{s(s+\lambda)}}\int_0^{\Lambda}\mquad dy_1dy_2\;\frac{(\sqrt{s+\lambda+y_2}-\sqrt{s+\lambda})}{y_1 y_2^2 (y_1-y_2)}\nn \\[1mm] \times &\Big[(\sqrt{s+y_1}-\sqrt{s})y_2-(\sqrt{s+y_2}-\sqrt{s})y_1\Big],
\end{align*}
where in the expression for $\widetilde{B}_2$ we exchanged the dummy variables $y_1$ and $y_2$.

Sum of the two amplitudes has a simpler expression, given by
\begin{align*}
\widetilde{B}(\lambda,s)&=\widetilde{B}_1(\lambda,s)+\widetilde{B}_2(\lambda,s)\cr 
&=\frac{1}{D}\times \frac{x_0^2}{\sqrt{s(s+\lambda)}}\times b(s,s+\lambda),
\end{align*}
where we define
\begin{align*}
 b(s_1,s_2)&=2\int_0^{\Lambda}\frac{dy_1dy_2}{y_1^2 y_2^2(y_1-y_2)} \times \cr
\bigg[&(\sqrt{s_1+y_1}-\sqrt{s_1}) (\sqrt{s_2+y_1}-\sqrt{s_2})y_2^2 \cr &-
(\sqrt{s_1+y_2}-\sqrt{s_1})(\sqrt{s_2+y_2}-\sqrt{s_2})y_1^2\bigg].
\end{align*}
In terms of re-scaled variables this result gives \Eq{eq:B tilde sub final expression rescaled}.
\begin{remark}
We have numerically verified the asymptotic divergence for large $\Lambda$, 
\begin{align}
b(s_1,s_2)= & -2 \ln^2\left(\Lambda \right)+\frac{2 \ln \Lambda}{\sqrt{s_1s_2}}\bigg[2\left(\sqrt{s_1}+\sqrt{s_2}\right)^2\times \nn \\[1mm]
& \ln\left( \sqrt{s_1}+\sqrt{s_2}\right) - s_1 \ln (s_1) - s_2\ln(s_2)\nn \\[1mm]
& + 2\sqrt{s_1 s_2}\left(1-2\ln 2 \right)\bigg]+\cdots .
\end{align}
\end{remark}

\subsubsection{Diagrams $G_1$ and $G_2$}

For $t_{\rm max}$, amplitude of $G_1$ and $G_2$ in \fref{fig:diagram} are
\begin{align*}
G_1(\tau,T)=&\frac{1}{2D}\int_0^{\Lambda}dy_1 dy_2 \int_{0}^{\infty}dm_1 dm_2\int_0^{\tau}dr_1\int_{r_1}^{\tau}ds \nn \\[1mm]
\times & \int_{s}^{\tau}dr_2  \, e^{y_1(r_1-s)}  e^{y_2(s-r_2)}\llangle[\Big]\dot{x}(r_1)\dot{x}(r_2)\rrangle[\Big]_{(m_1,m_2)}
\end{align*}
and
\begin{align*}
G_2(\tau,T)=&\frac{1}{2D} \int_0^{\Lambda}dy_1 dy_2 \int_{0}^{\infty}dm_1 dm_2\int_{\tau}^{T}dr_1\int_{r_1}^{T}ds  \nn \\[1mm]
\times & \int_{s}^{T}dr_2 \, e^{y_1(r_1-s)}  e^{y_2(s-r_2)}\llangle[\Big]\dot{x}(r_1)\dot{x}(r_2)\rrangle[\Big]_{(m_1,m_2)}
\end{align*}
with the angular brackets defined in \Eq{eq:angular bracket max}.

These expressions can be written as
\begin{align*}
G_1(\tau,T)=&\frac{1}{2D}\int_0^{\Lambda}dy_1dy_2 \int_{0}^{\infty}dm_1 dm_2 \nn \\[1mm] \times &  \mathscr{L}_{\tau}^{+}(m_1,x_0;-y_1,y_1-y_2,y_2)  Z^{+}_{T-\tau}(x_0,m_2) 
\end{align*}
and
\begin{align*}
G_2(\tau,T)=&\frac{1}{2D}\int_0^{\Lambda}dy_1dy_2 \int_0^{\infty}dm_1 dm_2 \nn \\[1mm] 
\times & Z^{+}_{\tau}(m_1,x_0) \mathscr{L}^{+}_{T-\tau}(x_0,m_2;-y_1,y_1-y_2,y_2),
\end{align*}
where $Z_t^{+}$ is in \Eq{eq:propagator free absorbing} and $\mathscr{L}^{+}$ is an analogue of \eqref{eq:L prop} in presence of absorbing boundary.

A double Laplace transformation \eqref{eq:D transformation} of the amplitudes are
\begin{align}
\widetilde{G}_1(\lambda,&s)=\frac{1}{2D}\int_0^{\Lambda}dy_1dy_2 \int_{0}^{\infty}dm_1 dm_2 \nn \\[1mm] 
\times & \widetilde{\mathscr{L}}_{s+\lambda}^{+}(m_1,x_0;-y_1,y_1-y_2,y_2) \widetilde{Z}_s^{+}(x_0,m_2) \label{eq:G1 max formal expression}
\end{align}
and
\begin{align}
\widetilde{G}_2(\lambda,&s)=\frac{1}{2D}\int_0^{\Lambda}dy_1dy_2\int_0^{\infty}dm_1 dm_2 \nn \\[1mm] 
\times & \widetilde{Z}^{+}_{s+\lambda}(m_1,x_0) \widetilde{\mathscr{L}}_{s}^{+}(x_0,m_2;-y_1,y_1-y_2,y_2),\label{eq:G2 max formal expression}
\end{align}
where the Laplace transformation of $\mathscr{L}^{+}$ is expressed in terms of $\widetilde{J}^{+}$ in an analogous relation of \Eq{eq:mathscr L in J}. From this relation and using the results in Eqs.~\eqref{eq:J2 + tilde integral final full} and \eqref{eq:B27}) we see that
\begin{align*}
\int_0^{\infty}dm_2 &\;\widetilde{\mathscr{L}}_{s}^{+}(x_0,m_2;-y_1,y_1-y_2,y_2)  \cr
=& \int_{0}^{\infty}dm_1 \widetilde{\mathscr{L}}_{s}^{+}(m_1,x_0;-y_1,y_1-y_2,y_2)
\end{align*}
with an expression for the latter in \Eq{eq:L scr 3 tilde integral right}. This gives
\begin{align}
\int_0^{\infty}\mquad dm \; \widetilde{\mathscr{L}}_{s}^{+}&(x_0, m;-y_1,y_1-y_2,y_2) \cr
&=\frac{2 D}{\sqrt{s}\; y_1^2y_2^2} \times h^{+}_{s}\left(\frac{x_0}{\sqrt{D}},y_1,y_2 \right)
\end{align}
with $h_s^{+}$ in \Eq{eq:h + full expr}.

Result for the integral of $\widetilde{Z}^{+}$ is in \Eq{eq:id Z plus int}. Using these results in \Eq{eq:G1 max formal expression} we get
\begin{align}
\widetilde{G}_{1}(\lambda,s)&=\mathfrak{h}(s,s+\lambda), \\
\widetilde{G}_{2}(\lambda,s)&=\mathfrak{h}(s+\lambda,s),
\end{align}
where
\begin{align}
\mathfrak{h}(s_1,s_2) = & \frac{\left(1-e^{-x_0\sqrt{\frac{s_1}{D}}}\right) }{s_1\sqrt{s_2}}\cr
 & \times  \int_0^{\Lambda}\frac{dy_1dy_2}{y_1^2 y_2^2} \; h^{+}_{s_2}\left(\frac{x_0}{\sqrt{D}},y_1,y_2 \right) .
\end{align}

For small $x_0$, using the asymptotic \Eq{eq:h+ h asymptotics} we get
\begin{equation}
\widetilde{G}_{1}(\lambda,s)\simeq \frac{x_0^2}{D \sqrt{s(s+\lambda)}}\times g(s+\lambda)
\label{eq:G1 max final explicit expressions}
\end{equation}
and
\begin{equation}
\widetilde{G}_{2}(\lambda,s)\simeq \frac{x_0^2}{D \sqrt{s(s+\lambda)}}\times g(s)\label{eq:G2 max final explicit expressions}
\end{equation}
with $g(x)$ defined in \Eq{eq:small g definition}. Beside the $x_0^2$ pre-factor, amplitudes are similar to asymptotics in  Eqs.~\eqref{eq:G1 max final explicit expressions} and \eqref{eq:G2 max final explicit expressions}  for $t_\text{last}$.

In terms of re-scaled variables, we get \Eq{eq:F tilde final max}.

\section{Expression for $\Psi^{\rm max}$}
\label{app:R max}
The expression for $\Psi^{\rm max}$ in \Eq{eq:F2 max exact integral} can be written as
\begin{align}
\Psi^{\rm max}\left(y_1,y_2,z\right)&= \mathfrak{d} + \mathfrak{b} + \mathfrak{c} - \mathfrak{a},\label{eq:Psimax final explicit expression}
\end{align}
where the terms on the right hand side are associated to the amplitudes in \Eq{eq:F2 tilde max final} and given by
\begin{widetext}
\begin{align}
\mathfrak{d}=&\sqrt{y_1+y_2+1} \left(\sqrt{y_1+y_2+1}-\sqrt{y_1+1}-\sqrt{y_2+1}+1\right) \cr
&\times\bigg\{-\sqrt{y_1+y_2+1} \left(\sqrt{y_1+y_2+1}-\sqrt{y_1+1}-\sqrt{y_2+1}+1\right)+\sqrt{| y_1+y_2-z| } \cr
&\times \bigg[\sqrt{| z-y_1| }\; \Big(\Theta (z-y_1)-\Theta (y_1+y_2-z)\Big)+\sqrt{| z-y_2| } \;\Big(\Theta (z-y_2)-\Theta (y_1+y_2-z)\Big)\cr
&\qquad+\sqrt{| y_1+y_2-z| } \; \Big(\Theta (y_1+y_2-z)-\Theta (z-y_1-y_2)\Big)+\sqrt{| z| } \;\Big(\Theta (y_1+y_2-z)-\Theta (z)\Big)\bigg]\bigg\},
\end{align}
\begin{align}
\mathfrak{b} = \frac{1}{y_1-y_2}\bigg\{&y_2^2 \left(\sqrt{y_1+1}-1\right) \left(\sqrt{y_1-z}\; \Theta (y_1-z)-\sqrt{y_1+1}+1\right)\cr
&-y_1^2 \left(\sqrt{y_2+1}-1\right) \left(\sqrt{y_2-z}\; \Theta (y_2-z)-\sqrt{y_2+1}+1\right)\bigg\},
\end{align}
\begin{align}
\mathfrak{c} = & \sqrt{y_1+1} \left(\sqrt{y_1+y_2+1}-\sqrt{y_1+1}-\sqrt{y_2+1}+1\right)^2 \left(-\sqrt{y_1-z} \;\Theta (y_1-z)+\sqrt{y_1+1}-1\right)\cr
& - \left(1-\sqrt{y_1+1}\right) \bigg\{\sqrt{y_1+1} \left(\sqrt{y_1+y_2+1}-\sqrt{y_1+1}-\sqrt{y_2+1}+1\right)^2 + \sqrt{| y_1-z| }\; \Theta (z-y_1) \cr
&\times \Theta (y_2-z)\; \Theta (y_1+y_2-z)\bigg[\left(\sqrt{y_1+y_2-z}-\sqrt{z-y_1}-\sqrt{y_2-z}+\sqrt{z}\right)^2-2 \left(\sqrt{z}-\sqrt{z-y_1}\right)^2\bigg] \cr
& + \sqrt{| y_1-z| } \; \Theta (y_1+y_2-z) \bigg[\Theta (y_1-z) \Theta (y_2-z) \left(z-\left(\sqrt{y_1+y_2-z}-\sqrt{y_1-z}-\sqrt{y_2-z}\right)^2\right)\cr
&+\Theta (z-y_1)\;  \Theta (y_2-z) \left(\left(\sqrt{z}-\sqrt{z-y_1}\right)^2-\left(\sqrt{y_2-z}-\sqrt{y_1+y_2-z}\right)^2\right)\nn \\[1mm]
&-2 \Theta (z-y_1)\;  \Theta (z-y_2) \sqrt{y_1+y_2-z} \left(\sqrt{z-y_1}+\sqrt{z-y_2}-\sqrt{z}\right)\nn \\[1mm]
&+\Theta (y_1-z)\; \Theta (z-y_2) \left(\left(\sqrt{z}-\sqrt{z-y_2}\right)^2-\left(\sqrt{y_1-z}-\sqrt{y_1+y_2-z}\right)^2\right) \bigg] \bigg\},
\end{align}
and
\begin{align}
\mathfrak{a} =& \left(1-\sqrt{y_1+1}\right) \left(1-\sqrt{y_2+1}\right)^2 \bigg[1-\sqrt{y_1+1}+\sqrt{y_1-z} \;\Theta (y_1-z)\bigg] +\left(1-\sqrt{{y_1}+1}\right) \left(1-\sqrt{y_2+1}\right) \cr
& \times \bigg[\sqrt{y_1-z} \;\sqrt{y_2-z} \;\Theta (y_1-z)\; \Theta (y_2-z)-\Big(\sqrt{z}-\sqrt{z-y_1} \;\Theta (z-y_1)\Big)\Big(\sqrt{z}-\sqrt{z-y_2}\; \Theta (z-y_2)\Big)\bigg]\cr
& +\left(1-\sqrt{y_1+1}\right) \bigg[y_2 \sqrt{y_1-z}\; \Theta (y_1-z)- 2 \sqrt{z} \sqrt{y_1-z} \;\Theta (y_1-z) \Big(\sqrt{z}-\sqrt{z-y_2} \;\Theta (z-y_2)\Big)\cr
& \qquad \qquad -2 \sqrt{z} \sqrt{y_2-z}\; \Theta (y_2-z)\Big(\sqrt{z}-\sqrt{z-y_1}\; \Theta (z-y_1)\Big)\bigg].
\end{align}
Here $\Theta(x)$ is the Heaviside step function. These expressions are also given in the supplemental Mathematica notebook \cite{Mathematica} for their numerical evaluation.
\end{widetext}

\section{Two-loop diagrams for distribution of $t_{\rm pos}$}
\label{app:two loop positive}
\begin{figure}
\includegraphics[width=0.43\textwidth]{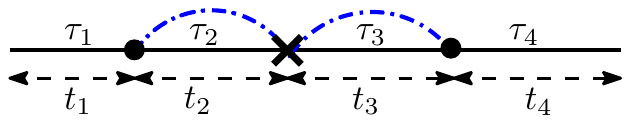}
\caption{An illustration for a change of variables in the amplitude \Eq{eq:D positive app intermediate} of the diagram $D$ in  \fref{fig:diagramPositive} to write the expression in \Eq{eq:D psitive in Zbb}. Inside each time window $t_i$ the process is conditioned to stay a net $\tau_i$ amount of time on positive side. \label{fig:D t illustration positive}}
\end{figure}

Among the two diagrams in \fref{fig:diagramPositive} which contribute to second order, the diagram $D$ is simpler to evaluate. Corresponding amplitude is in \Eq{eq:D positive formal}, which can be expressed in terms of conditional propagator \Eq{eq:conditional propagator definition} using the correlation in 
\Eq{eq:two-point correlation conditional}. 
\begin{align}
D(\tau,&T)=\frac{2^2 D^2}{2D} \int_0^{\Lambda} dy_1 dy_2 \int_0^T dr_1 \int_{r_1}^T  ds
\int_{s}^T dr_2 \nn \\[1mm]
\times & \int\limits_0^{r_1}d\tau_1\int\limits_0^{r_2-r_1}  d\tau_2
\int\limits_0^{T-r_2}  d\tau_3 \int\limits_{-\infty}^{\infty}  dx_1\, dx_2\, dm \nn \\[2mm]
\times & \delta(\tau-\tau_1-\tau_2-\tau_3)\, e^{y_1(r_1-s)} e^{y_2(s-r_2)}\, \mathbb{Z}_{r_1}(0,x_1\vert \tau_1) \nn \\[2mm]
\times &   \partial_{x_1}\mathbb{Z}_{r_2-r_1}(x_1,x_2\vert \tau_2)\, \partial_{x_2}\mathbb{Z}_{T-r_2}(x_2,m\vert \tau_3).\label{eq:D positive app intermediate}
\end{align}

For reasons that will be clear shortly, we make a change of variables (see illustration in \fref{fig:D t illustration positive}), and write
\begin{align}
D(\tau,T)&=2D\int_0^{\Lambda}dy_1 dy_2\int_0^{\infty}dt_1dt_2 dt_3dt_4\cr
\times \int_0^{t_1} & d\tau_1  \int_0^{t_2}d\tau_2 \int_0^{t_3}d\tau_3 \int_0^{t_4}d\tau_4 \int_{-\infty}^{\infty}dx_1 dx_2 dx_3 dm \nn \\[2mm]
\times &\delta(T-t_1-t_2-t_3-t_4)\;\delta(\tau-\tau_1-\tau_2-\tau_3-\tau_4) \nn \\[2mm]
\times &e^{-y_1 t_2 -y_2 t_3}\; \mathbb{Z}_{t_1}(0,x_1 \vert \tau_1) \; \partial_{x_1}\mathbb{Z}_{t_2}(x_1,x_2\vert \tau_2)  \nn \\[2mm]
\times & \mathbb{Z}_{t_3}(x_2,x_3 \vert \tau_3) \; \partial_{x_3}\mathbb{Z}_{t_4}(x_3,m \vert \tau_4),
\label{eq:D psitive in Zbb}
\end{align}
where in the last two lines of the expression we used $\mathbb{Z}_{t_2+t_3}(x_1,x_3\vert \tau_2+\tau_3)=\int dx_2 \mathbb{Z}_{t_2}(x_1,x_2\vert \tau_2)\mathbb{Z}_{t_3}(x_2,x_3 \vert \tau_3)$.

A double Laplace transformation \Eq{eq:D transformation} of the amplitude gives a simpler expression
\begin{align*}
\widetilde{D}(\lambda,s)=& 2D\int_{0}^{\Lambda}dy_1dy_2\int_{-\infty}^{\infty}dx_1 dx_2 dx_3 dm\;\widetilde{\mathbb{Z}}_{s}(0,x_1 \vert \lambda)  \nn \\[2mm]
\times \partial_{x_1} &\widetilde{\mathbb{Z}}_{s+y_1}(x_1,x_2 \vert \lambda)\;\widetilde{\mathbb{Z}}_{s+y_2}(x_2,x_3\vert \lambda)\; \partial_{x_3}\widetilde{\mathbb{Z}}_{s}(x_3,m \vert \lambda)
\end{align*}
with $\widetilde{\mathbb{Z}}$ defined in \Eq{eq:Z tilde conditional final}.

Results for spatial integration of $\widetilde{\mathbb{Z}}_s$ are derived in \aref{app:identities for conditional propagator} and successively using them we get (a lengthy but straightforward algebra) an explicit expression for the amplitude.
\begin{align}
\widetilde{D}(\lambda,s)&=\frac{2}{\sqrt{s(s+\lambda)}\left(\sqrt{s}+\sqrt{s+\lambda}\right)}\int_0^{\Lambda} \frac{dy_1dy_2}{y_1 y_2}\nn \\[1mm]
& \qquad \times \left\{\frac{y_2\, \mathfrak{h}(1,z,y_1)}{(y_2-y_1)}+\frac{y_1\, \mathfrak{h}(1,z,y_2)}{(y_1-y_2)}\right\},\label{eq:D tilde positive final}
\end{align}
where $\mathfrak{h}(s_1,s_2,y)$ is defined in \Eq{eq:h positive}. In terms of re-scaled variables, \Eq{eq:D tilde positive final} gives \Eq{eq:D tilde positive final rescaled}.

\begin{figure}
\includegraphics[width=0.47\textwidth]{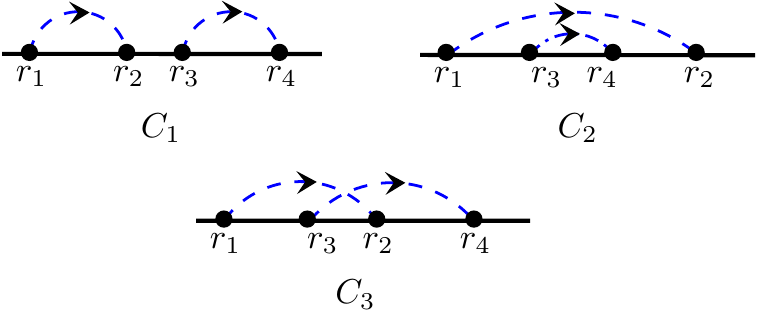}
\caption{The diagram $C$ in \fref{fig:diagramPositive} is split into three parts according to relative position of the loops. For these diagrams we choose $r_2>r_1$ and $r_4>r_3$, as indicated by the arrowheads. \label{fig:C123 positive}}
\end{figure}

For the diagram $C$ in \fref{fig:diagramPositive}, we write the amplitude \eqref{eq:C positive formal} in three parts according to the order of time variables (associated diagrams are indicated in \fref{fig:C123 positive}). For example, amplitude of diagram $C_1$ is
\begin{align}
C_1(\tau,T)=&\frac{2}{8D^2} \int_{-\infty}^{\infty} dm \int_0^{\Lambda} dy_1 dy_2 \int_0^T dr_1 \nn \\[1mm]
\times & \int_{r_1}^T  dr_2  \int_{r_2}^T dr_3 \int_{r_3}^T  dr_4 \; e^{y_1(r_1-r_2)} e^{y_2(r_3-r_4)} \nn \\[2mm]
&\times \llangle[\big] \dot{x}(r_1)\dot{x}(r_{2})\dot{x}(r_3)\dot{x}(r_{4}) \rrangle[\big]_{(0,m)},
\label{eq:C1 positive formal}
\end{align}
where the pre-factor $2$ is the degeneracy for exchange of pairs $(r_1,r_2)$ and $(r_3,r_4)$ for the diagram $C_1$ in \fref{fig:C123 positive}.

Similar to the diagram $D$, these amplitudes can be expressed in terms of conditional propagator \eqref{eq:conditional propagator definition}. The four point correlation in the conditional case is given by, for $r_1<  r_2 < r_3 <r_4< T$,
\begin{align}
&\llangle \dot{x}(r_1)\dot{x}(r_{2})\dot{x}(r_{3})\dot{x}(r_{4}) \rrangle_{(m_1,m_2)}= 2^4D^4 
\int_{-\infty}^{\infty}\mquad dx_1dx_2  \times \cr & dx_3dx_4\int_0^{r_1}d\tau_1 \int_0^{r_2-r_1}\mquad d\tau_2 \int_0^{r_3-r_2}\mquad d\tau_3 \int_0^{r_4-r_3}\mquad d\tau_4 \int_0^{T-r_4}\mquad d\tau_5 \nn \\[2mm]
& \qquad\times  \delta(\tau-\tau_1-\tau_2-\tau_3-\tau_4-\tau_5)\mathbb{Z}_{r_1}(m_1,x_1\vert \tau_1)\nn \\[2mm]
&\qquad\times  \partial_{x_1}\mathbb{Z}_{r_2-r_1}(x_1,x_2\vert \tau_2)\partial_{x_2}\mathbb{Z}_{r_3-r_2}(x_2,x_3\vert \tau_3) \nn \\[2mm]
& \qquad\times \partial_{x_3}\mathbb{Z}_{r_4-r_3}(x_3,x_4\vert \tau_4)\partial_{x_4}\mathbb{Z}_{T-r_4}(x_4,m_2\vert \tau_5),
\label{eq:four-point correlation conditional}
\end{align}
where the conditional average is defined in \Eq{eq:average +}.  This is analogous to \Eq{eq:id xxxx dot} without a condition on positive time  and can be derived following a similar analysis given in \sref{sec:correlation x dot}.

Following this result \eqref{eq:four-point correlation conditional} and the amplitude in \Eq{eq:C1 positive formal} we write the 
\begin{align}
C_1(\tau,T)&=2\times \frac{2^4D^4}{8D^2}\int_{0}^{\Lambda}\mquad dy_1 dy_2\int_0^{\infty}\mquad dt_1 dt_2 dt_3 dt_4 dt_5\nn \\[2mm]
&\times  \int_0^{t_1}d\tau_1 \int_0^{t_2}d\tau_2 \int_0^{t_3}d\tau_3 \int_0^{t_4}d\tau_4 \int_0^{t_5}d\tau_5\nn \\[2mm]
\times \int_{-\infty}^{\infty}&dx_1 dx_2 dx_3 dx_4 dm\;\delta(T-t_1-t_2-t_3-t_4-t_5)\nn \\[2mm]
&\times \delta(\tau-\tau_1-\tau_2-\tau_3-\tau_4-\tau_5) \; e^{-y_1 t_2-y_2 t_4}  \nn \\[2mm]
\times \mathbb{Z}_{t_1}(&0,x_1\vert \tau_1)\partial_{x_1}\mathbb{Z}_{t_2}(x_1,x_2\vert \tau_2)\partial_{x_2}\mathbb{Z}_{t_3}(x_2,x_3\vert \tau_3) \nn \\[2mm]
&\times \partial_{x_3}\mathbb{Z}_{t_4}(x_3,x_4\vert \tau_4)\partial_{x_4}\mathbb{Z}_{t_4}(x_4,m\vert \tau_4),\label{eq:C1 positive formal two}
\end{align}
where we have made a change of integration variables similar to that used for the diagram $D$ in \Eq{eq:D psitive in Zbb}.

Following a very similar analysis we find that amplitude of other two diagrams in \fref{fig:C123 positive} are almost same as in \Eq{eq:C1 positive formal two}, with only the term $e^{-y_1 r_2-y_2 r_4}$ replaced by $e^{-y_1 (t_2+t_3+t_4)-y_2 t_3}$ for $C_2$ and by $e^{-y_1 (t_2+t_3)-y_2 (t_3+t_4)}$ for $C_3$.

A double Laplace transformation \eqref{eq:D transformation} of the amplitudes integrates the delta functions and lead to a simpler formula,
\begin{align}
\widetilde{C}_1(\lambda,s)=& 4D^2\int_{0}^{\Lambda}dy_1dy_2\int_{-\infty}^{\infty}dx_1 dx_2 dx_3 dx_4 dm\nn \\[2mm]
\times \widetilde{\mathbb{Z}}_{s}&(0,x_1 \vert \lambda)\partial_{x_1}\widetilde{\mathbb{Z}}_{s+y_1}(x_1,x_2 \vert \lambda)\partial_{x_2} \widetilde{\mathbb{Z}}_{s}(x_2,x_3\vert \lambda)\nonumber \\[2mm] 
&\times  \partial_{x_3}\widetilde{\mathbb{Z}}_{s+y_2}(x_3,x_4\vert \lambda) \partial_{x_4}\widetilde{\mathbb{Z}}_{s}(x_4,m \vert \lambda),\label{eq:C1 app intermediate}
\end{align}
with $\widetilde{\mathbb{Z}}$ defined in \Eq{eq:Z tilde conditional final}. The other two amplitudes
\begin{align}
\widetilde{C}_2(\lambda,s)=&4D^2\int_{0}^{\Lambda}dy_1dy_2\int_{-\infty}^{\infty}dx_1 dx_2 dx_3 dx_4 dm \nn \\[2mm]
\widetilde{\mathbb{Z}}_{s}(0,x_1 &\lambda)\partial_{x_1}\widetilde{\mathbb{Z}}_{s+y_1}(x_1,x_2 \vert \lambda)\partial_{x_2}\widetilde{\mathbb{Z}}_{s+y_1+y_2}(x_2,x_3 \vert \lambda )\nn \\[2mm]
&\partial_{x_3}\widetilde{\mathbb{Z}}_{s+y_1}(x_3,x_4 \vert \lambda)\partial_{x_4}
\widetilde{\mathbb{Z}}_{s}(x_4,m \vert \lambda),\label{eq:C2 app intermediate}
\end{align}
and
\begin{align}
\widetilde{C}_3&(\lambda,s)=4D^2\int_{0}^{\Lambda}dy_1dy_2\int_{-\infty}^{\infty}dx_1 dx_2 dx_3 dx_4 dm \nn \\[2mm]
&\widetilde{\mathbb{Z}}_{s}(0,x_1 \vert \lambda)\partial_{x_1}\widetilde{\mathbb{Z}}_{s+y_1}(x_1,x_2 \vert \lambda)\partial_{x_2}\widetilde{\mathbb{Z}}_{s+y_1+y_2}(x_2,x_3  \vert \lambda)\nn \\[2mm] & \qquad \qquad \partial_{x_3}\widetilde{\mathbb{Z}}_{s+y_2}(x_3,x_4  \vert \lambda)\partial_{x_4}
\widetilde{\mathbb{Z}}_{s}(x_4,m \vert \lambda).\label{eq:C3 app intermediate}
\end{align}
Difference in Eqs.~\eqref{eq:C2 app intermediate} and \eqref{eq:C3 app intermediate} are in the subscript of a single $\widetilde{\mathbb{Z}}$ term.

Spatial integrals in these amplitudes can be evaluated by successively applying results from Appendix \ref{app:identities for conditional propagator}. It follows a lengthy but straightforward algebra. We write their final expression as follows.
\begin{align}
& \widetilde{C}_1(\lambda,s)=\frac{4}{\sqrt{s(s+\lambda)}\; (\sqrt{s}+\sqrt{s+\lambda})} \label{eq:C1 tilde positive final} \\[1mm] &
\times \int_{0}^{\Lambda}\frac{dy_1dy_2}{y_1y_2}\bigg[\mathfrak{f}(s,s+\lambda,y_1,y_2)+\mathfrak{f}(s+\lambda,s,y_1,y_2)\bigg] \nn
\end{align}
with $\mathfrak{f}$ in \Eq{eq:f positive}. Amplitudes of $C_2$ and $C_3$ are similar, 
\begin{align}
& \widetilde{C}_2(\lambda,s)+\widetilde{C}_3(\lambda,s)=\frac{4}{\sqrt{s(s+\lambda)}\;(\sqrt{s}+\sqrt{s+\lambda})}\nn \\[1mm] & 
\times\int_{0}^{\Lambda}\frac{dy_1dy_2}{y_1^2y_2}\bigg[\mathfrak{g}(s,s+\lambda,y_1,y_2)+\mathfrak{g}(s+\lambda,s,y_1,y_2)\bigg] \label{eq:C2C3 tilde positive final},\nn\\
\end{align}
with $\mathfrak{g}$ in \Eq{eq:g positive}. Writing them together in terms of re-scaled variables we get \Eq{eq:C tilde final expression scaled}.

\begin{remark}
We have verified the expression in Eqs.~\eqref{eq:D tilde positive final}, \eqref{eq:C1 tilde positive final}, and \eqref{eq:C2C3 tilde positive final} using the formula \Eq{eq:Z tilde conditional final} in Eqs.~\eqref{eq:C1 app intermediate}, \eqref{eq:C2 app intermediate}, and \eqref{eq:C3 app intermediate} and then numerically integrating in Mathematica.
\end{remark}

\section{Expression for $\Psi^{\rm pos}$}
\label{app:Psi pos}
Similar to the \Eq{eq:Psimax final explicit expression} for $t_{\rm max}$ we write $\Psi^{\rm pos}$ in \Eq{eq:F2 pos final exact} as a combination of three term.
\begin{align}
\Psi^{\rm pos}\left(y_1,y_2,z\right)&= \mathfrak{c} + \mathfrak{d} - \frac{1}{4}\mathfrak{a},\label{eq:Psipos final explicit expression}
\end{align}
where the terms on the right hand side corresponds to amplitudes in \Eq{eq:F2 pos final before transformation}.  Expression for $\mathfrak{c}(y_1,y_2,z)$ is cumbersome to write here and it is given in the supplemental Mathematica notebook \cite{Mathematica}.  In comparison,   
$\mathfrak{d}$ and $\mathfrak{a}$ have simpler expression,  given below.  Their numerical verification is also given in the Mathematica notebook.
\begin{align}
\mathfrak{d}(y_1,y_2,z)=& y_1 y_2+\frac{y_2^2 \; r(y_1,z)-y_1^2 \; r(y_2,z)}{y_1-y_2},
\end{align}
with
\begin{align}
&r(y,z)=-\sqrt{z(z-y)}\; \Theta (z-y)+\\[1mm]
&\left(\sqrt{y+1}-1\right) \sqrt{y-z} \;\Theta (y-z)+\sqrt{y+1}+z-1.\nn
\end{align}
Here $\Theta(x)$ is the Heaviside step  function.  
\begin{equation}
\mathfrak{a}(y_1,y_2,z)=u(y_1,y_2,z)+u(y_2,y_1,z),
\end{equation}
with
\begin{widetext}
\begin{align}
u(y_1,y_2,&z)=2-4 z+\frac{z^2}{2}+\frac{y_1 y_2}{2} -y_1 z+2 y_1+\bigg(6 z-4-2y_2 +2 (1-z) \sqrt{y_2+1}\bigg)\sqrt{y_1+1}\;+\cr
&| y_1-z| \bigg (\Theta(y_1-z)-\Theta(z-y_1)\bigg) \bigg(2+y_2-z-2 \sqrt{y_2+1} -2 \sqrt{y_2+1}\sqrt{y_2-z} \;\Theta(y_2-z)\bigg)+\cr
& 2 \sqrt{| y_2-z| } \bigg\{\Theta(y_2-z) \bigg[z \left(2 \sqrt{y_1+1}+\sqrt{y_2+1}-3\right)+y_1 \left(1-\sqrt{y_2+1}\right)+\cr
&\qquad \qquad \qquad \qquad \qquad \qquad 2 \left(1-\sqrt{y_1+1}-\sqrt{y_2+1}+\sqrt{(y_1+1) (y_2+1)}\right)\bigg]+ \cr
&\qquad \qquad \qquad \qquad \quad \sqrt{z} \;\Theta(z-y_2) \left(2 \sqrt{(y_1+1) (y_2+1)}-4 \sqrt{y_1+1}-2 \sqrt{y_2+1}+y_1-z+4\right)
\bigg\}+\cr
& 2 | y_1-z|  \sqrt{| y_2-z| } \bigg(\Theta(y_1-z)-\Theta(z-y_1)\bigg) \bigg(\Theta(y_2-z)+\sqrt{z} \;\Theta(z-y_2)\bigg)+\cr
& \frac{1}{2} | y_1-z|  | y_2-z|  \bigg(\Theta(y_1-z)-\Theta(z-y_1)\bigg) \bigg(\Theta(y_2-z)-\Theta(z-y_2)\bigg)-\cr
& 2 \sqrt{| y_1-z| } \sqrt{| y_2-z| }\bigg\{\sqrt{z} \left(\sqrt{y_1+1}+\sqrt{y_2+1}-2\right) \bigg(\Theta(y_1-z) \Theta(z-y_2)+\Theta(z-y_1) \Theta(y_2-z)\bigg)+\cr
&\left(z-1+\sqrt{y_1+1}+\sqrt{y_2+1}-\sqrt{(y_1+1) (y_2+1)}\right) \bigg(\Theta(y_1-z) \Theta(y_2-z)-\Theta(z-y_1) \Theta(z-y_2)\bigg) \bigg\}.
\end{align}
\end{widetext}

\section{A list of integrals for the Brownian propagator}
\label{app:Identities}
The Brownian propagator $Z_t(m_1,m_2)$ in \Eq{eq:W0 Z} is symmetric under exchange of $m_1$ and $m_2$, and therefore
\begin{equation}
\partial_{m_1}Z_{t}(m_1,m_2)=-\partial_{m_2}Z_t(m_1,m_2)\label{eq:Z der exchange}
\end{equation}
and its Laplace transformation \eqref{eq:L of Z Z+}
\begin{equation}
\partial_{m_1}\widetilde{Z}_s(m_1,m_2)=-\partial_{m_2}\widetilde{Z}_s(m_1,m_2). \label{eq:Z tilde der exchange}
\end{equation}
There is an analogous formula for the propagator $\widetilde{Z}^{+}_s$ in presence of absorbing line.
\begin{equation}
\partial_{m_1}\widetilde{Z}^{+}_s(m_1,m_2)=-\partial_{m_2}\widetilde{Z}^{+}_s(m_1,m_2)+\frac{1}{D}e^{-\sqrt{\frac{s}{D}}(m_1+m_2)}.
\label{eq:id swap}
\end{equation}

We list the following results for the integral of the propagators, which are frequently used in this paper. They can be numerically verified in Mathematica.
\begin{equation}
\int_{-\infty}^{\infty}dm_2\widetilde{Z}_{s}(m_1,m_2)=\frac{1}{s}
\end{equation}
and its analogue with absorbing boundary
\begin{equation}
\int_{0}^{\infty}dm_2\widetilde{Z}^{+}_{s}(m_1,m_2)=\frac{1}{s}\left(1-e^{-m_1\sqrt{\frac{s}{D}}}\right).
\label{eq:id Z plus int}
\end{equation}

Another useful result
\begin{align}
\int_{0}^{\infty}dm_2&\widetilde{Z}^{+}_{s+y}(m_1,m_2)e^{-m_2\sqrt{\frac{s}{D}}}\cr
&=\frac{1}{y}\left(e^{-m_1\sqrt{\frac{s}{D}}}-e^{-m_1\sqrt{\frac{s+y}{D}}}\right).
\label{eq:A10}
\end{align}
Due to a symmetry $\widetilde{Z}_s(m_1,m_2)=\widetilde{Z}_s(m_2,m_1)$ an integral over $m_1$ yields the same results as above.

For product of two propagators we get
\begin{align}
\int_{-\infty}^{\infty}dx &\widetilde{Z}_{r}(m_1,x)\widetilde{Z}_{s}(x,m_2)=\label{eq:int z z}\\[2mm]
&\begin{cases}\frac{\widetilde{Z}_{s}(m_1,m_2)}{r-s}+\frac{\widetilde{Z}_{r}(m_1,m_2)}{s-r} & \mbox{if } s\ne r, \nn\\[2mm] \frac{1+\sqrt{\frac{s}{D}}\vert m_1-m_2\vert}{2s}\;\widetilde{Z}_{s}(m_1,m_2) & \mbox{if } r=s.\end{cases}\nn
\end{align}
and for its analogue with absorbing boundary
\begin{align}
\int_{0}^{\infty}dx &\widetilde{Z}^{+}_{r}(m_1,x)\widetilde{Z}^{+}_{s}(x,m_2)=\label{eq:A11}\\[2mm]
&\begin{cases}
\displaystyle\frac{\widetilde{Z}^{+}_{s}(m_1,m_2)}{r-s}+\frac{\widetilde{Z}^{+}_{r}(m_1,m_2)}{s-r} & \mbox{if } s\ne r, \cr & \cr
\displaystyle
\frac{1+\sqrt{\frac{s}{D}}\vert m_1-m_2\vert}{2s}\;\widetilde{Z}^{+}_{s}(m_1,m_2)& \mbox{if } r=s,
\nn\\[2mm] \displaystyle\quad -\frac{\min\{m_1,m_2\}}{2s}e^{-(m_1+m_2)\sqrt{\frac{s}{D}}} \end{cases}\nn
\end{align}
For product of three propagators, corresponding formula is
\begin{widetext}
\begin{equation}
\int_{-\infty}^{\infty}dx\int_{-\infty}^{\infty}dy \widetilde{Z}_{r}(m_1,x)\widetilde{Z}_{s}(x,y)\widetilde{Z}_{t}(y,m_2)=\begin{cases}
\displaystyle
\frac{\widetilde{Z}_{r}(m_1,m_2)}{(s-r)(t-r)}+\frac{\widetilde{Z}_{s}(m_1,m_2)}{(r-s)(t-s)}+\frac{\widetilde{Z}_{t}(m_1,m_2)}{(r-t)(s-t)} 
& 
\mbox{if} \quad r\ne s\ne t \cr & 
\cr \cr
\displaystyle
\left[\frac{1+\sqrt{\frac{r}{D}}\vert m_1-m_2\vert}{2r}-\frac{1}{s-r}\right]\frac{\widetilde{Z}_{r}(m_1,m_2)}{s-r}
\cr 
\displaystyle
\qquad \qquad +\frac{\widetilde{Z}_{s}(m_1,m_2)}{(r-s)^2} & \mbox{if}\quad r=t\ne s  
\cr \cr
& 
\cr 
\displaystyle\left[\frac{1+\sqrt{\frac{r}{D}}\vert m_1-m_2\vert}{2r}-\frac{1}{t-r}\right]\frac{\widetilde{Z}_{r}(m_1,m_2)}{t-r}\cr 
\displaystyle
\qquad \qquad +\frac{\widetilde{Z}_{t}(m_1,m_2)}{(r-t)^2} & \mbox{if}\quad r=s\ne t\end{cases}
\label{eq:int z z z}
\end{equation}
and its counterpart in presence of absorbing line,
\begin{equation}
\int_{0}^{\infty}\!\!\!\!dx\int_{0}^{\infty}\!\!\!\!dy \widetilde{Z}^{+}_{r}(m_1,x)\widetilde{Z}^{+}_{s}(x,y)\widetilde{Z}^{+}_{t}(y,m_2)=
\begin{cases}
\displaystyle
\frac{\widetilde{Z}^{+}_{r}(m_1,m_2)}{(s-r)(t-r)}+\frac{\widetilde{Z}^{+}_{s}(m_1,m_2)}{(r-s)(t-s)}+\frac{\widetilde{Z}^{+}_{t}(m_1,m_2)}{(r-t)(s-t)} & \mbox{if} \quad r\ne s\ne t 
\cr\cr\cr
\displaystyle
\left[\frac{1+\sqrt{\frac{r}{D}}\vert m_1-m_2\vert}{2r}-\frac{1}{s-r}\right]\frac{\widetilde{Z}^{+}_{r}(m_1,m_2)}{s-r}
\cr 
\displaystyle
\quad  +\frac{\widetilde{Z}^{+}_{s}(m_1,m_2)}{(r-s)^2}- \frac{\min(m_1,m_2)}{2r(s-r)}e^{-(m_1+m_2)\sqrt{\frac{r}{D}}} & \mbox{if}\quad r=t\ne s  
\cr\cr\cr
\displaystyle
\left[\frac{1+\sqrt{\frac{r}{D}}\vert m_1-m_2\vert}{2r}-\frac{1}{t-r}\right]\frac{\widetilde{Z}^{+}_{r}(m_1,m_2)}{t-r}
\cr 
\displaystyle
\quad  +\frac{\widetilde{Z}^{+}_{t}(m_1,m_2)}{(r-t)^2} - \frac{\min(m_1,m_2)}{2r(t-r)}e^{-(m_1+m_2)\sqrt{\frac{r}{D}}}& \mbox{if}\quad r=s\ne t\end{cases}
\label{eq:int z z z plus}
\end{equation}
\end{widetext}

\section{Time-correlation of Brownian velocities}
\label{sec:correlation x dot}
Here, we derive multi-time correlations of velocity $\dot{x}(t)$ for a standard Brownian motion with diffusivity $D$. The first moment is defined by
\begin{equation}
\langle \dot{x}(t) \rangle =\int_{x(0)=m_1}^{x(T)=m_2}\mathcal{D}[x]e^{-\frac{S_0}{D}} \dot{x}(t),
\label{eq:angular Brownian}
\end{equation}
where the angular brackets denote average with a Brownian measure of diffusivity $D$ starting at position $x(0)=m_1$ and finishing at time $T$ at position $x(T)=m_2$. For evaluating the average we consider a small window between time $t$ and $t+\Delta t$ such that
\begin{align}
\langle \dot{x}(t) &\rangle=\lim_{\Delta t\to 0} \int_{-\infty}^{\infty} dx dy \; Z_{t}(m_1,x)\times \cr
&\left[\frac{e^{-\frac{(y-x)^2}{4D\Delta t}}}{\sqrt{4\pi D \Delta t}} \left(\frac{y-x}{\Delta t}\right) \right] Z_{T-t-\Delta t}(y,m_2),
\end{align}
where the Brownian propagator $Z$ is in \Eq{eq:W0 Z} and we use \Eq{eq:S0} for small $\Delta t$. Writing
\begin{equation*}
 \frac{e^{-\frac{(y-x)^2}{4D\Delta t}}}{\sqrt{4\pi D \Delta t}} \left(\frac{y-x}{\Delta t}\right) =-2D \; \partial_y \left[\frac{e^{-\frac{(y-x)^2}{4D\Delta t}}}{\sqrt{4\pi D \Delta t}} \right]
\end{equation*}
and using integration by parts for $y$ variable, we get
\begin{align*}
\langle \dot{x}(t) &\rangle=2D \int_{-\infty}^{\infty} dx dy  \; Z_{t}(m_1,x)\times \cr
&\lim_{\Delta t\to 0} \left[\frac{e^{-\frac{(y-x)^2}{4D\Delta t}}}{\sqrt{4\pi D \Delta t}} \right] \partial_y Z_{T-t-\Delta t}(y,m_2).
\end{align*}
In the $\Delta t \to 0$ limit, it gives an expression
\begin{align}
\langle \dot{x}(t)\rangle =2D\int_{-\infty}^{\infty} dx \;  Z_{t}(m_1,x) \partial_{x}\; Z_{T-t}(x,m_2),
\label{eq:id x dot}
\end{align}
which can be explicitly evaluated using \Eq{eq:W0 Z}.

For two-time correlation one can similarly show that 
\begin{align}
\langle\dot{x}(r_1)\dot{x}(r_2)\rangle=& 2^2D^2\mathscr{C}(r_1,r_2)\cr
&+2D \; \delta(r_1-r_2)Z_T(m_1,m_2),
\label{eq:id xx dot}
\end{align}
where $\mathscr{C}(r_1,r_2)$ is a symmetric function given by
\begin{align}
\mathscr{C}(r_1,r_2)&=\int_{-\infty}^{\infty}\mquad dx_1dx_2 Z_{r_1}(m_1,x_1)\times \cr & \partial_{x1}Z_{r_2-r_1}(x_1,x_2) \partial_{x2}Z_{T-r_2}(x_2,m_2),
\end{align}
for $r_2>r_1$. The integral remains finite for $r_1\to r_2$ limit.

A generalization of \Eq{eq:id xx dot} in an analogy of Wick's theorem gives multi-time correlations. For example, we get  
\begin{align}
\langle&\dot{x}(r_1)\dot{x}(r_2)\dot{x}(r_3)\rangle=\label{eq:id xxx dot}\\
&\qquad 2^3D^3\mathscr{C}(r_1,r_2,r_3)+2D\sum_{\text pairs}\delta(r_i-r_j) \langle \dot{x}(r_k)\rangle, \nn
\end{align}
where $\mathscr{C}(r_1,r_2,r_3)$ is a symmetric function under permutation of its arguments and given by
\begin{align}
\mathscr{C}(r_1,r_2,r_3)=&\int_{-\infty}^{\infty}dx_1dx_2dx_3  Z_{r_1}(m_1,x_1)\\[1mm] \times &  \partial_{x_1}Z_{r_2-r_1}(x_1,x_2)\;\partial_{x_2}Z_{r_3-r_2}(x_2,x_3) \nn\\[2mm] \times & \partial_{x_3}Z_{T-r_3}(x_3,m_2),\nn
\end{align}
for $r_3>r_1>r_1$. 

For the four-time correlation, we get
\begin{align}
\langle \dot{x}(r_1)\dot{x}(r_{2})\dot{x}(r_{3})&\dot{x}(r_{4}) \rangle= 2^4D^4\mathscr{C}(r_1,r_2,r_3,r_4)\nn \\[2mm] &+ 2D\sum_{pairs}\delta(r_i-r_j)\langle \dot{x}(r_k)\dot{x}(r_{\ell}) \rangle\label{eq:id xxxx dot}
\end{align}
with
\begin{align}
\mathscr{C}(r_1,r_2,r_3,r_4) =&\int_{-\infty}^{\infty}dx_1dx_2dx_3dx_4\times  \\   Z_{r_1}(m_1,x_1)&\partial_{x_1}Z_{r_2-r_1}(x_1,x_2)\partial_{x_2}Z_{r_3-r_2}(x_2,x_3) \nn\\[2mm] \times \partial_{x_3}&Z_{r_4-r_3}(x_3,x_4)\partial_{x_4}Z_{T-r_4}(x_4,m_2),\nn
\end{align}
for $r_1<  r_2 < r_3 <\cdots< r_4$.

Expression for these correlations can be further simplified. For the first moment \Eq{eq:id x dot}, using \Eq{eq:Z der exchange} and then integrating over $x$, we get
\begin{align}
\langle \dot{x}(t)\rangle =(-2D \partial_{m_2})Z_{T}(m_1,m_2). \label{eq:x dot simple}
\end{align}
Similarly, from \Eq{eq:id xx dot} we get 
\begin{align}
\langle \dot{x}(r_1)\dot{x}(r_2)\rangle =\left[2^2D^2 \partial_{m_2}+2D \delta(r_1-r_2)\right]Z_{T}(m_1,m_2), \label{eq:xx dot simple}
\end{align}
and for three-time correlation in \Eq{eq:id xxx dot} we get 
\begin{align}
\langle \dot{x}(r_1)\dot{x}(r_{2})\dot{x}(r_{3}&) \rangle  = (-2D\; \partial_{m_2})\bigg[2^2D^2\partial_{m_2}^2+\cr
& 2D\sum_{\text pairs}\delta(r_i-r_j)\bigg] Z_T(m_1,m_2). \label{eq:xxx dot simple}
\end{align}

More generally, for $r_1< r_2 < \cdots < r_{2n}$ we see that
\begin{equation}
\langle \dot{x}(r_1)\cdots \dot{x}(r_{2n}) \rangle=2^{2n}D^{2n}\partial^{2n}_{m_2}Z_T(m_1,m_2),
\label{eq:main identity}
\end{equation}
which is used for a derivation of \Eq{eq:GH der m}.

\begin{remark}
Formulas in Eqs.~\eqref{eq:xx dot simple} and \eqref{eq:xxx dot simple} are mentioned earlier in Eqs.~\eqref{eq:two-point correlation} and \eqref{eq:four-point correlation}.
\end{remark}
\begin{remark}
In presence of an absorbing wall, correlations have a very similar formula as in Eqs.~\eqref{eq:id x dot}, \eqref{eq:id xx dot}, \eqref{eq:id xxx dot}, and \eqref{eq:id xxxx dot}, where one need to substitute the propagator $Z$ by $Z^+$. However, they can not be simplified like in Eqs.~(\ref{eq:x dot simple}\;-\;\ref{eq:xxx dot simple}).
\end{remark}

\section{Identities for $J_t$ in \Eq{eq:J2}} 
\label{app:B}
In this section, we give a list of results for $J_t$ in \Eq{eq:J2} and its analogue $J_t^{+}$ with absorbing boundary.
These results are used in our analysis. 

\subsection{$J_t(m_1,m_2;y)$}
Using \eqref{eq:x dot simple} in \Eq{eq:J2} we write
\begin{equation*}
J_t(m_1,m_2;y)=2D\; \partial_{m_2}Z_t(m_1,m_2)\left\{\frac{e^{-yt}-1}{y}\right\}.
\end{equation*}
It's Laplace transform is
\begin{equation*}
\widetilde{J}_{s}(m_1,m_2;y)=\frac{2D}{y}\partial_{m_2}\left(\widetilde{Z}_{s+y}(m_1,m_2)-\widetilde{Z}_{s}(m_1,m_2)\right)
\end{equation*}
and using \Eq{eq:Z tilde final} it leads to
\begin{align}
\widetilde{J}_{s}(m_1,m_2&;y)=\frac{\sgn(m_1-m_2)}{y}\times \cr& \left[e^{-\vert m_1-m_2\vert\sqrt{\frac{s+y}{D}}}-e^{-\vert m_1-m_2\vert\sqrt{\frac{s}{D}}}\right]. \label{eq:J1 tilde final}
\end{align}
Here $\sgn(x)$ gives the sign of $x$.

\subsection{$J_t^{+}(m_1,m_2;y)$}
An analogue of $J_t$ in presence of absorbing line is 
\begin{equation}
J_{t}^{+}(m_1,m_2;y)=\int_0^{t}dr \; e^{-y r}\langle \dot{x}(r)\rangle^{+},
\end{equation}
with the average $\langle \cdot \rangle^+$ defined as in \Eq{eq:angular Brownian} with absorbing boundary at origin. 
Using the analogous formula of \Eq{eq:id x dot} for absorbing boundary and taking Laplace transformation we get
\begin{align}
\widetilde{J}_{s}^{+}(m_1,m_2;y)=2D\int_{0}^{\infty}dx\widetilde{Z}_{s+y}^{+}(m_1,x) \partial_{x}\widetilde{Z}_{s}^{+}(x,m_2).
\label{eq: app J+ 1y formal}
\end{align}
Further, using \Eq{eq:id swap} and \Eq{eq:A11} leads
\begin{align}
&\widetilde{J}^{+}_s(m_1,m_2;y)= -2D\partial_{m_2}\Bigg[\int_{0}^{\infty}dx\widetilde{Z}^{+}_{s+y}(m_1,x)\times \cr &\widetilde{Z}^{+}_{s}(x,m_2)\Bigg] + 2 \int_{0}^{\infty}dx\widetilde{Z}^{+}_{s+y}(m_1,x)e^{-\sqrt{\frac{s}{D}}(x+m_2)}.\label{eq:J+ tilde 1 in Z}
\end{align}

Invoking the explicit expression of $\widetilde{Z}^{+}$ in \Eq{eq:Z + tilde final} leads to a small $x_0$ asymptotic,
\begin{equation}
\widetilde{J}^{+}_s(m_1,x_0;y)\simeq \frac{2x_0 \sqrt{s}}{y\sqrt{D}}\left\{e^{-m_1\sqrt{\frac{s+y}{D}}}-e^{-m_1\sqrt{\frac{s}{D}}}\right\},\label{eq:J + tilde 1 asymptotic}
\end{equation}
which has been used many times in our analysis.

Another useful result is for integrals of $\widetilde{J}^{+}_s$. It is straightforward to see that an integration over $m_1$ gives
\begin{align}
\int_{0}^{\infty}&dm_1\widetilde{J}^{+}_{s}(m_1,m_2;y)\cr
&=\frac{2\sqrt{D}}{y\sqrt{(s+y)}}\left\{e^{-m_2\sqrt{\frac{s+y}{D}}}-e^{-m_2\sqrt{\frac{s}{D}}}\right\},
\label{eq:J1IntL}
\end{align}
and an integration over $m_2$ gives
\begin{align}
\int_{0}^{\infty}dm_2&\widetilde{J}^{+}_{s}(m_1,m_2;y)\cr &=\frac{2\sqrt{D}}{y\sqrt{s}}\left\{e^{-m_1\sqrt{\frac{s}{D}}}-e^{-m_1\sqrt{\frac{s+y}{D}}}\right\},
\label{eq:J1IntR}
\end{align}
where we used \Eq{eq:A10}. The same result can also be derived using a symmetry
\begin{equation*}
\widetilde{J}^{+}_{s}(m_1,m_2;y)=-\widetilde{J}^{+}_{s+y}(m_2,m_1;-y),
\end{equation*}
which is evident from \Eq{eq:J+ tilde 1 in Z} and the symmetry of $\widetilde{Z}^{+}_t$.

\subsection{$J_t(m_1,m_2;y_1,y_2)$}
For $J_t(m_1,m_2;y_1,y_2)$ in \Eq{eq:J2} using the correlation \Eq{eq:xx dot simple} with the choice of integration \Eq{eq:int dr1dr2 delta} we get
\begin{align*}
&J_t(m_1,m_2;y_1,y_2)\nn \\[2mm] &=2^{2}D^{2}\frac{\partial^{2}_{m_2}Z_{t}(m_1,m_2)}{y_1y_2}\left[\frac{y_1+y_2e^{-t(y_1+y_2)}}{y_1+y_2}-e^{-y_2 t}\right].
\end{align*}
A Laplace transformation gives
\begin{align*}
&\widetilde{J}_{s}(m_1,m_2;y_1,y_2)=\frac{2^{2}D^{2}}{y_1y_2(y_1+y_2)}\bigg[y_1\partial^{2}_{m_2}\widetilde{Z}_{s}(m_1,m_2)+\cr &y_2\partial^{2}_{m_2}\widetilde{Z}_{s+y_1+y_2}(m_1,m_2) -(y_1+y_2)\partial^{2}_{m_2}\widetilde{Z}_{s+y_2}(m_1,m_2)\bigg].
\end{align*}
The explicit formula of $\widetilde{Z}$ in \Eq{eq:Z tilde final} leads to the result given in \Eq{eq:J2 tilde final full}.

A special case of \Eq{eq:J2 tilde final full}, used earlier for deriving the result \Eq{eq:chi2 last}, is
\begin{align}
\widetilde{J}_{s}(0,x_0;-y&,y)=\frac{\sqrt{D}}{y^2 }\Bigg[2\sqrt{s+y}e^{-x_0\sqrt{\frac{s+y}{D}}}\cr &-\frac{e^{-x_0\sqrt{\frac{s}{D}}}}{\sqrt{s}}\bigg(2s+y-yx_0\sqrt{\frac{s}{D}}\bigg)\,\Bigg].
\label{eq:J2 0}
\end{align}

\subsection{$J_t^{+}(m_1,m_2;y_1,y_2)$}
Starting with the definition 
\begin{align}
J_{t}^{+}&(m_1,m_2;y_1,y_2)\cr &=\int_0^{t}dr_1\int_{r_1}^{t}dr_2 \; e^{-y_1 r_1-y_2 r_2}\langle \dot{x}(r_1)\dot{x}(r_2)\rangle^{+} \label{eq:J + 2 definition}
\end{align}
with the convention in \Eq{eq:int dr1dr2 delta} for time-integrals and using an analogue of \Eq{eq:id xx dot} for correlations  with absorbing boundary, we write
\begin{align*}
J_{t}^{+}(m_1,&m_2;y_1,y_2)=\int_{-\infty}^{\infty}\mquad dx_1dx_2\int_0^{t}dr_1\int_{r_1}^{t}dr_2 \; e^{-y_1 r_1-y_2 r_2}\nn \\[2mm] & \times Z_{r_1}(m_1,x_1) \partial_{x1}Z_{r_2-r_1}(x_1,x_2) \partial_{x2}Z_{T-r_2}(x_2,m_2) .
\end{align*}
It's Laplace transformation (in $t\to s$ variable) is
\begin{align}
\widetilde{J}_{s}^{+}&(m_1,m_2;y_1,y_2) = 2^2D^2\int_{0}^{\infty}dx_1
dx_2\times  \label{eq:J+ tilde 2 in Z}\\[2mm]  &\widetilde{Z}^{+}_{s+y_1+y_2}(m_1,x_1)\partial_{x_1}\widetilde{Z}^{+}_{s+y_2}(x_1,x_2)\partial_{x_2}\widetilde{Z}^{+}_{s}(x_2,m_2).\nn
\end{align}
An explicit expression can be derived using the result in \eq{eq:Z + tilde final}. 

Analysis gets simplified realizing that
\begin{align}
\widetilde{J}_{s}^{+}&(m_1,m_2;y_1,y_2)\label{eq:J + tilde 2 in Z J 1}\\ & =  2D\int_{0}^{\infty}dx
\widetilde{Z}^{+}_{s+y_1+y_2}(m_1,x)\partial_{x}\widetilde{J}_{s}^{+}(x,m_2,y_2),\nn
\end{align}
with $\widetilde{J}_{s}^{+}(x,m_2,y_2)$ in \Eq{eq: app J+ 1y formal}.
Using this, for example, one can derive a useful asymptotic for small $x_0$ by using \Eq{eq:J + tilde 1 asymptotic} and \Eq{eq:Z + tilde final}, which gives
\begin{align}
\widetilde{J}_{s}^{+}&(m_1,x_0;y_1,y_2)\cr \simeq &\; 4 x_0\Bigg\{\frac{s}{y_2(y_1+y_2)}\left(e^{-m_1\sqrt{\frac{s}{D}}}-e^{-m_1\sqrt{\frac{s+y_1+y_2}{D}}}\right)\cr - & \frac{\sqrt{s(s+y_2)}}{y_1y_2}\left(e^{-m_1\sqrt{\frac{s+y_2}{D}}}-e^{-m_1\sqrt{\frac{s+y_1+y_2}{D}}}\right)\Bigg\}.
\label{eq:J + tilde 2 asymptotics}\qquad
\end{align}

For an analogous formula of \Eq{eq:J1IntR} we evaluate the integration in \Eq{eq:J + tilde 2 in Z J 1} using \Eq{eq:id Z plus int}, a symmetry $\widetilde{Z}^{+}_s(m_1,m_2)=\widetilde{Z}^{+}_s(m_2,m_1)$, the results in Eqs.~\eqref{eq:Z tilde der exchange}, \eqref{eq:A10}, \eqref{eq:A11}, and using integration by parts. This way it is straightforward to get the result in \Eq{eq:J2 + tilde integral final full}.

In a similar way we derive the integral over $m_1$, and the result is given in \Eq{eq:B27}.
Alternatively, one can use a symmetry
\begin{equation*}
\widetilde{J}_{s}^{+}(m_1,m_2;y_1,y_2)=\widetilde{J}_{s+y_1+y_2}^{+}(m_2,m_1;-y_2,-y_1),
\end{equation*}
which is evident from \Eq{eq:J+ tilde 3 in Z} using the symmetry $\widetilde{Z}_{s}^{+}(x_1,x_2)=\widetilde{Z}_{s}^{+}(x_2,x_1)$.

A special case of \Eq{eq:J2 + tilde integral final full}, which is used for deriving \Eq{eq:chi2 last}, is
\begin{align}
\int_{0}^{\infty}&dm_2\widetilde{J}_{s}^{+}(m_1,m_2;-y,y)= \frac{2\sqrt{D}}{y^2\sqrt{s}}\bigg\{ 2\sqrt{(s+y)}\times\cr &\sqrt{D}  \left(e^{-m_1\sqrt{\frac{s}{D}}}-e^{-m_1\sqrt{\frac{s+y}{D}}}\right)-m_1 y e^{-m_1\sqrt{\frac{s}{D}}}\bigg\}.\quad
\label{eq:IJPm1}
\end{align}

\subsection{$J_t^{+}(m_1,m_2;y_1,y_2,y_3)$}
Similar to \Eq{eq:J + 2 definition} we define $J_t^{+}(m_1,m_2;y_1,y_2,y_3)$. Using the analogue of \Eq{eq:id xxx dot} with an absorbing boundary and then taking a Laplace transformation (in $t\to s$ variable) we write
\begin{align}
\widetilde{J}_{s}^{+}(m_1,m_2;&y_1,y_2,y_3)=2^3D^3\int_{0}^{\infty}dx_1dx_2dx_3 \nn\\[2mm]
&\times \widetilde{Z}_{s+y_1+y_2+y_3}^{+}(m_1,x_1)\partial_{x_1}\widetilde{Z}_{s+y_2+y_3}^{+}(x_1,x_2)\nn\\[2mm]
& \times \partial_{x_2}\widetilde{Z}_{s+y_3}^{+}(x_2,x_3)\partial_{x_3}\widetilde{Z}_{s}^{+}(x_3,m_2).
\label{eq:J+ tilde 3 in Z}
\end{align}
For an explicit result we note that
\begin{align}
\widetilde{J}_{s}^{+}(m_1,m_2;&y_1,y_2,y_3)= 2D\int_{0}^{\infty} dx \label{eq:B32}\\ &
\times \widetilde{Z}^{+}_{s+y_1+y_2+y_3}(m_1,x)\partial_{x}\widetilde{J}^{+}_{s}(x,m_2;y_2,y_3)\nn
\end{align}
with \Eq{eq:J+ tilde 3 in Z}. Then, Eqs.~\eqref{eq:Z + tilde final} and \eqref{eq:J + tilde 2 asymptotics} can be used to get an asymptotic for small $x_0$.

Integral of $\widetilde{J}_{s}^{+}(m_1,x_0;y_1,y_2,y_3)$ analogous to \Eq{eq:J1IntL} is also straightforward to derive using \Eq{eq:B32}. For small $x_0$,
\begin{align}
\int_0^\infty  dm_1&\widetilde{J}_{s}^{+}(m_1,x_0;y_1,y_2,y_3)\simeq 8 D x_0\nn \\[2mm]
\bigg\{ &\frac{\left(s+y_3\right) \left(\sqrt{s+y_3}-\sqrt{s+y_1+y_2+y_3}\right)}{\sqrt{s} \; y_2 \left(y_1+y_2\right) y_3}+\nn \\[2mm]
& \frac{\sqrt{s+y_2+y_3} \left[\left(y_2+y_3\right) \sqrt{s+y_3}-y_2\sqrt{s} \right] }{\sqrt{s}\; y_1 y_2 y_3 \left(y_2+y_3\right)}\times \nn \\[2mm]
&\left(\sqrt{s+y_1+y_2+y_3}-\sqrt{s+y_2+y_3}\right)+\nn \\[2mm]
&\frac{\sqrt{s} \left(\sqrt{s+y_1+y_2+y_3}-\sqrt{s}\right)}{y_3 \left(y_2+y_3\right) \left(y_1+y_2+y_3\right)}\bigg\},\label{eq:left integral J 3 small x0}
\end{align}
which is used for a derivation of \Eq{eq:154}.

For an integral over $m_2$ variable, one can use a symmetry
\begin{align}
\widetilde{J}_{s}^{+}&(m_1,m_2;y_1,y_2,y_3)\cr & =-\widetilde{J}_{s+y_1+y_2+y_3}^{+}(m_2,m_1;-y_3,-y_2,-y_1),
\label{eq:sym J3}
\end{align}
which is evident from \Eq{eq:J+ tilde 3 in Z} and $\widetilde{Z}^{+}_s(m_1,m_2)=\widetilde{Z}^{+}_s(m_2,m_1)$. The result is useful for a derivation of \Eq{eq:G17}.

\subsection{$J_t^{+}(m_1,m_2;y_1,y_2,y_3,y_4)$}
Similar to Eqs.~\eqref{eq:J + tilde 2 in Z J 1} and \eqref{eq:B32},
\begin{align}
J_{t}^{+}(m_1,&m_2;y_1,y_2,y_3,y_4)=\int_{0}^{t}dr_1\int_{r_1}^{t}dr_2\int_{r_2}^{t}dr_3\times \cr  \int_{r_3}^{t}dr_4 &e^{-y_1 r_1- y_2r_2-y_3 r_3- y_4r_4}\langle\dot{x}(r_1)\dot{x}(r_2)\dot{x}(r_3)\dot{x}(r_4)\rangle^{+} \label{eq:J + 4 definition}
\end{align}
follows a hierarchy where
\begin{align}
\widetilde{J}_{s}^{+}(&m_1,m_2;y_1,y_2,y_3,y_4)= 2D\int_{0}^{\infty}dx\times 
\cr & \widetilde{Z}_{s+y_1+y_2+y_3+y_4}^{+}(m_1,x)\;\partial_{x}\widetilde{J}^{+}_{s}(x,m_2;y_2,y_3,y_4),\qquad \quad 
\label{eq:B39}
\end{align}
which leads to explicit results explicit result for \Eq{eq:J + 4 definition}. 

For example, an integral over $m_2$ variable is given in \Eq{eq:J4 + integrate}. From this one can derive also the integral over $m_1$ variable using a symmetry relation
\begin{align}
\widetilde{J}_{s}^{+}&(m_1,m_2;y_1,y_2,y_3,y_4) \cr = & \widetilde{J}_{s+y_1+y_2+y_3+y_4}^{+}(m_2,m_1;-y_4,-y_3,-y_2,-y_1),\qquad 
\end{align}
which is evident from \Eq{eq:B39} and a symmetry $\widetilde{Z}^{+}_s(m_1,m_2)=\widetilde{Z}^{+}_s(m_2,m_1)$.

\section{Identities for $\mathscr{I}_{\tau}^{+}$ in \Eq{eq:I3+}}
Using \Eq{eq:150} we get a relation for their Laplace transformation
\begin{align}
\widetilde{\mathscr{I}}_s^{+}&(m_1,m_2;y_1,y_2,y_3)=\widetilde{J}_s^{+}(m_1,m_2;y_1,y_2,y_3)\cr \quad +&\widetilde{J}_s^{+}(m_1,m_2;y_1,y_3,y_2)+\widetilde{J}_s^{+}(m_1,m_2;y_3,y_1,y_2).\qquad \label{eq:N1}
\end{align}

This leads to the results we need, namely,
\begin{align*}
\int_{0}^{\infty}dm_1\widetilde{\mathscr{I}}_{s+\lambda+y_2}^{+}&(m_1,x_0;-y_1,y_1,-y_2) \cr =\int_{0}^{\infty}dm_1 \bigg\{& \widetilde{J}_{s+\lambda+y_2}^{+}(m_1,x_0;-y_1,y_1,-y_2)\cr &+\widetilde{J}_{s+\lambda+y_2}^{+}(m_1,x_0;-y_1,-y_2,y_1)\cr & +\widetilde{J}_{s+\lambda+y_2}^{+}(m_1,x_0;-y_2,-y_1,y_1) \bigg\},
\end{align*}
which using \Eq{eq:left integral J 3 small x0} for small $x_0$ limit gives \Eq{eq:154}. A analogous integral
\begin{align}
\int_{0}^{\infty}dm_2&\widetilde{\mathscr{I}}_s^{+}(x_0,m_2;-y_2,y_2,y_1)\simeq \frac{4D x_0 \sqrt{s+y_1}}{y_1 y_2^2\sqrt{s}}\times \cr &\bigg(\sqrt{s{+}y_1{+}y_2}-\sqrt{s{+}y_1}-\sqrt{s{+}y_2}+\sqrt{s}\bigg)^2,
\end{align}
for small $x_0$, is derived using Eqs.~\eqref{eq:N1}, \eqref{eq:left integral J 3 small x0}, and \eqref{eq:sym J3}. It is used for a derivation of  \Eq{eq:G17}.

\section{Identities for conditional propagator $\mathbb{Z}_t$} \label{app:identities for conditional propagator}
In this section we give a list of identities for conditional Brownian propagator $\mathbb{Z}_t$ in \Eq{eq:conditional propagator definition}. These identities are often used for our analysis in \sref{sec:positive}.

In \Eq{eq:Za final} we see that
\begin{equation}
\widetilde{\mathbb{A}}_{s}(0,x\vert \lambda )=0=\widetilde{\mathbb{A}}_{s}(x,0\vert \lambda).
\end{equation}
Substituting this and \Eq{eq:Zb final} in \Eq{eq:Z tilde conditional final} we get
\begin{align}
\widetilde{\mathbb{Z}}_{s}(0,x\vert \lambda )&=\widetilde{\mathbb{B}}_{s}(0,x\vert \lambda)\cr &=\frac{\sqrt{s+\lambda}-\sqrt{s}}{\lambda \sqrt{D}}\;e^{-\vert x \vert \sqrt{\frac{s+\lambda \Theta(x)}{D}}}.\label{eq:Zbb x1 0}
\end{align}
The result is used for the zeroth order amplitude in \Eq{eq:W0 final positive} and also appears in the linear order amplitude \Eq{eq:W1 tilde pos some intermediate expression}.

For results about integrals of $\widetilde{\mathbb{Z}}_{s}$ we use that for $\widetilde{\mathbb{A}}$ in \Eq{eq:Za final},
\begin{equation*}
\int_{-\infty}^{\infty}\mquad dx_2\;\widetilde{\mathbb{A}}_{s}(x_1,x_2\vert \lambda)=\frac{  1-e^{-\vert x_1 \vert \sqrt{\frac{s+\lambda\Theta(x_1)}{D}}} }{s+\lambda\Theta(x_1)}
\end{equation*}
and for $\widetilde{\mathbb{B}}$ in \Eq{eq:Zb final},
\begin{equation*}
\int_{-\infty}^{\infty}dx_2\;\widetilde{\mathbb{B}}_{s}(x_1,x_2\vert \lambda)=\frac{e^{-\vert x_1 \vert \sqrt{\frac{s+\lambda\Theta(x_1)}{D}}}}{\sqrt{s(s+\lambda)}}.
\end{equation*}
Then \Eq{eq:Z tilde conditional final} leads to
\begin{align}
\int_{-\infty}^{\infty}&\mquad dx_2\;\widetilde{\mathbb{Z}}_{s}(x_1,x_2\vert \lambda)\cr
&=\frac{  1-e^{-\vert x_1 \vert \sqrt{\frac{s+\lambda\Theta(x_1)}{D}}} }{s+\lambda\Theta(x_1)}+\frac{e^{-\vert x_1 \vert \sqrt{\frac{s+\lambda\Theta(x_1)}{D}}}}{\sqrt{s(s+\lambda)}}.\label{eq:Zbb tilde integral Right}
\end{align}

For a related result, we use
\begin{equation*}
\int_{-\infty}^{\infty}\mquad dx_2\partial_{x_1}\widetilde{\mathbb{A}}_s(x_1,x_2\vert \lambda)=\frac{\sgn(x_1)e^{-\vert x_1 \vert \sqrt{\frac{s+\lambda\Theta(x_1)}{D}}}}{\sqrt{D(s+\lambda\Theta(x_1))}},
\end{equation*}
and
\begin{align*}
 \int_{-\infty}^{\infty}dx_2\partial_{x_1}&\widetilde{\mathbb{B}}_s(x_1,x_2\vert \lambda)\cr & =\frac{\sgn(-x_1)\sqrt{s+\lambda\Theta(x_1)}}{\sqrt{Ds(s+\lambda)}}\; e^{-\vert x_1 \vert \sqrt{\frac{s+\lambda\Theta(x_1)}{D}}},
\end{align*}
to get
\begin{equation}
\int_{-\infty}^{\infty}\mquad dx_2\partial_{x_1}\widetilde{\mathbb{Z}}_s(x_1,x_2\vert \lambda)=\frac{e^{-\vert x_1 \vert \sqrt{\frac{s+\lambda\Theta(x_1)}{D}}}\left[\sqrt{s}-\sqrt{s+\lambda}\right]}{\sqrt{Ds(s+\lambda)}},\label{eq:Zbb tilde integral der right}
\end{equation}
which appears in the amplitudes  \Eq{eq:W1 tilde pos some intermediate expression} and \Eq{eq:W2 pos first equation}.

In the rest we list a few more identities which frequently appear for calculating the amplitude \Eq{eq:W2 pos first equation}. Their derivation is similar to those shown for Eqs.~\Eq{eq:Zbb tilde integral Right} and \Eq{eq:Zbb tilde integral der right}. They can be verified numerically in Mathematica using the expressions in Eqs.~\eqref{eq:Za final},\,\eqref{eq:Zb final}, and \eqref{eq:Z tilde conditional final}. 

These are as follows
\begin{align}
\int_{-\infty}^{\infty}\mquad dx_2\widetilde{\mathbb{Z}}_{s_2}&(x_1,x_2 \vert \lambda)e^{-\vert x_2 \vert \sqrt{\frac{s_1+\lambda \Theta(x_2)}{D}}}\\[2mm]
&=\frac{e^{-\vert x_1 \vert \sqrt{\frac{s_1+\lambda\Theta(x_1)}{D}}}}{s_2-s_1} -\left(\frac{\sqrt{s_2}-\sqrt{s_2+\lambda}}{\sqrt{s_1}-\sqrt{s_1+\lambda}} \right)\nn \\[2mm] &\qquad\qquad\times \frac{e^{-\vert x_1 \vert \sqrt{\frac{s_2+\lambda\Theta(x_1)}{D}}}}{s_2-s_1}\nn
\end{align}
and
\begin{align}
\int_{-\infty}^{\infty}&dx_2\widetilde{\mathbb{Z}}_{s}(0,x_2\vert \lambda)\partial_{x_2}e^{-\vert x_2 \vert \sqrt{\frac{s_1+\lambda \Theta(x_2)}{D}}} \label{eq:Zbb exp der int}\\[2mm]  =&\frac{1}{\sqrt{D}}\times\frac{\sqrt{s_1(s+\lambda)}-\sqrt{s(s_1+\lambda)}}{(\sqrt{s+\lambda}+\sqrt{s})(\sqrt{s}+\sqrt{s_1})(\sqrt{s+\lambda}+\sqrt{s_1+\lambda})}.\nn
\end{align}
An analogous result (difference with \Eq{eq:Zbb exp der int} is in a space derivative)
\begin{align}
& \int_{-\infty}^{\infty}\mquad dx_1\left[\partial_{x_1}\widetilde{\mathbb{Z}}_{s_1}(0,x_1\vert \lambda )\right]e^{-\vert x_1 \vert \sqrt{\frac{s_2+\lambda \Theta(x_1)}{D}}}\label{eq:B49}\\[2mm]&=\frac{1}{\sqrt{D}}
\times \frac{\sqrt{s_1(s_2+\lambda)}-\sqrt{(s_1+\lambda)s_2}}{(\sqrt{s_1+\lambda}+\sqrt{s_1})(\sqrt{s_1}+\sqrt{s_2})(\sqrt{s_1+\lambda}+\sqrt{s_2+\lambda})}.\nn
\end{align}

More identities involving products of $\mathbb{Z}$ are as follows. 
\begin{align}
\int_{-\infty}^{\infty}\mquad & dx_2dm\widetilde{\mathbb{Z}}_{s_2}(x_1,x_2\vert \lambda)\partial_{x_2}\widetilde{\mathbb{Z}}_{s_1}(x_2,m\vert \lambda)=\frac{1}{s_2-s_1}\times \cr
&\frac{1}{\sqrt{D\;s_1(s_1+\lambda)}}\Bigg[e^{-\vert x_1 \vert \sqrt{\frac{s_1+\lambda \Theta(x_1)}{D}}} \left(\sqrt{s_1}-\sqrt{s_1+\lambda}\right) \cr 
&\qquad -e^{-\vert x_1 \vert \sqrt{\frac{s_2+\lambda \Theta(x_1)}{D}}}\left(\sqrt{s_2}-\sqrt{s_2+\lambda}\right)\Bigg],
\label{eq:B48}
\end{align}
and
\begin{align}
 &\int_{-\infty}^{\infty} dx_1
\widetilde{\mathbb{Z}}_{s}(0,x_1\vert \lambda)\partial_{x_1}\widetilde{\mathbb{Z}}_{s+y_1}( x_1,x_2\vert \lambda) \nn \\[2mm]
&\qquad =\frac{1}{D y_1}\times\frac{1}{\left(\sqrt{\lambda +s}+\sqrt{s}\right)}\times \cr
&\qquad \quad \Bigg\{\sgn(x_2) \sqrt{\lambda  \Theta (x_2)+s} \; e^{-\left| x_2\right|  \sqrt{\frac{\lambda  \Theta (x_2)+s}{D}}}+\cr
\Bigg[&\frac{\left(\sqrt{s+y_1}-\sqrt{s}\right) \left(\sqrt{(\lambda +s) (s+y_1)}-\sqrt{s (\lambda +s+y_1)}\right)}{\left(\sqrt{\lambda +s}+\sqrt{\lambda +s+y_1}\right) \left(\sqrt{\lambda +s+y_1}+\sqrt{s+y_1}\right)}\cr & \qquad -\sgn(x_2) \sqrt{\lambda  \Theta (x_2)+s}\Bigg] e^{-\left| x_2\right|  \sqrt{\frac{\lambda  \Theta (x_2)+s+y_1}{d}}}\Bigg\}.
\end{align}
A last one involving products of four $\mathbb{Z}$,
\begin{align}
\int_{-\infty}^{\infty}dx_1dx_2 &dx_3dm\;\widetilde{\mathbb{Z}}_{s}(0,x_1\vert \lambda)\partial_{x_1}\widetilde{\mathbb{Z}}_{s+y_1}(x_1,x_2 \vert \lambda)\nn \\[2mm]
 \times  \widetilde{\mathbb{Z}}_{s+y_2}&(x_2,x_3\vert \lambda )\partial_{x_3}\widetilde{\mathbb{Z}}_{s}(x_3,m\vert \lambda)\nn \\[2mm]
=&\frac{1}{D\sqrt{s(s+\lambda)}}\times \frac{1}{(\sqrt{s}+\sqrt{s+\lambda})}\nn \\[2mm]
& \times \left[\frac{h(s,s+\lambda,y_1)}{y_1(y_2-y_1)}+\frac{h(s,s+\lambda,y_2}{y_2(y_1-y_2)}\right],
\end{align}
where $h(s_1,s_2,y)$ is defined in \Eq{eq:h positive}. This is used for the amplitude of diagram $D$ in \Eq{eq:D tilde positive final rescaled}.

\begin{figure}
\includegraphics[width=0.48\textwidth]{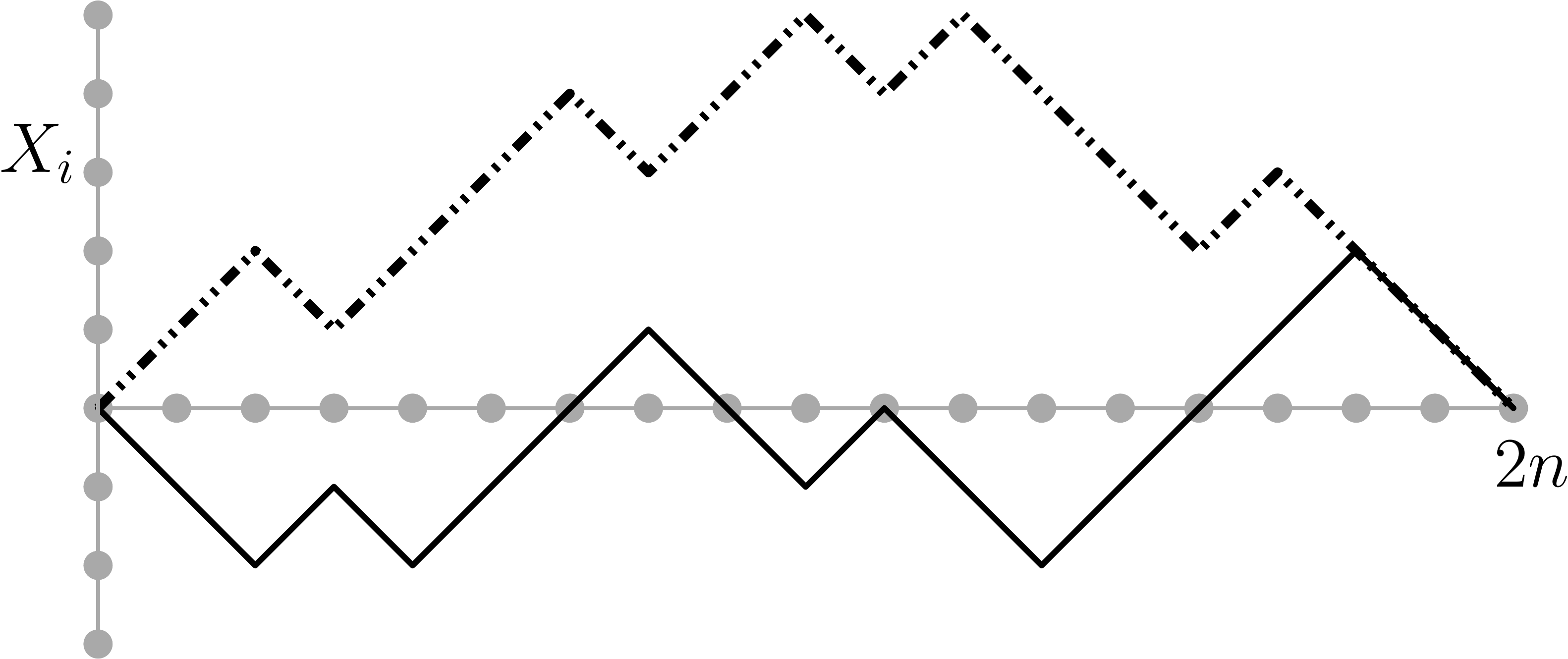}
\caption{The zigzag solid line shows a random walk bridge of $2n=18$ steps that spends $2m=6$ steps on the positive side. The dashed line shows an excursion of $2n$ steps that is conditioned to stay positive, for the entire duration. \label{fig:Random walk bridge}}
\end{figure}

\section{Uniform distribution of $t_{\rm pos}$ for a Brownian Bridge} 
\label{app:Brownian Bridge}
In \sref{sec:conditional propagator} we used a result that for a Brownian bridge, time spent on positive half has a uniform distribution. Here, we give a derivation of this result.

Our derivation is for a random walk of total $2n$ steps on an infinite chain. The walker is conditioned to take equal number of positive and negative \textit{steps} such that at the final step the walker returns to the starting point, which we choose to be the origin. Continuous limit of the process is a Brownian bridge, and the distribution of positive time for the Random walk gives the distribution for Brownian bridge in the continuous limit. 

For our derivation, we define a generating function
\begin{equation}
G(\kappa,\rho)=1+\sum_{n=1}^{\infty}\sum_{m=0}^{n}\kappa^{n}\rho^{m}\frac{1}{2^{2n}}N(2n,2m),
\label{eq:G def}
\end{equation}
where $(\kappa,\rho)$ are parameters and $N(2n,2m)$ gives the total number of Random walk bridges of length $2n$ with $2m$ number of steps spent on the positive side of the chain (see illustration in \fref{fig:Random walk bridge}).

We define a second generating function 
\begin{equation}
g(\kappa)=\sum_{n=1}^{\infty}\frac{\kappa^n}{2^{2n}}N^{+}(2n),
\label{eq:g kappa}
\end{equation}																					
where $N^{+}(2n)$ gives the number of random bridges that stay on the positive side of the chain for the entire duration $2n$ (Random walk excursion. See illustration in \fref{fig:Random walk bridge}). 

Using method of images it is straightforward to show that
\begin{equation}
N^{+}(2n)=\binom{2n-2}{n-1}-\binom{2n-2}{n}=\frac{(2n-2)!}{n!(n-1)!},
\end{equation}
leading to 
\begin{equation}
g(\kappa)=\frac{1}{2}\left(1-\sqrt{1-\kappa}\right).
\end{equation}

\begin{figure}[t]
\includegraphics[width=0.47\textwidth]{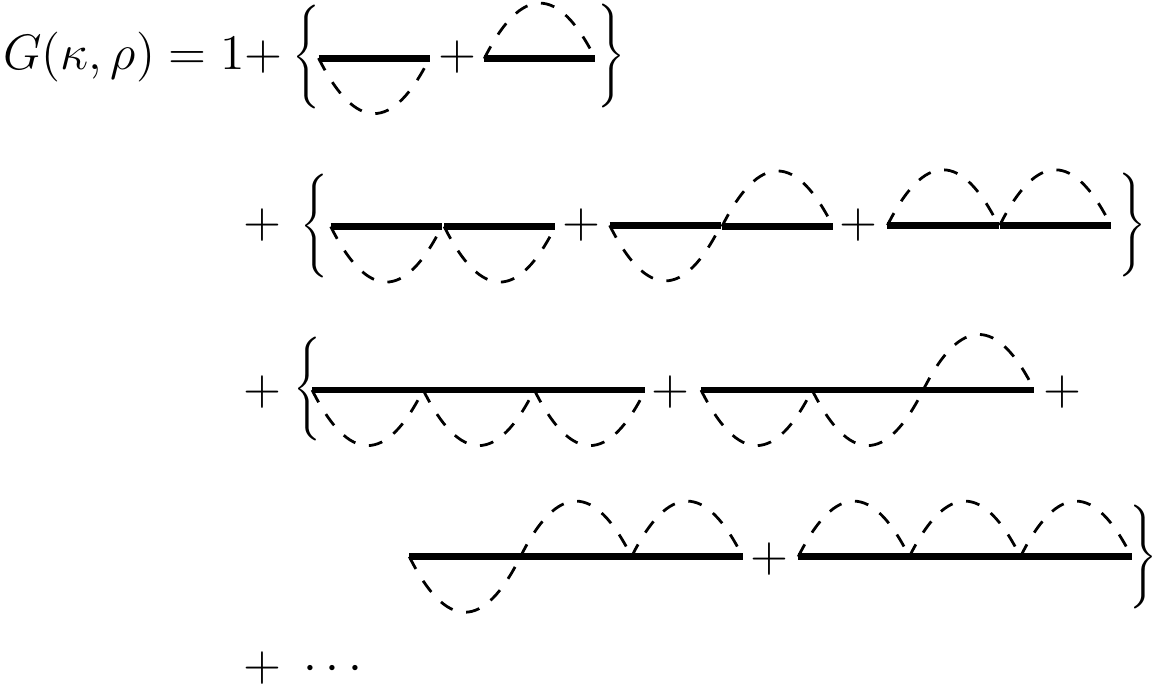}
\caption{A graphical representation of the infinite summation in \Eq{eq:G g sum}. Down-sided excursions represent $g(\kappa)$ and up-sided excursions represent $g(\kappa\,\rho)$. Relative order of excursions give the degeneracy in \Eq{eq:G g sum}. }\label{fig:bridge expansion}
\end{figure}

To calculate $G(\kappa,\rho)$ we use a relation
\begin{align}
G(\kappa,\rho)=&1+\left[g(\kappa)+g(\kappa\rho) \right]+\cr
&\left[g(\kappa)^2+2g(\kappa)g(\kappa\rho)+g(\kappa\rho)^2\right]+\cr
&\left[g(\kappa)^3+3g(\kappa)^2g(\kappa\rho)+3g(\kappa)g(\kappa\rho)^2+g(\kappa\rho)^3\right]\cr
&+\cdots, \label{eq:G g sum}
\end{align}
which can be seen by the graphical illustration in \fref{fig:bridge expansion}. Completing the summation we get
\begin{equation}
G(\kappa,\rho)=\frac{1}{1-g(\kappa)-g(\kappa\rho)}.
\end{equation}
Using the formula for $g(\kappa)$ in \Eq{eq:g kappa} we write
\begin{equation}
G(\kappa, \rho)=1+\sum_{n=1}^{\infty}\sum_{m=0}^n\kappa^n\rho^m\frac{(2n)!}{2^{2n}(n+1)(n!)^2}.
\end{equation}
Comparing with \Eq{eq:G def} it is evident that $N(2n,2m)$ is independent of $m$. Equivalently, there are equal number of paths for all values of $m$ in a random walk bridge of length $2n$. In the continuous limit, this means that for a Brownian bridge, all values of fractional positive time are equally probable.

\bibliographystyle{KAY-hyper}
\bibliography{citation,citation2}

\ifx\doi\undefined
\providecommand{\doi}[2]{\href{http://dx.doi.org/#1}{#2}}
\else
\renewcommand{\doi}[2]{\href{http://dx.doi.org/#1}{#2}}
\fi
\providecommand{\link}[2]{\href{#1}{#2}}
\providecommand{\arxiv}[1]{\href{http://arxiv.org/abs/#1}{#1}}
\providecommand{\hal}[1]{\href{https://hal.archives-ouvertes.fr/hal-#1}{hal-#1}}
\providecommand{\mrnumber}[1]{\href{https://mathscinet.ams.org/mathscinet/search/publdoc.html?pg1=MR&s1=#1&loc=fromreflist}{MR#1}}

\begin{thebibliography}{10}

\bibitem{DecreusefondUstunel1998}
L.~Decreusefond and A.S. \"Ust\"unel,
\newblock {\em Fractional {Brownian} motion: Theory and applications},
\newblock in {\em ESAIM: PROC}, pages 75--86, EDP Sciences, 1998.

\bibitem{fbmReview2}
H.~Qian,
\newblock {\em Fractional {B}rownian motion and fractional gaussian noise},
\newblock in {\em Processes with Long-Range Correlations}, pages 22--33,
  Springer, 2003.

\bibitem{NourdinBook}
I.~Nourdin,
\newblock \doi{10.1007/978-88-470-2823-4}{\rm {\em Selected aspects of
  fractional {B}rownian motion}}\null,
\newblock Bocconi \& Springer Series, 2012.

\bibitem{Shevchenko2015}
G.~Shevchenko,
\newblock {\em Fractional {B}rownian motion in a nutshell},
\newblock \doi{10.1142/S2010194515600022}{\rm Int. J. Mod. Phys. {\bf 36}
  (2015)   1560002}\null.

\bibitem{Kolmogorov1940}
A.N. Kolmogorov,
\newblock {\em {Wienersche spiralen und einige andere interessante Kurven im
  Hilbertschen Raum}},
\newblock CR(Dokl.) Acad. Sci. URSS {\bf 26} (1940)   115--118.

\bibitem{MandelbrotVanNess1968}
B.B. Mandelbrot and J.W. {Van~Ness},
\newblock {\em Fractional {B}rownian motions, fractional noises and
  applications},
\newblock \doi{10.1137/1010093}{\rm SIAM Review {\bf 10} (1968)
  422--437}\null.

\bibitem{MetzlerKlafter2000}
R.~Metzler and J.~Klafter,
\newblock {\em The random walk's guide to anomalous diffusion: a fractional
  dynamics approach},
\newblock \doi{10.1016/S0370-1573(00)00070-3}{\rm Phys. Rep. {\bf 339} (2000)
  1--77}\null.

\bibitem{CohenIstas2013}
S.~Cohen and J.~Istas,
\newblock \doi{10.1007/978-3-642-36739-7}{\rm {\em Fractional Fields and
  Applications}}\null,
\newblock Springer, 2013.

\bibitem{KuklaKornatowskiDemuthGirnusal1996}
V.~Kukla, J.~Kornatowski, D.~Demuth, I.~Girnus, H.~Pfeifer, L.V.C. Rees,
  S.~Schunk, K.K. Unger  and J.~Karger,
\newblock {\em {NMR} studies of single-file diffusion in unidimensional channel
  zeolites},
\newblock \doi{10.1126/science.272.5262.702}{\rm Science (1996)   702
  --704}\null.

\bibitem{WeiBechingerLeiderer2000}
Q.-H. Wei, C.~Bechinger  and P.~Leiderer,
\newblock {\em Single-file diffusion of colloids in one-dimensional channels},
\newblock \doi{10.1126/science.287.5453.625}{\rm Science {\bf 287} (2000)
  625--627}\null.

\bibitem{LizanaAmbjornssonTaloniBarkaiLomholt2010}
L~Lizana, T.~Ambj\"ornsson, A.~Taloni, E.~Barkai  and M.A. Lomholt,
\newblock {\em Foundation of fractional {Langevin} equation: Harmonization of a
  many-body problem},
\newblock \doi{10.1103/PhysRevE.81.051118}{\rm Phys. Rev. E {\bf 81} (2010)
  051118}\null.

\bibitem{KrapivskyMallickSadhu2015}
P.L. Krapivsky, K.~Mallick  and T.~Sadhu,
\newblock {\em Dynamical properties of single-file diffusion},
\newblock \doi{10.1088/1742-5468/2015/09/P09007}{\rm J. Stat. Mech. (2015)
  P09007}\null,
\newblock \arxiv{arXiv:1505.01287}.

\bibitem{SadhuDerrida2015}
T.~Sadhu and B.~Derrida,
\newblock {\em Large deviation function of a tracer position in single file
  diffusion},
\newblock \doi{10.1088/1742-5468/2015/09/P09008}{\rm J. Stat. Mech. {\bf 2015}
  (2015)   P09008}\null.

\bibitem{SadhuDerrida2016}
T.~Sadhu and B.~Derrida,
\newblock {\em Correlations of the density and of the current in
  non-equilibrium diffusive systems},
\newblock \doi{10.1088/1742-5468/2016/11/113202}{\rm J. Stat. Mech. {\bf 2016}
  (2016)   113202}\null.

\bibitem{ZoiaRossoMajumdar2009}
A.~Zoia, A.~Rosso  and S.N. Majumdar,
\newblock {\em Asymptotic behavior of self-affine processes in semi-infinite
  domains},
\newblock \doi{10.1103/PhysRevLett.102.120602}{\rm Phys. Rev. Lett. {\bf 102}
  (2009)   120602}\null.

\bibitem{DubbeldamRostiashvili2011}
J.L.A. Dubbeldam, V.G. Rostiashvili, A.~Milchev  and T.A. Vilgis,
\newblock {\em Fractional {B}rownian motion approach to polymer translocation:
  The governing equation of motion},
\newblock \doi{10.1103/PhysRevE.83.011802}{\rm Phys. Rev. E {\bf 83} (2011)
  011802}\null.

\bibitem{PalyulinAlaNissilaMetzler2014}
V.~Palyulin, T.~Ala-Nissila  and R.~Metzler,
\newblock {\em Polymer translocation: the first two decades and the recent
  diversification},
\newblock \doi{10.1039/C4SM01819B}{\rm Soft Matter {\bf 10} (2014)
  9016--9037}\null.

\bibitem{BouchaudGeorges1990}
J.-P. Bouchaud and A.~Georges,
\newblock {\em Anomalous diffusion in disordered media: statistical mechanisms,
  models and physical applications},
\newblock \doi{10.1016/0370-1573(90)90099-N}{\rm Phys. Rep. {\bf 195} (1990)
  127--293}\null.

\bibitem{Peters1996}
E.E. Peters,
\newblock {\em Chaos and order in the capital markets},
\newblock Wiley finance editions,
\newblock Wiley, New York, 2 edition, 1996.

\bibitem{CutlandKoppWillinger1995}
N.J. Cutland, P.E. Kopp  and W.~Willinger,
\newblock {\em Stock price returns and the {Joseph} effect: A fractional
  version of the {Black-Scholes} model},
\newblock in E.~Bolthausen, M.~Dozzi  and F.~Russo, editors, {\em Seminar on
  Stochastic Analysis, Random Fields and Applications}, {\em {\em Volume}~36}
  of {\em Progress in Probability}, pages 327--351, Birkh{\"a}user Basel, 1995.

\bibitem{Biagini2008}
F.~Biagini, Y~Hu, B.~Oksendal  and T.~Zhang,
\newblock {\em Stochastic Calculus for Fractional Brownian Motion and
  Applications},
\newblock Springer Verlag, London, 2008.

\bibitem{Sottinen2001}
T.~Sottinen,
\newblock {\em Fractional {Brownian} motion, random walks and binary market
  models},
\newblock \doi{10.1007/PL00013536}{\rm Finance and Stochastics {\bf 5} (2001)
  343--355}\null.

\bibitem{Hurst1951}
H.E. Hurst,
\newblock {\em Long-term storage capacity of reservoirs},
\newblock Transactions of the American Society of Civil Engineers {\bf 116}
  (1951)   770--799.

\bibitem{MandelbrotWallis1968}
B.B. Mandelbrot and J.R. Wallis,
\newblock {\em Noah, {J}oseph, and operational hydrology},
\newblock \doi{10.1029/WR004i005p00909}{\rm Water Resources Research {\bf 4}
  (1968)   909--918}\null.

\bibitem{GuptaRossoTexier2013}
S.~Gupta, A.~Rosso  and C.~Texier,
\newblock {\em Dynamics of a tagged monomer: Effects of elastic pinning and
  harmonic absorption},
\newblock \doi{10.1103/PhysRevLett.111.210601}{\rm Phys. Rev. Lett. {\bf 111}
  (2013)   210601}\null.

\bibitem{Moreno2014}
E.~Monte-Moreno and M.~Hern{\'{a}}ndez-Pajares,
\newblock {\em Occurrence of solar flares viewed with gps: Statistics and
  fractal nature},
\newblock \doi{10.1002/2014JA020206}{\rm J. Geophys. Res. {\bf 119} (2014)
  9216--9227}\null.

\bibitem{Simonsen2003}
I.~Simonsen,
\newblock {\em Measuring anti-correlations in the nordic electricity spot
  market by wavelets},
\newblock \doi{10.1016/S0378-4371(02)01938-6}{\rm Physica A {\bf 322} (2003)
  597--606}\null.

\bibitem{Norros2006}
I.~Norros,
\newblock {\em On the use of fractional {B}rownian motion in the theory of
  connectionless networks},
\newblock \doi{10.1109/49.400651}{\rm IEEE J. Sel. A. Commun. {\bf 13} (2006)
  953--962}\null.

\bibitem{Burnecki2012}
K.~Burnecki, E.~Kepten, J.~Janczura, I.~Bronshtein, Y.~Garini  and A.~Weron,
\newblock {\em Universal algorithm for identification of fractional brownian
  motion. a case of telomere subdiffusion},
\newblock \doi{10.1016/j.bpj.2012.09.040}{\rm Biophysical Journal {\bf 103}
  (2012)   1839--1847}\null.

\bibitem{JeonTejedorBurovBarkaiSelhuber-UnkelBerg-SorensenOddershedeMetzler2011}
J.H. Jeon, V.~Tejedor, S.~Burov, E.~Barkai, C.~Selhuber-Unkel,
  K.~Berg-S\o{}rensen, L.~Oddershede  and R.~Metzler,
\newblock {\em In vivo anomalous diffusion and weak ergodicity breaking of
  lipid granules},
\newblock \doi{10.1103/PhysRevLett.106.048103}{\rm Phys. Rev. Lett. {\bf 106}
  (2011)   048103}\null.

\bibitem{Ernst2012}
D.~Ernst, M.~Hellmann, K.~J\"{u}rgen  and M.~Weiss,
\newblock {\em {Fractional Brownian motion in crowded fluids}},
\newblock \doi{10.1039/C2SM25220A}{\rm Soft Matter {\bf 8} (2012)
  4886--4889}\null.

\bibitem{Molchan1999}
G.~M. Molchan,
\newblock {\em Maximum of a fractional {B}rownian motion: Probabilities of
  small values},
\newblock \doi{10.1007/s002200050669}{\rm Comm. Math. Phys. {\bf 205} (1999)
  97--111}\null.

\bibitem{KrugKallabisMajumdarCornellBraySire1997}
J.~Krug, H.~Kallabis, S.N. Majumdar, S.~J. Cornell, A.~J. Bray  and C.~Sire,
\newblock {\em Persistence exponents for fluctuating interfaces},
\newblock \doi{10.1103/PhysRevE.56.2702}{\rm Phys. Rev. E {\bf 56} (1997)
  2702--2712}\null.

\bibitem{GuerinLevernierBenichouVoituriez2016}
T.~Gu{\'e}rin, N.~Levernier, O.~B{\'e}nichou  and R.~Voituriez,
\newblock {\em Mean first-passage times of non-{Markovian} random walkers in
  confinement},
\newblock \doi{10.1038/nature18272}{\rm Nature {\bf 534} (2016)
  356--359}\null.

\bibitem{Majumdar2006}
S.N. Majumdar,
\newblock {\em Brownian functionals in physics and computer science},
\newblock in {\em The Legacy of Albert Einstein.}, {\em {\em Volume}~89}, pages
  93--129, World scientific, 2005,
\newblock \arxiv{arXiv:cond-mat/0510064}.

\bibitem{Hida1977}
T.~Hida,
\newblock {\em Functionals of {B}rownian motion},
\newblock in J.~Ko{\v z}e{\v s}nik, editor, {\em Transactions of the Seventh
  Prague Conference on Information Theory, Statistical Decision Functions,
  Random Processes and of the 1974 European Meeting of Statisticians: held at
  Prague, from August 18 to 23, 1974}, pages 239--243, Springer Netherlands,
  Dordrecht, 1977.

\bibitem{Ruelle1999}
D.~Ruelle,
\newblock {\em Smooth dynamics and new theoretical ideas in nonequilibrium
  statistical mechanics},
\newblock \doi{10.1023/A:1004593915069}{\rm J. Stat. Phys. {\bf 95} (1999)
  393--468}\null.

\bibitem{Ruelle2004}
D.~Ruelle,
\newblock {\em {Conversations on Nonequilibrium Physics With an
  Extraterrestrial}},
\newblock \doi{10.1063/1.1768674}{\rm Phys. Today {\bf 57} (2004)
  48--53}\null.

\bibitem{LebowitzSpohn1999}
J.L. Lebowitz and H.~Spohn,
\newblock {\em A {Gallavotti--Cohen}-type symmetry in the large deviation
  functional for stochastic dynamics},
\newblock \doi{10.1023/A:1004589714161}{\rm J. Stat. Phys. {\bf 95} (1999)
  333--365}\null.

\bibitem{Levy1940}
P.~L{\'e}vy,
\newblock {\em Sur certains processus stochastiques homog\`enes},
\newblock Compositio Mathematica {\bf 7} (1940)   283--339.

\bibitem{FellerBook}
W.~Feller,
\newblock {\em Introduction to Probability Theory and Its Applications},
\newblock John Wiley \& Sons, 1950.

\bibitem{Morters2010}
P.~M{\"{o}}rters and Y.~Peres,
\newblock {\em Brownian Motion},
\newblock Cambridge University Press, 2010.

\bibitem{Yen2013}
J.Y. Yen and M.~Yor,
\newblock {\em Paul L{\'e}vy's Arcsine Laws}, pages 43--54,
\newblock Springer International Publishing, Cham, 2013.

\bibitem{MajumdarRandon-FurlingKearneyYor2008}
S.N. Majumdar, J.~Randon-Furling, M.J. Kearney  and M.~Yor,
\newblock {\em On the time to reach maximum for a variety of constrained
  {Brownian} motions},
\newblock \doi{10.1088/1751-8113/41/36/365005}{\rm J. Phys. A {\bf 41} (2008)
  365005}\null.

\bibitem{Bouchaud2008}
S.~N. Majumdar and J.~P. Bouchaud,
\newblock {\em Optimal time to sell a stock in the black–scholes model:
  comment on {{`}Thou shalt buy and hold{'}, by A. Shiryaev, Z. Xu and X. Y.
  Zhou}},
\newblock \doi{10.1080/14697680802569093}{\rm Quantitative Finance {\bf 8}
  (2008)   753--760}\null.

\bibitem{Randon-FurlingMajumdar2007}
J.~Randon-Furling and S.N. Majumdar,
\newblock {\em Distribution of the time at which the deviation of a {B}rownian
  motion is maximum before its first-passage time},
\newblock \doi{10.1088/1742-5468/2007/10/p10008}{\rm J. Stat. Mech. {\bf 2007}
  (2007)   P10008}\null.

\bibitem{Majumdar2010}
S.N. {Majumdar},
\newblock {\em {Universal first-passage properties of discrete-time random
  walks and {L{\'e}vy} flights on a line: Statistics of the global maximum and
  records}},
\newblock \doi{10.1016/j.physa.2010.01.021}{\rm Physica A {\bf 389} (2010)
  4299--4316}\null.

\bibitem{SchehrLeDoussal2010}
G.~Schehr and P.~Le Doussal,
\newblock {\em Extreme value statistics from the real space renormalization
  group: {Brownian} motion, {Bessel} processes and continuous time random
  walks},
\newblock \doi{10.1088/1742-5468/2010/01/P01009}{\rm J. Stat. Mech. {\bf 2010}
  (2010)   P01009}\null,
\newblock \arxiv{arXiv:0910.4913}.

\bibitem{HOCHBERG1994}
K.~J. Hochberg and E.~Orsingher,
\newblock {\em The arcsine law and its analogs for processes governed by signed
  and complex measures},
\newblock \doi{http://dx.doi.org/10.1016/0304-4149(94)90029-9}{\rm Stoch. Proc.
  App. {\bf 52} (1994)   273--292}\null.

\bibitem{Pitman1992}
J.~Pitman and M.~Yor,
\newblock {\em Arcsine laws and interval partitions derived from a stable
  subordinator},
\newblock Proc. London Math. Soc (1992)   65--326.

\bibitem{Carmona1994}
P.~Carmona, F.~Petit  and M.~Yor,
\newblock {\em Some extensions of the arcsine law as partial consequences of
  the scaling property of {B}rownian motion},
\newblock \doi{10.1007/BF01204951}{\rm Prob. Theo. Rel. Fields {\bf 100} (1994)
    1--29}\null.

\bibitem{Lamperti1958}
J.~Lamperti,
\newblock {\em An occupation time theorem for a class of stochastic processes},
\newblock \doi{10.2307/1993222}{\rm Trans. Amer. Math. Soc. {\bf 88} (1958)
  380--387}\null.

\bibitem{BARLOW1989}
M.~Barlow, J.~Pitman  and M.~Yor,
\newblock {\em Une extension multidimensionnelle de la loi de tare sinus},
\newblock in {\em Siminaire de Probabilites XXIII}, pages 294--314, Springer,
  Berlin, 1989.

\bibitem{Bingham1994}
N.~H. Bingham and R.~A. Doney,
\newblock {\em On higher-dimensional analogues of the arcsine law},
\newblock \doi{http://dx.doi.org/10.2307/3214239}{\rm J. App. Prob. {\bf 25}
  (1988)   120--131}\null.

\bibitem{Ernst2017}
P.~A. Ernst and L.~Shepp,
\newblock {\em On occupation times of the first and third quadrants for planar
  {B}rownian motion},
\newblock \doi{10.1017/jpr.2016.104}{\rm J. Appl. Prob. {\bf 54} (2017)
  337--342}\null.

\bibitem{Dale1980}
D.~Charles and W.~Rosemarie,
\newblock {\em The arcsine law and the treasury bill futures market},
\newblock \doi{10.2469/faj.v36.n6.71}{\rm Financial Analysts Journal {\bf 36}
  (1980)   71--74}\null.

\bibitem{Baz2004}
J.~Baz and G.~Chacko,
\newblock {\em Financial Derivatives: Pricing, Applications, and Mathematics},
\newblock Cambridge University Press, 2004.

\bibitem{Clauset2015}
A.~Clauset, M.~Kogan  and S.~Redner,
\newblock {\em Safe leads and lead changes in competitive team sports},
\newblock \doi{10.1103/PhysRevE.91.062815}{\rm Phys. Rev. E {\bf 91} (2015)
  062815--062826}\null.

\bibitem{WieseMajumdarRosso2010}
K.J. Wiese, S.N. Majumdar  and A.~Rosso,
\newblock {\em Perturbation theory for fractional {Brownian} motion in presence
  of absorbing boundaries},
\newblock \doi{10.1103/PhysRevE.83.061141}{\rm Phys. Rev. E {\bf 83} (2011)
  061141}\null,
\newblock \arxiv{arXiv:1011.4807}.

\bibitem{DelormeWiese2015}
M.~Delorme and {K.J.} Wiese,
\newblock {\em Maximum of a fractional {Brownian} motion: Analytic results from
  perturbation theory},
\newblock \doi{10.1103/PhysRevLett.115.210601}{\rm Phys. Rev. Lett. {\bf 115}
  (2015)   210601}\null,
\newblock \arxiv{arXiv:1507.06238}.

\bibitem{DelormeWiese2016b}
M.~Delorme and K.J. Wiese,
\newblock {\em Extreme-value statistics of fractional {Brownian} motion
  bridges},
\newblock \doi{10.1103/PhysRevE.94.052105}{\rm Phys. Rev. E {\bf 94} (2016)
  052105}\null,
\newblock \arxiv{arXiv:1605.04132}.

\bibitem{DelormeWiese2016}
M.~Delorme and K.J. Wiese,
\newblock {\em Perturbative expansion for the maximum of fractional {Brownian}
  motion},
\newblock \doi{10.1103/PhysRevE.94.012134}{\rm Phys. Rev. E {\bf 94} (2016)
  012134}\null,
\newblock \arxiv{arXiv:1603.00651}.

\bibitem{DelormeRossoWiese2017}
M.~Delorme, A.~Rosso  and K.J. Wiese,
\newblock {\em Pickands' constant at first order in an expansion around
  {Brownian} motion},
\newblock \doi{10.1088/1751-8121/aa5c98}{\rm J. Phys. A {\bf 50} (2017)
  16LT04}\null,
\newblock \arxiv{arXiv:1609.07909}.

\bibitem{DelormeThesis}
M.~Delorme,
\newblock {\em Stochastic processes and disordered systems, around {Brownian}
  motion},
\newblock PhD thesis, PSL Research University, 2016.

\bibitem{Wiese2018}
{K.J.} Wiese,
\newblock {\em First passage in an interval for fractional {Brownian} motion},
\newblock \doi{10.1103/PhysRevE.99.032106}{\rm Phys. Rev. E {\bf 99} (2018)
  032106}\null,
\newblock \arxiv{arXiv:1807.08807}.

\bibitem{ArutkinWalterWiese2020}
M.~Arutkin, B.~Walter  and K.J. Wiese,
\newblock {\em Extreme events for fractional {Brownian} motion with drift:
  Theory and numerical validation},
\newblock \doi{10.1103/PhysRevE.102.022102}{\rm Phys. Rev. E {\bf 102} (2020)
  022102}\null,
\newblock \arxiv{arXiv:1908.10801}.

\bibitem{SadhuDelormeWiese2017}
T.~Sadhu, M.~Delorme  and {K.J.} Wiese,
\newblock {\em Generalized arcsine laws for fractional {Brownian} motion},
\newblock \doi{10.1103/PhysRevLett.120.040603}{\rm Phys. Rev. Lett. {\bf 120}
  (2018)   040603}\null,
\newblock \arxiv{arXiv:1706.01675}.

\bibitem{Kac1949}
M.~Kac,
\newblock {\em On distributions of certain {Wiener} functionals},
\newblock \doi{10.1090/S0002-9947-1949-0027960-X}{\rm Trans. Amer. Math. Soc.
  {\bf 65} (1949)   1--13}\null.

\bibitem{Zinn}
J.~Zinn-Justin,
\newblock \doi{10.1093/acprof:oso/9780199227198.001.0001}{\rm {\em Quantum
  Field Theory and Critical Phenomena}}\null,
\newblock Oxford University Press, Oxford, 1989.

\bibitem{DaviesHarte1987}
R.B. Davies and D.S. Harte,
\newblock {\em Tests for hurst effect},
\newblock \doi{10.1093/biomet/74.1.95}{\rm Biometrika {\bf 74} (1987)
  95--101}\null.

\bibitem{DiekerPhD}
A.B. Dieker,
\newblock {\em \link{http://www.columbia.edu/~ad3217/fbm/thesis.pdf}{Simulation
  of fractional {B}rownian motion}},
\newblock PhD thesis, University of Twente, 2004.

\bibitem{WalterWiese2019a}
B.~Walter and K.J. Wiese,
\newblock {\em Monte {Carlo} sampler of first passage times for fractional
  {Brownian} motion using adaptive bisections: Source code},
\newblock \link{https://hal.archives-ouvertes.fr/hal-02270046}{hal-02270046}
  (2019).

\bibitem{WalterWiese2019b}
B.~Walter and K.J. Wiese,
\newblock {\em Sampling first-passage times of fractional {Brownian} motion
  using adaptive bisections},
\newblock \doi{10.1103/PhysRevE.101.043312}{\rm Phys. Rev. E {\bf 101} (2020)
  043312}\null,
\newblock \arxiv{arXiv:1908.11634}.

\bibitem{Dhar1999}
A.~Dhar and S.N. Majumdar,
\newblock {\em Residence time distribution for a class of gaussian markov
  processes},
\newblock \doi{10.1103/PhysRevE.59.6413}{\rm Phys. Rev. E {\bf 59} (1999)
  6413--6418}\null.

\bibitem{Owen2014}
D.~B. Owen,
\newblock {\em Orthant probabilities},
\newblock in {\em Wiley StatsRef: Statistics Reference Online}, American Cancer
  Society, 2014.

\bibitem{Mathematica}
T.~Sadhu and K.J. Wiese,
\newblock {\em Supplemental mathematica notebook which describes certain steps
  to evaluate expresion for $\mathcal{F}_2$.}

\bibitem{MajumdarComtet2005}
S.N. Majumdar and A.~Comtet,
\newblock {\em {Airy Distribution Function: From the Area Under a {B}rownian
  Excursion to the Maximal Height of Fluctuating Interfaces}},
\newblock \doi{10.1007/s10955-005-3022-4}{\rm J. Stat. Phys. {\bf 119} (2005)
  777--826}\null.

\bibitem{Coeurjolly2000}
J.F. Coeurjolly,
\newblock {\em Simulation and identification of the fractional {Brownian}
  motion: a bibliographical and comparative study},
\newblock \doi{10.18637/jss.v005.i07}{\rm Journal of Statistical Software {\bf
  05} (2000)   i07}\null.

\bibitem{WoodChan}
A.T.A. Wood and G.~Chan,
\newblock {\em Simulation of stationary gaussian processes in {$[0,1]^d$}},
\newblock \doi{10.2307/1390903}{\rm Journal of Computational and Graphical
  Statistics {\bf 3} (1994)   409--432}\null.

\end{thebibliography}

\end{document}